\documentclass{emulateapj}

\usepackage[]{amsmath}
\usepackage{amssymb}
\usepackage{natbib}
\usepackage{apjfonts}
\usepackage{graphicx}
\usepackage{multirow}
\usepackage{booktabs}
\usepackage{bm}
\usepackage{xspace}
\usepackage{afterpage}
\usepackage{gensymb}

\shorttitle{Spatially Resolved SFR in  galaxies in different environments}
\shortauthors{Vulcani et al.}

\usepackage{color}			      
\definecolor{midgray}{gray}{0.4}		
\definecolor{orange}{rgb}{1,0.5,0}    

\usepackage[colorlinks=true,citecolor=midgray,linkcolor=midgray,urlcolor=midgray]{hyperref}        

\newcommand{\HST}{\emph{HST}}

\newcommand{\Ha}{H$\alpha$\xspace}

\begin{document}

\title{THE GRISM LENS-AMPLIFIED SURVEY FROM SPACE (GLASS). VII. The diversity of the distribution of star formation 
 in cluster  and field galaxies at $0.3\leq \lowercase{z}\leq0.7$} 

\author{Benedetta Vulcani\altaffilmark{1}}
\author{Tommaso Treu\altaffilmark{2}}
\author{Kasper B. Schmidt\altaffilmark{3}}
\author{Takahiro Morishita\altaffilmark{2,4,5}}
\author{Alan Dressler\altaffilmark{6}}
\author{Bianca M. Poggianti\altaffilmark{7}}
\author{Louis Abramson\altaffilmark{2}}
\author{Marusa Brada\v{c}\altaffilmark{8}}
\author{Gabriel B. Brammer\altaffilmark{9}}
\author{Austin Hoag\altaffilmark{8}}
\author{Matthew Malkan\altaffilmark{2}}
\author{Laura Pentericci\altaffilmark{10}}
\author{Michele Trenti\altaffilmark{1}}

\affil{\altaffilmark{1}School of Physics, University of Melbourne, VIC 3010, Australia}
\affil{\altaffilmark{2}Department of Physics and Astronomy, University of California, Los
Angeles, CA, USA 90095-1547}
\affil{\altaffilmark{3}Leibniz-Institut für Astrophysik Potsdam (AIP), An der Sternwarte 16, 14482 Potsdam, 
Germany}
\affil{\altaffilmark{4} Astronomical Institute, Tohoku University, Aramaki, Aoba, Sendai 980-8578, Japan}
\affil{\altaffilmark{5} Institute for International Advanced Research and Education, Tohoku University, Aramaki, Aoba, Sendai 980-8578, Japan}
\affil{\altaffilmark{6}The Observatories of the Carnegie Institution for Science, 813 Santa Barbara St., 
Pasadena, CA 91101, USA}
\affil{\altaffilmark{7}INAF-Astronomical Observatory of Padova, Italy}
\affil{\altaffilmark{8}Department of Physics, University of California, Davis, CA, 95616, USA}
\affil{\altaffilmark{9}Space Telescope Science Institute, 3700 San Martin Drive, Baltimore, MD, 21218, USA}
\affil{\altaffilmark{10}INAF - Osservatorio Astronomico di Roma Via Frascati 33 - 00040 Monte Porzio Catone, I}

\begin{abstract}
Exploiting the slitless spectroscopy taken as part of the Grism
Lens-Amplified Survey from Space (GLASS), we present an extended
analysis of the spatial distribution of star formation in 76 galaxies
in 10 clusters at 0.3$<z<$0.7.  We use 85 foreground and background
galaxies in the same redshift range as a field sample. The samples are
well matched in stellar mass (10$^8$-10$^{11}$ M$_\odot$)
and star formation rate (0.5-50 M$_\odot \, yr^{-1}$).  We visually
classify galaxies in terms of broad-band morphology, \Ha morphology
and likely physical process acting on the galaxy.
Most \Ha emitters have a spiral morphology
(41$\pm8$\% in clusters, 51$\pm8$\% in the field), followed by
mergers/interactions (28$\pm8$\%, 31$\pm7$\%, respectively) and early-type
galaxies (remarkably as high as 29$\pm$8 in clusters and 15$\pm$6\% in
the field).  A diversity of \Ha morphologies is detected, suggesting  
a diversity of physical processes.  In clusters,  30$\pm$8\% of the
galaxies present a regular morphology, mostly consistent  with star formation
diffused uniformly across the stellar population (mostly in the disk
component, when present). The second most common morphology
(28$\pm$8\%) is asymmetric/jellyfish, consistent with ram pressure
stripping or other non-gravitational processes in 18$\pm$8\% of the cases. 
Ram pressure stripping appears
significantly less prominent in the field (2$\pm$2\%), where the most common
morphology/mechanism appears to be consistent with minor gas rich
mergers or clump accretion. This work demonstrates that while environment
specific mechanisms affect galaxy evolution at this redshift,  they
are diverse and their effects subtle. A full understanding of this
complexity requires larger samples and detailed and spatially resolved physical models.
\end{abstract}


\section{Introduction}

Over the last four decades, several studies have shown that galaxy
properties correlate with their environment, and these correlations
vary as a function of redshfit
\citep[e.g.][]{butcher84, dressler80, dressler97, poggianti99, ellis97, lewis02, treu03, gomez03, goto03, postman05,  kauffmann04, grutzbauch11}.
The discovery
of these correlations raised a fundamental question that remains at
center stage to this date. How much is galaxy evolution driven by
internal processes as opposed to collective phenomena found only in
specific environments? This question has sometimes been phrased in
terms of nature vs nurture, even though the distinction is not clear
cut: today's clusters correspond to some of the most overdense regions
in the early universe and therefore we expect their evolution to be
accelerated with respect to average or underdense region, even if
cluster-specific mechanisms were not at all relevant \citep{dressler80, abramson16, lilly16, morishita16}. 
Thus, one of the
key challenges consists of finding observational signatures that
uniquely point to cluster specific mechanism and then characterize
their overall importance.

For example, the fraction of star forming galaxies has been found to
significantly decrease going from the dense cluster centers to the
field. In clusters at 0$<z<$0.1 the star formation rate (SFR) in star
forming galaxies also declines with decreasing  radius
\citep{vonderlinden10, paccagnella16}, as would be expected if
cluster-specific environmental processes act to impede star
formation. However, there is no consensus between the relative
importance of the specific mechanisms at play. It is not even clear
whether star formation is actively quenched by the cluster environment
or whether it occurs prior to the galaxy entering the cluster sphere
of influence. For example, \cite{lewis02} found that the fraction of
star forming galaxies strongly decreases with declining radius, 
but, the same correlation holds for galaxies more
than two virial radii from the cluster center. They concluded that
this rules out cluster specific processes being solely responsible for
the declining fraction of star-forming galaxies towards smaller
radii. However, e.g. \cite{haines13} at $z<0.15$ and
\cite{patel09, vulcani10, koyama13} at 0.4$<z<$0.8,   \cite{muzzin12} at 0.85$<z<$1.20 found that within the virial
radius the SFRs of star forming galaxies are significantly lower than
those found in field regions, at any given mass  In contrast, e.g. \cite{sobral11} found that at 
$z\sim 0.8$ differences hold only at low masses, 
\cite{koyama13, darvish16} showed a lack of environmental 
dependence of the SFR-Mass relation out to z$\sim 1-2$.

\cite{haines13} also found that the SFRs of the galaxies in the infall
regions are indistinguishable from those in the field. They argue that
the processes which suppress the star formation within infalling
galaxies must be related to processes occurring within the cluster,
likely due to interactions with the intra-cluster medium (ICM).

 \cite{muzzin12} also argue that  the lack of strong  correlation of the properties of star-forming and quiescent galaxies 
with their environment can be understood if the environmental- quenching timescale is rapid, and that the evolution of the internal-quenching and 
environmental-quenching rates mirrors each other, regardless of which processes dominate the overall quenching process. 

 We note that while many studies have so far been mostly confined to  field versus clusters,  
intermediate environments such as galaxy groups, outskirts of clusters and filaments  are equally 
important to shed light on the processes inducing galaxy transformations 
\citep[e.g., see][]{kodama01, porter07, porter08, sobral11, coppin12, darvish14}.

The picture is further complicated by the diversity of physical
processes that might transform the morphology and star formation
properties of galaxies. 
 Internal feedback,  likely driven  by either supernovae or  active galactic nucleus (AGN), 
and/or low escape velocities might be the responsible for outflows from both star formation 
and AGN activity to be much more efficient and induce faster transformations \citep[see, e.g.,][]{mcgee14, sobral15}. 
At the same time, many external physical mechanisms can take place.
Gravitational effects can distort a galaxy and
tear away stars and gas \citep{bekki99}. Rapid and frequent
galaxy-galaxy encounters induce gravitational perturbations which can
greatly affect the stellar and gas components of cluster galaxies
\citep{moore96}.  Gas falling onto a cluster is heated by shocks
leading to a hot, diffuse ICM which permeates the space between the
galaxies in clusters. In turn, the ICM can impact the gas within a
galaxy by either compressing it, leading to triggered star formation
\citep{bekkicouch03, stroe14, stroe15}, or by removing the galaxy gas which is required
to fuel star formation and thus quenching  star formation.
The high speed relative motion between the galaxy and ICM can induce
ram-pressure stripping
\citep{gunngott72}.

Although all these process are observed to be at work at some level, a
number of studies have shown that their  overall effects are
subtle. Controlling for a galaxy stellar mass and color, environmental
differences are small and difficult to detect even in the best
spatially unresolved data \citep{morishita16}.

Spatially resolved data, preferably spectroscopic, is necessary to make
progress, since each process is expected to leave some signature on
the spatial distribution of star formation within a galaxy. Broadly
speaking, ram pressure is expected to partially or completely strip
layers of gas from a galaxy, leaving a recognizable pattern of star
formation with truncated disks smaller than the undisturbed stellar
disk \citep[e.g.,][]{yagi15}. The morphology of the gas should also
reflect the motion of the galaxy through the ICM. Strangulation, which
is the removal of the hot gas halo surrounding the galaxy either via
hydrodynamical or tidal effects \citep{larson80, balogh00}, should
deprive the galaxy of its outer gas reservoir but have a much gentler
effect on the deeply embedded interstellar medium (ISM) and diffuse star formation. Wet
major and minor mergers, cold gas accretion and harassment should
also leave relatively significant signatures in the gas dynamics
\citep{moore96}. 

Recent progress in hydrodynamical numerical
simulations \citep[e.g. FIRE,][]{hopkins14} that are starting to resolve individual
star forming regions within galaxies suggests that in the near
future it will be possible to go beyond these qualitative statements
to full quantitive characterizations of the signatures of these
processes.

The most commonly adopted tracer of instantaneous star formation is
the \Ha line emission as it scales with the quantity of ionizing
photons produced by hot young stars \citep{kennicutt98}.  In the local
universe, a number of studies have focused on the analysis of the \Ha
spatial distribution of a limited number of systems in clusters
clearly presenting signs of stripping (e.g.,
\citealt{merluzzi13, fumagalli14, fossati16, merluzzi16}, Poggianti et al. in prep.)

 While several spatially resolved studies of \Ha have been conducted beyond the local universe from the ground
\citep[e.g.][]{yang08, goncalves10, sobral13b, swinbank12, wisnioski15, stott16}, very few studies  have been conducted  at $z = 0-1$ from space \citep[e.g.,][]{atek10, straughn11, livermore12}.
Spatially resolved star formation maps at $z\sim$1 have been obtained
for field galaxies using both the Advanced Camera for Surveys (ACS) I
band and the G141 grism on the Wide-Field Camera 3 (WFC3) on board the Hubble Space
Telescope (HST) as part of the 3D-HST Survey \citep{vandokkum11,
brammer12, schmidt13, momcheva15}. \cite{nelson12, nelson13, nelson15}
mapped the \Ha and stellar continuum with high resolution showing that
star formation broadly follows the rest-frame optical light, but is
slightly more extended. By stacking galaxies in bins of stellar mass, they
found that star formation has  to occur in
approximately exponential distributions on average and is enhanced at all
radii above the star formation main sequence \citep[SFMS,][]{noeske07}
and suppressed at all radii below the SFMS.

\cite{wuyts12,wuyts13} characterized the resolved stellar populations
with multi-wavelength broadband imaging from CANDELS \citep{postman12}
and \Ha surface brightness profiles from 3D-HST \citep{brammer12} at $z=0.7-1.5$
at the same kiloparsec
resolution. They found that \Ha morphologies resemble more
closely those observed in the ACS F814W band than in the WFC3/F160W band,
especially for the larger systems. In addition, they showed how the
rate of ongoing star formation per unit area tracks the amount of
stellar mass assembled over the same area. Off-center clumps are
characterized by enhanced \Ha equivalent widths, bluer broad- band
colors, and correspondingly higher specific SFRs than the underlying
disk, suggesting that the ACS clump selection preferentially 
picks up those regions of elevated star formation activity that are the least obscured by dust.

In this paper, we exploit the unprecedented depth and angular
resolution of WFC3 G102 grism dataset obtained as part of the Grism
Lens-Amplified Survey from Space (GLASS; GO-13459; PI:
Treu,\footnote{\url{http:// glass.astro.ucla.edu} }
\citealt{schmidt14, treu15})  to carry out a large and detailed study of the \Ha morphology in
clusters and field galaxies at $0.3<z<0.7$.  Building on the
techniques developed in our pilot study
\citep{vulcani15}, we increase the sample size from 25 and 17 to 76
and 85 galaxies in the clusters and the field, respectively. We define
and apply a new morphological classification scheme for \Ha emission
to this full statistical sample with the goal of identifying potential
signatures of the underlying physical processes and assess their
frequency across environments, and across the SFR-mass plane.
 
In a companion paper (Vulcani et al. 2016, submitted; hereafter Paper
VIII), we investigate trends with cluster properties, such as the hot
gas density as traced by the X-ray emission, the total surface mass density
as inferred from gravitational lens models, and the local number density, to
inspect whether local cluster conditions have an impact on the extent
and location of the star formation.
 
The paper is structured as follows. \S\ref{sec:glass} introduces the
dataset, along with the data reduction and the redshift
determinations.  \S\ref{sec:sample} describes the sample and how
galaxy properties have been determined. \S\ref{sec:results} presents
our results, mainly focussing on the morphology of \Ha emitters at
different wavelengths and on the physical processes that are likely to
have induced these morphologies. We also quantify the position and
extension of the \Ha emission within the galaxies and characterize the
SFR-mass relation for the different types of \Ha emitters. Focusing on
spirals, we test the hypothesis that star formation is mainly
occurring in the disk.  In \S\ref{sec:d_c} we discuss our results and
conclude.
 
Throughout the paper, we assume $H_{0}=70 \, \rm km \, s^{-1} \,
Mpc^{-1}$, $\Omega_{0}=0.3$, and $\Omega_{\Lambda} =0.7$.  The adopted
initial mass function (IMF) is that of \cite{chabrier03} in the mass
range 0.1--100 $\textrm{M}_{\odot}$.

\section{The Grism Lens-Amplified Survey from Space data set}
\label{sec:glass}
\begin{table*}[!t]
\caption{Cluster properties \label{tab:clus}}
\centering
\begin{tabular}{llccccccccc} 
\hline
\hline
\bf{cluster} 	&\bf{short}		& \bf{RA}  & \bf{DEC}  & \bf{z} 	&  \bf{phys scale}&\bf{L$_{\rm X}$} 	& \bf{M$_{\rm 500}$}   & \bf{r$_{\rm 500}$}  & {\bf PA1} & {\bf PA2}    \\
			& \bf{name}	&(J2000) &(J2000) &				&  (kpc/$^{\prime\prime}$)& (10$^{44}$erg s$^{-1}$)	& \	  (10$^{14}$M$_\odot$) & (Mpc) & &    \\
\hline
Abell2744       		& A2744 	&00:14:21.2 	&-30:23:50.1 	&0.308  		&4.535	&15.28$\pm$0.39& 17.6$\pm$2.3 &1.65$\pm$0.07 & 135 & 233\\ 
RXJ2248.7-4431  	& RXJ2248 &22:48:44.4 	&-44:31:48.5 	&0.348  	&4.921	&30.81$\pm$1.57& 22.5$\pm$3.3 & 1.76$\pm$0.08& 053 & 133\\
Abell370        		& A370 	&02:39:52.9 	&-01:34:36.5	&0.375      		&5.162	&8.56$\pm$0.37& 11.7$\pm$2.1 &1.40$\pm$0.08 & 155 &253 \\
MACS0416.1-2403 	& MACS0416	& 04:16:08.9 	&-24:04:28.7 	&0.420    &5.532	&8.11$\pm$0.50&  9.1$\pm$2.0 &1.27$\pm$0.09 & 164 &247\\
RXJ1347.5-1145  	& RXJ1347 &13:47:30.6 	&-11:45:10.0  	&0.451 		&5.766	&47.33$\pm$1.2& 21.7$\pm$3.0 & 1.67$\pm$0.07& 203 & 283\\
MACS1423.8+2404   & MACS1423 	& 14:23:47.8  	& +24:04:40 	& 0.543   	& 6.382	&13.96$\pm$0.52& 6.64$\pm$0.88  & 1.09$\pm$0.05 & 008& 088 \\
MACS1149.6+2223	& MACS1149 	&11:49:36.3 	&+22:23:58.1 	&0.544 	&6.376	&17.25$\pm$0.68& 18.7$\pm$3.0 &1.53$\pm$0.08 & 032 & 125\\
MACS0717.5+3745   & MACS0717	& 07:17:31.6  	& +37:45:18 	& 0.546    & 6.400 	&24.99$\pm$0.92& 24.9$\pm$2.7 & 1.69$\pm$0.06 & 020& 280 \\
MACS2129.4-0741 	& MACS2129 	&21:29:26.0 	&-07:41:28.0 	&0.589   	&6.524	&13.69$\pm$0.57& 10.6$\pm$1.4 &1.26$\pm$0.05 & 050 & 328\\
MACS0744.9+3927  & MACS0744	& 07:44:52.8 	&+39:27:24.0 	&0.686    		&7.087	&18.94$\pm$0.61& 12.5$\pm$1.6 &1.27$\pm$0.05 & 019 & 104\\  
\hline
\end{tabular}
 \tablecomments{J2000 coordinates, redshift, physical scale, X-ray luminosity, M$_{500}$ \citep[from][]
{mantz10}, r$_{500}$  and the two position angles. }
\end{table*}

GLASS is a 140 orbit slitless spectroscopic survey conducted with
\HST{} in cycle 21. It has observed the cores of 10 massive galaxy
clusters with the WFC3 NIR grisms G102 and G141 providing an
uninterrupted wavelength coverage from 0.8$\mu$m to 1.7$\mu$m.  The 10
clusters are listed in Table \ref{tab:clus}.  Observations for GLASS were
completed in January 2015, and the first public data release was
completed in March 2016.  Prior to each grism exposure, imaging
through either F105W or F140W was obtained to assist the extraction of
the spectra and the modeling of contamination from nearby objects on
the sky.  The total exposure time per cluster was 10 orbits in G102
(with either F105W or F140W) and 4 orbits in G141 with F140W. Each
cluster was observed at two position angles (PAs) approximately 90
degrees apart to facilitate clean extraction of the spectra for
objects in crowded cluster fields.  6 GLASS clusters are imaged by the
Hubble Frontier Fields (HFF; P.I. Lotz, \citealt{lotz16}) and 8 by the
Cluster Lensing And Supernova survey with Hubble (CLASH; P.I. Postman,
\citealt{postman12}), providing excellent multiband data.

\subsection{Data reduction}
\label{sec:reduction}
The GLASS observations follow the 
updated version of the 3D- HST
reduction pipeline\footnote{\url{http://
code.google.com/p/threedhst/}} described by \cite{brammer12,
momcheva15}. The updated pipeline combines the individual exposures
into mosaics by interlacing them. 

Each exposure was interlaced to a final G102 (G141) grism
mosaic. Before sky-subtraction and interlacing, each individual
exposure was checked and corrected for elevated backgrounds due to the
He Earth-glow described by \cite{brammer14}.  From the final mosaics,
the spectra of individual objects are extracted by predicting the
position and extent of each two-dimensional spectrum based on the
SExtractor \citep{bertin96} segmentation map combined with deep mosaic
of the direct NIR GLASS and CLASH images. As this is done for each
object, the contamination, i.e., the dispersed light from neighboring
objects in the direct image field-of-view, can be estimated and
accounted for. Full details on the observations and data reduction are
given in \cite{treu15}, while a complete description of the 3D-HST
pipeline, spectral extractions, and spectral fitting, is provided by
\citet{momcheva15}.

The spectra analyzed in this study were all visually inspected with
the publicly available GLASS inspection GUI,
GiG\footnote{\href{https://github.com/kasperschmidt/GLASSinspectionGUIs}{github.com/kasperschmidt/GLASSinspectionGUIs}}
\citep{treu15},  to identify and flag systematic errors in the contamination model, assess the degree of contamination in the spectra, and identify strong emission lines and the presence of a
continuum.

\subsection{Redshift determinations}
To determine redshifts, templates were compared to each of the four available grism spectra 
independently (G102 and G141 at two PAs each) to compute a posterior 
distribution function for the redshift. If available, photometric redshift distributions were used as input priors 
to the grism fits  to reduce computational time. Then, with the help 
of the publicly available GLASS inspection GUI for redshifts \citep[GiGz,][]{treu15}, we flagged which grism 
fits are reliable or alternatively entered a redshift by hand if the redshift 
was misidentified by the automatic procedure.
Using GiGz we assigned a quality $Q_z$ to the redshift (4=secure; 3=probable; 2=possible; 1=tentative, but 
likely an artifact; 0=no-$z$). These quality criteria take into account the 
signal to noise ratio of the detection, the possibility that the line is a contaminant, and the identification of 
the feature with a specific emission line. This procedure was carried out 
independently by at least two inspectors per cluster  \citep[see][for 
details]{treu15}.

The full redshift catalogs from the inspection of the 10 GLASS clusters are available at \url{https://archive.stsci.edu/prepds/glass/}.

\section{The sample}
\label{sec:sample}

We make use of all the 10 GLASS clusters. Virial radii 
$r_{\rm 500}$ have been computed from virial masses M$_{\rm 500}$ taken from \cite{mantz10}: 
$$r_{500}=\sqrt[3]{\frac{3}{4\pi}\frac{M_{500}}{500\rho_{cr}}}$$
where $\rho_{cr}=\frac{3H^2}{8\pi G}= 
\frac{3H_0^2}{8\pi G}\times\left[\Omega_\Lambda+\Omega_0\times
\left(1+z)^3\right)\right]$, with G being the gravitational constant = $4.29 \times 10^{-9} {\rm (km/s)^2\,  Mpc 
\, M}_\odot$.  

From the redshift catalogs, we extract galaxies with reliable redshift
($Q_z\geq$2.5) within $\pm$0.03 of the cluster redshift and consider them
as cluster members. The redshift uncertainty is larger than the
expected range of cluster velocities in order to account for
uncertainties in the low resolution grism data. Our sample members
might therefore include a fraction of interlopers, but we expect it to
be extremely small, considering that we are looking at the cores of
rich clusters that are highly overdense \citep[see, e.g.,][]{morishita16}.

We then select galaxies with detected \Ha emission. We exclude the
Brightest Cluster Galaxies (BCGs) from our analysis, which are not
representative of the general cluster galaxy population. Given the
cluster redshifts, \Ha is found at the observed wavelength range
$8500\leq \lambda \leq11100$
\AA, and we therefore mostly exploit the G102 grism data in our analysis, but we check whether
 \Ha is detected in the G141 grism for the higher redshift galaxies
 ($z\geq0.67$, i.e. members of MACS0744 and a few field galaxies).

\begin{figure}
\centering
\includegraphics[scale=0.45]{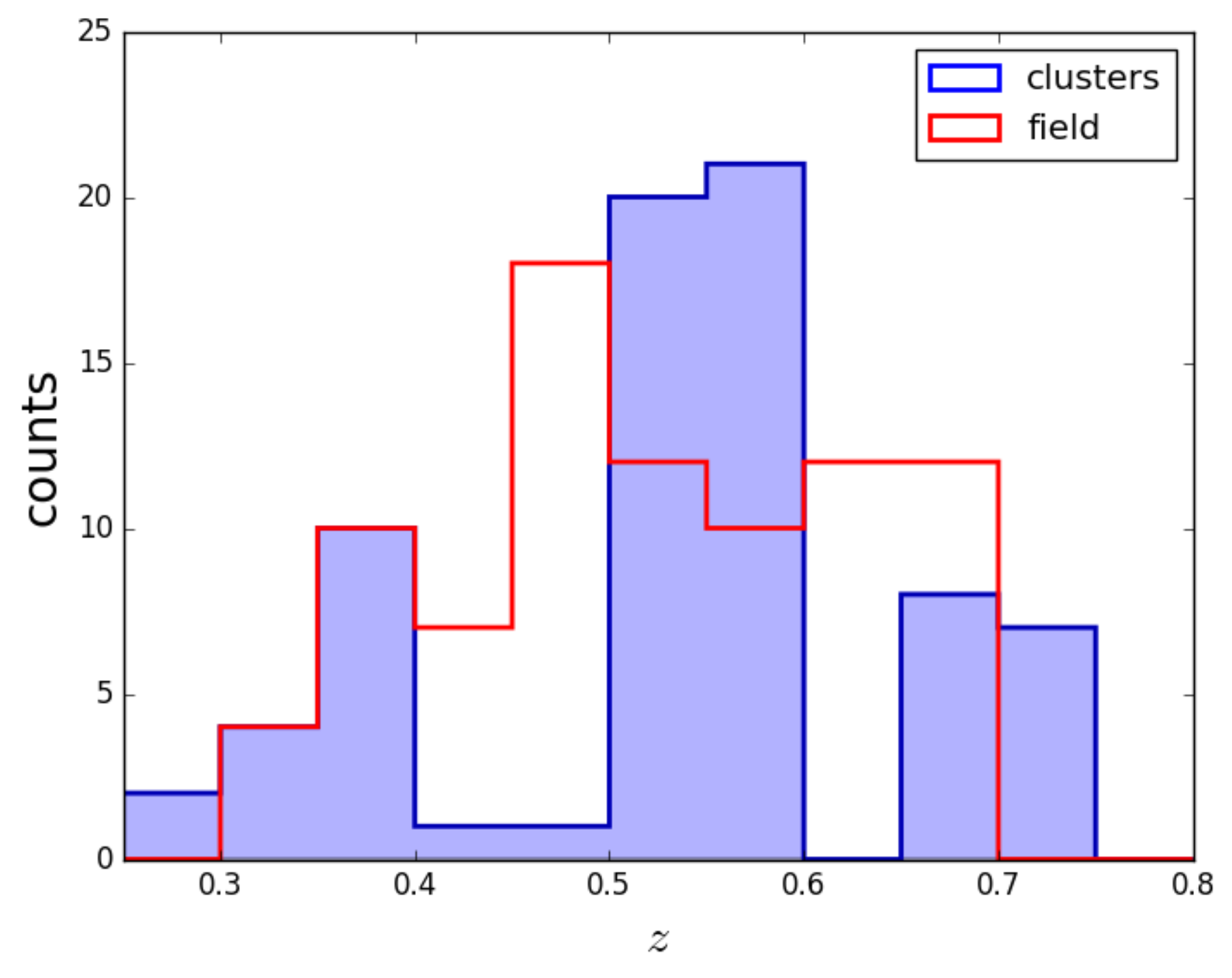}
\caption{Redshift distribution for cluster (blue) and field (red) galaxies in our sample.
\label{fig:z}}
\end{figure}

We  assemble a field sample consisting of all galaxies with reliable redshift, \Ha in emission  
in the G102 grism and redshift outside the cluster redshift intervals 
($z<z_{cl}-0.03$ or $z>z_{cl}+0.03$). Overall, we limit our field sample range to the redshift range spanned 
by cluster members:  0.3$\leq z \leq$0.7. Note that this cut was 
not adopted by \citet{vulcani15}, therefore some galaxies some galaxies from that work are not included in the current selection.
Figure \ref{fig:z} shows the redshift distribution of our samples. 
We  do not have additional information on the 
environments of the field galaxies; some of them might actually be located in groups, but certainly not in rich clusters.

Overall, our sample includes 76 cluster members and 85 field galaxies, distributed among the different 
clusters as summarized in Table \ref{tab:gals}.
\begin{table}
\caption{Number of galaxies with \Ha in emission  \label{tab:gals}}
\centering
\begin{tabular}{lcc} 
\hline
\hline
\bf{cluster} 	&\bf{cluster members}		& \bf{field galaxies}     \\
\hline
A2744 	&4&5\\
RXJ2248 & 3 &4 \\
A370    & 8&10 \\
MACS0416	&2&5\\
RXJ1347 &2&8\\
MACS1423 	&10&18\\
MACS1149 	&8&7\\
MACS0717	&16&8\\
MACS2129 	&8&12\\
MACS0744	&15&8\\
\hline
total &76&85\\
\hline
\end{tabular}
\end{table}

\subsection{Stellar masses} 

Stellar mass estimates are obtained as described  by \cite{morishita16}. 
  7 HST/WFC3+ACS bands (F435W, F606W, F814W, F105W, F125W,
F140W, F160W) have been used.
Briefly, the SED parameters for 
all galaxies have been derived  using FAST  v.1.0 \citep{kriek09}  
using the spectroscopic redshift of each object. CLASH \citep{postman12} or, when available, HFF 
photometry \citep{lotz16} have been adopted. 
The stellar population model of GALAXEV \citep{bc03}, solar metallicity and  a \cite{chabrier03} IMF have been 
adopted. The Calzetti dust law \citep{calzetti00} is restricted to the range 
$0\leq A_V \leq 4.0$ mag and the age can vary from  0.1 to the age of the universe at the galaxy redshift. 
Exponentially-declining star-formation history, where $SFR(t)\propto \exp(-t/
\tau)$ at the time t, with $\tau$ in the range of $8 \leq log\tau \leq 10$ is assumed. The errors in the SED 
parameters are 1$\sigma$ uncertainties derived by FAST. 

 We note that the templates for SED fitting used here do not include emission lines. 
In \cite{vulcani15}, where we analyzed only two clusters, we had used a SED fitting procedure that considered emission lines \citep[see][for details]{castellano14}. 
Comparing those mass estimates to the one used here for the subset of objects in common, 
after correcting for the IMF, we get that results are largely in agreement and the dispersion  is $<$0.1dex, which is smaller than the typical error on mass estimates (0.2-0.3 dex).

Stellar population properties  have not been fitted for A370, since the final HFF observations are 
not scheduled until September 2016.

\subsection{\Ha maps} \label{Hamaps}

The GLASS grism spectra have high spatial resolution and low spectral resolution, meaning that 
emission line structure reflects almost exclusively spatial structure 
(morphology) in contrast to data with high spectral resolution  where structure reflects velocity (rotation or 
dispersion). Spectra can be seen as images of a galaxy taken at $\sim$24 
\AA{} increments ($\sim$12 \AA{} after interlacing) and placed next to each other (offset by one pixel) on 
the detector. 

The details of the procedure we followed to make emission line maps of galaxies are described by 
\cite{vulcani15}. 
Briefly, we obtained the maps from the spectra coming from the different
exposures (one per PA) of each galaxy independently, and only in the
last step we combined them. 
From the flux-calibrated galaxy 2D continuum spectra we subtracted the sky background
and the contamination.  From two regions
contiguous to the \Ha emission we determined the $y$-position of the
peak of the continuum. This position was needed to measure the
offset in the $y$-direction of the \Ha emission with respect to the
galaxy center in the light of the continuum. Subsequently, we subtracted
the 2D stellar continuum model obtained by convolving the best-fit 1D continuum
 without emission lines with the actual 2D data, ensuring that all
model flux pixels are non-negative. We were therefore left with the surface brightness map of the \Ha line.

When spectra from both PAs are reliable, i.e. not contaminated by other sources or not at the edge of the grism, 
we  combined the maps obtained  for the two PAs.
When also the \Ha maps from the G141 filter are available, we combined them with the G102 maps, to 
increase the signal-to-noise ratio.  
Overall, for 80 galaxies \Ha maps have been reliably obtained in the spectra of both PAs in the G102 grism; for 9  
galaxies 
\Ha maps have been obtained also from the G141 spectra. 

As a final step, we superimposed the \Ha
map onto an image of the galaxy taken with the F475W filter
(rest-frame UV) and onto an image in the F140W (IR). Images are taken from the HFF photometry  \citep{lotz16} or CLASH HST 
photometry, \citep{postman12}. We use the F475W filter to map relatively recent ($\sim
$100 Myr)  star formation,
and the F140W to trace the older stellar population
as opposed to ongoing  ($\sim$10Myr) star formation traced by \Ha. 
Note that for A2744 we used the F435W filter, because the F475W 
filter is not available. 

To superimpose the \Ha maps to the images, we aligned each map to the continuum 
image of the galaxy, rotating them by the angle of its PA,
keeping the $y$-offset unaltered with  respect to the continuum. On
the $x$-axis, there is a degeneracy between the spatial dimension and
the wavelength uncertainty, it is therefore not possible to determine uniquely
the central position of the \Ha map for each PA separately. However, for the vast majority of cases in
which spectra from both PAs are reliable, we used the fact that
the 2 PAs differ by almost $90\degree$, therefore the $x$-direction of
one spectrum roughly corresponds to the $y$-direction of the second
spectrum and vice-versa. We shifted the two spectra
independently along their $x$-direction to maximize the cross correlation between the two maps
to get the intersect.
For the galaxies with reliable spectra in both PAs, we  also measured the real distance between the peak 
of the \Ha emission and the continuum emission, obtained as the quadratic 
sum of the two offsets.

\subsection{SFRs}

From the \Ha maps we derive SFR maps.
We use the conversion factor derived by \cite{kennicutt94}  and \cite{madau98}: 
$$\textrm{SFR} [\textrm{M}_\odot \, \textrm{yr}^{-1}] = 5.5 \times 10^{-42}\textrm{L}(H\alpha)[\textrm{erg}\,  
\textrm{s}^{-1}]
$$
valid for a \cite{kr01} IMF.  We then converted SFRs to our adopted \cite{chabrier03} IMF by adding 0.05 dex to the logarithmic values. 
We compute both the surface  SFR density ($\Sigma$\textrm{SFR}, $\textrm{M}_\odot \, \textrm{yr}^{-1}\, 
\textrm{kpc}^{-2}$) and the total SFRs ($\textrm{M}_\odot \, \textrm{yr}
^{-1}$), separately for the spectra coming from the two PAs and then  combine them taking the mean 
values. Errors are summed in quadrature and divided by two. The measurements from the two PAs are consistent within the uncertainty. 
The total SFRs are obtained 
summing the surface  SFR density  within the Kron radius measured by Sextractor 
from a combined NIR image  of the galaxy. 

There are two possible limitations when using \Ha as SFR estimator:
the contamination by the [NII] line doublet, and uncertainties in the
extinction corrections to be applied to each galaxy. We account for
both effects, even though they are both small and none of the
conclusions of this work hinge on the details of these corrections.

To correct for  the [NII] contamination,  we apply the locally calibrated correction factor 
given by \cite{james05}. As opposed to  previous works which considered 
only central regions, these authors developed a method which takes into account the variation of the \Ha-
[NII] with radial distance from the galaxy center, finding an average value of 
\Ha/(\Ha + [NII])= 0.823. This approach is appropriate given our goal to investigate extended emission. 

The second caveat is the effect of dust extinction. Star formation normally 
takes place in dense and dusty molecular clouds, so a significant fraction 
of the emitted light from young stars is absorbed by the dust and re-emitted at rest-frame IR wavelengths. 
 \cite{garnbest10} modeled a mass-dependent attenuation by dust, based 
on the \cite{calzetti00} dust attenuation law, of the form

\begin{equation}
A(H\alpha)=\sum_{i=0}^{3} B_i \cdot X^i
\end{equation}

with $X=\log_{10}(M_\ast/1.1\cdot 10^{10}M_\odot)$ (to take into account the different IMF) and $B_i$=$0.91$, $0.77$,  $0.11$, $-0.09$, respectively.

Even though the relationship was measured from observations of $z\sim0$ starburst galaxies, it has been shown to be appropriate for galaxies at redshifts up to $z\sim0.8$ \citep[see, e.g.,][]{garn10,sobral12, dominguez13, price14}.

According to \cite{garnbest10}, the equation allows to predict the extinction of a galaxy 
with a given stellar mass to within a typical error of 0.28 mag. This typical error is broadly comparable to the 
accuracy with which extinctions can be estimated from the Balmer decrement.

We use this correction to obtain the intrinsic SFRs:
\begin{eqnarray*}
\textrm{SFR}_{\textrm{int}} = \textrm{SFR}_{\textrm{obs}} \times 10^{A(H\alpha)\times 0.4}.
\end{eqnarray*}

We assume that there is no spatial variation in extinction across the galaxy, even though 
high-resolution imaging in multiple HST bands \citep{wuyts12} 
and analyses of such data in combination with \Ha maps extracted from grism spectroscopy 
\citep{wuyts13} indicate that such an assumption may be over-simplistic, 
particularly in the more massive galaxies where the largest spatial color variations are seen. It is hard to 
anticipate how corrections for non-uniform extinction might affect our 
conclusions, since the correction to the sizes will depend on the actual distribution of dust. 

However, we expect the effect of dust to be relatively small,
especially in a differential sense, i.e. it should not effect
significantly our comparison between cluster and field galaxies.  In
support of this we note that Wang et al. (2016; in prep.) measure dust
attenuation from the Balmer decrement for a subsample of field
galaxies at $1.3\leq z \leq 2.3$ in MACS1149. They find that the
measured attenuation $A_V$ is almost always
much smaller than 1 mag, and quite homogeneous across the galaxies, indicating
that the spatial variation of the dust is not very important.

\begin{figure}
\centering
\includegraphics[scale=0.45]{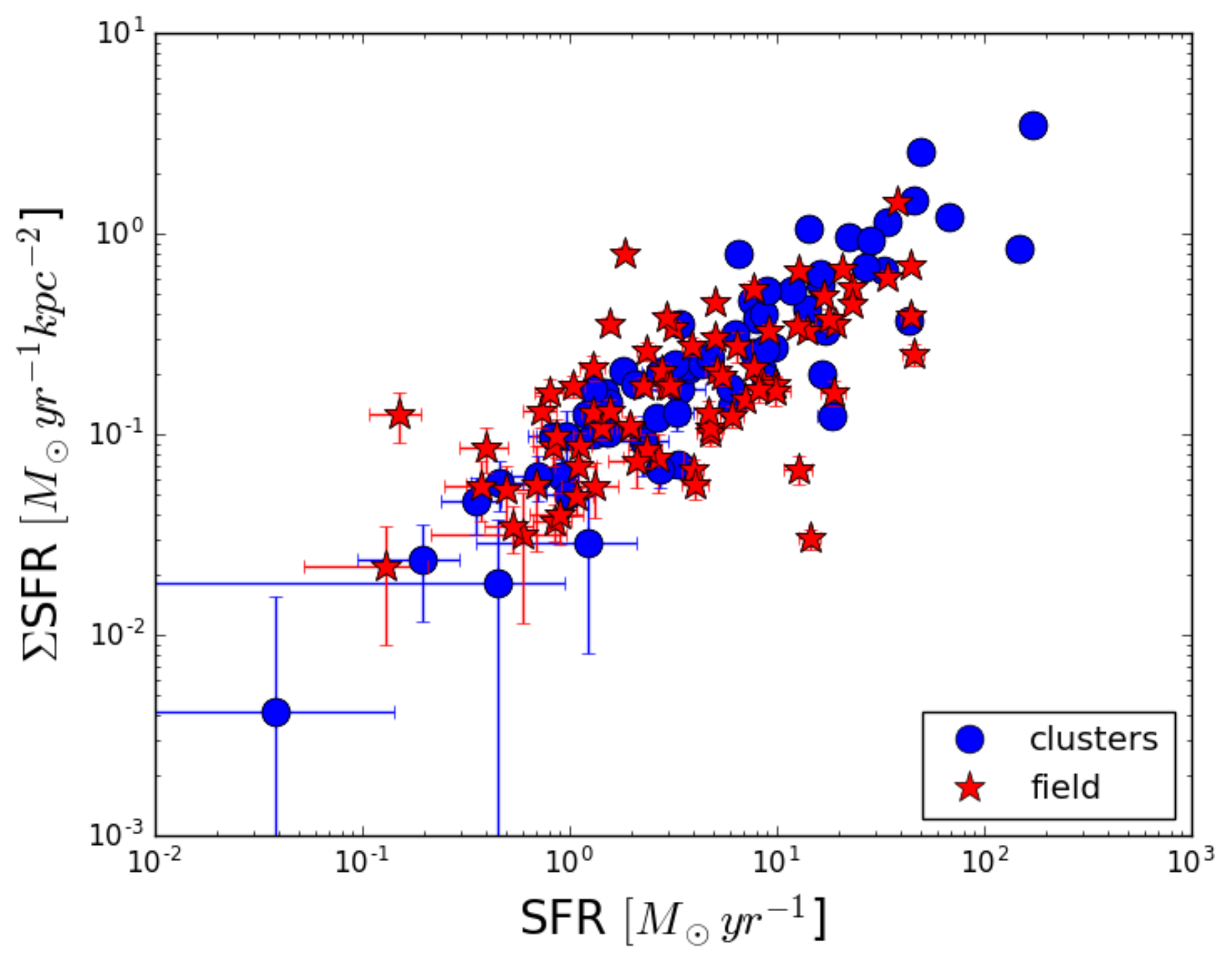}
\caption{$\Sigma$SFR-SFR for cluster (blue) and field (red) galaxies in our sample. Error bars show the typical uncertainties
on the measurements. Our  $\Sigma
$SFR limit is around $5\times 10^{-2}$ M$_\odot$ yr$^{-1}$ kpc$^{-2}$ for 
SFR$\sim1 M_\odot yr^{-1}$ and is independent on environment, suggesting that the 
physical conditions in star forming galaxies are similar in clusters and field.
\label{fig:SFR_SFRd}}
\end{figure}

Figure \ref{fig:SFR_SFRd} shows the correlation between   $\Sigma$SFRs and total SFR. 
Our $\Sigma $SFR limit is around $5\times 10^{-2}$
M$_\odot$ yr$^{-1}$ kpc$^{-2}$ for SFR$\sim1 M_\odot yr^{-1}$ both in
clusters and in the field, suggesting that poor sensitivity does not
vary between the field and cluster samples, as expected because we are
using the same dataset.  We use this value as indication of the
completeness limit of our samples.

\begin{figure*}
\centering
\includegraphics[scale=0.33]{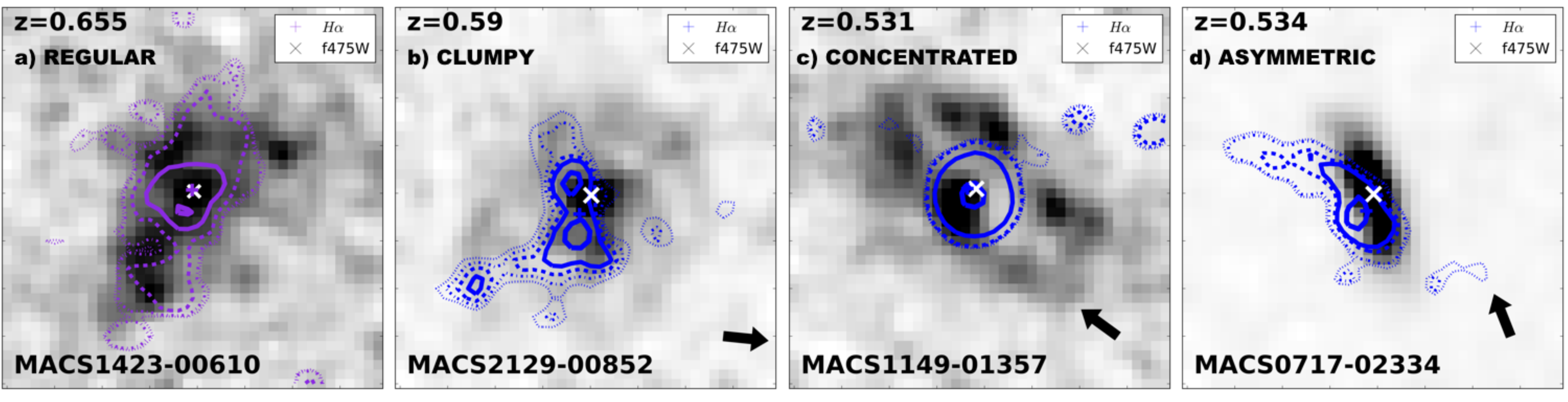}
\caption{Examples of \Ha maps with different morphologies, as indicated in the labels. Maps are 
superimposed on the image of the galaxy in the F475W filter. Contour levels 
represent the 35$^{th}$, 50$^{th}$, 65$^{th}$, 80$^{th}$, 95$^{th}$ percentiles of the light distribution, 
respectively. Blue contours indicate that the \Ha map is obtained just from one 
spectrum, purple contours indicate that the \Ha map is obtained  from the two orthogonal spectra. For 
galaxies in clusters, arrows on the bottom right corner indicate the direction of 
the cluster center. The redshift of the galaxy is indicated on the top left corner.
\label{Ha_morf}}
\end{figure*}

\begin{figure*}
\centering
\includegraphics[scale=0.33]{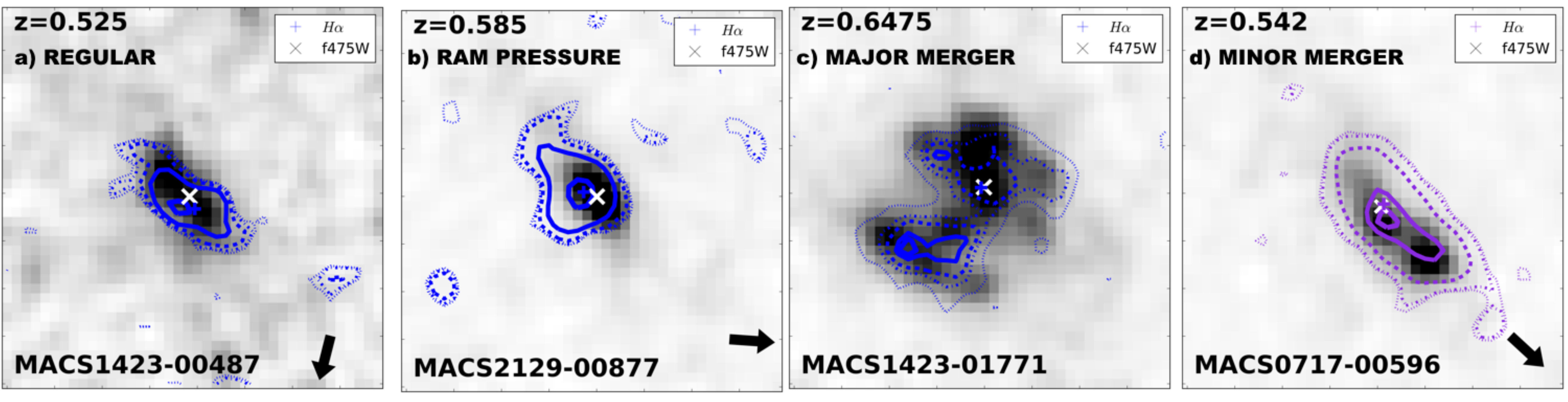}
\caption{Examples of  \Ha maps influenced by different physical processes, as indicated in the labels. 
Colors, lines and symbols are as in Fig.\ref{Ha_morf}.
\label{Ha_proc}}
\end{figure*}

\subsection{\Ha EWs}
Finally, we also compute \Ha equivalent widths EW(\Ha) from the collapsed 2D spectra.
We define the line profile by adopting a fixed rest frame wavelength range, centered on the theoretical 
wavelength, 6480-6650 \AA{}, and then  obtain the line flux, $f_{\textrm{line}}$, 
by summing the flux within the line. 
The continuum is defined by two regions of 100 \AA{} located at the two extremes of the line profile. We fit 
a straight line to the average continuum in the two regions and  sum the 
flux below the line, to obtain $f_{\textrm{cont}}$. The rest-frame  EW(\Ha) is therefore defined by
$$
\textrm{EW}(H\alpha) =  \frac{f_{\textrm{line}}}{f_{\textrm{cont}}\times(1+z)}
$$ Our approach ignores underlying \Ha absorption, assuming it to be
negligible for these strong emitters. As usual, when two spectra for
the same galaxy are reliable, the final value is given by the average
of the two EW estimates, and the error is obtained by summing in
quadrature the individual errors. Overall, the EW measurements from the
two PAs are consistent within the uncertainties.

\subsection{Sizes and B/T ratios}\label{sec:sizes}

Morphological parameters (sizes, bulge to total (B/T) ratios) for all the galaxies have been obtained using  GALFIT \citep{peng02}, as 
described by \cite{morishita16}. Measurements have been performed 
using the F140W and F375W for  HFF photometry \citep{lotz16} or the F140W and F475W for CLASH HST 
photometry \citep{postman12}. 

The fitting was performed assuming a double profile, to separate the
different components of the galaxies (i.e. bulge+disk).  Following a common choice \citep[e.g.][]{graham05, meert15}, for bulges a
Sersic index n=4 has been adopted, while for disks the typical
exponential profile has been chosen (n=1). Such decomposition allows
us to inspect whether the \Ha flux traces the disk. A detailed
description of the fitting procedure--though only for a single component analysis-- can be found in
\cite{morishita16}.

To estimate the \Ha  size of the galaxies, we choose to fix the Sersic index to n=1. This assumption is valid
only for galaxies where star formation is taking place in disks, as we will discuss in \S\ref{sec:spirals}.  Therefore, we run 
GALFIT with the same parameters as done for the disk component  in the  continuum.

\subsection{Visual morphologies}\label{sec:visinsp}

Galaxies in our sample have been visually inspected by a subset of authors (B.V., T.T., A.D. and T.M.)  to 
classify their morphology.  Visual inspection was carried out with the publicly available Graphic User Interface GIG 
for morphologies (GiGm) described in Appendix \ref{GIGm}.
We classified both the broad-band morphology  and the morphology of the \Ha emission. 
For the former, we followed the standard Hubble  classification, 
subdividing galaxies into Ellipticals (E), Lenticulars (S0), Spirals (Sp) and Irregulars (Irr). In 
addition, we flagged galaxies undergoing a either minor or major merger (Mer). 
For the \Ha morphology, we introduce a new  scheme (illustrated in Fig. \ref{Ha_morf}):

\begin{itemize}
\item {\it regular}, when the \Ha light distribution is smooth and follows the galaxy disk (Fig. \ref{Ha_morf}a);
\item {\it clumpy}, when multiple peaks in the \Ha light  distribution have been detected, possibly a 
consequence of a  merger where the core of the two galaxies are still 
distinguishable (Fig. \ref{Ha_morf}b);
\item {\it concentrated}, when the \Ha light distribution appears more concentrated than the underlying disk or overall continuum if a disk is no clearly identifiable (Fig. 
\ref{Ha_morf}c);
\item {\it asymmetric/jellyfish}, when the emission appears torqued and extends mainly on one side of the galaxy (Fig. \ref{Ha_morf}d).
\end{itemize}

In addition, we also assign a label to each galaxy to attempt to classify the most likely physical processes responsible for altering its continuum and \Ha morphology. This is clearly a qualitative and approximate classification scheme, considering that multiple processes might be simultaneously at work and that the mapping between morphology and process is not always unique and unambiguous.
In spite of the uncertainties, we believe there is merit in categorizing in a self consistent manner the diversity of morphological features across environments. In the future, this classification scheme might be replaced with full 2D comparisons with numerical simulations. However, this is not currently possible and a qualitative classification appears to be a useful first step.

We adopt the following scheme
(illustrated in Fig. \ref{Ha_proc}):
\begin{itemize}
\item {\it regular}, when the \Ha light distribution appears regular and undisturbed (Fig. \ref{Ha_proc}a);
\item {\it ram pressure}, when an asymmetry in the \Ha distribution or in the surface brightness is detected
(Fig. \ref{Ha_proc}b). We are not able to detect weak cases of ram pressure stripping, for example 
when a galaxy is at its second or third passage toward the galaxy center, but only the strongest ones, when large quantities of gas are still available, 
and the ionized gas is stripped away in a line that approximately points away from the cluster center. Even though we know the environment 
in which galaxies are embedded, in clusters we did not explicitly use the direction of the cluster center to characterize this process. Appendix \ref{RP}
shows all the galaxies in our sample we think are mainly affected by this process;
\item {\it major merger}, when the constituents of the mergers are both visible both in the F140W and F475W filters, suggesting
they have comparable stellar mass and luminosity (Fig. \ref{Ha_proc}c);
\item {\it minor merger/interaction}, when the F475W filter shows the presence of material 
infalling onto the main galaxies that is not detected in the F140W filter, suggesting a low mass to light ratio and a low stellar mass  (Fig. \ref{Ha_proc}d);
\item {\it other}, when none of the above applies. 
\end{itemize}

To determine whether our \Ha emitters are characterized by a distinctive morphology with respect 
to the general galaxy population, we also visually classified a control 
sample of non-\Ha emitter galaxies. 
This sample was drawn from the GLASS subsample for which photometric redshift have been 
determined by  \cite{morishita16}. This sample includes  four HFF
clusters A2744, MACS0416, MACS0717, MACS1149.  For each galaxy in our \Ha emitter sample, we 
extracted two galaxies from the photo$-z$ sample with masses similar 
 to the reference galaxy (usually one slightly less massive and one slightly more massive), in the 
same environment and possibly at similar redshift. The latter criterion had 
to be relaxed for cluster galaxies at $z>0.55$ due to the lack of cluster galaxies in this redshift range in the photo$-z$ catalog. 

The classifications obtained from the different inspectors have been combined, adopting the most common 
classification. In case of broad disagreement ($<25\%$ of the cases), the galaxies were re-inspected and discussed to reach an agreement.

\subsection{Summary of the samples}
In summary we  have four galaxy sub-samples with the following quantities estimated that we will use in our analysis:
\begin{enumerate}
\item \Ha emitters, clusters: Continuum morphology, \Ha morphology, main process, SFR, $\Sigma$SFR, $M_\ast$, offsets between emissions, sizes;
\item \Ha emitters, field: Continuum morphology, \Ha morphology, main process, SFR, $\Sigma$SFR, $M_\ast$, offsets between emissions, sizes;
\item non-\Ha emitters  clusters: Continuum morphology, $M_\ast$;
\item non-\Ha emitters field: Continuum morphology, $M_\ast$;
\end{enumerate}
The first two samples constitute the main sample, the other our control sample. 
In the next section we present extensive comparisons of these four samples and their derived characteristics.

\section{Results}\label{sec:results}
The  properties of galaxies in our main sample are given in Table \ref{tab:clu_gal}  (the total 
sample is given in the online version of the article). They include 
galaxy positions, redshifts, environments,  magnitudes, stellar masses, EW(\Ha)s, SFRs, $\Sigma$SFRs,  
the offset between the peak of the light in \Ha and in the rest-frame UV 
continuum, broad-band morphologies   and \Ha morphologies and main processes acting on the galaxies.

\begin{figure}
\centering
\includegraphics[scale=0.45]{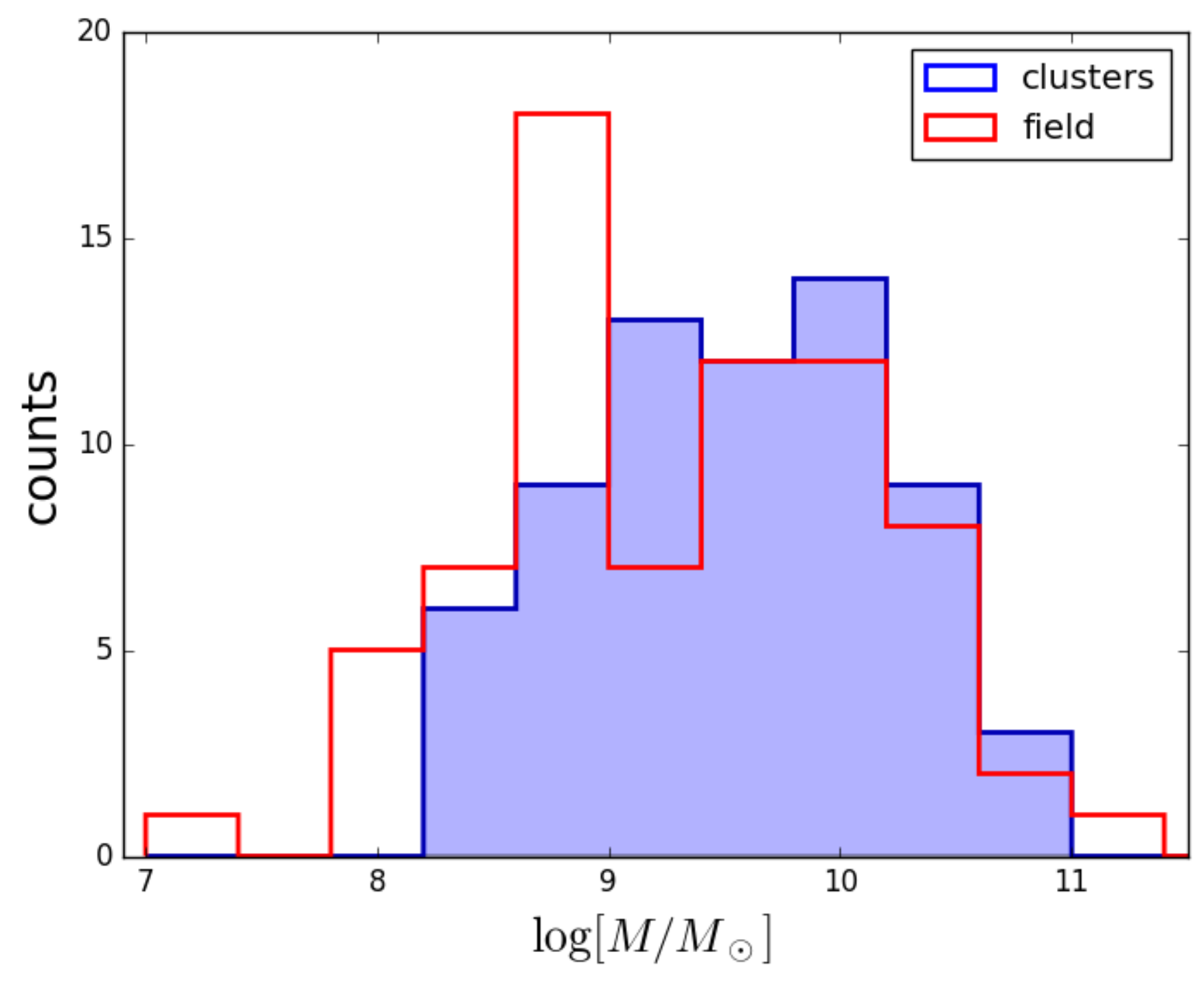}
\caption{Stellar mass distribution for cluster (blue) and field (red) galaxies in our sample. Galaxies in the two environments are 
characterized by very similar mass distributions. Therefore any
differences in star formation properties between the two samples
are not driven by differences in stellar mass.
\label{fig:mass}}
\end{figure}

Figure \ref{fig:mass} shows the distribution of stellar masses for our
main sample. Galaxies in the two environments are
characterized by very similar mass distributions \citep[in agreement
with][] {vulcani13}. Therefore any differences in star formation
properties  are not driven by
differences in stellar mass.

\begin{turnpage}
\begin{table*}
\caption{Properties of galaxies
\label{tab:clu_gal}}
\centering
\setlength{\tabcolsep}{3pt}
\begin{tabular}{lcccccccccccccc}
\hline
\hline
  \multicolumn{1}{c}{objname} &
  \multicolumn{1}{c}{RA} &
  \multicolumn{1}{c}{DEC} &
  \multicolumn{1}{c}{z$^1$} &
  \multicolumn{1}{c}{env} &
  \multicolumn{1}{c}{MAG\_AUTO} &
  \multicolumn{1}{c}{$\log M_\ast/M_\sun$} &
  \multicolumn{1}{c}{EW} &
  \multicolumn{1}{c}{SFR} &
  \multicolumn{1}{c}{$\Sigma$SFR} &
  \multicolumn{1}{c}{offsetPA1$^2$} &
  \multicolumn{1}{c}{offsetPA2$^2$} &
  \multicolumn{1}{c}{morph$^3$} &
  \multicolumn{1}{c}{\Ha morph$^4$} &
  \multicolumn{1}{c}{proc \Ha$^5$} \\
  \multicolumn{1}{c}{} &
  \multicolumn{1}{c}{(J2000)} &
  \multicolumn{1}{c}{(J2000)} &
  \multicolumn{1}{c}{} &
  \multicolumn{1}{c}{} &
  \multicolumn{1}{c}{ABmag} &
  \multicolumn{1}{c}{} &
  \multicolumn{1}{c}{\AA{}} &
  \multicolumn{1}{c}{($M_\odot \, yr^{-1}$)} &
  \multicolumn{1}{c}{($M_\odot \, yr^{-1} \, kpc^2$)} &
  \multicolumn{1}{c}{(kpc)} &
  \multicolumn{1}{c}{(kpc)} &
  \multicolumn{1}{c}{} &
  \multicolumn{1}{c}{} &
  \multicolumn{1}{c}{} \\
\hline
  A2744-00065 & 3.57700058 & -30.37947795 & 0.496 & field & 20.4 & 10.3 & 30.7$\pm$0.6 & 2.7$\pm$0.9 & 0.08$\pm$0.03 & - & -0.1$\pm$0.1 & Ell & Reg & RP\\
  A370-00765 & 39.95471146 & -1.55565658 & 0.576 & field & 21.0 & - & 39$\pm$2 & - & - & 0.1$\pm$0.2 & -  & Mer & Cl & MaM\\
  MACS0416-00286 & 64.032599 & -24.063501 & 0.312 & field & 22.4 & 8.7 & 70$\pm$20 & 1.3$\pm$0.4& 0.05$\pm$0.01 & 0.2$\pm$0.6 & - & Unc & Cl& MiM\\
  MACS0717-00173 & 109.39848457 & 37.766442326 & 0.556 & cluster & 22.1 & 9.3 & 30$\pm$2 & 3.7$\pm$0.3 & 0.22$\pm$0.01 & - & -0.0$\pm$0.1 & S0 & Asy & MiM\\
  MACS0744-00175 & 116.22258717 & 39.47348495 & 0.488 & field & 21.7 & 9.3 & 54$\pm$4 & 4.9$\pm$0.7& 0.11$\pm$0.01& -0.9$\pm$0.2 & 0.0$\pm$0.2 & Spir & Conc & Reg\\
  MACS1149-00063 & 177.397721 & 22.415211 & 0.536 & cluster & 21.8 & 9.9 & 9.7$\pm$0.5 & 4.5$\pm$0.7 & 0.18$\pm$0.03 & 0.0$\pm$0.2 & - & S0 & Conc & RP\\
  MACS1423-01729 & 215.94220899 & 24.0667282 & 0.460 & field & 22.6 & 8.9 & 42$\pm$4 & 1.6$\pm$0.2& 0.13$\pm$0.02 & 0.0$\pm$0.1 & -0.8$\pm$0.2 & Spir & Asy & MiM\\
  MACS2129-00163 & 322.35546305 & -7.67532645 & 0.466 & field & 21.4 & 9.6 & 15.2$\pm$0.4 & 4.0$\pm$0.4 & 0.28$\pm$0.03 & - & -0.2$\pm$0.1 & Spir & Conc & Reg\\
  RXJ1347-01406 & 206.89375736 & -11.76375986 & 0.538 & field & 21.4 & 9.6 & 65$\pm$2 & 13.2$\pm$0.6 & 0.67$\pm$0.03 & 0.08$\pm$0.04 & 0.49$\pm$0.04 & Mer & Ass & MaM\\
  RXJ2248-00104 & 342.16731251 & -44.51393496 & 0.343 & cluster & 19.7 & 9.9 & 17.4$\pm$0.5 & 28$\pm$2 & 0.70$\pm$0.04 & 1.1$\pm$0.1 & -0.3$\pm$0.2 & Spir & Ass & RP\\
\hline\end{tabular}
 \tablecomments{J2000 coordinates, redshift, environment, magnitude, stellar mass, \Ha equivalent width, SFR, $\Sigma$SFR,  offsets between the \Ha emission and the continuum 
emission (as measured form the F475W filter) along the two directions, broad-band continuum morphologies, \Ha morphologies and main processes acting on the galaxy (see text for details). The 
whole table is available online. \\
 $^1$redshifts slightly changed from the GLASS catalog to better match the center of the \Ha maps. \\
 $^2$Reported offsets are along the y-direction of the corresponding PA. The orientation of the offset  (counterclockwise from North) is $\theta=-PA1-44.69$.\\
 $^3$Ell=elliptical, S0=lenticular, Spir=spiral, Irr=Irregular, Mer=merger \\
 $^4$Reg=regular, Cl=Clumpy, Conc=concentrated, Asy=asymmetric/jellyfish \\
 $^5$Reg=nothing, RP=ram pressure stripping, MaM=Major Merger, MiM=Major Merger, other=other } 
\end{table*}
\end{turnpage}

In the next subsections we will use all the data at our disposal   to investigate a number of aspects related 
to galaxy evolution in the different environments. We will first  focus on the 
morphologies of our galaxies and contrast the morphological distribution of \Ha emitters to that of the 
control sample of non \Ha emitters.  Then we will focus only on \Ha emitters and consider also the morphology 
of the \Ha emission and characterize the possible processes responsible 
for such morphologies in the different environments. We will also characterize the SFR-mass relation, 
to see whether galaxies with different   \Ha properties and are experiencing different physical processes 
 are located in different regions of this plane. Finally, for spiral galaxies - for which disk galaxy sizes
are meaningful - we will  
investigate if the  \Ha emission follows the extension of the disk, indicative of galaxies forming 
stars in the secondary component. 

Given the GLASS strategy (spectroscopic redshifts mainly for galaxies with emission lines)
and the way our samples have been selected (visual selection), a statistical analysis on the incidence of \Ha emitters in 
the different environments is not currently possible, and  we therefore do not investigate the frequency of the \Ha emitters with respect to the overall population in clusters and field. 
 
\subsection{The peculiar morphology of  \Ha emitters}
 
\begin{figure}
\centering
\includegraphics[scale=0.28]{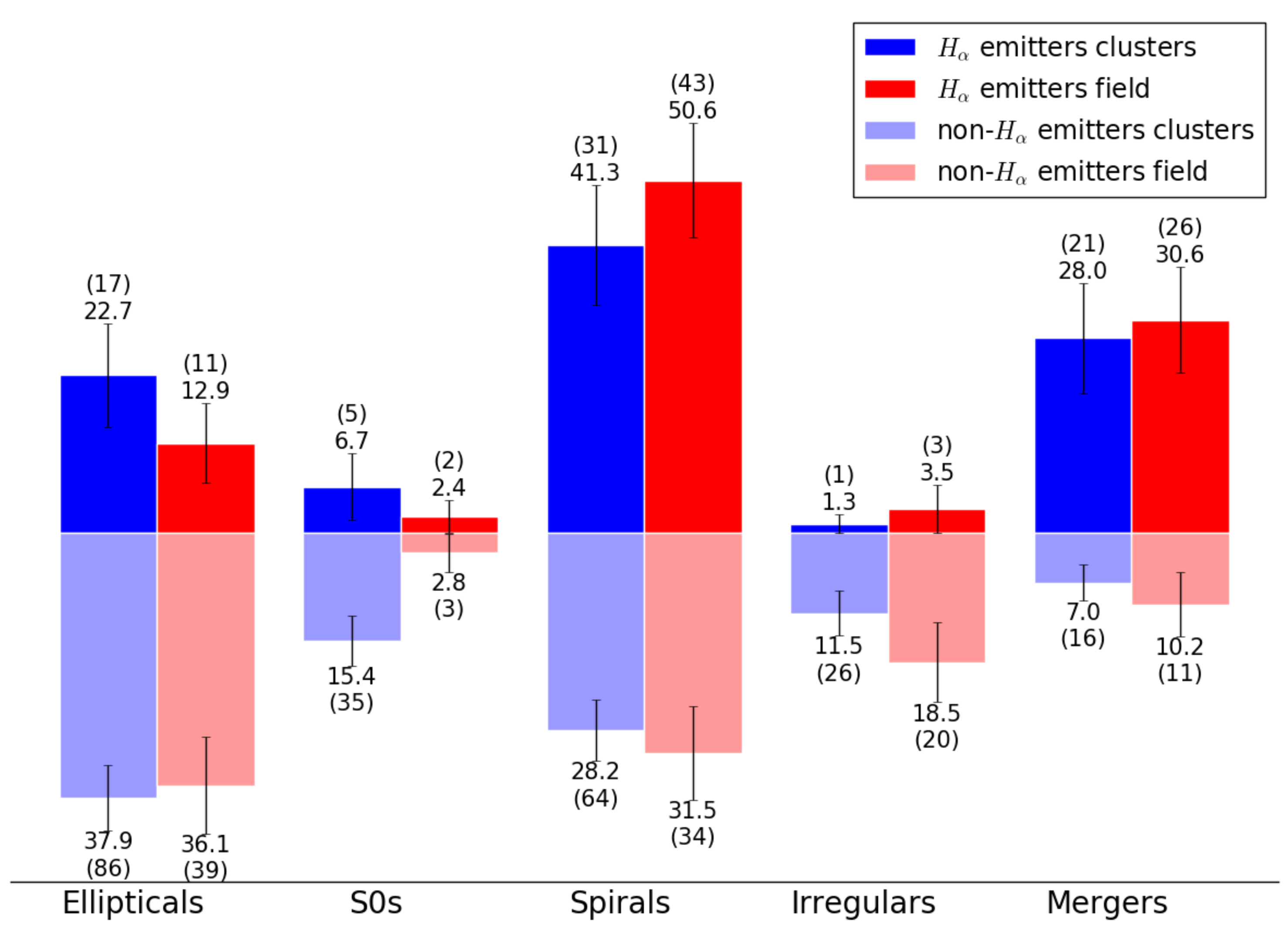}
\caption{Morphological percentages of galaxies in clusters (blue) and in the field (red) for \Ha emitters 
(darker colors, upper histograms) and  non-\Ha emitters (lighter colors, bottom histograms). Numbers represent 
percentages, error bars represent 1$\sigma$ binomial errors \citep{gehrels86}.  Numbers in brackets give the number of objects in each class.
Spiral galaxies dominate the population of
\Ha emitters, followed by merging systems. Interestingly, many \Ha emitters have an early-type (S0+elliptical) morphology. 
In the control sample, there is no predominance of one morphological types, and the fraction of passive spirals
is surprisingly high, compared to previous works \citep[e.g.][]{poggianti09}.
\label{fig:cfr_morph}}
\end{figure}

Figure \ref{fig:cfr_morph} summarizes the morphological percentages for galaxies in clusters and in the 
field, for \Ha emitter and non-\Ha emitter galaxies.  The two samples present morphological 
distributions which are all consistent within the 1$\sigma$ errors: the \Ha emitters are  dominated by spiral galaxies, both in 
clusters ($41\pm8\%$) and in the field ($51\pm8\%$), followed by 
merging systems ($28\pm8\%$ in clusters, $31\pm7\%$ in the field). Ellipticals constitute 22$\pm$7\% of the \Ha emitter population in clusters, and 13$\pm$6\% in the 
field. In both environments, S0s represent less than 7\% of the entire 
population and they are slightly more numerous in clusters, in agreement with \cite{dressler97, postman05}. Irregulars represent just a few percent of the total population.
We stress that among the \Ha emitters, the fraction of both mergers and early-type (elliptical+S0) galaxies is remarkably high. Although this was somewhat expected for mergers, it is perhaps surprising for the early-type galaxies.
Understanding the origin of star forming ellipticals is beyond the scope of this work and will be revisited in future work. 

In the control sample of non \Ha emitters, there is no predominance of a unique morphological 
class, with spirals constituting 28$\pm$4\% of the total population in clusters and  31$\pm$7\% in the field, 
ellipticals 38$\pm$5\% in clusters,  36$\pm$7\% in the field. The incidence of merging systems is significantly
less  than among \Ha emitters (7$\pm$3\% in clusters, 10$\pm$3\% in the field), while irregulars are  more 
numerous (12$\pm$3\% in clusters, 19$\pm$6\% in the field). However, we caution the reader that in many cases it has been very difficult to distinguish 
between these two classes. 
It is very interesting to note the presence of  such a high number of spiral and irregular galaxies that do not present 
\Ha in emission above our \Ha surface brightness detection limit of $5\times 10^{-2}$ M$_\odot$ yr$^{-1}$ kpc$^{-2}$, 
indicating that, despite their morphology, these object are already passive, perhaps on their way to becoming lenticular galaxies \citep[e.g.][]{moran05, ellis97, treu03}.

\begin{figure}
\centering
\includegraphics[scale=0.28]{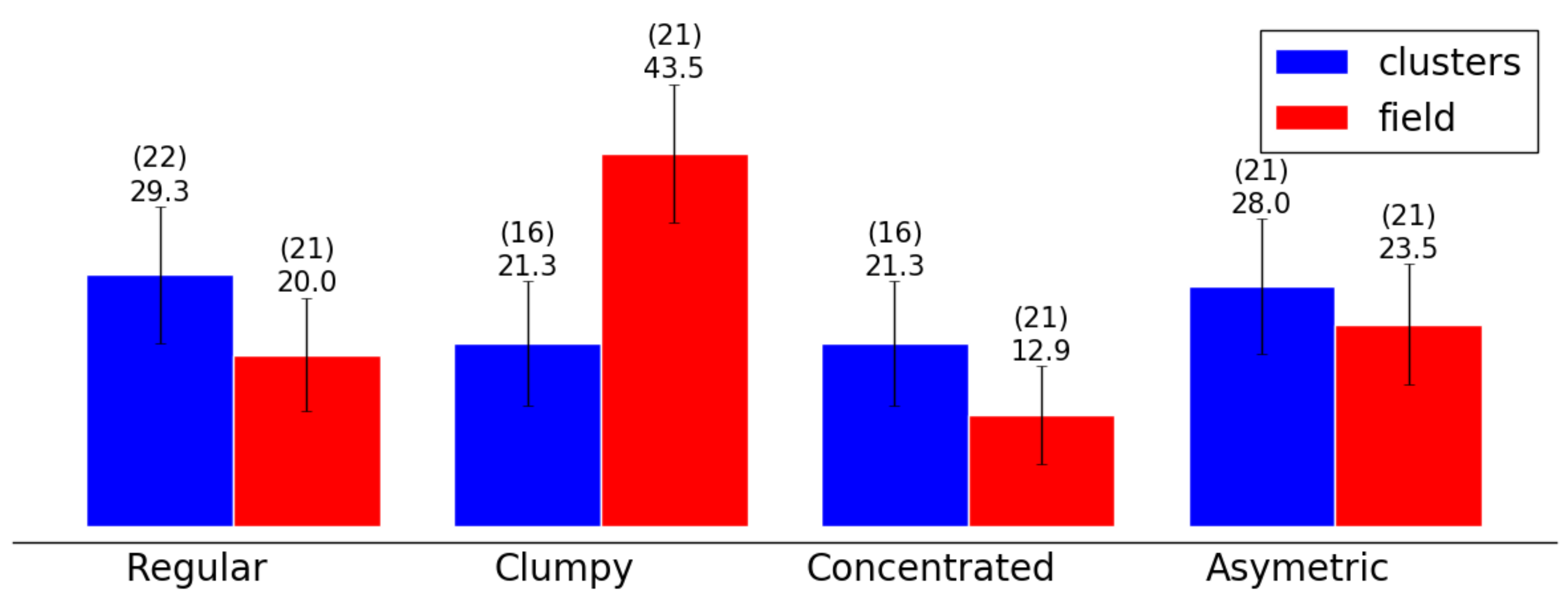}
\includegraphics[scale=0.28]{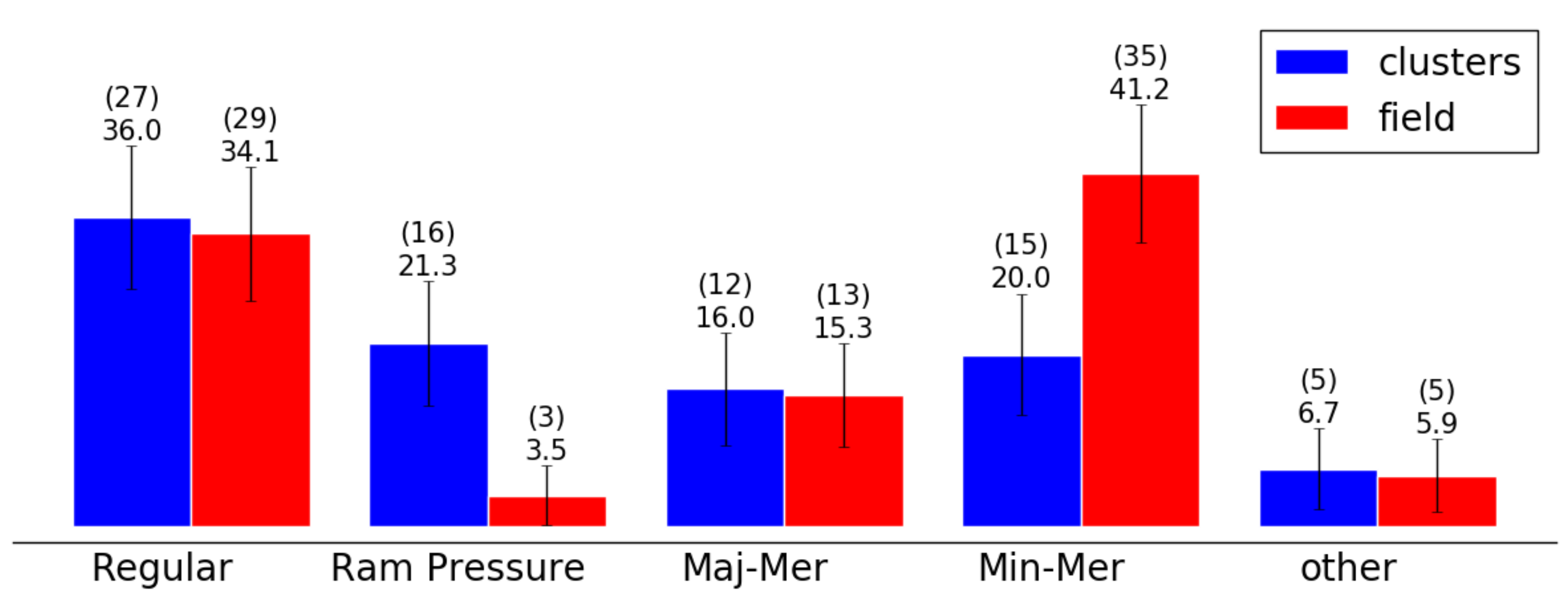}
\caption{\Ha Morphology (upper panel) and proposed main process  responsible for such 
morphology (bottom panel) percentages for \Ha emitters in clusters (blue) and in the field 
(red). Numbers represent 
percentages, error bars represent 1$\sigma$ binomial errors \citep{gehrels86}.  Numbers in brackets give the number of objects in each class.
All morphologies and processes are  well 
represented, consistent with a diversity of mechanisms being responsible for the morphology of \Ha in these systems.
\label{fig:morphs_emitters}}
\end{figure}

\subsection{The diversity of the  \Ha morphologies}
\begin{figure*}[!t]
\centering
\includegraphics[scale=0.49]{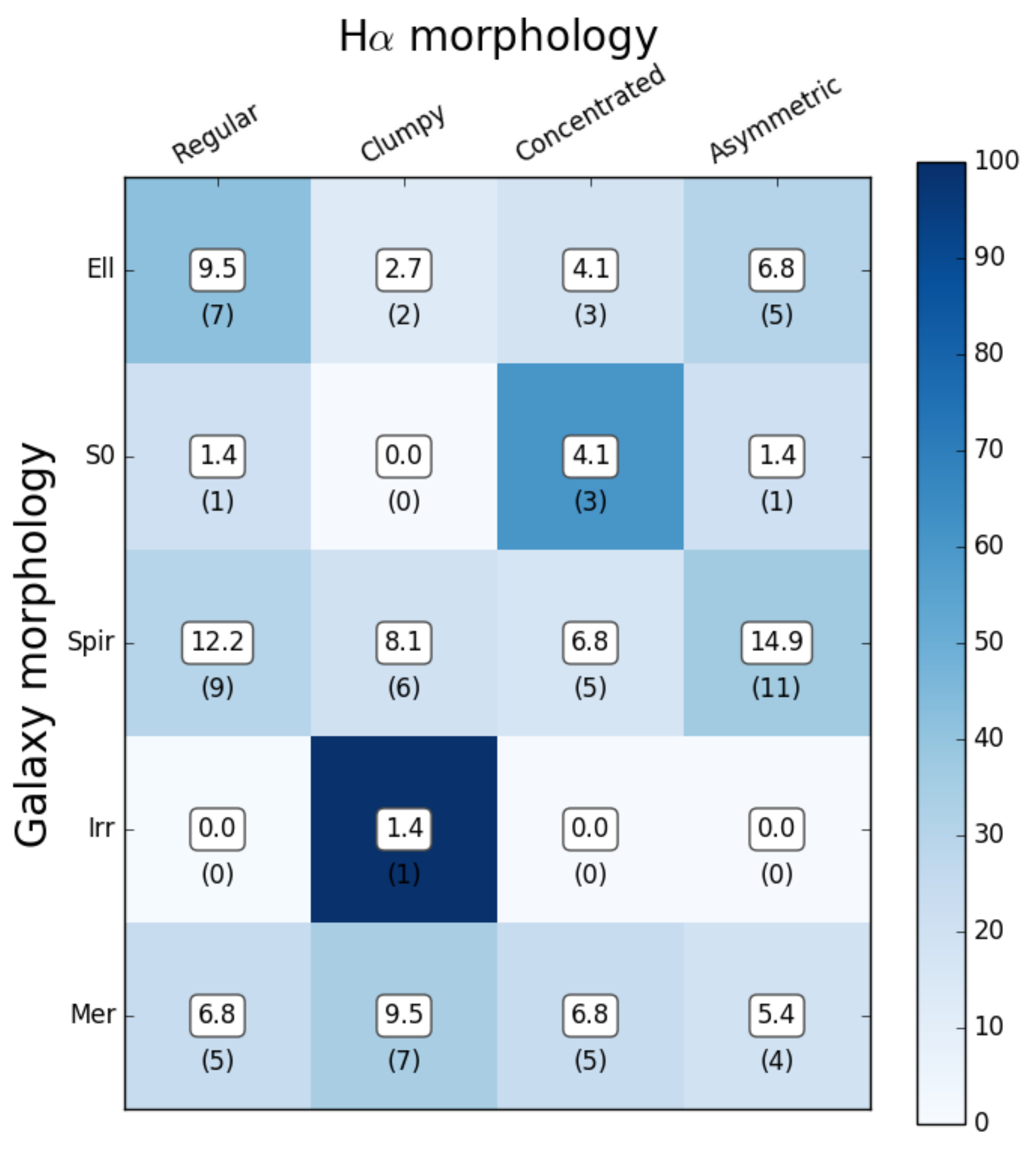}
\includegraphics[scale=0.49]{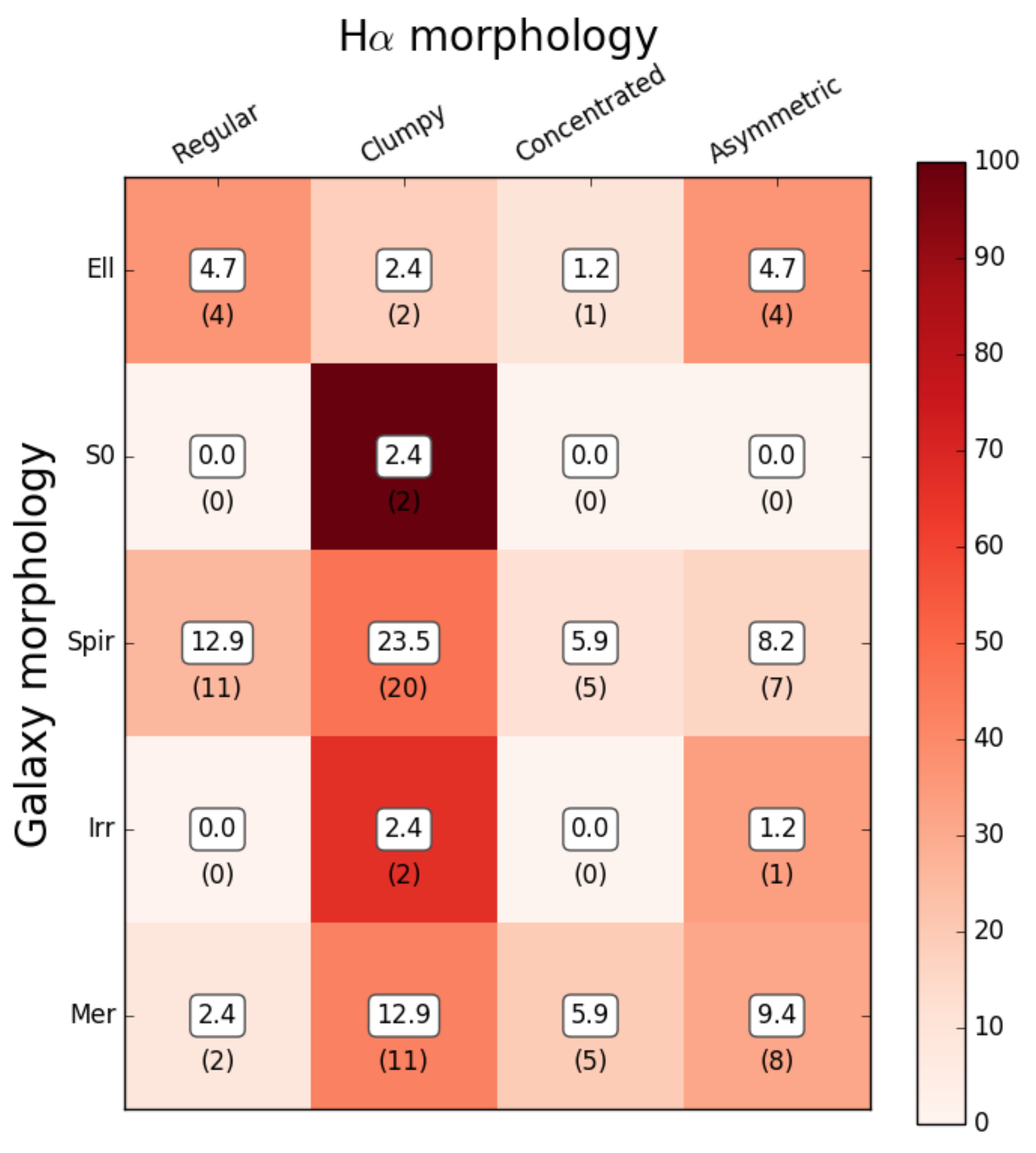}
\caption{Percentages of galaxies of a given broad-band morphology and \Ha morphology in clusters (
left panel) and in the field (right panel). The shown numbers represent the fraction of objects of a given \Ha-galaxy morphology combination, i.e. the sum to 100.   Numbers in brackets give the number of objects in each class. The color coding in the rows represents the percentages of a given \Ha morphology for each morphological class separately.
For instance, from the color bar, we see that $\sim 40\%$ of spirals in the clusters have asymmetric \Ha morphologies and that this combination makes up 14.7\% of the full sample. 
The same class of objects make up 8.2\% of the field galaxies, but only represent $\sim10\%$ of all spirals, as clumpy \Ha dominates the spiral field sample.
\label{fig:morph_Hamorph}}
\end{figure*}

\begin{figure*}[!t]
\centering
\includegraphics[scale=0.49]{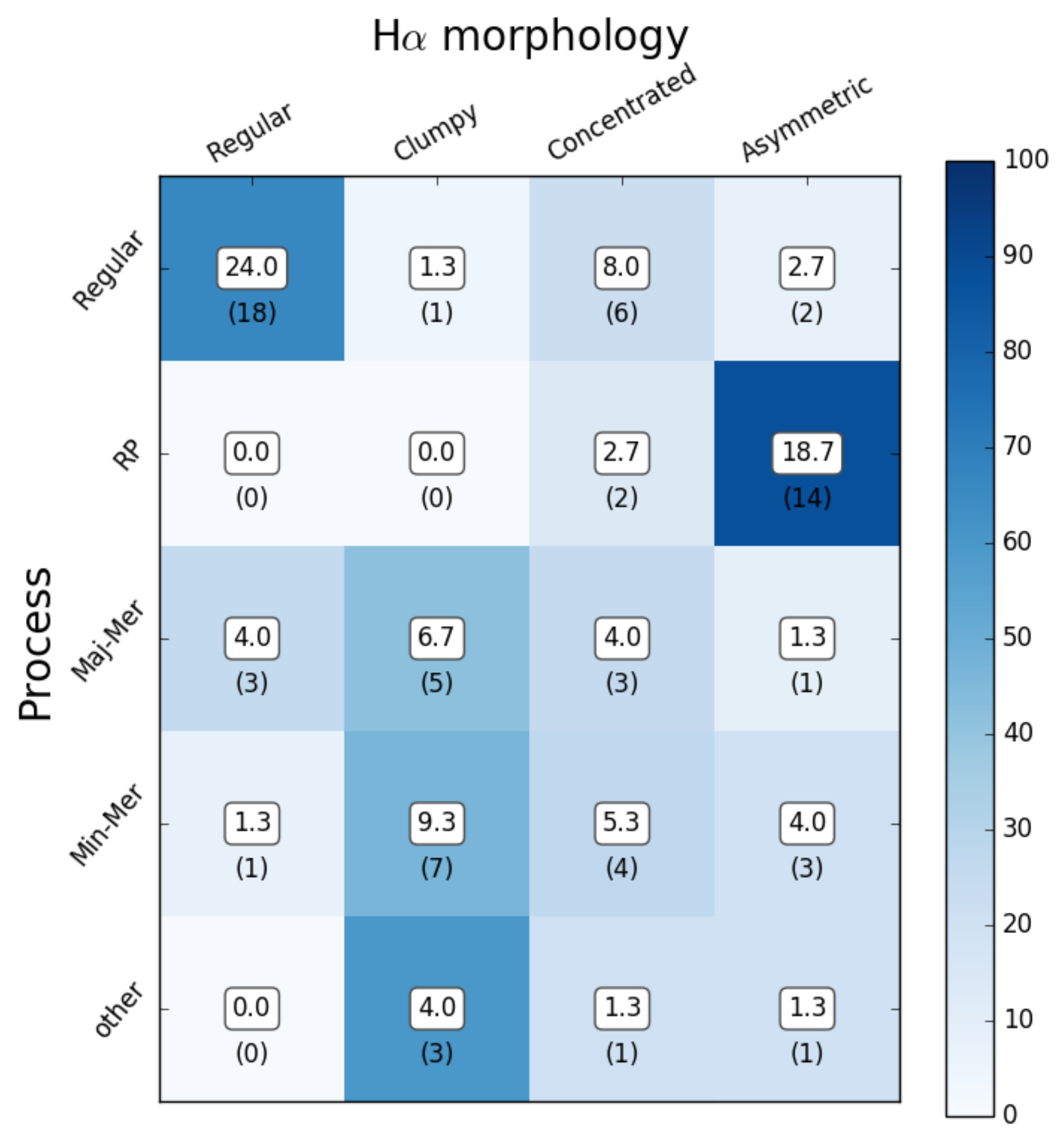}
\includegraphics[scale=0.49]{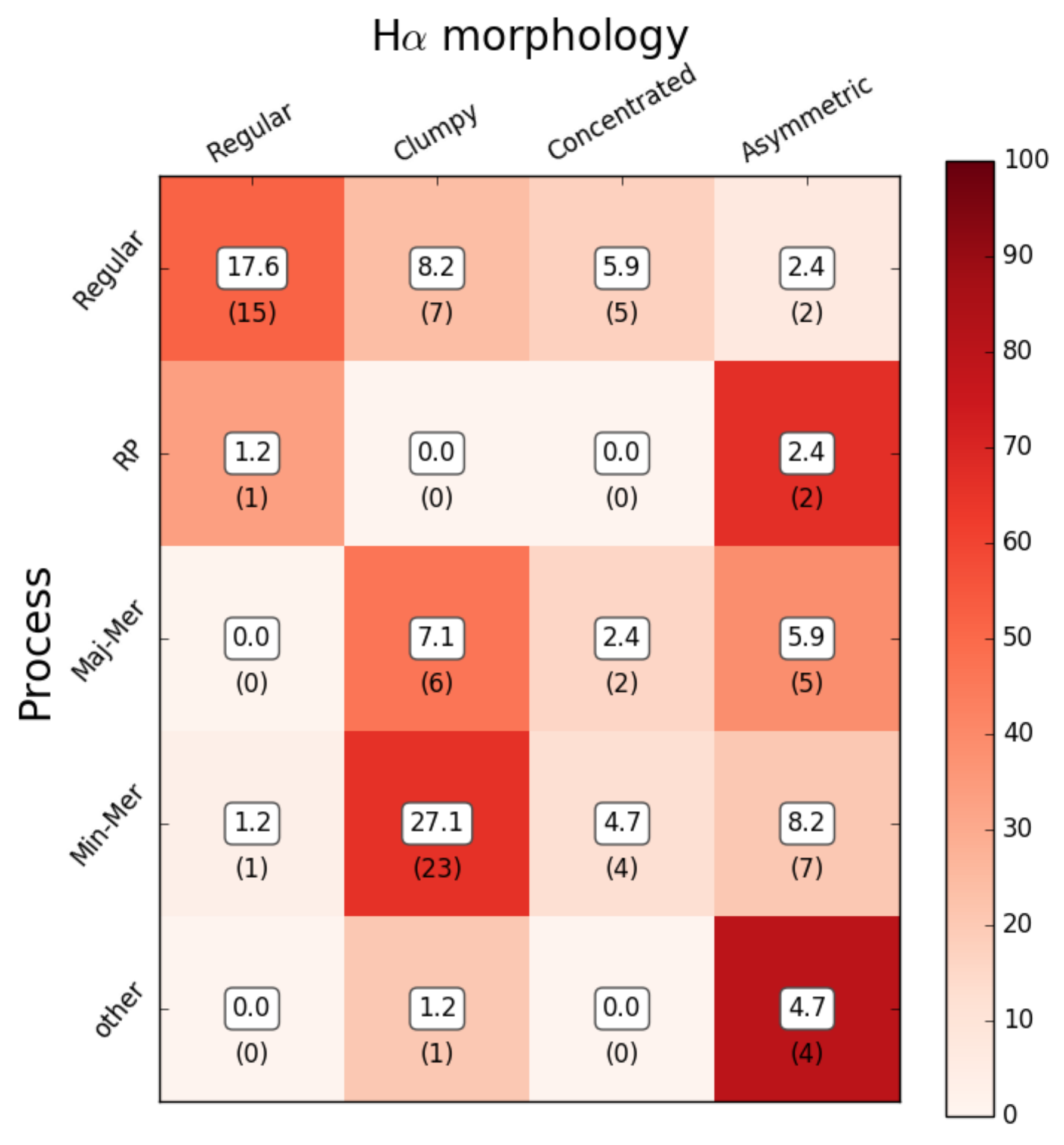}
\caption{
Figures similar to Figure \ref{fig:morph_Hamorph}, but comparing \Ha morphology to the responsible processes. 
Again the left panel represents the cluster sample and the right panel the field sample.
\label{fig:Hamorph_proc}}
\end{figure*}

Having established that \Ha emitters are a peculiar subsample of the entire galaxy population,  we 
focus on the properties of these galaxies, and compare the broad-band morphology
 to the morphology of the \Ha line. We will pay particular attention to spiral galaxies, which 
represent the most common class of emitters, both in clusters and in the field.  

Figure \ref{fig:morphs_emitters} presents the incidence of each \Ha
morphological class (upper panel) and process likely to be responsible
for the observed \Ha morphology (bottom panel), in both environments
separately.  We stress that that classification of the process label
should be interpreted as qualitative and the adopted label is
short-hand for what it is likely to be complex and not-necessarily
unique mechanisms.

We find that there is no dominant morphology for the \Ha emission, and
all the classes are almost uniformly populated.  The only exception is
the predominance of the
clumpy \Ha morphology in the field ($44\pm8\%$). 
Overall, this result is
consistent with a diversity of factors impacting galaxy star forming
regions and therefore on the \Ha morphology.  The other percentages agree within the errors,
but in clusters there seems to be a larger number of galaxies with
regular (30$\pm$8\% vs 20$\pm$7\% in the field), asymmetric (28$\pm$8\% vs 23$\pm$7\%
in the field) and more concentrated (21$\pm$7\% vs 13$\pm$6\% in the field)
\Ha morphology.

Different physical processes might be acting on the different galaxies and the bottom panel of Fig. 
\ref{fig:morphs_emitters} shows that in both environments star formation seems to proceed regularly and more or less undisturbed  on $\sim35\%$ of the 
galaxies and none of the proposed process seems to be the responsible
for the \Ha morphology for $\sim 6\%$ the galaxies.  The most common
classification in clusters seems to be ram pressure stripping, with
$21\pm7\%$ of all galaxies falling in this category. In contrast, in the
field, only $4\pm 3$\% of galaxies show signs of possible stripping. Major mergers
seem to have a similar incidence in both environments ($15\pm7$\%),
while minor mergers are by far more common in the field (41$\pm8$\%) than
in clusters (20$\pm$7\%).

We note that fractions do not depend on stellar mass: the mass distributions of the different \Ha morphological classes
are very comparable, as are also those of galaxies classified as undergoing different physical processes. 

It is interesting to note how the fraction of galaxies classified as
minor mergers in the two environments is the same as the fraction of
galaxies with a clumpy \Ha morphology, suggesting a physical link
between minor merging and \Ha morphology.  However, to understand the
connection between process label and \Ha appearance we must inspect
the galaxy properties simultaneously, as done in Figures
\ref{fig:morph_Hamorph} and \ref{fig:Hamorph_proc}.

Figure \ref{fig:morph_Hamorph} presents the incidence of galaxies of a
given broad-band morphology and given \Ha morphology, for the two
environments separately.  Numbers refer to the percentages with
respect to the total population of \Ha emitters in a given
environment, colors provide the information on the percentages of a
given \Ha morphology for each morphological class separately; e.g. the
fraction of cluster galaxies with regular \Ha morphology among
elliptical galaxies. We warn the reader that when splitting the samples in many subgroups, 
uncertainties become important and prevent us from reaching solid conclusions.

Some trends emerge and there are  interesting differences between the two environments. 
In clusters, elliptical galaxies most likely present a regular \Ha morphology (9$\pm$5\% of all galaxies, $\sim40\%$ of elliptical galaxies), but in a 
few cases they present also the other \Ha morphologies. In the field, ellipticals have the same chance of having a regular 
or asymmetric \Ha morphology (5$\pm$4\% of the total, $\sim40\%$ of all ellipticals). 
Besides being a small fraction of the total sample, S0s in clusters most likely have concentrated \Ha, while in the field they have a clumpy \Ha distribution. 
Spiral galaxies have all kinds of \Ha morphologies, but in clusters they have mostly asymmetric \Ha (15$\pm6$\%), while in the field by far clumpy \Ha (24$\pm$7\%). 
As expected, in both environments irregular galaxies do not have regular or concentrated \Ha disk, but it is either clumpy or asymmetric. 
Finally, merging systems have all kinds of \Ha morphologies, with a preference for clumpy morphology in the field (13$\pm$6\%), followed by the asymmetric one (9$\pm$5\%). 
It is interesting to note though that, especially in clusters merging systems can maintain a regular \Ha morphology ($\sim$30\% of all mergers).

Similarly, Figure \ref{fig:Hamorph_proc} presents the incidence of galaxies of a  given \Ha morphology classified as experiencing a given process, for the two environments separately. 
Again numbers refer to the percentages with respect to the total population of \Ha emitters in a given environment, colors provide the information on 
the percentages of a given \Ha morphology for each physical process separately; e.g. the fraction of galaxies with asymmetric \Ha morphology 
among galaxies that are classified as likely experiencing ram pressure stripping.

Overall, 25$\pm$7\% of galaxies in clusters present a regular \Ha morphology and have regular star formation. This percentage is 18$\pm$6\% in the field. 
In both  environments, regular label is also associated with a concentrated \Ha morphology, perhaps suggestive of a nuclear starburst without major outside disturbances, even though this case is not very common ($\sim$6-8\% of all galaxies, $<30$\% of galaxies classified as regular). In clusters, the second most populated class is that of galaxies with asymmetric \Ha classified as affected by ram pressure stripping (18$\pm$7\% of the total population and $>$90\% of galaxies  likely experiencing this process). In a few cases ram pressure seems  to produce a concentrated \Ha morphology (3$\pm$3\%). In this case, we could be witnessing the final stage of a stripping event, when the galaxy is left with a truncated \Ha disk. As expected, ram pressure is not very effective in the field (<4\%). In clusters, major and minor mergers seem to produce all the different \Ha morphologies,  with a light preference for the clumpy class. In contrast, in the field, 27$\pm$7\% of the total population have been classified as having clumpy \Ha morphologies undergoing a minor merger and almost all minor mergers produce a clumpy \Ha morphology (>90\%). In the field, major mergers are less common events, and  produce both clumpy (7$\pm4$\%) and asymmetric  (6$\pm4$\%) \Ha morphologies. 
Unidentified processes produce mainly clumpy \Ha morphologies in clusters ($4\pm4$\%), and asymmetric \Ha distribution in the field (5$\pm$4\%). 

Unfortunately, due to small number statistic, we can not look for trends taking into account simultaneously all the three parameters (broad-band morphology, \Ha morphology, and process).

From the analysis above, based on our qualitative classification, it appears evident that our sample includes a large variety of objects with different 
properties and that different processes, representative of the different environments, can indeed produce similar features.  Interestingly, 
in many cases the effect of the environment is hardly detectable, even with these sensitive tools and prior knowledge. This is consistent with the idea that environment-specific processes play a secondary role in galaxy evolution, 
at least for galaxies with $M_\ast>10^8 M_\sun$.
We now proceed by discussing in detail the most interesting classes. 

\subsection{Classes of peculiar objects}
\begin{figure*}[!t]
\centering
\includegraphics[scale=0.14]{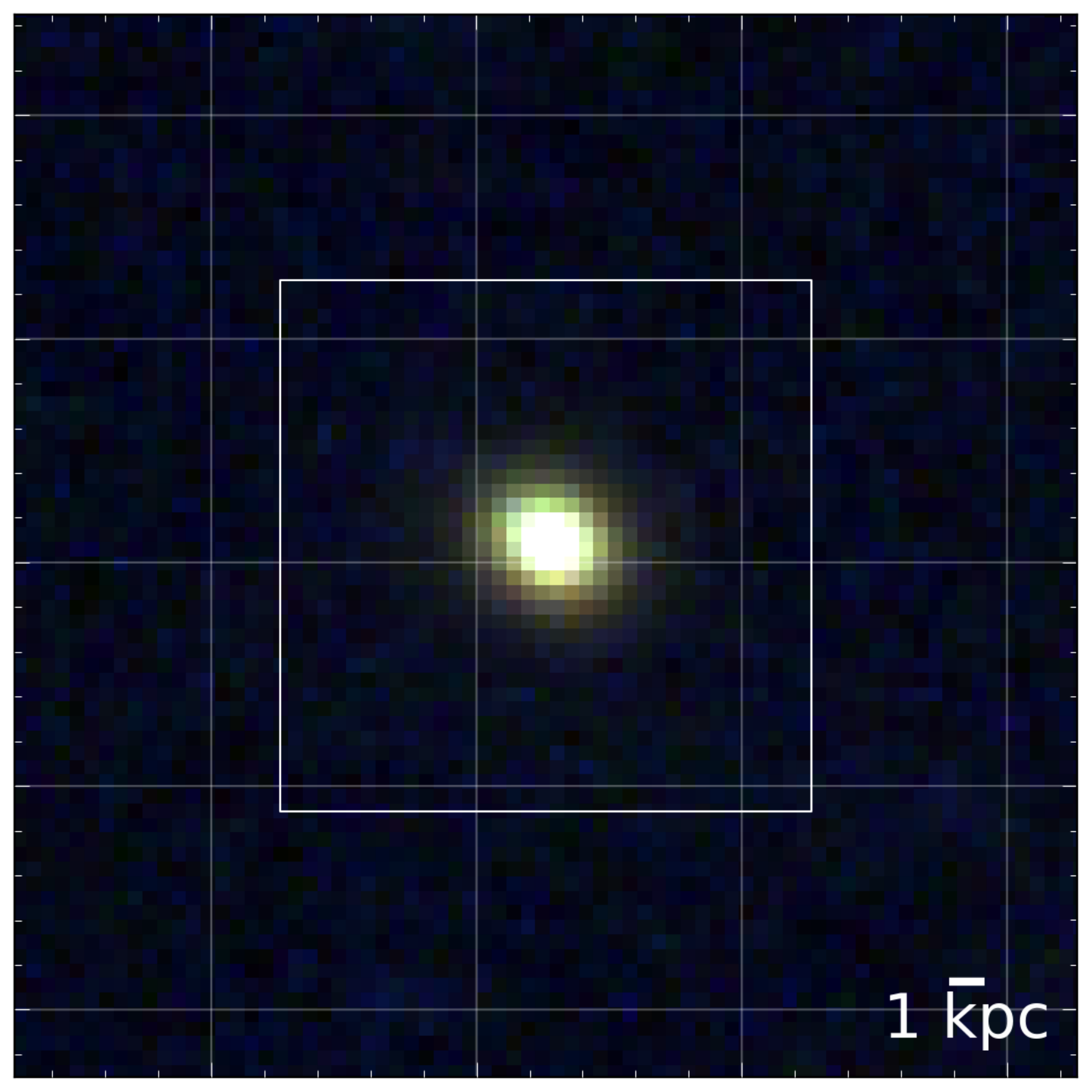}
\includegraphics[scale=0.14]{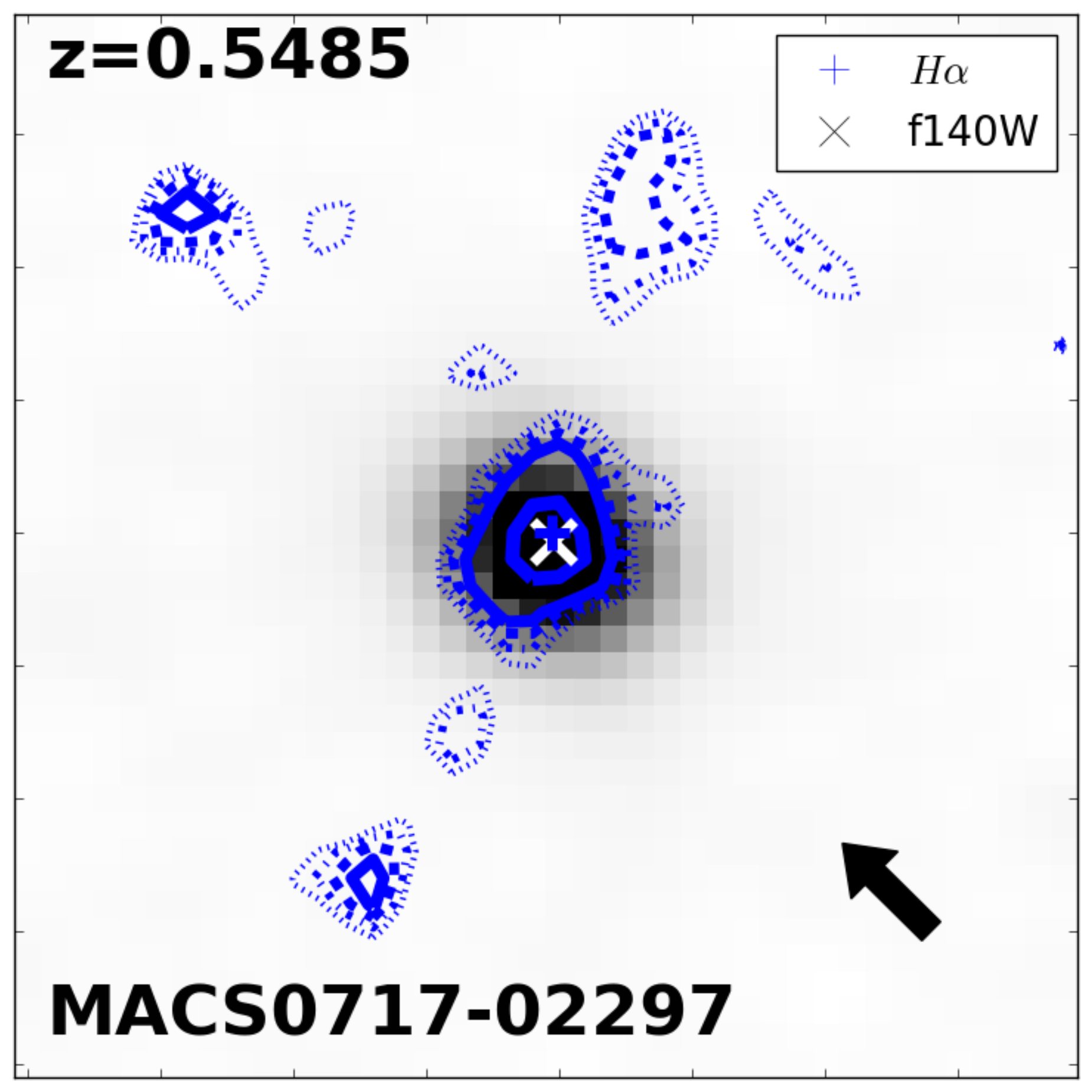}
\includegraphics[scale=0.14]{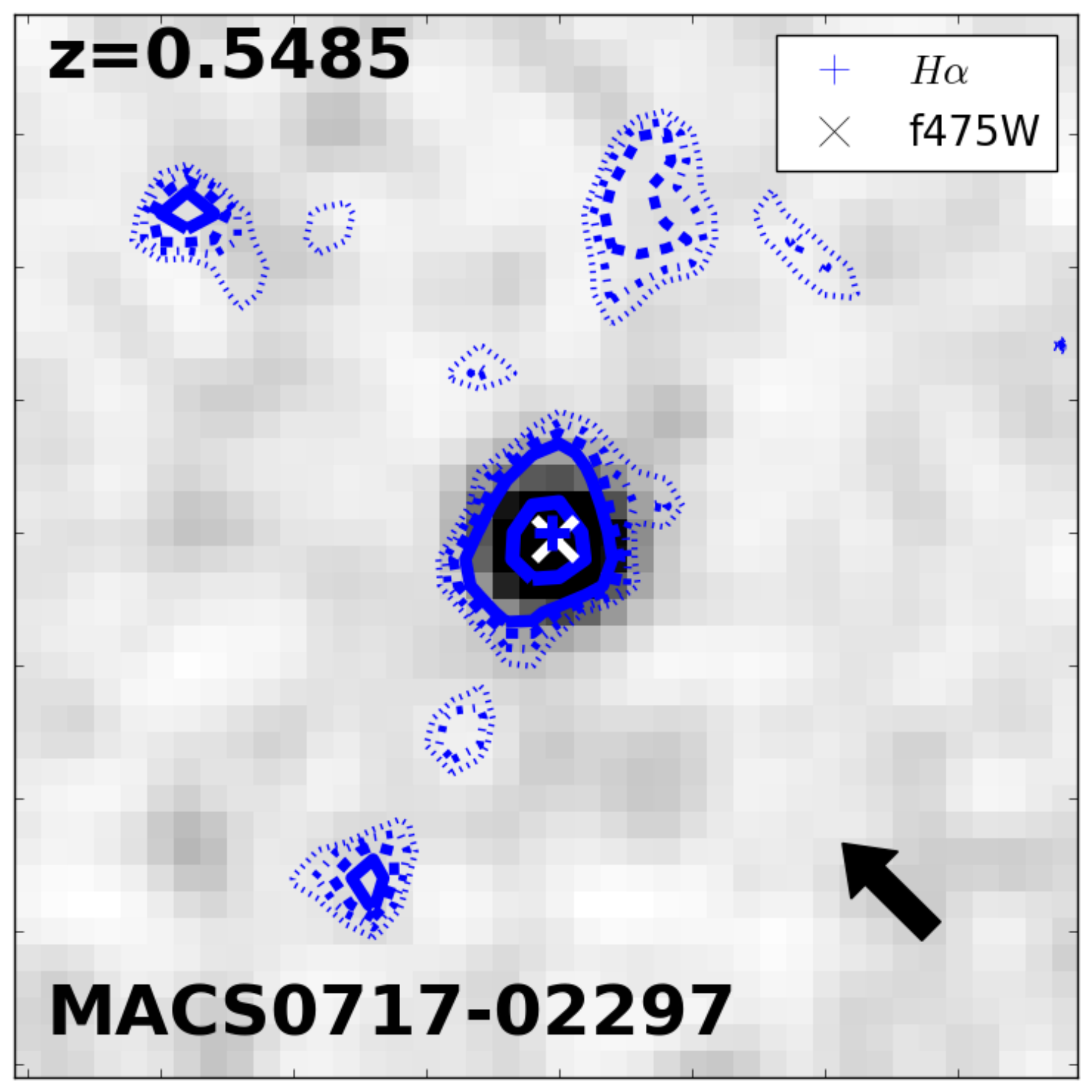}
\includegraphics[scale=0.14]{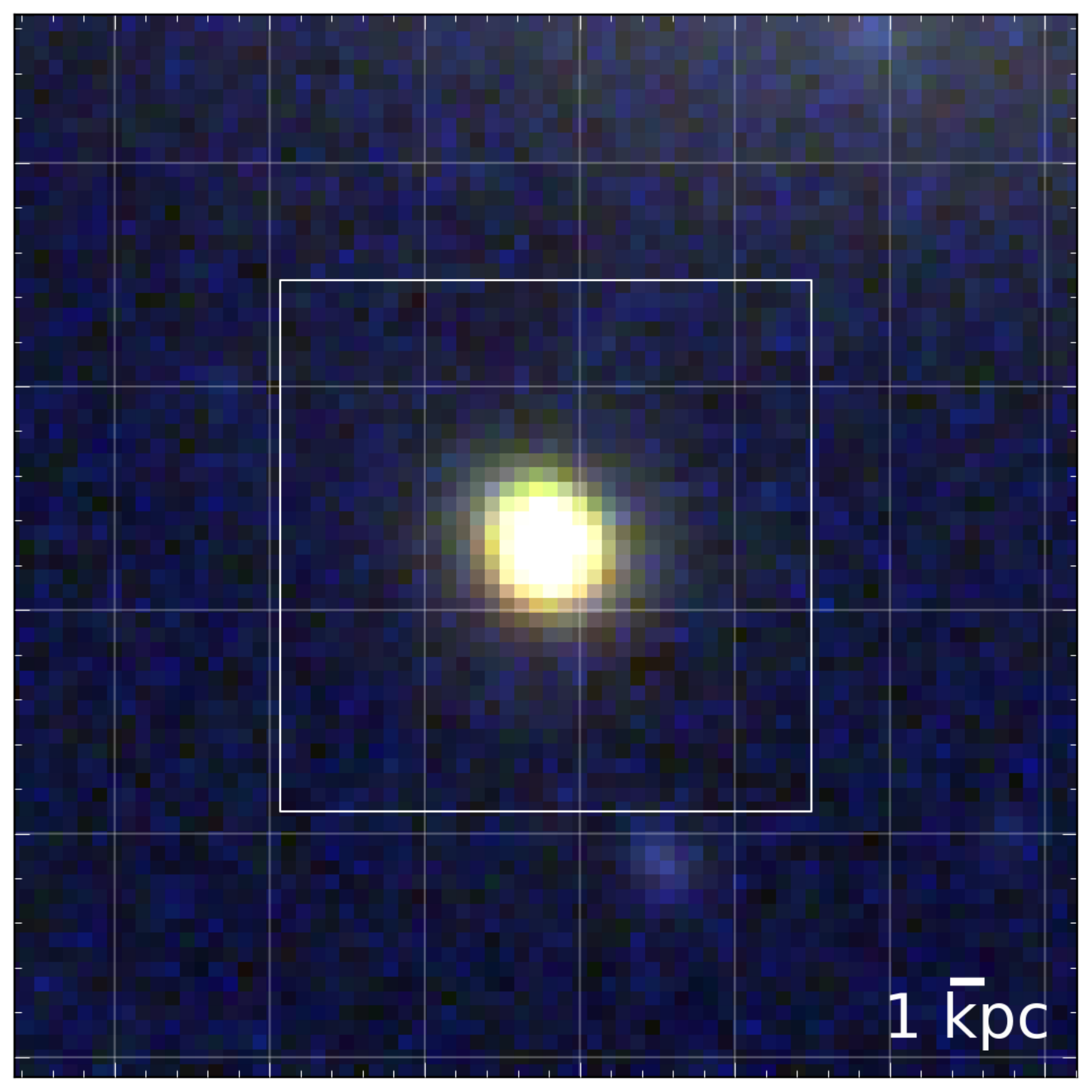}
\includegraphics[scale=0.14]{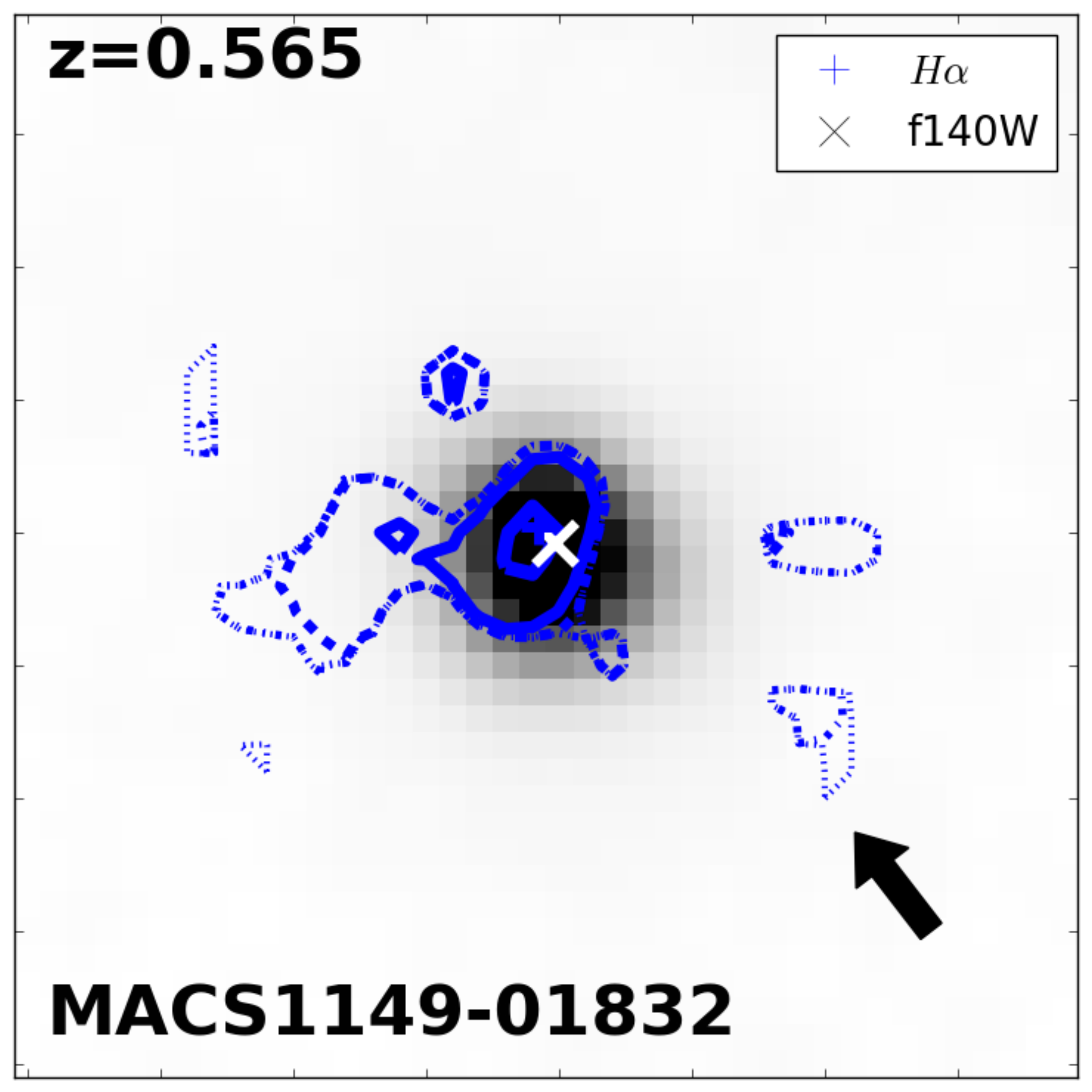}
\includegraphics[scale=0.14]{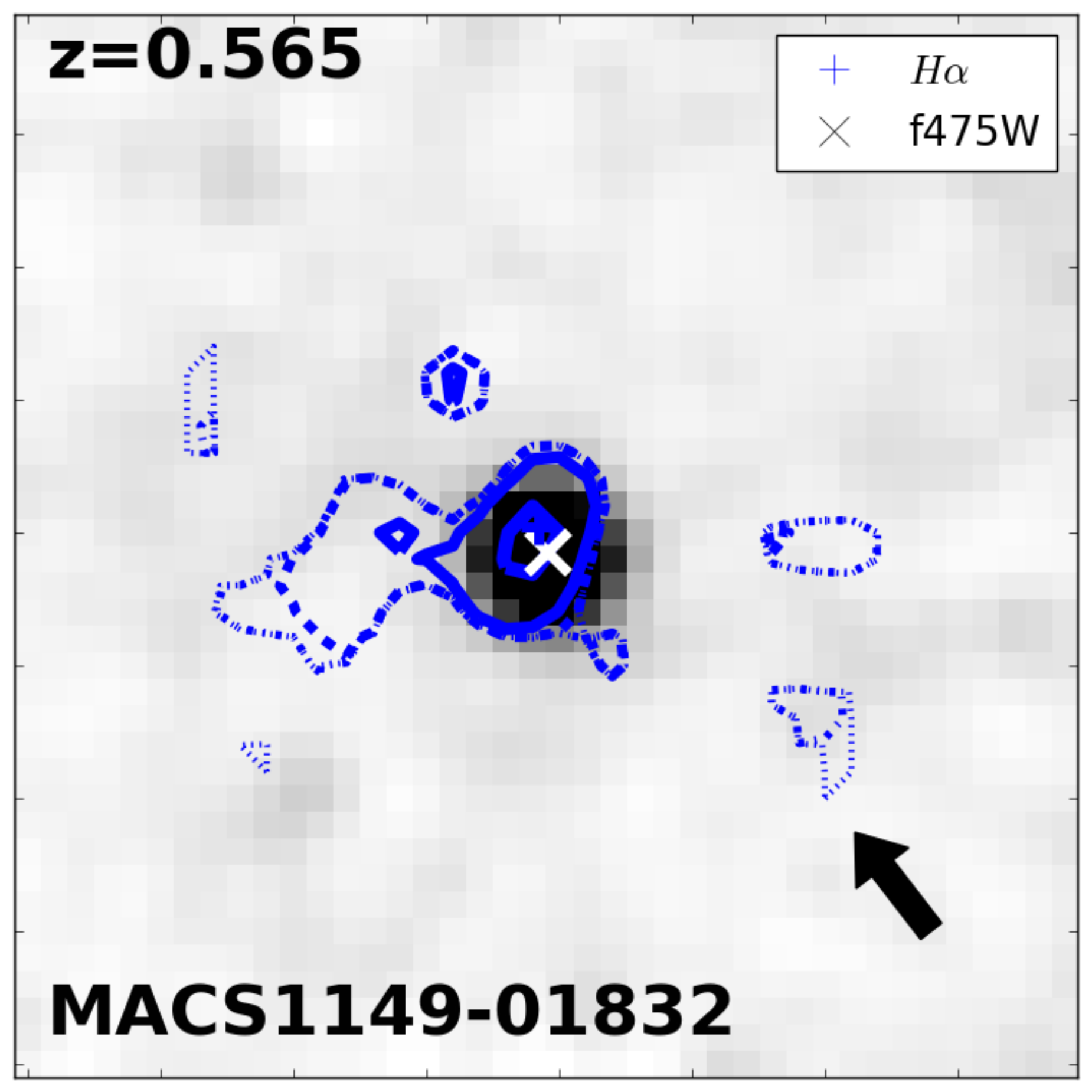}
\includegraphics[scale=0.14]{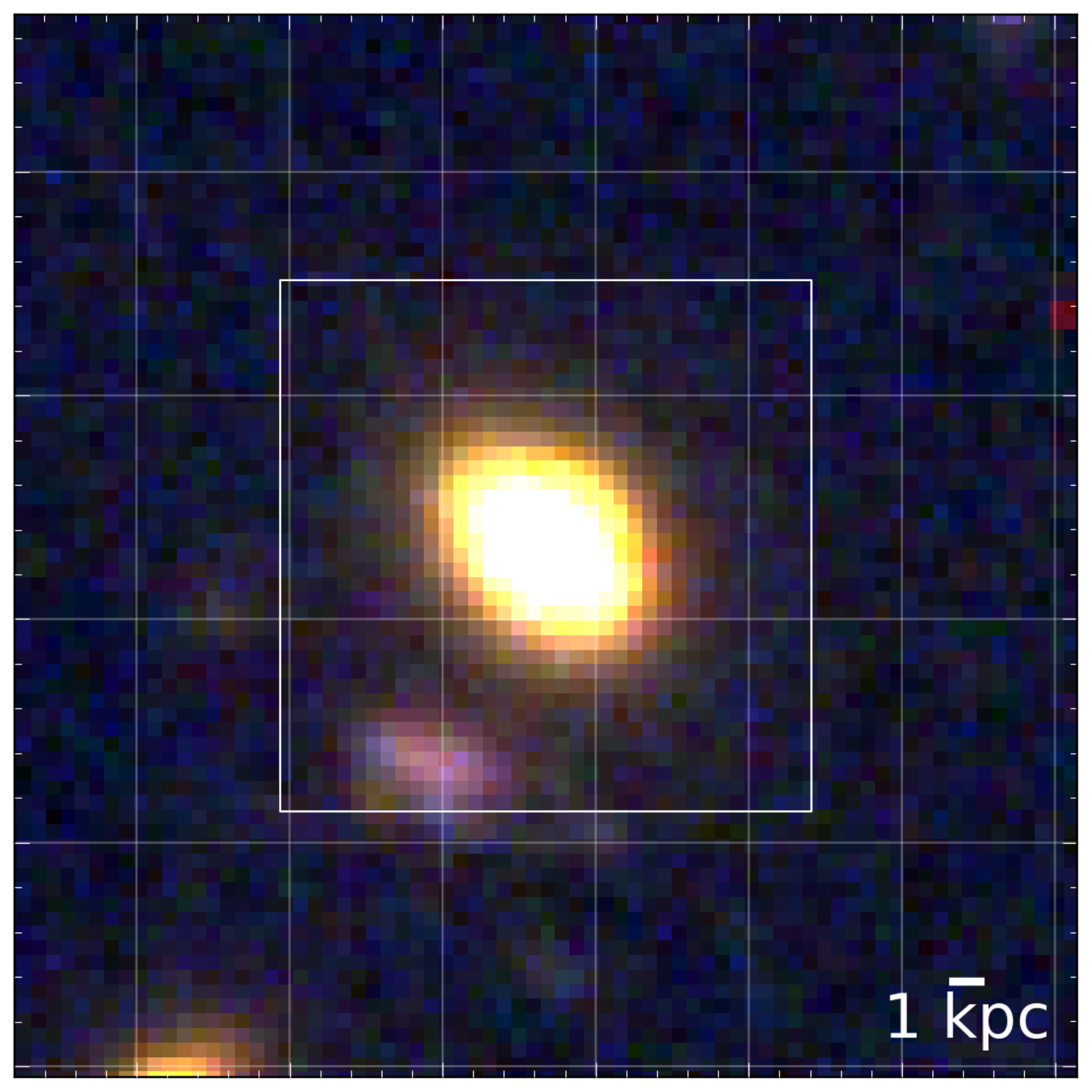}
\includegraphics[scale=0.14]{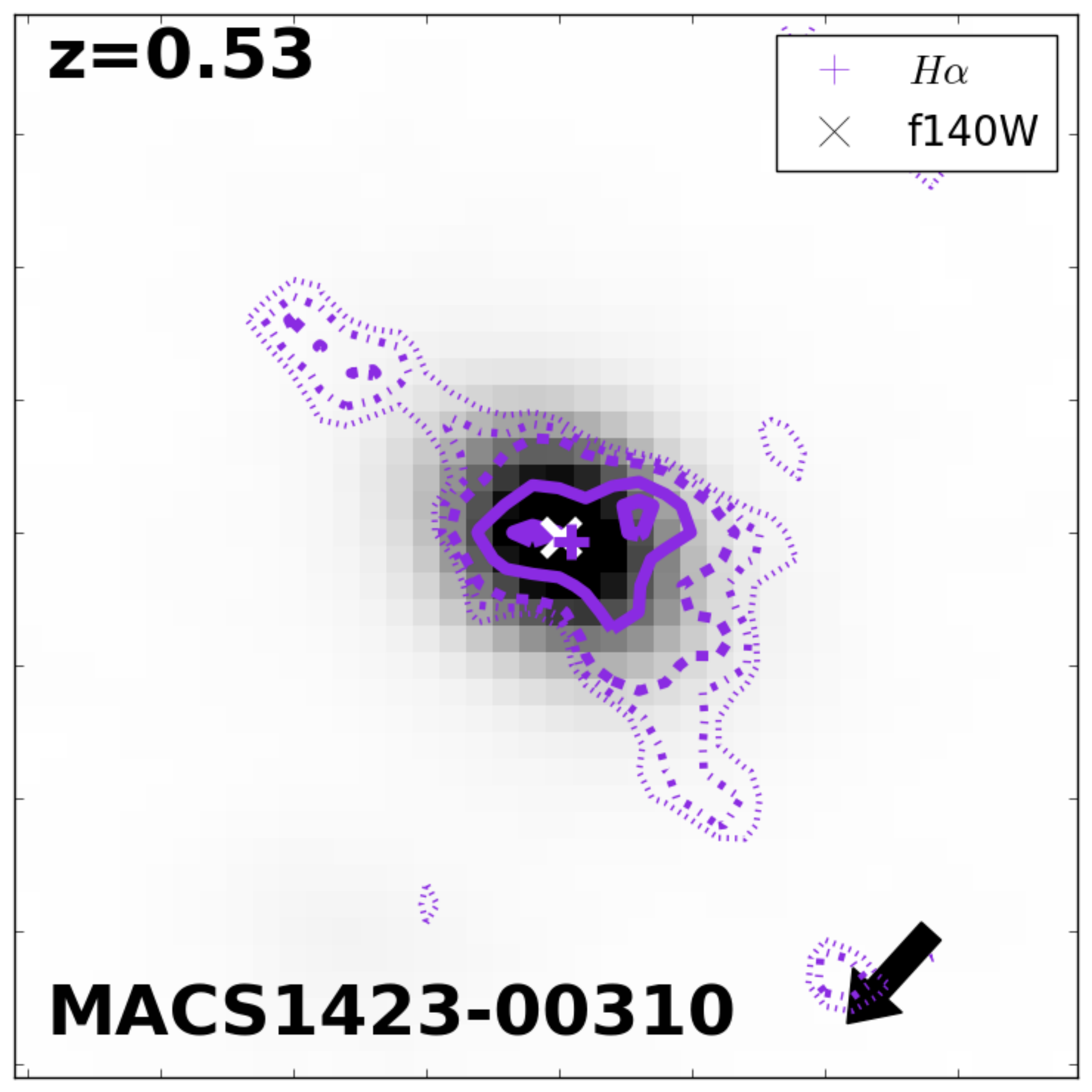}
\includegraphics[scale=0.14]{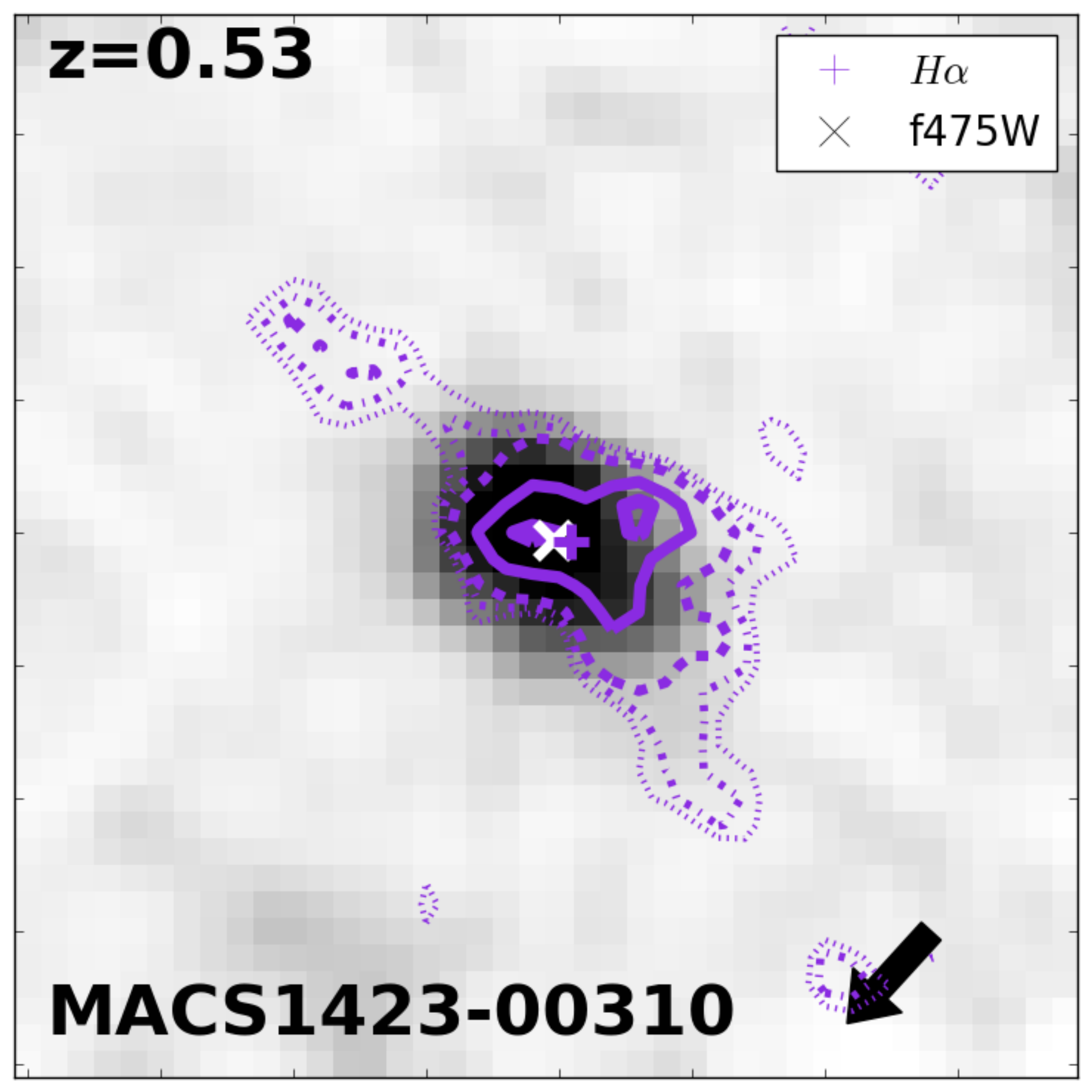}
\includegraphics[scale=0.14]{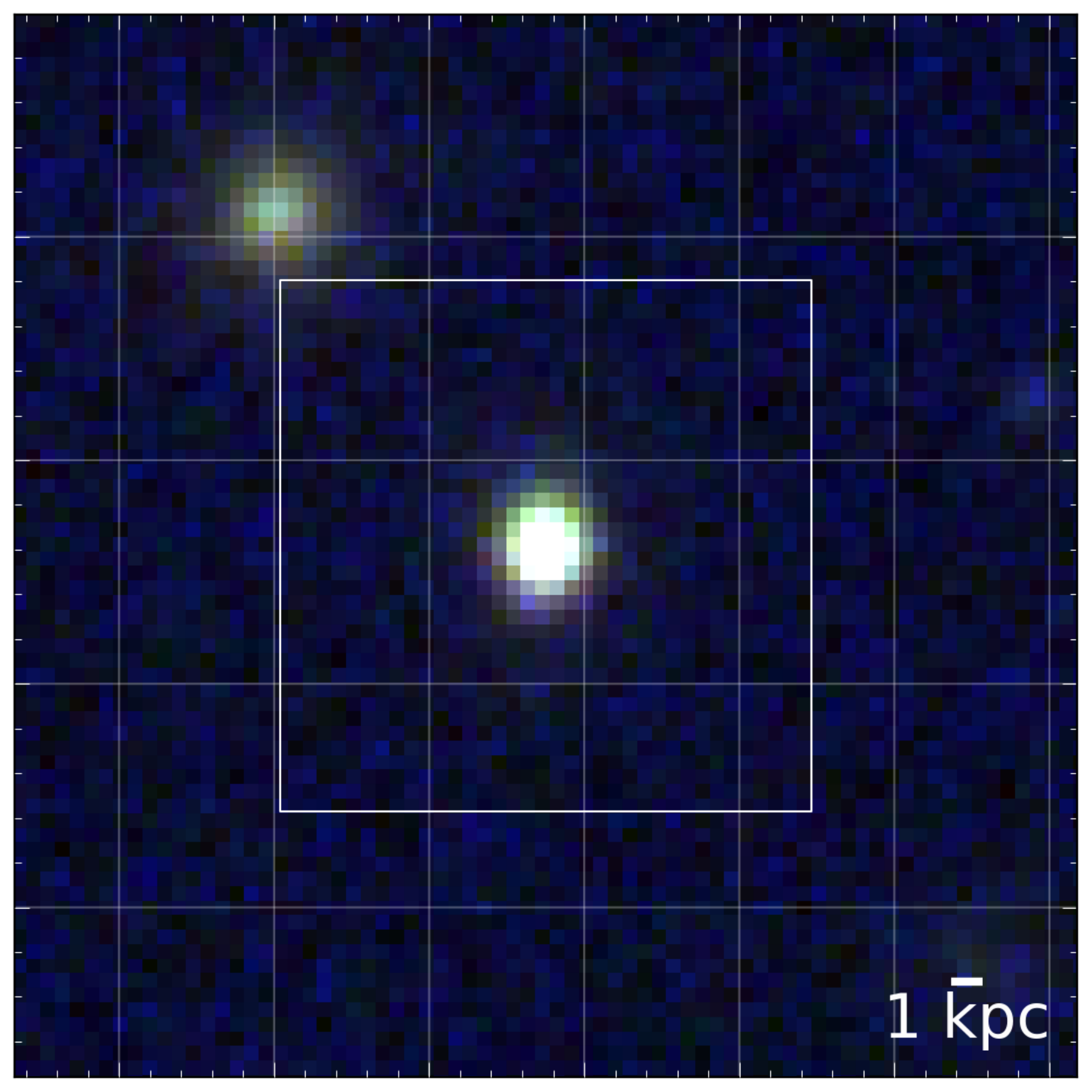}
\includegraphics[scale=0.14]{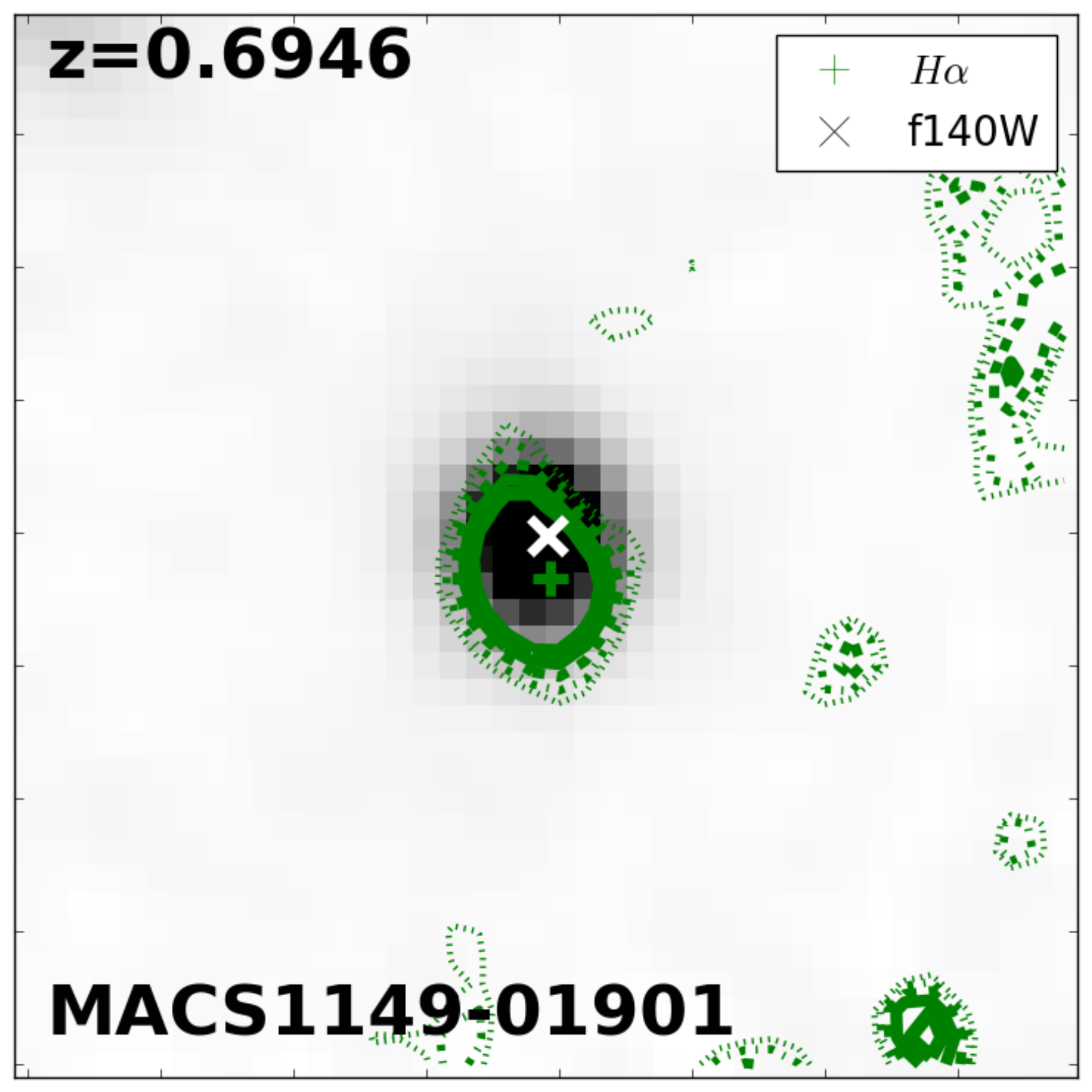}
\includegraphics[scale=0.14]{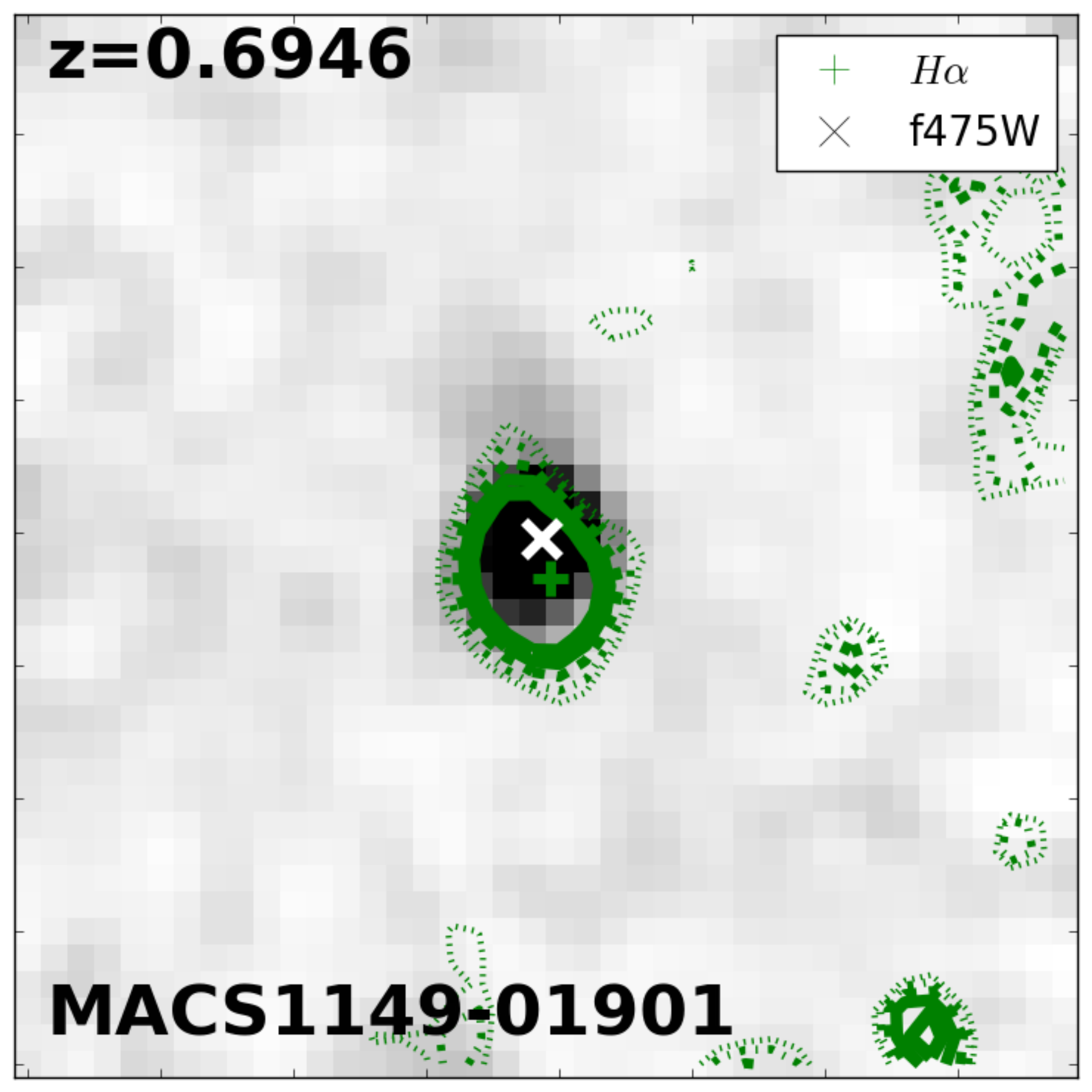}
\caption{Example of galaxies with elliptical broad-band morphology and detected \Ha emission. For each galaxy, the 
left panel shows the color composite image of the galaxy based on the 
CLASH \citep{postman12} or HFF \citep{lotz16} HST data. The blue channel is composed by the F435W, F475W, F555W, 
F606W, and F625W filters, the green by the F775W, F814W, F850lp, F105W, 
F110W filters, and the red by the F125W, F140W, F160W filters. The central panel 
shows the \Ha map superimposed on the image of the galaxy in the F140W filter 
and the right panel shows the \Ha map superimposed on the image of the galaxy in the F475W filter. 
Contour levels represent the 35$^{th}$, 50$^{th}$, 65$^{th}$, 80$^{th}$, 95$^{th}
$ percentiles of the light distribution, respectively. Blue contours indicate that \Ha maps are obtained  
from one spectrum, purple contours indicate that \Ha maps are obtained  
from  two orthogonal spectra, and green contours indicate that \Ha maps are obtained combining both the 
G102 and G141 grisms (only for $z>0.67$). In the color composite image, 
the Field of View is twice as big as the single band images. A smoothing filter has been applied to the 
maps and an arbitrary stretch  to the images for display purposes.  For galaxies 
in clusters, arrows on the bottom right corner indicate the direction of the cluster center. The redshift of the 
galaxy is indicated on the top left corner.
 \label{ell_ha}}
\end{figure*}

\subsubsection{Elliptical galaxies with extended \Ha in emission}
\begin{figure*}
\centering
\includegraphics[scale=0.14]{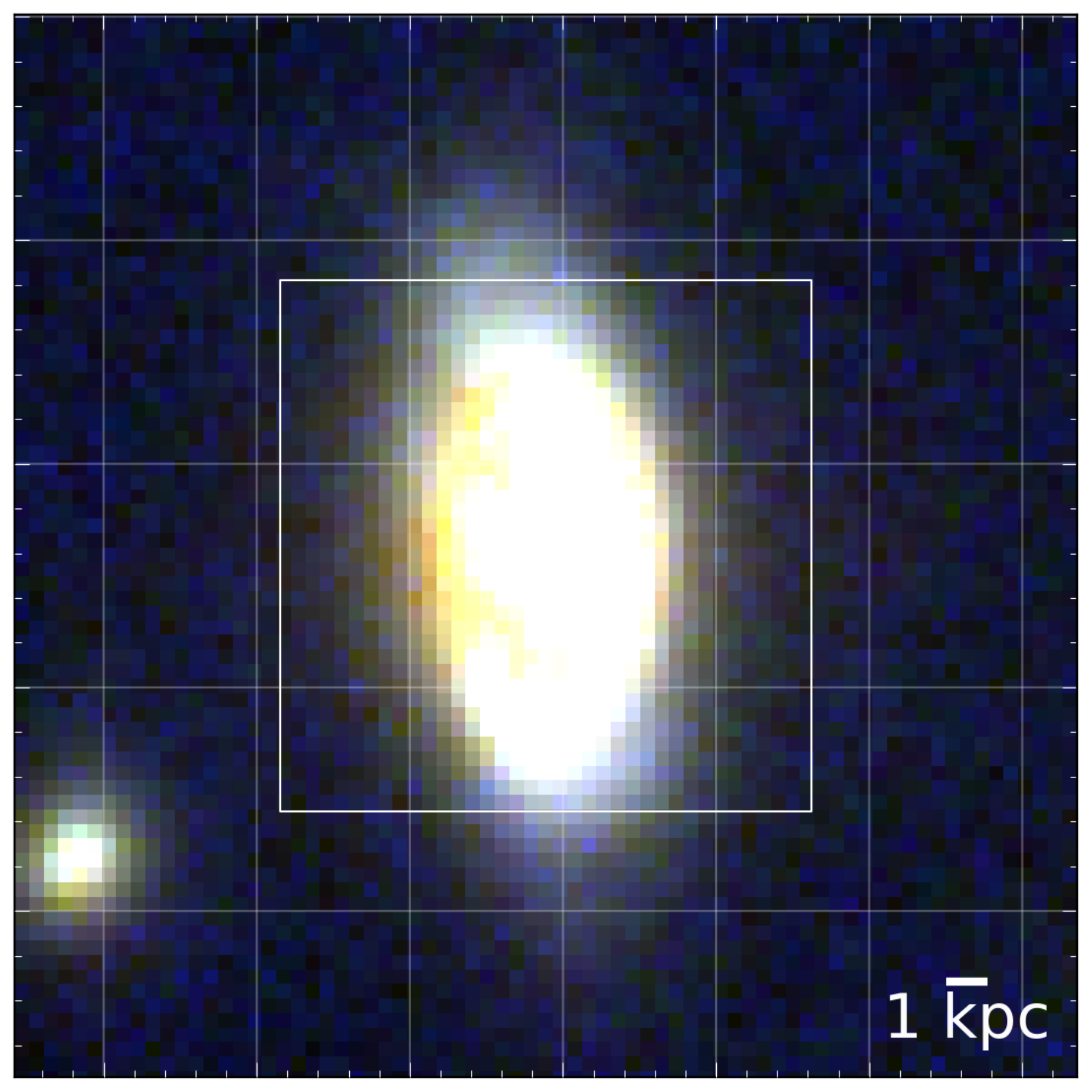}
\includegraphics[scale=0.14]{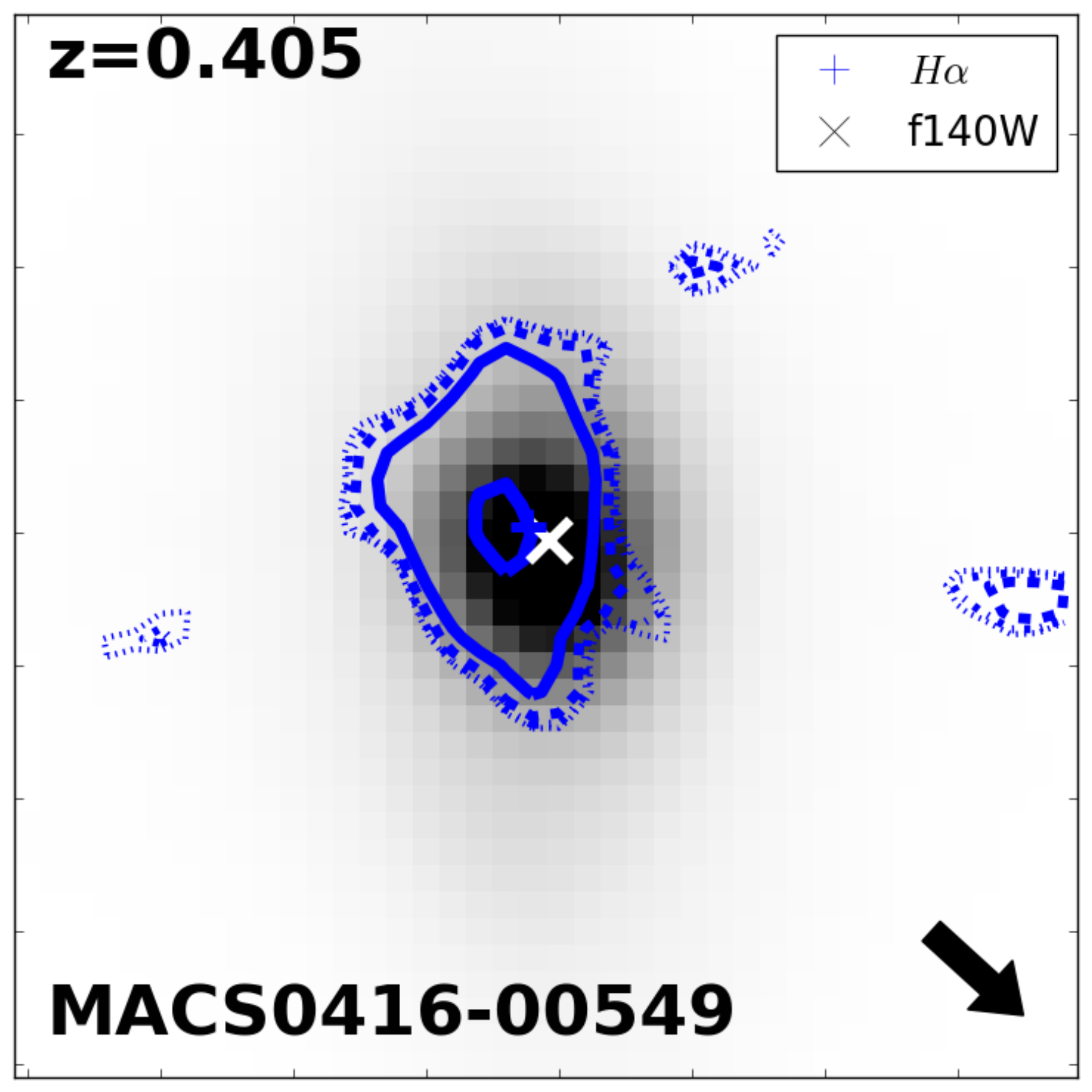}
\includegraphics[scale=0.14]{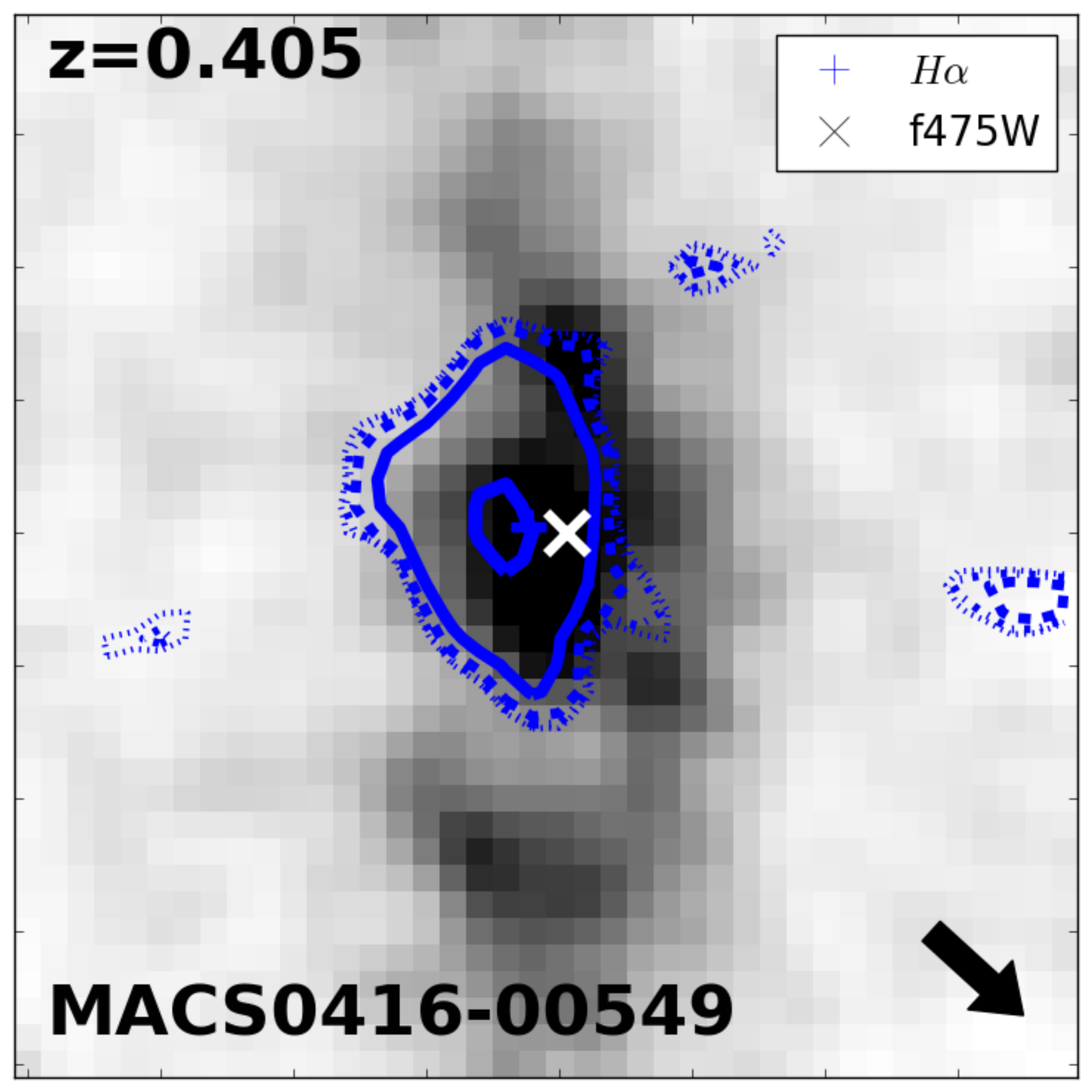}
\includegraphics[scale=0.14]{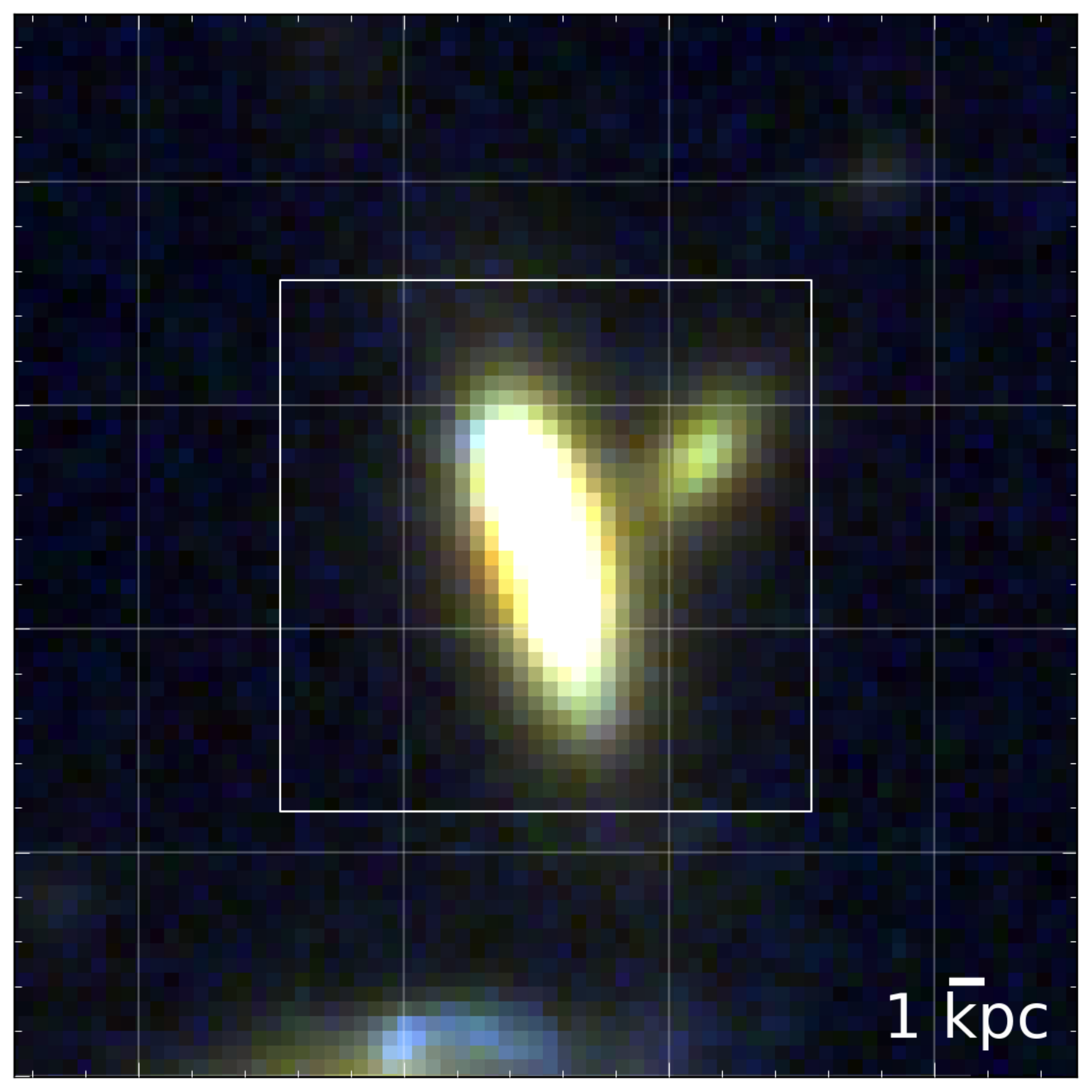}
\includegraphics[scale=0.14]{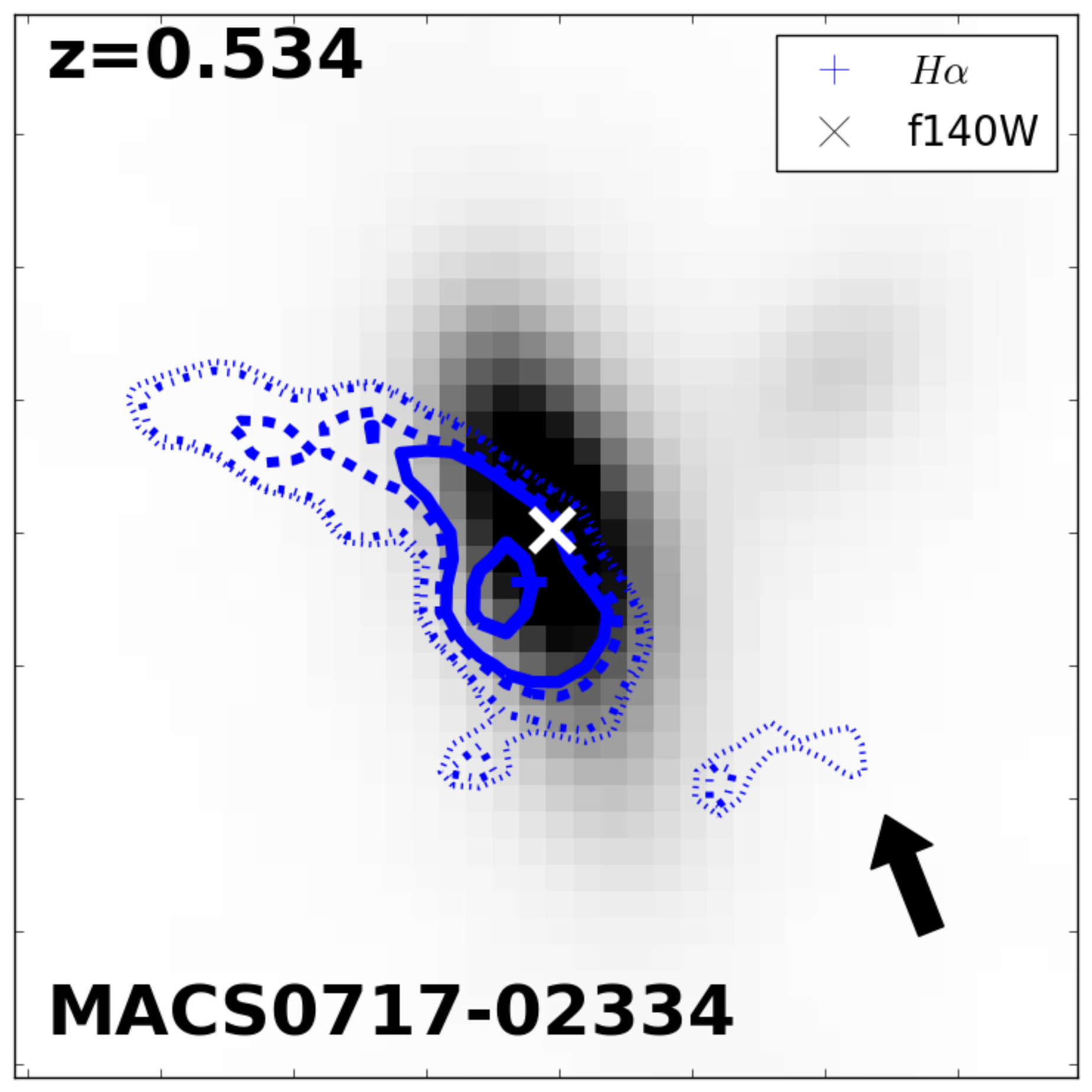}
\includegraphics[scale=0.14]{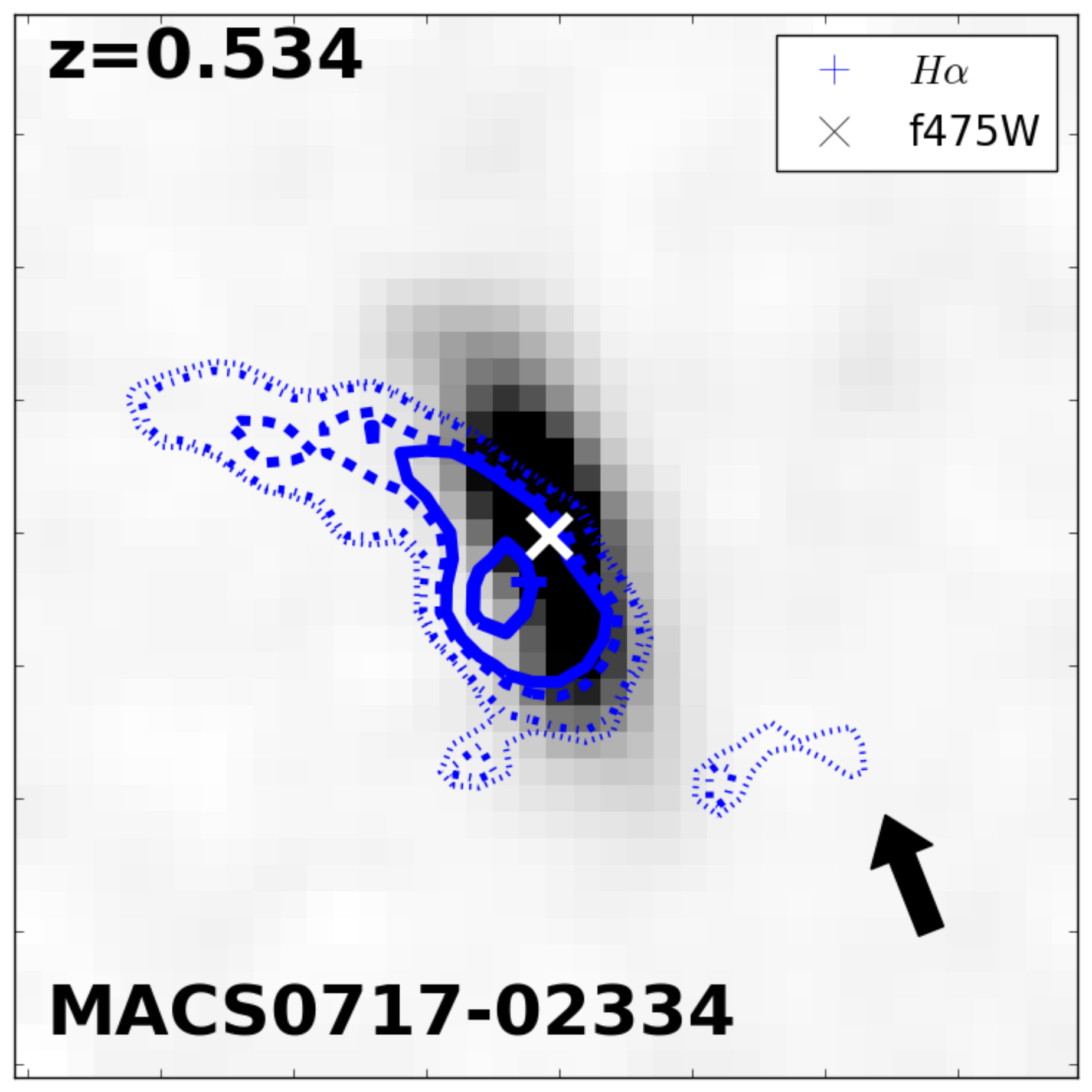}
\includegraphics[scale=0.14]{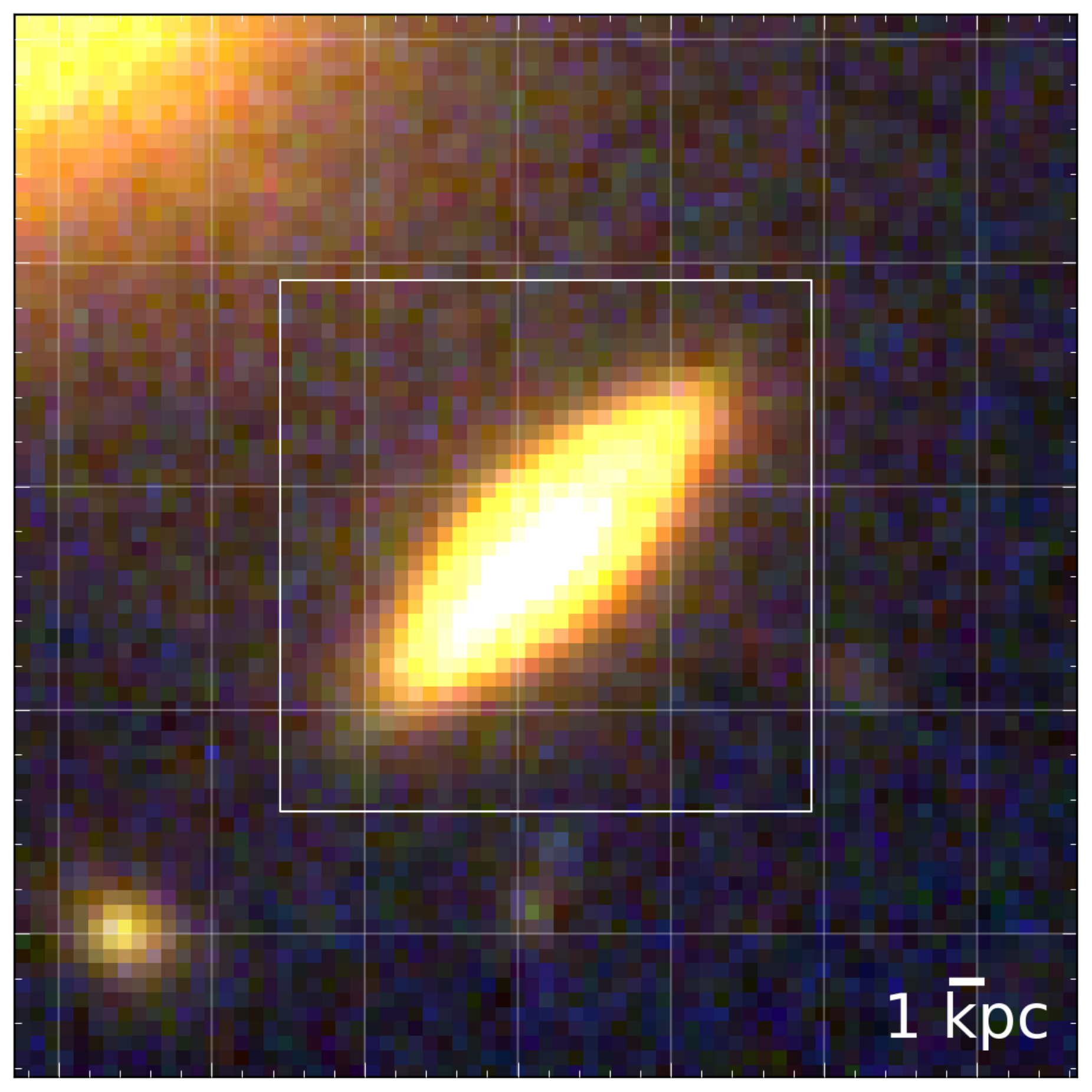}
\includegraphics[scale=0.14]{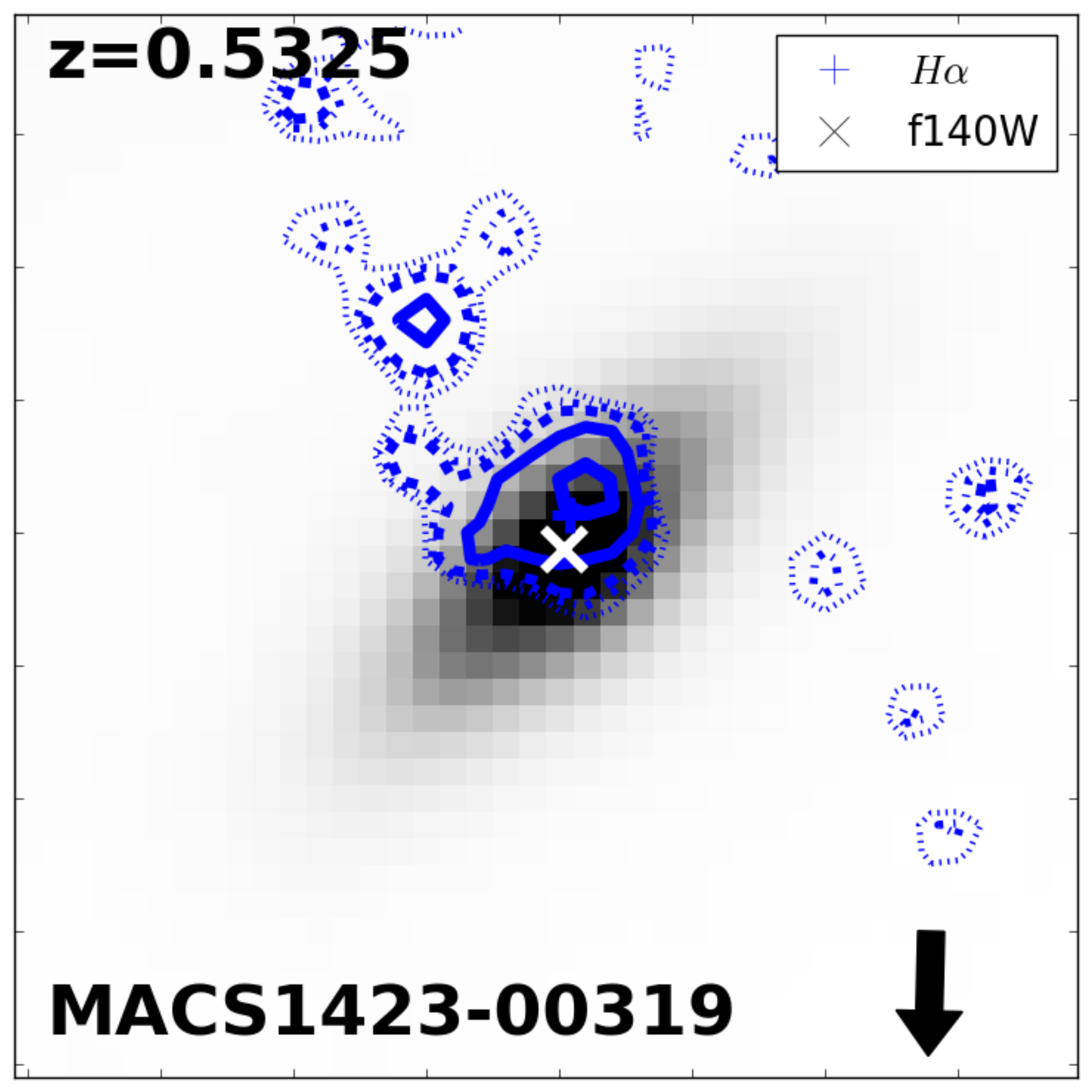}
\includegraphics[scale=0.14]{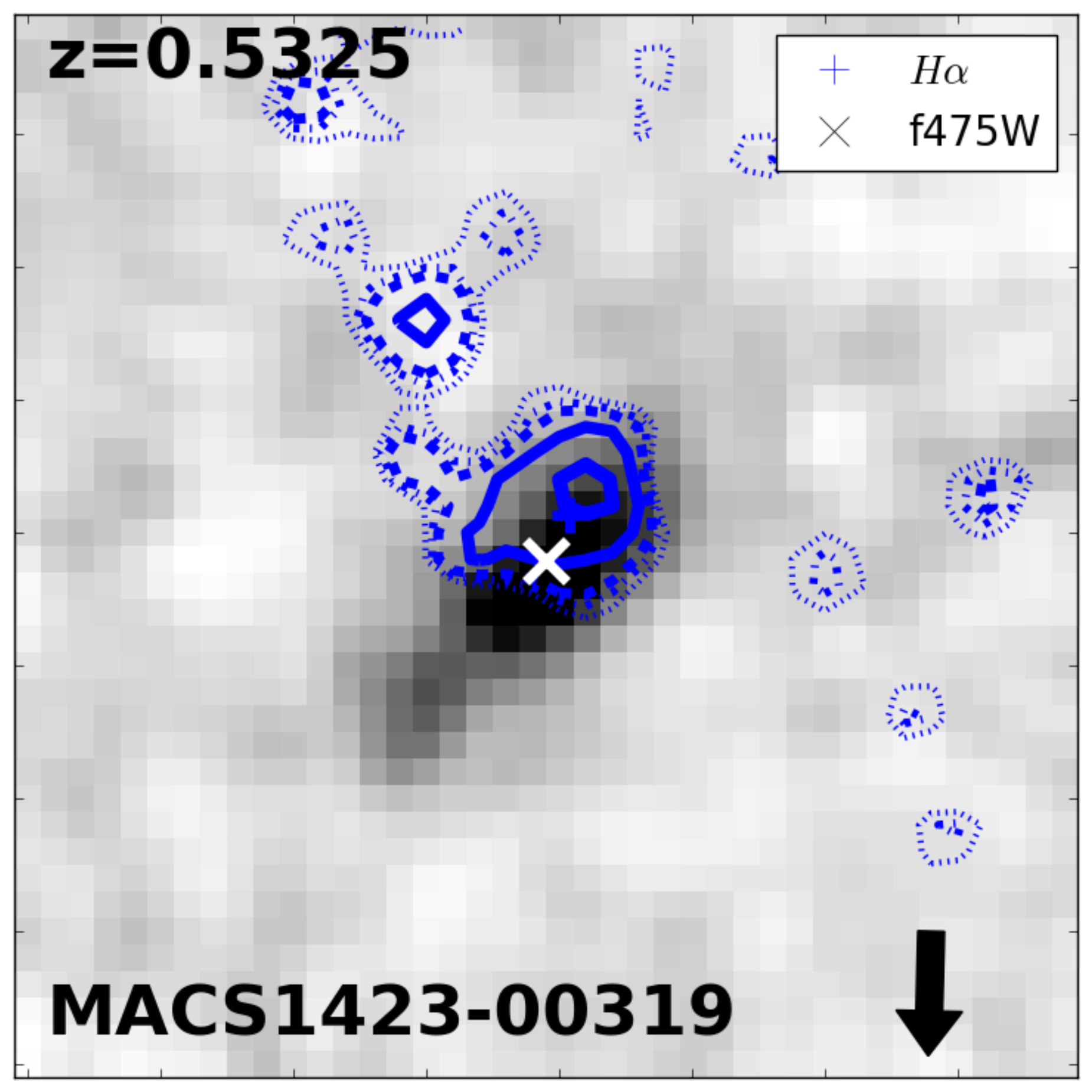}
\includegraphics[scale=0.14]{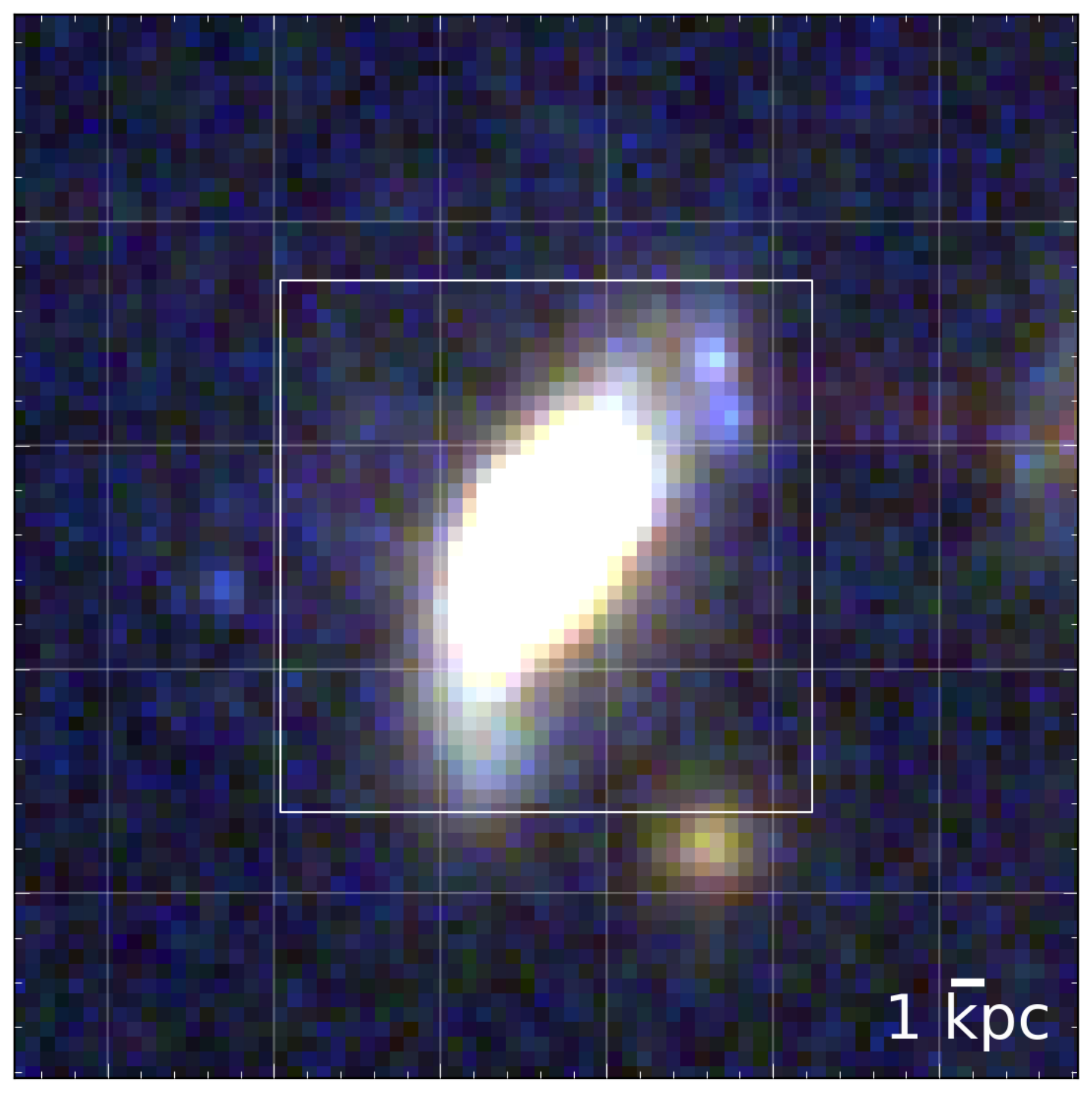}
\includegraphics[scale=0.14]{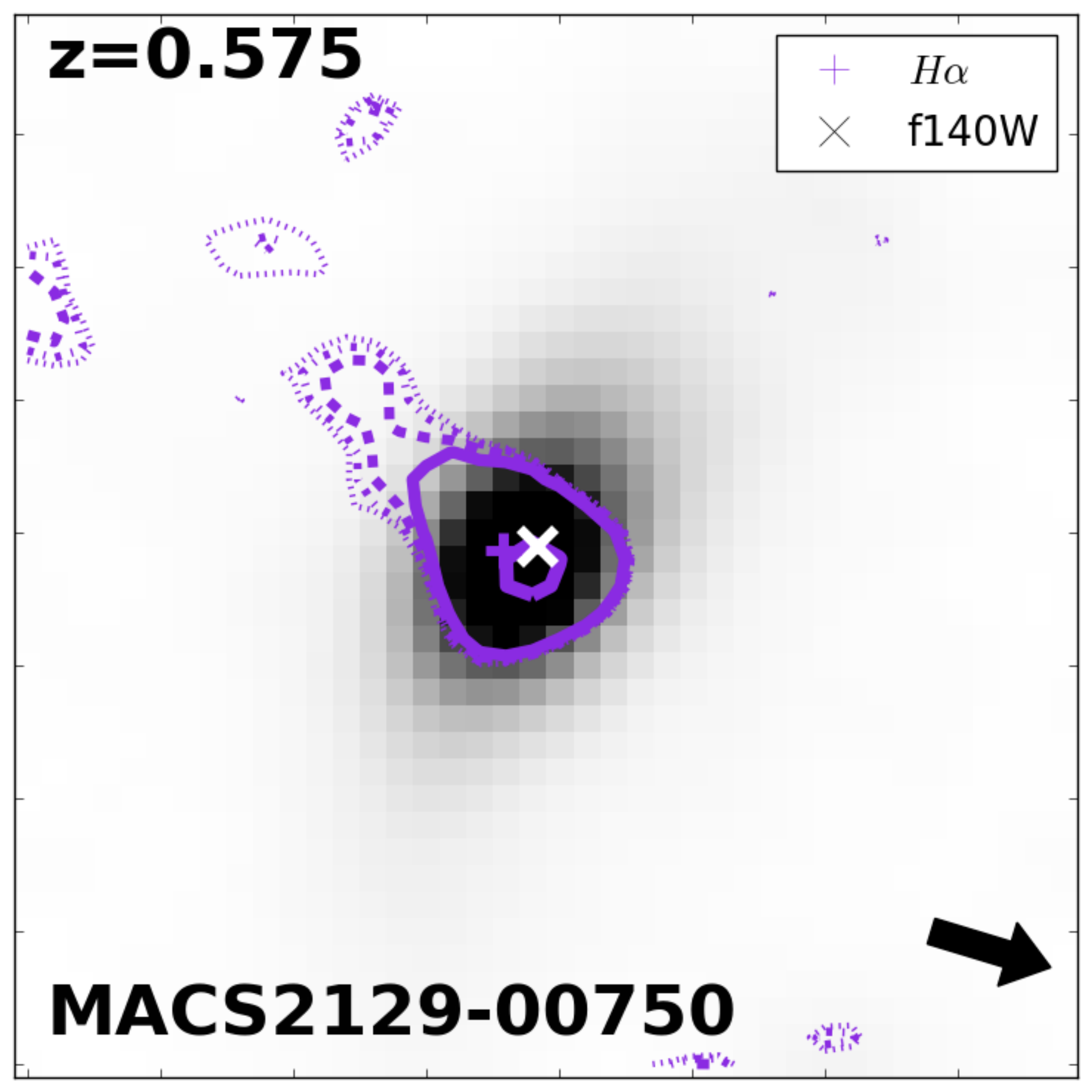}
\includegraphics[scale=0.14]{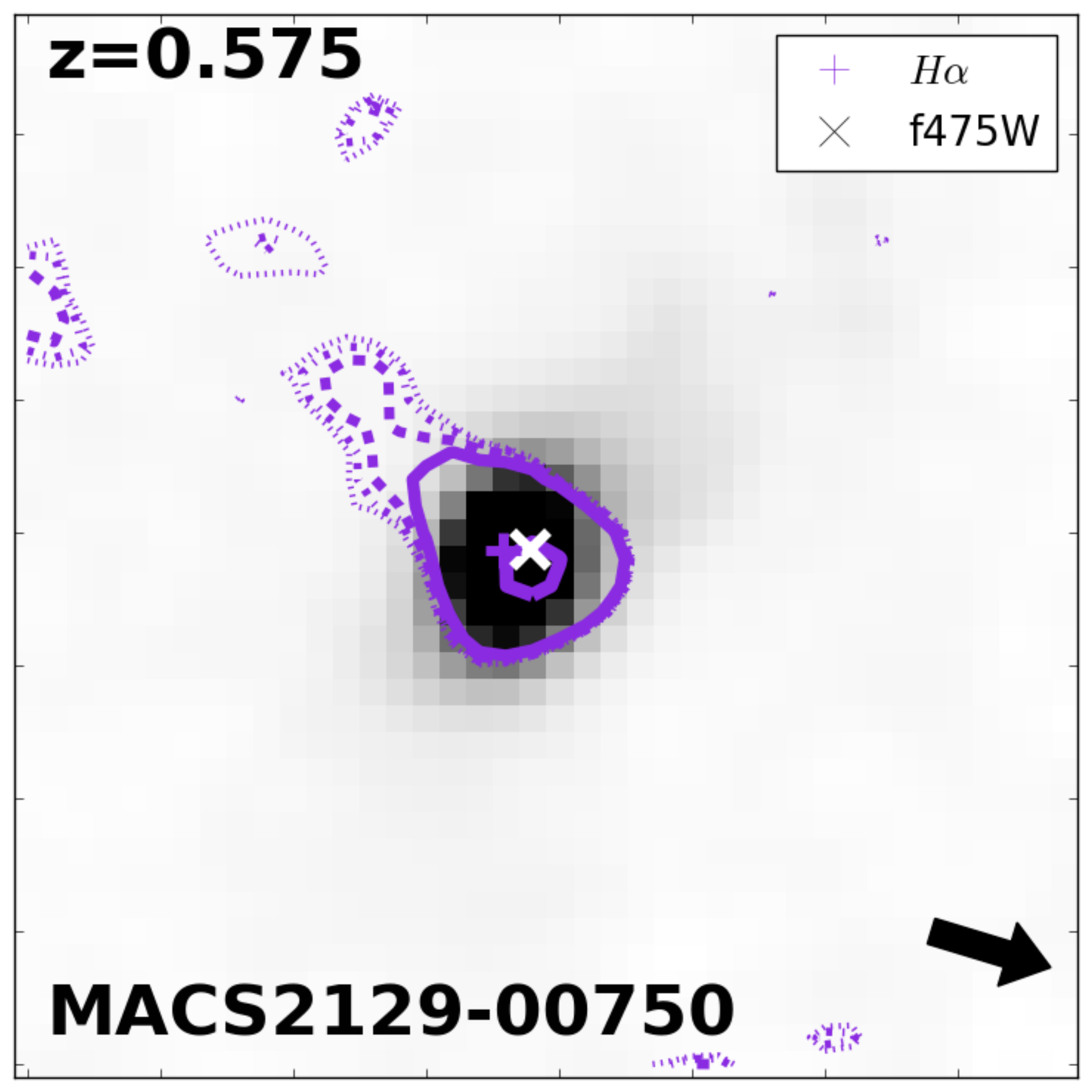}
\includegraphics[scale=0.14]{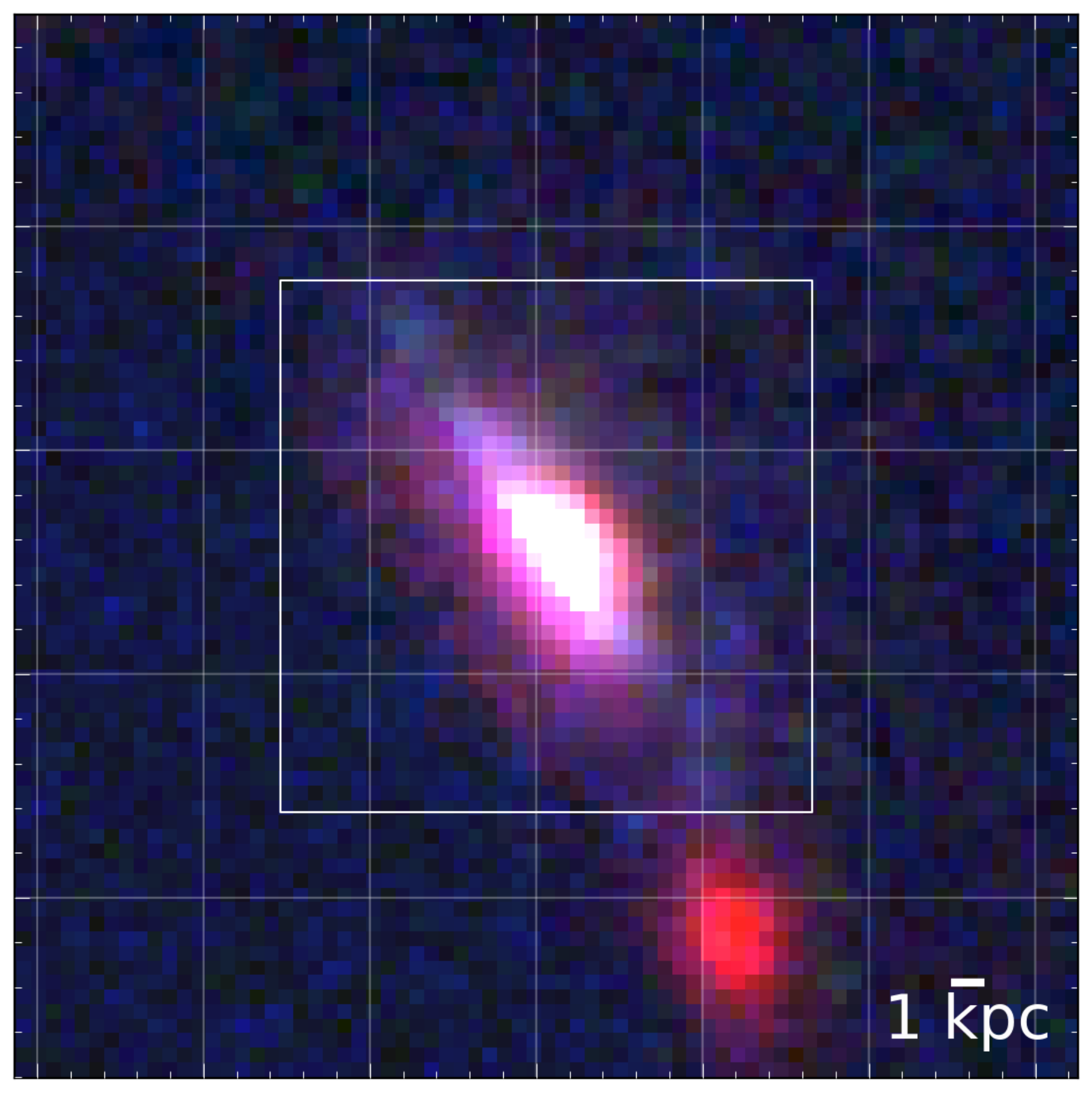}
\includegraphics[scale=0.14]{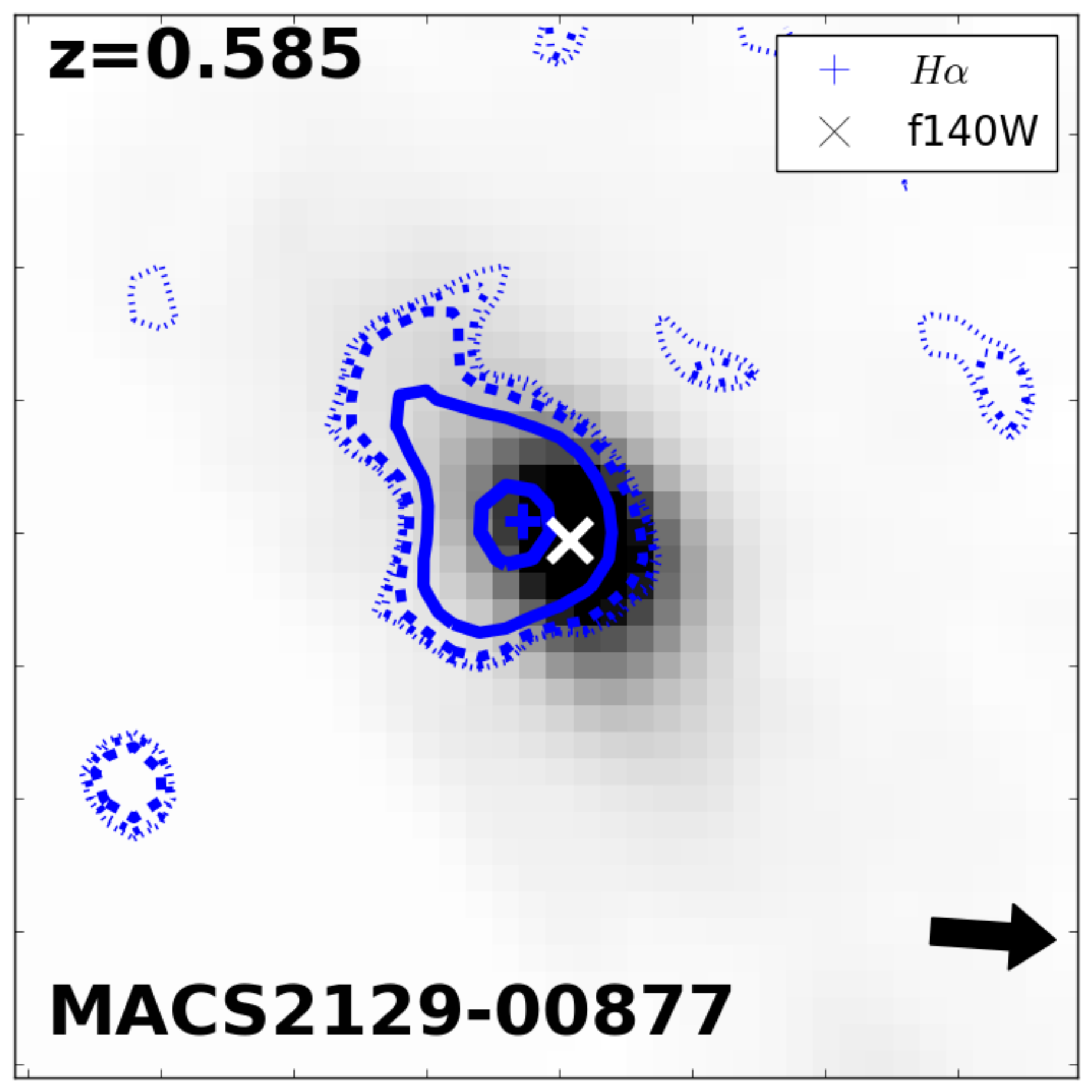}
\includegraphics[scale=0.14]{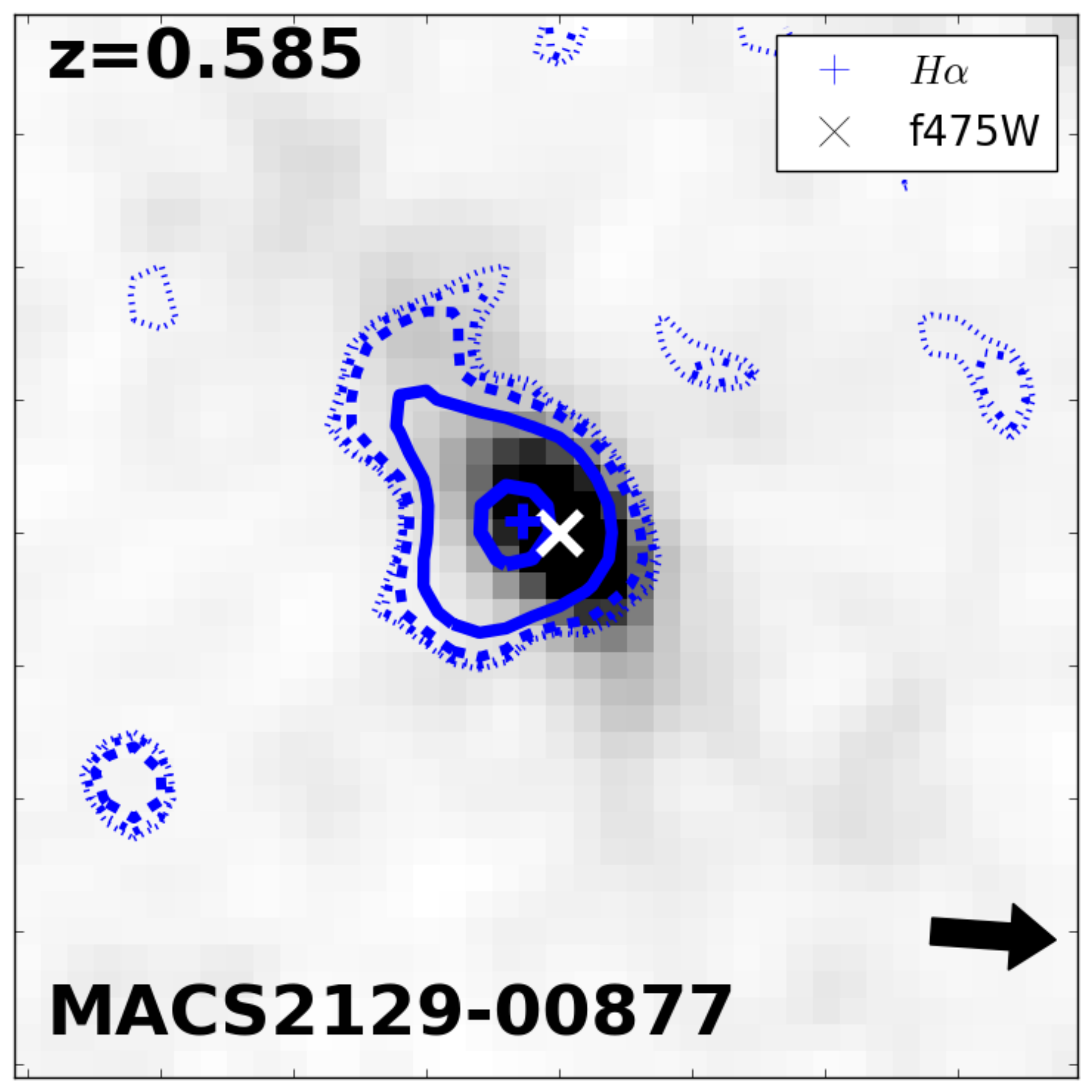}
\includegraphics[scale=0.14]{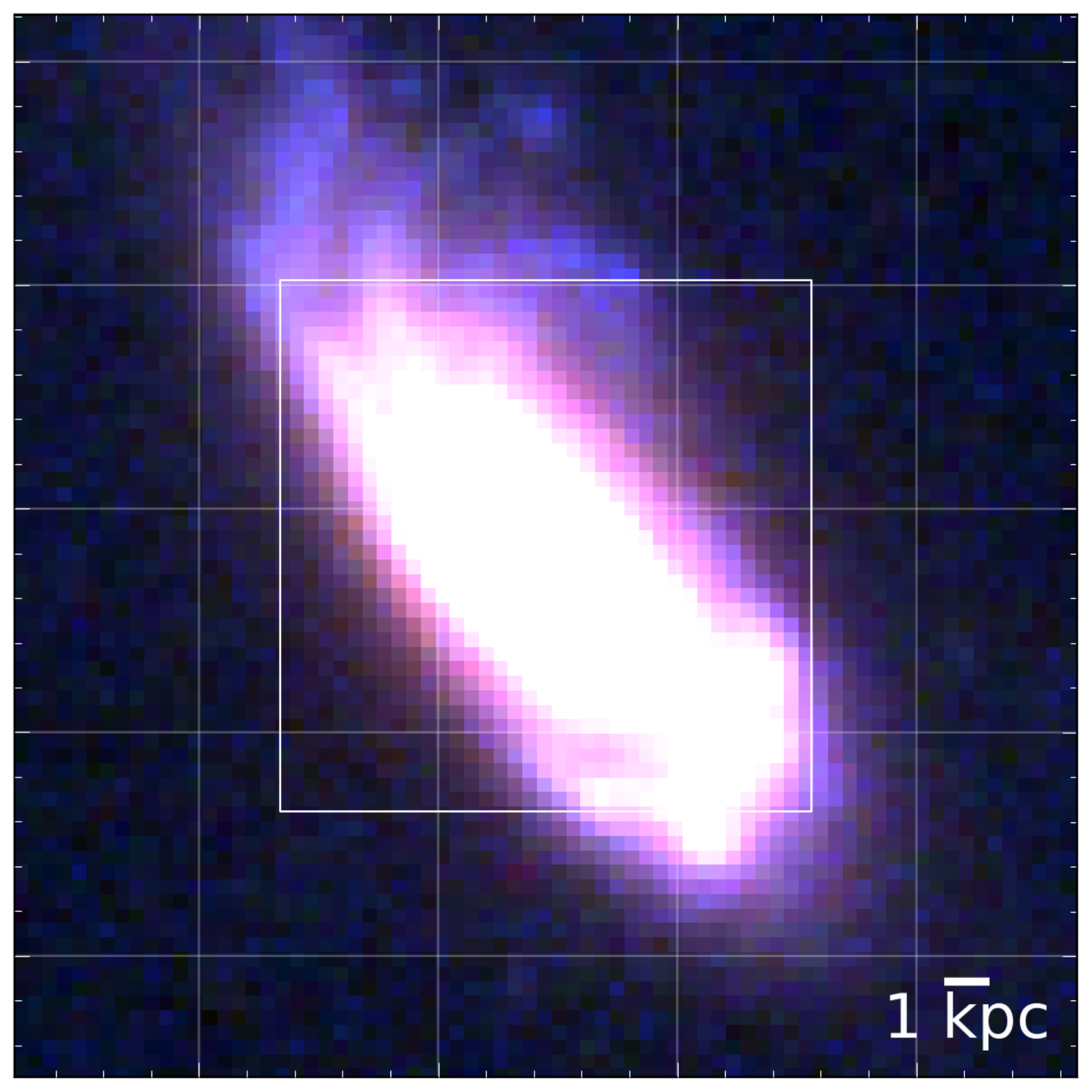}
\includegraphics[scale=0.14]{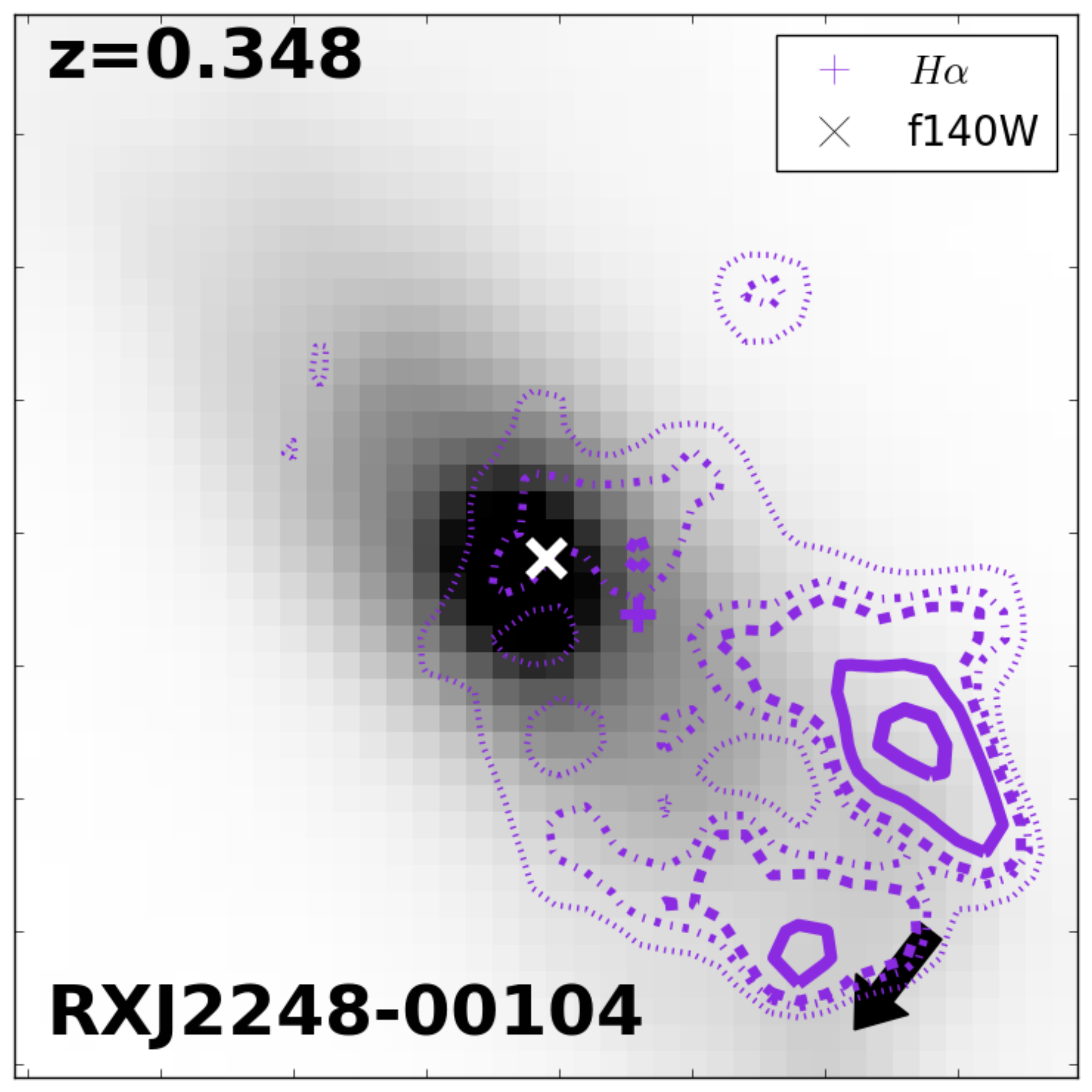}
\includegraphics[scale=0.14]{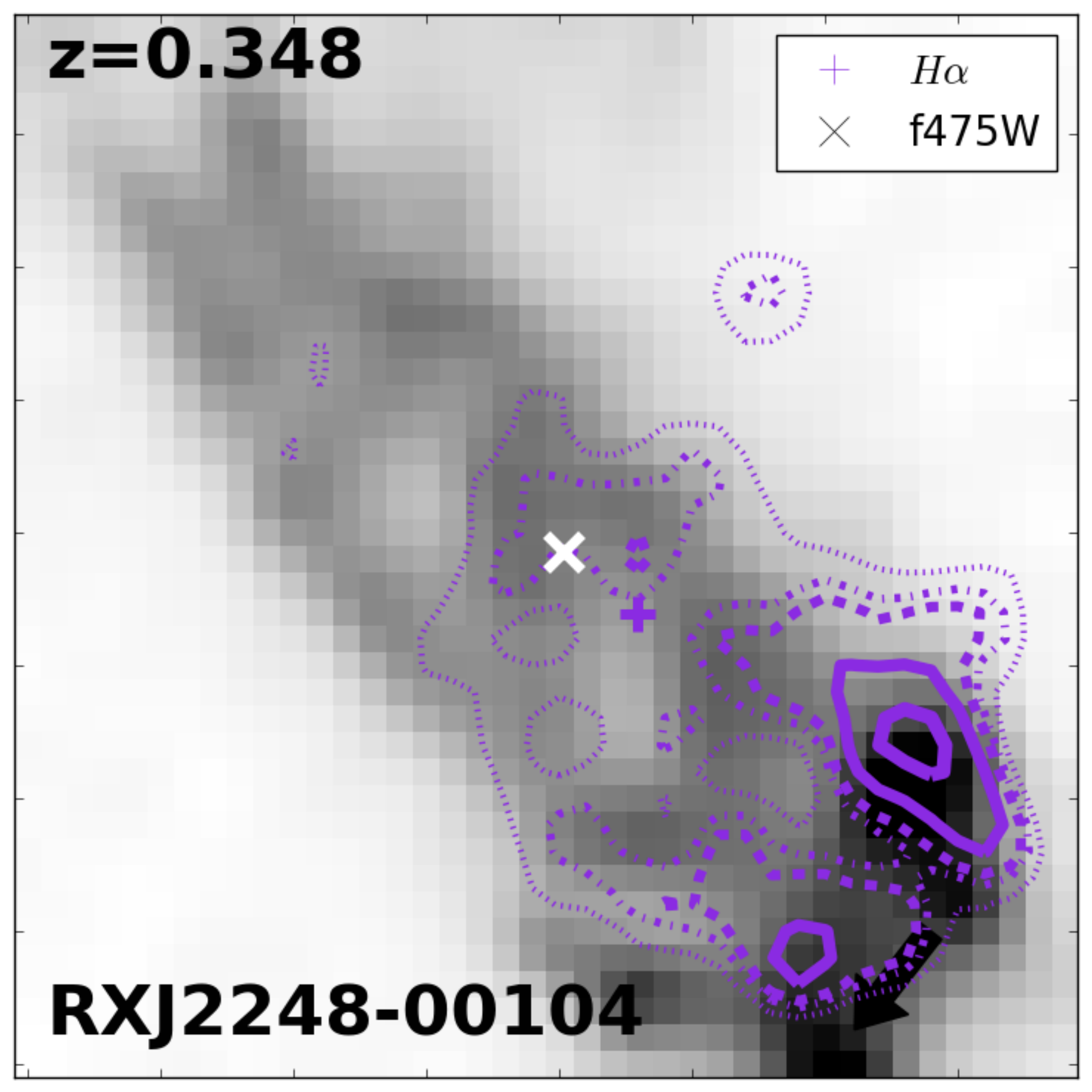}
\caption{Example of galaxies with spiral morphology and asymmetric \Ha mainly due to ram pressure 
stripping. Panels, colors, lines and labels are as in Fig. \ref{ell_ha}. 
 \label{spir_as}}
\end{figure*}

As mentioned above, 23(13)\% of \Ha emitters in clusters (field) present an elliptical morphology. This 
fraction is quite surprising since elliptical galaxies are generally thought to be quiescent \citep[although blue cores have been seen in the past; Menanteau et al. 2005a,][]{menanteau05b, treu05, schawinski09}.
Of these, in clusters $\sim$40\% present a regular \Ha morphology, $\sim
$10\% a clumpy \Ha morphology, $\sim$20\% a concentrated \Ha 
morphology and $\sim$30\% an asymmetric morphology.  The corresponding percentages for galaxies in 
the field are: $\sim$ 35\%, $\sim20\%$, $\sim$10\%, $\sim$35\%, 
respectively. Figure \ref{ell_ha} shows some examples of these galaxies. For each object, a composite 
image of the galaxy based on the CLASH \citep{postman12} or HFF 
\citep{lotz16} HST images is shown, along with the \Ha map superimposed on both the image of the galaxy in 
the F140W filter and in the F475W filter. 
As can be seen from the images, both the broad band morphology and the \Ha emission are highly certain and therefore these are not spurious detections or misclassified objects.
In clusters, 4 cases show clear signs of ram pressure stripping (like MACS1149-01832, in the upper right 
panels of Fig.\ref{ell_ha}), while there is only a tentative case of stripping  in 
the field.

In both environments, these galaxies have a median SFR of $\sim3 M_\sun \, yr^{-1}$; field galaxies have 
a median mass of$\sim 10^{8.8} M_\sun$, cluster galaxies of $\sim10^{9.7} 
M_\sun$, therefore they are not the most massive objects in our sample, but have high SFR, suggesting they might actually be nuclear star bursts 
 or AGNs. However, these galaxies do not all have concentrated \Ha, indicating that even if present, the AGN can not be entirely responsible for the 
 detected \Ha emission.  In addition, the X-ray  maps show that  these elliptical galaxies are not strongly emitting in X-ray, therefore  the
 presence of  strong AGNs is excluded.
 
 The origin of these objects is unclear, they might be disk+bulge galaxies which have had 
 their disk destroyed or removed by some process or very dusty objects. This class of objects will be revisited in a future work.

\subsubsection{Spiral galaxies with asymmetric \Ha  likely due to ram pressure}
\begin{figure*}
\centering
\includegraphics[scale=0.14]{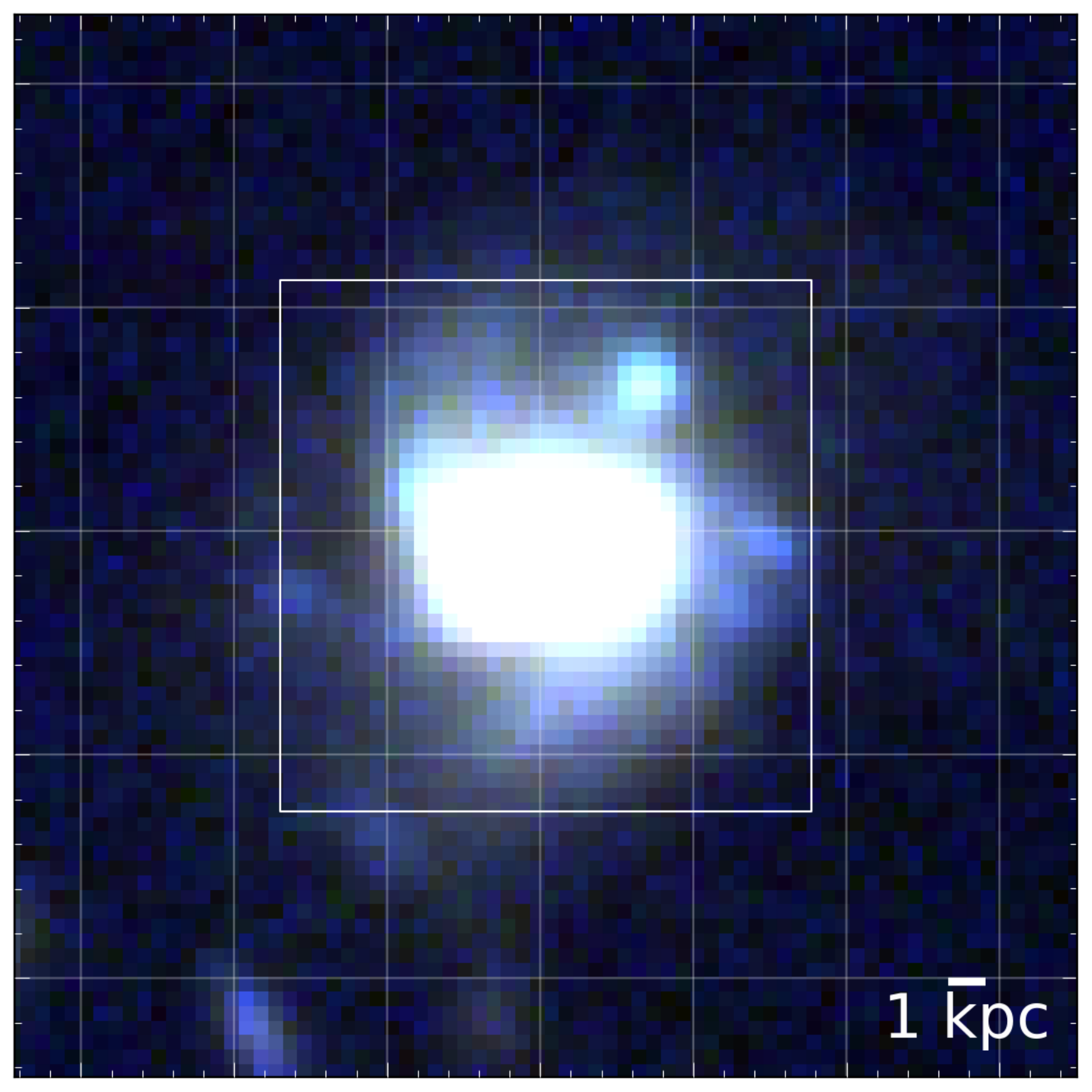}
\includegraphics[scale=0.14]{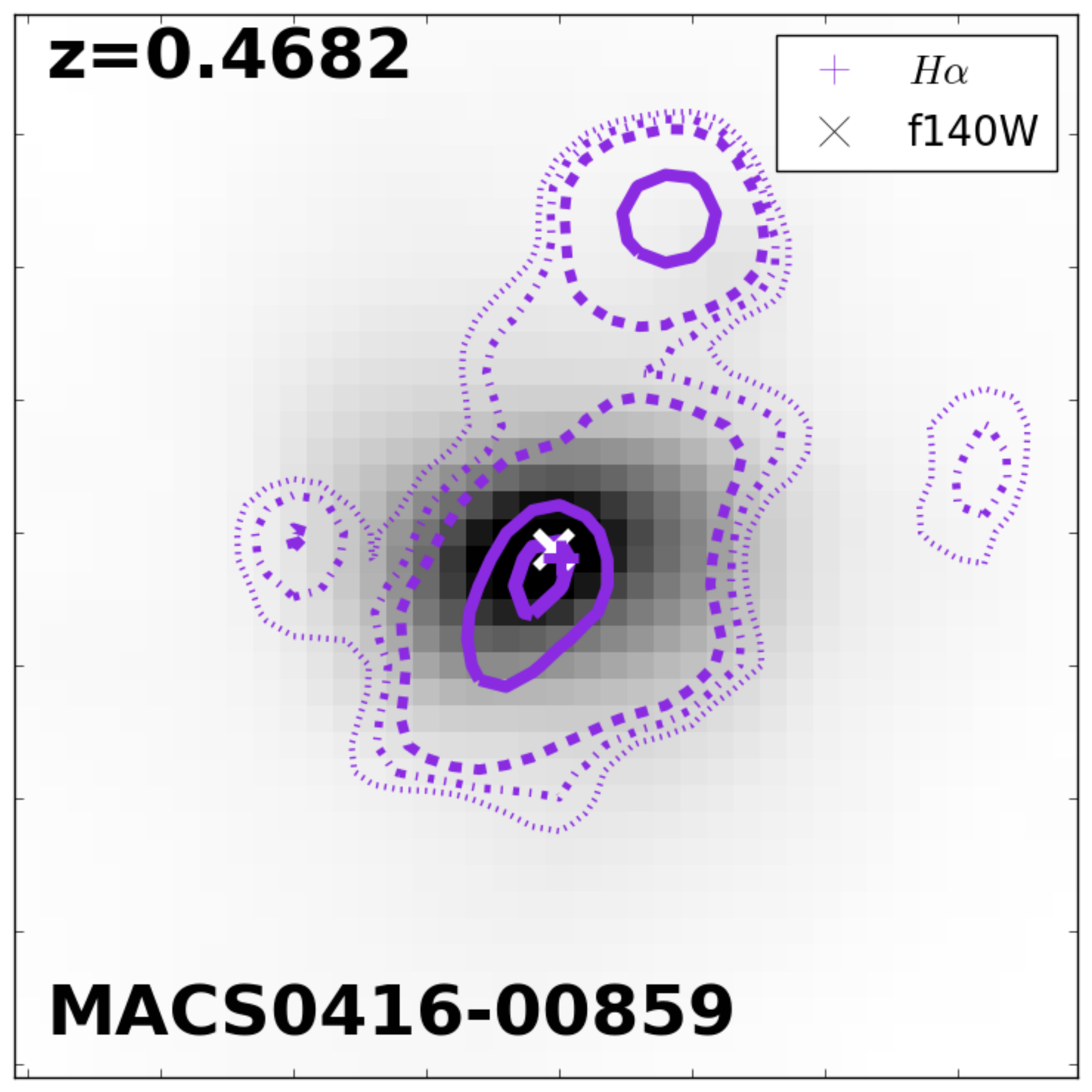}
\includegraphics[scale=0.14]{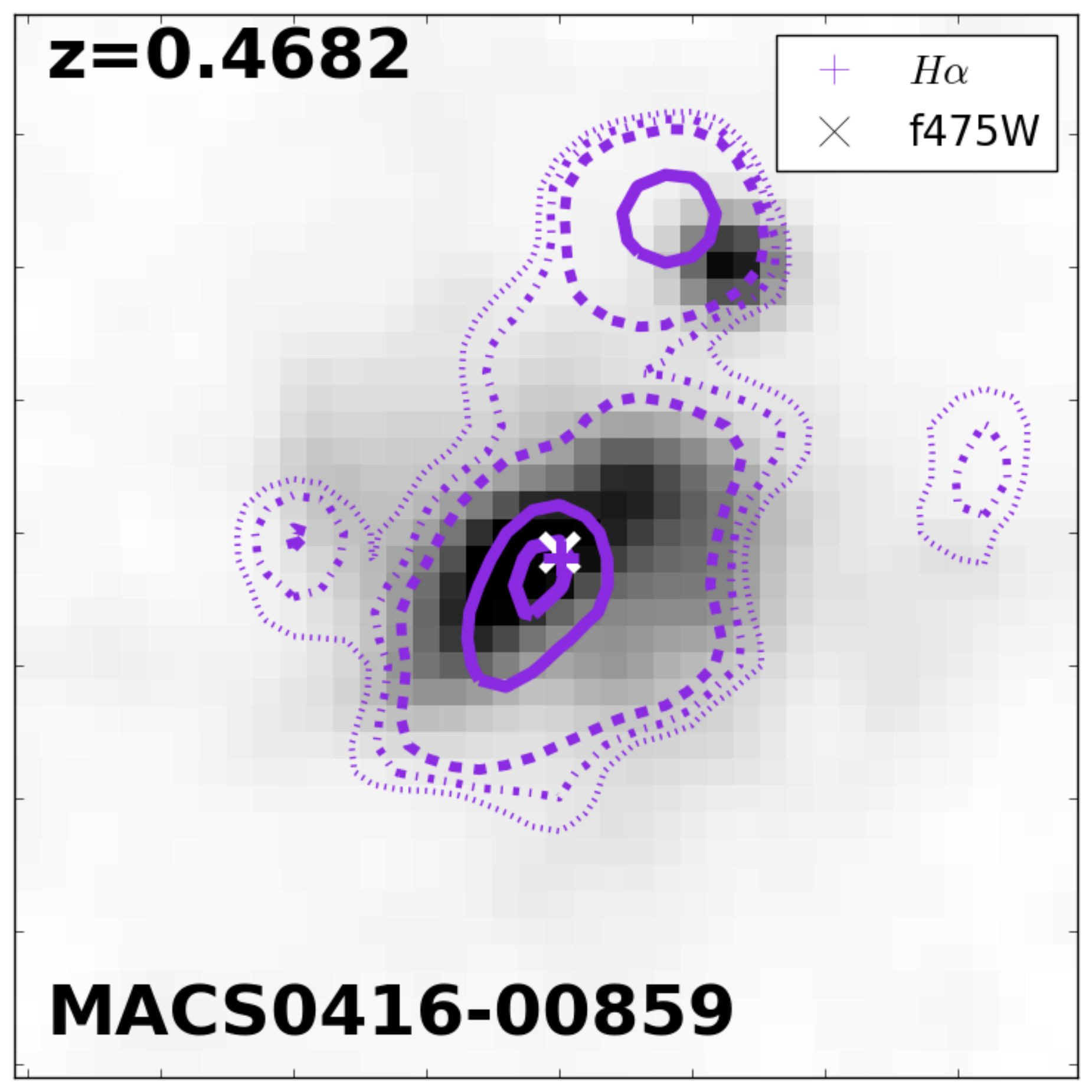}
\includegraphics[scale=0.14]{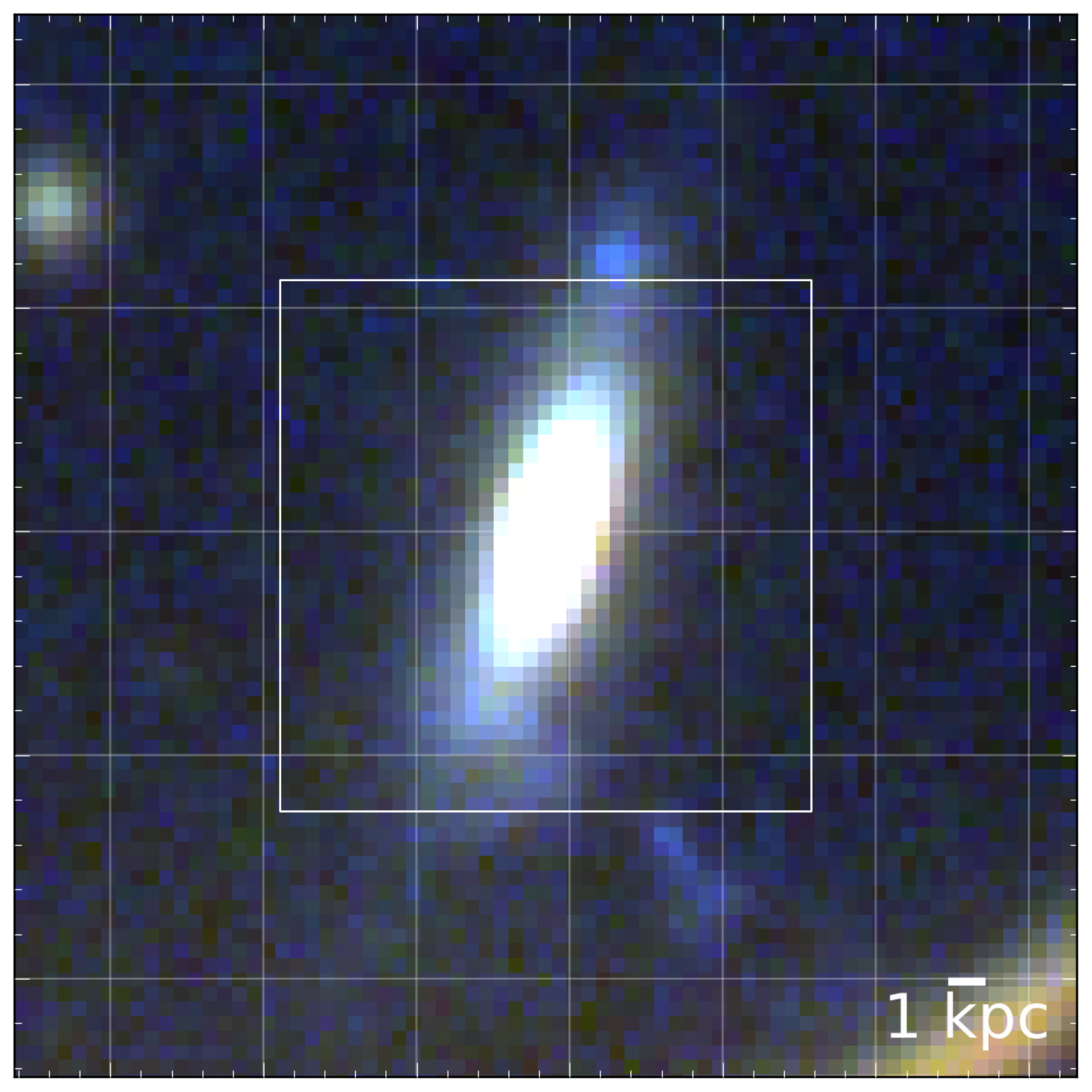}
\includegraphics[scale=0.14]{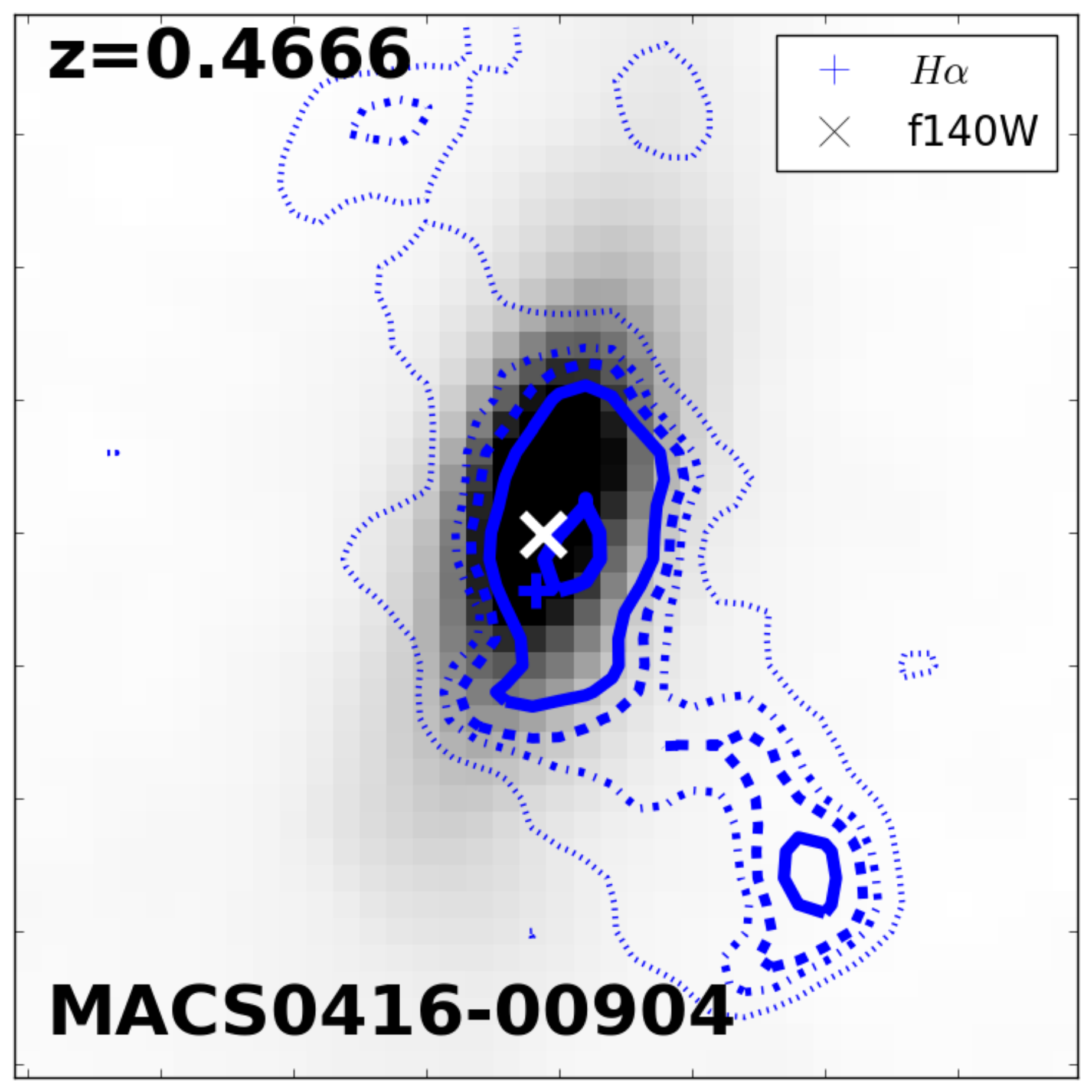}
\includegraphics[scale=0.14]{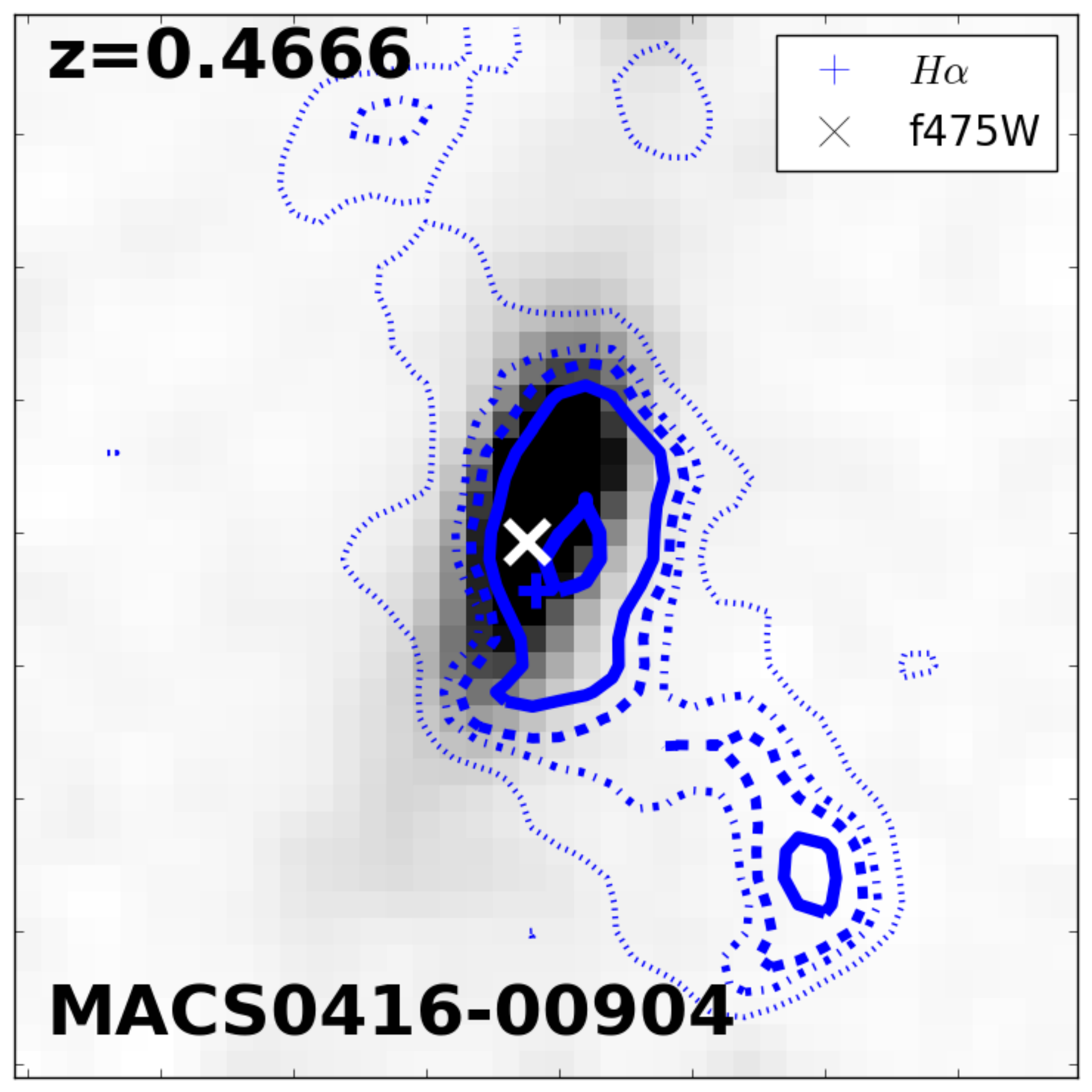}
\includegraphics[scale=0.14]{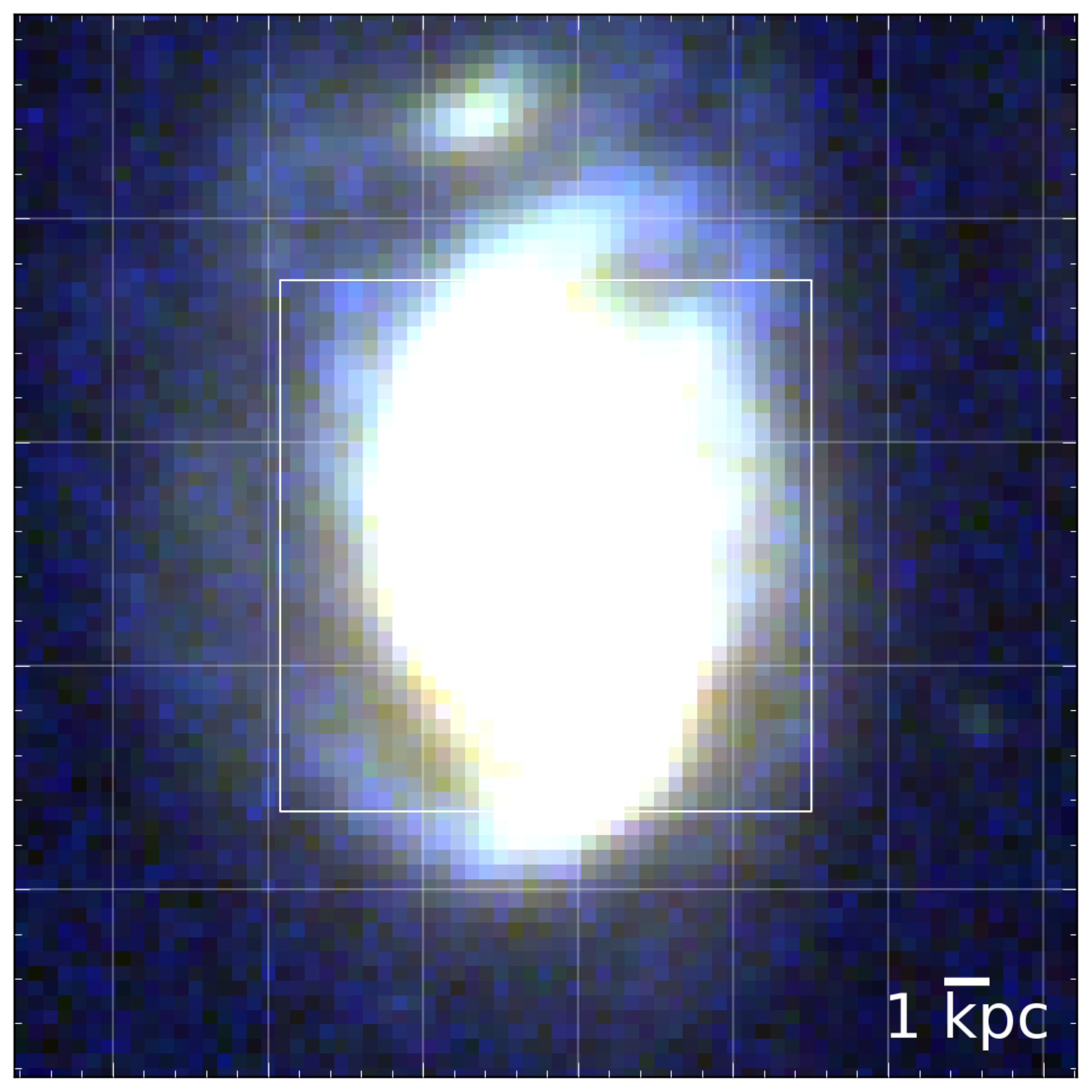}
\includegraphics[scale=0.14]{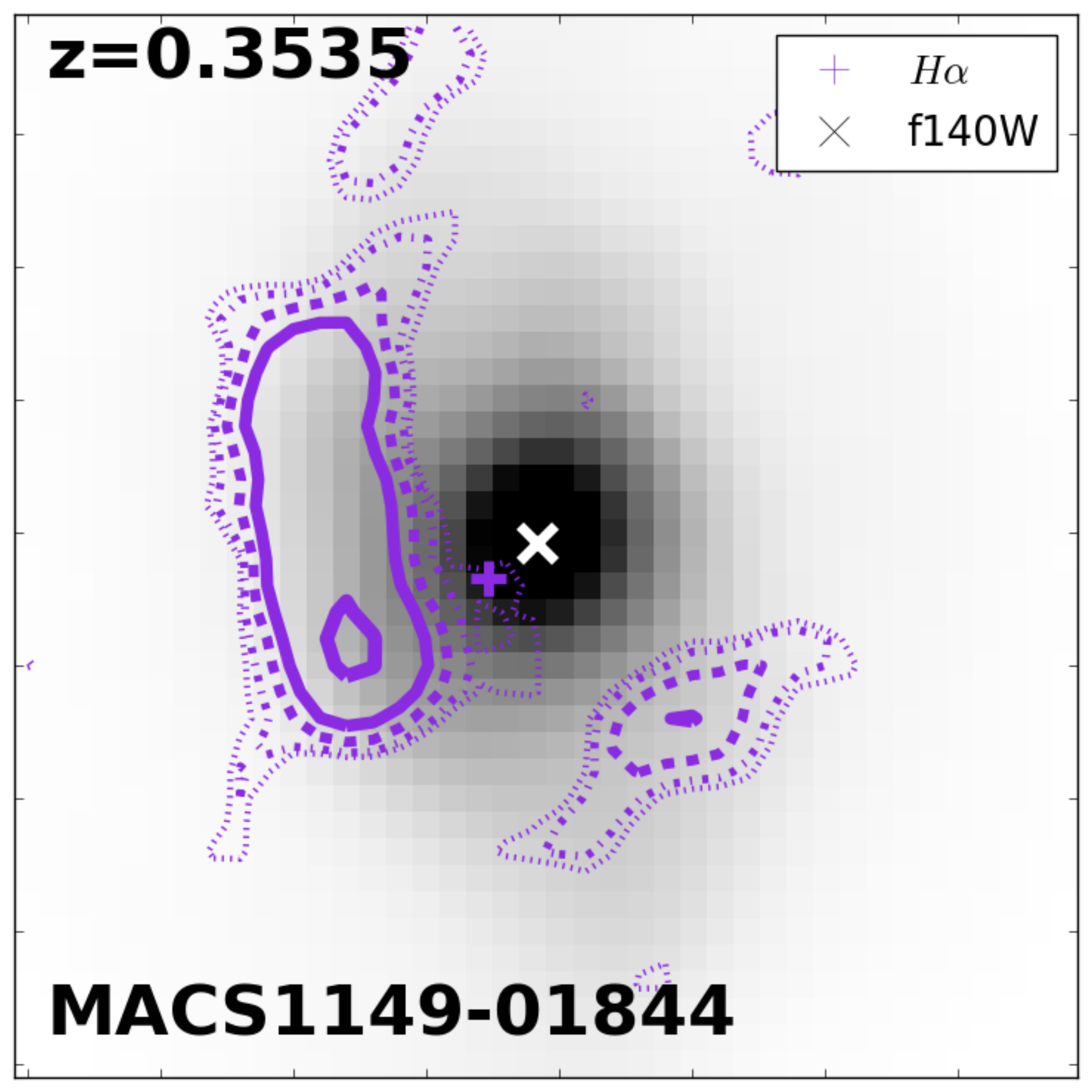}
\includegraphics[scale=0.14]{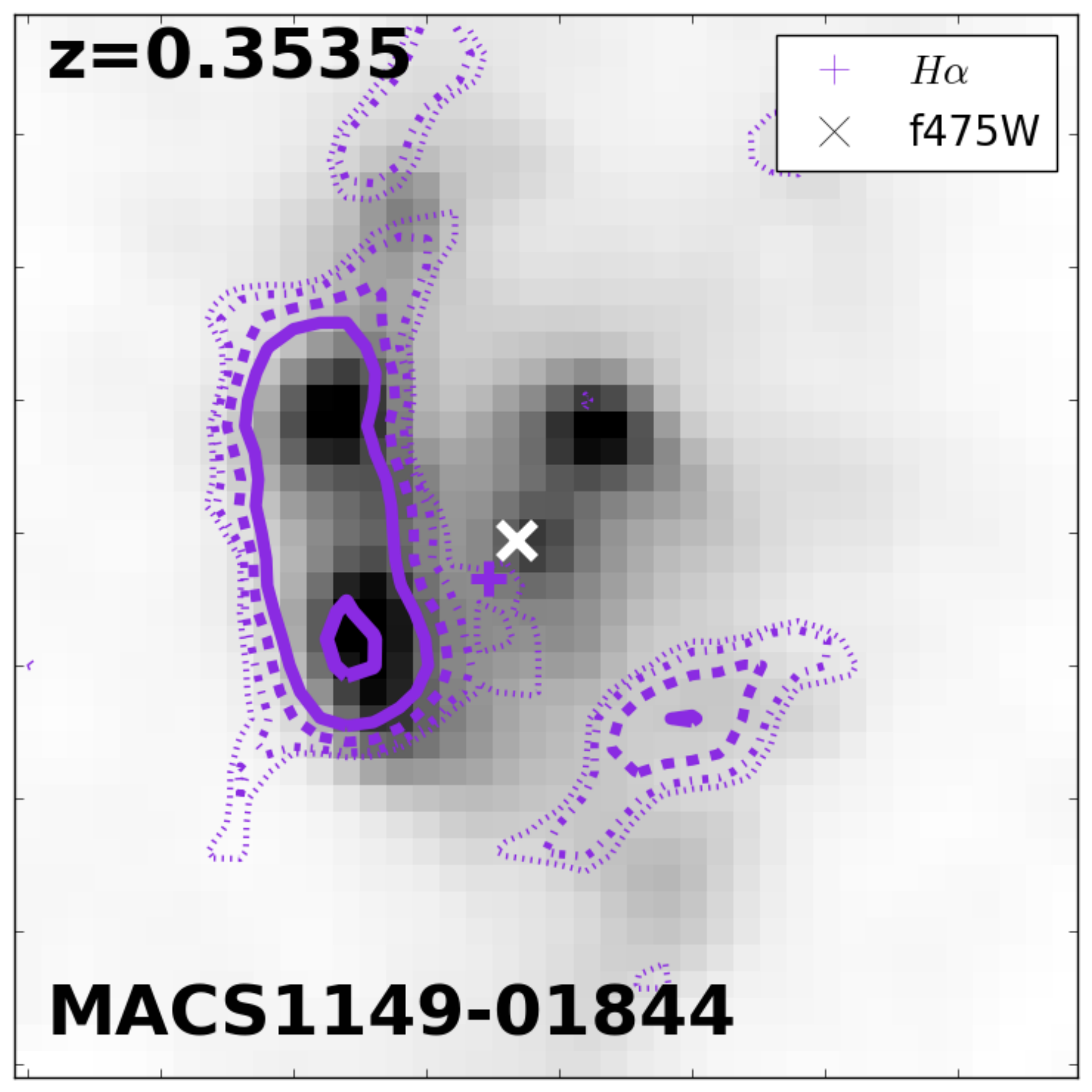}
\includegraphics[scale=0.14]{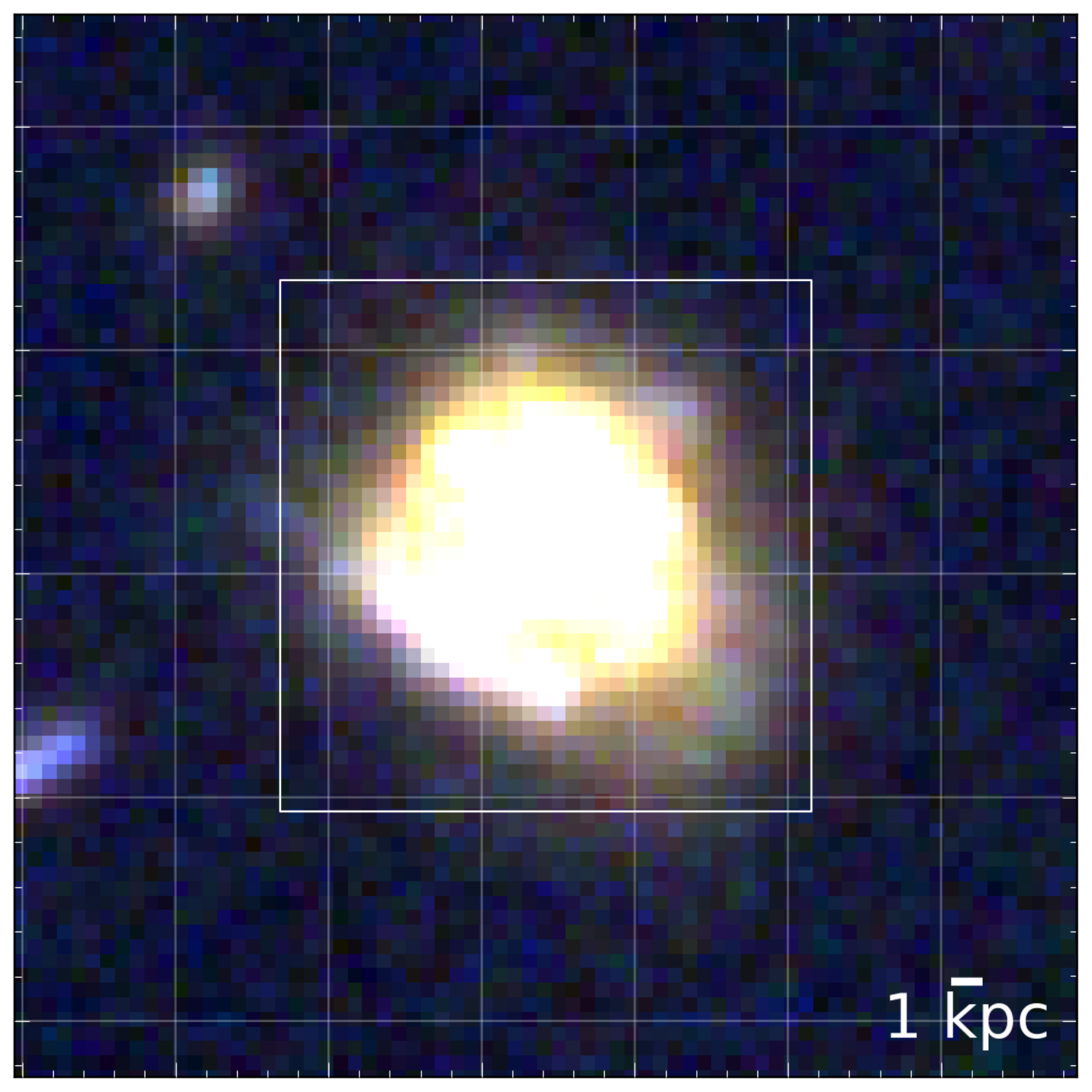}
\includegraphics[scale=0.14]{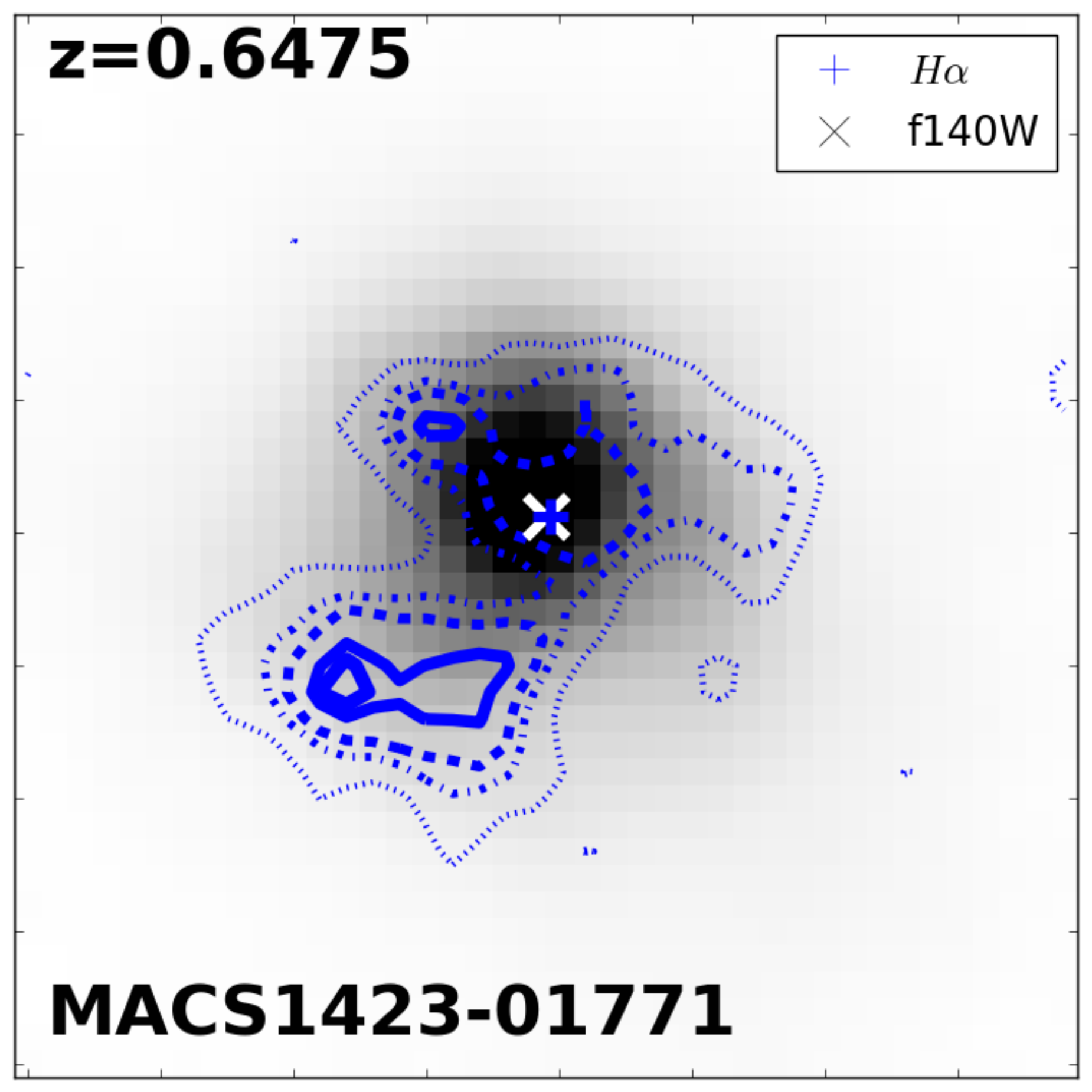}
\includegraphics[scale=0.14]{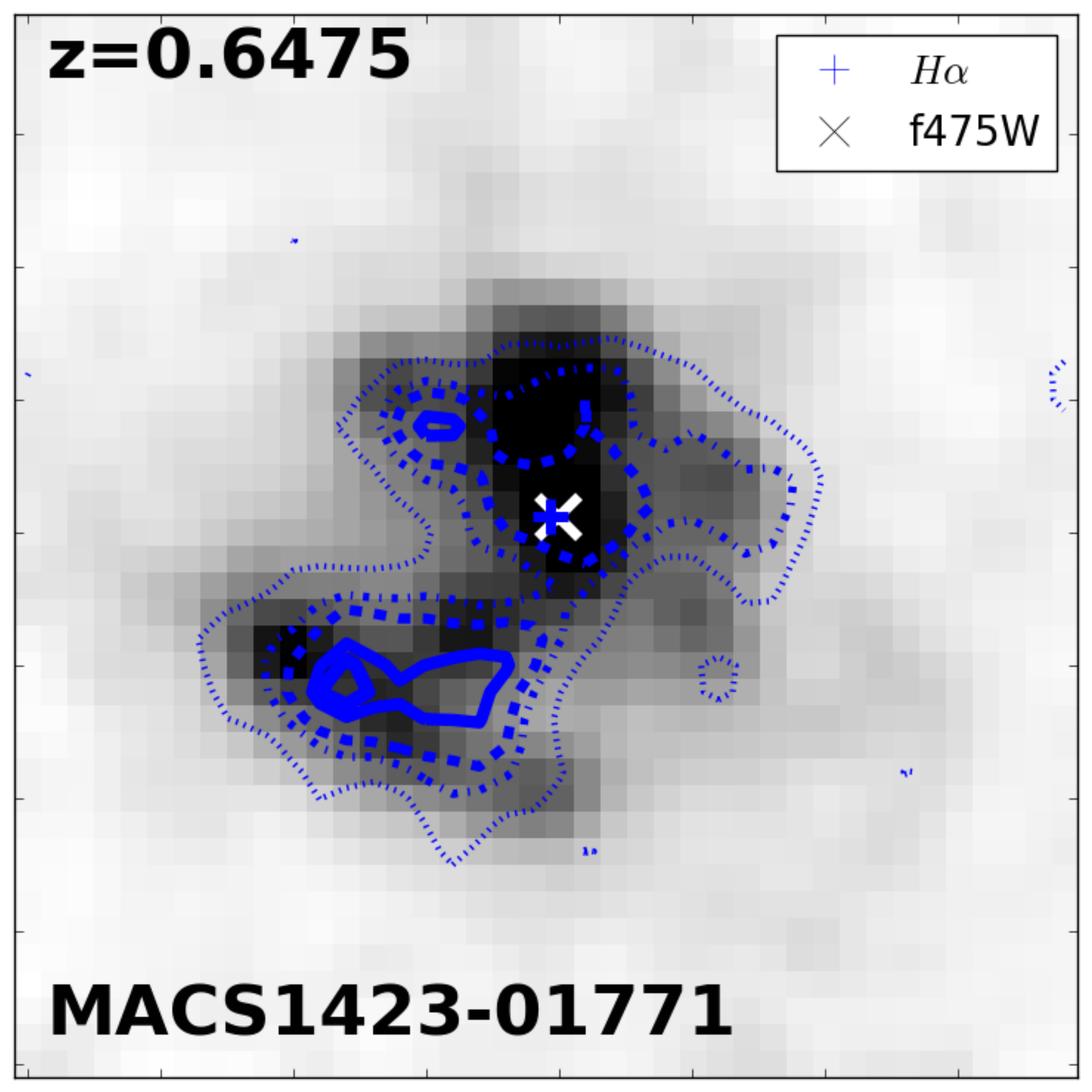}
\includegraphics[scale=0.14]{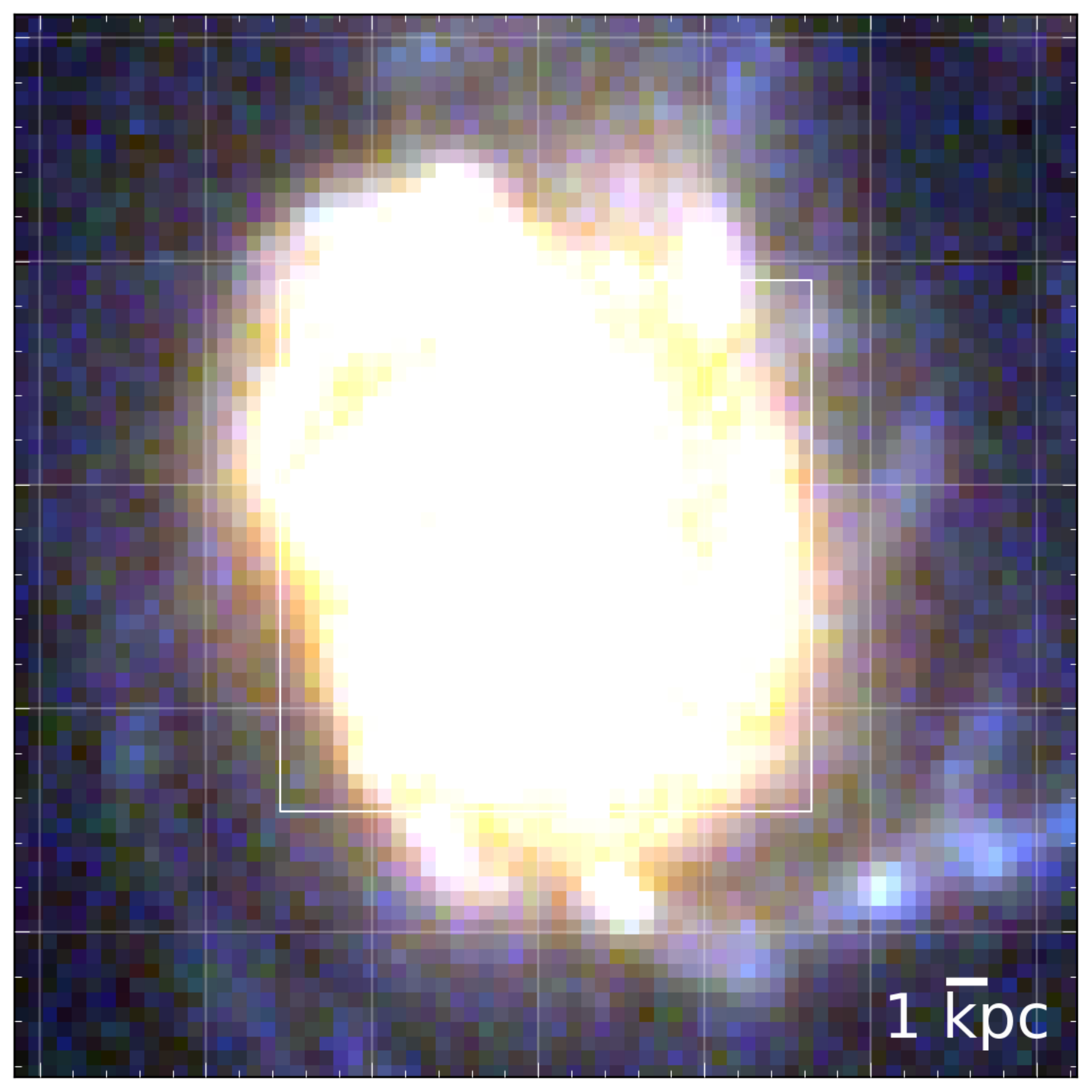}
\includegraphics[scale=0.14]{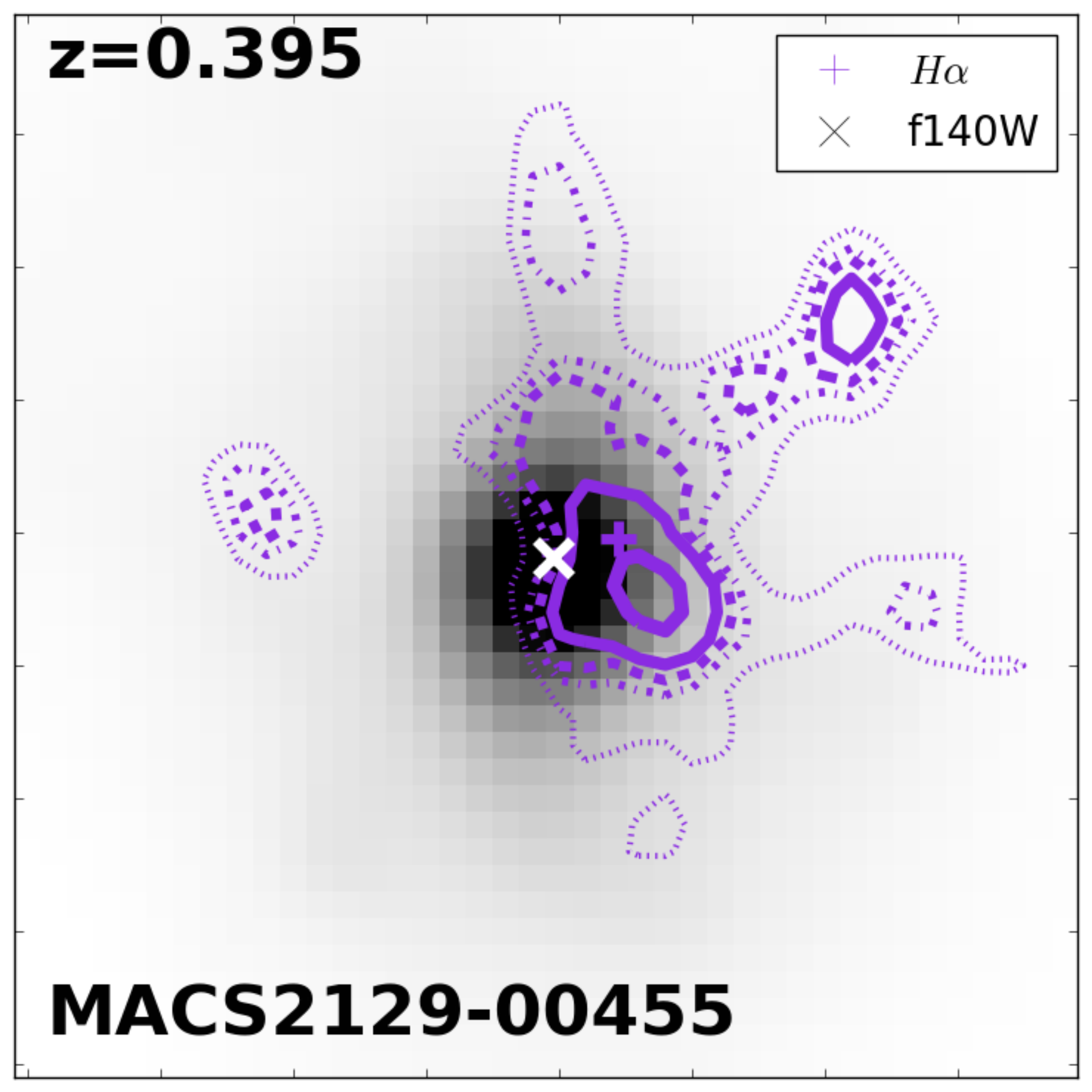}
\includegraphics[scale=0.14]{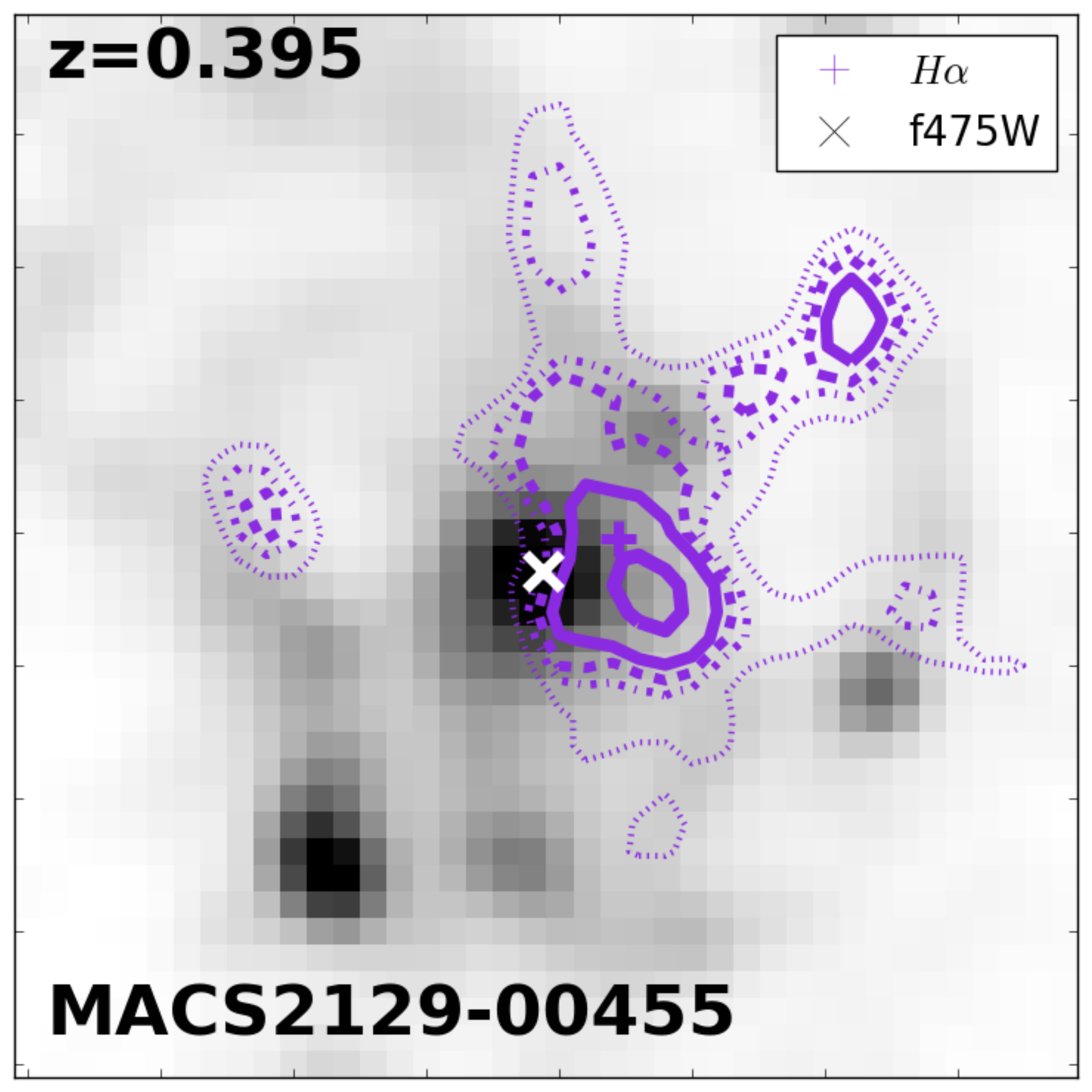}
\includegraphics[scale=0.14]{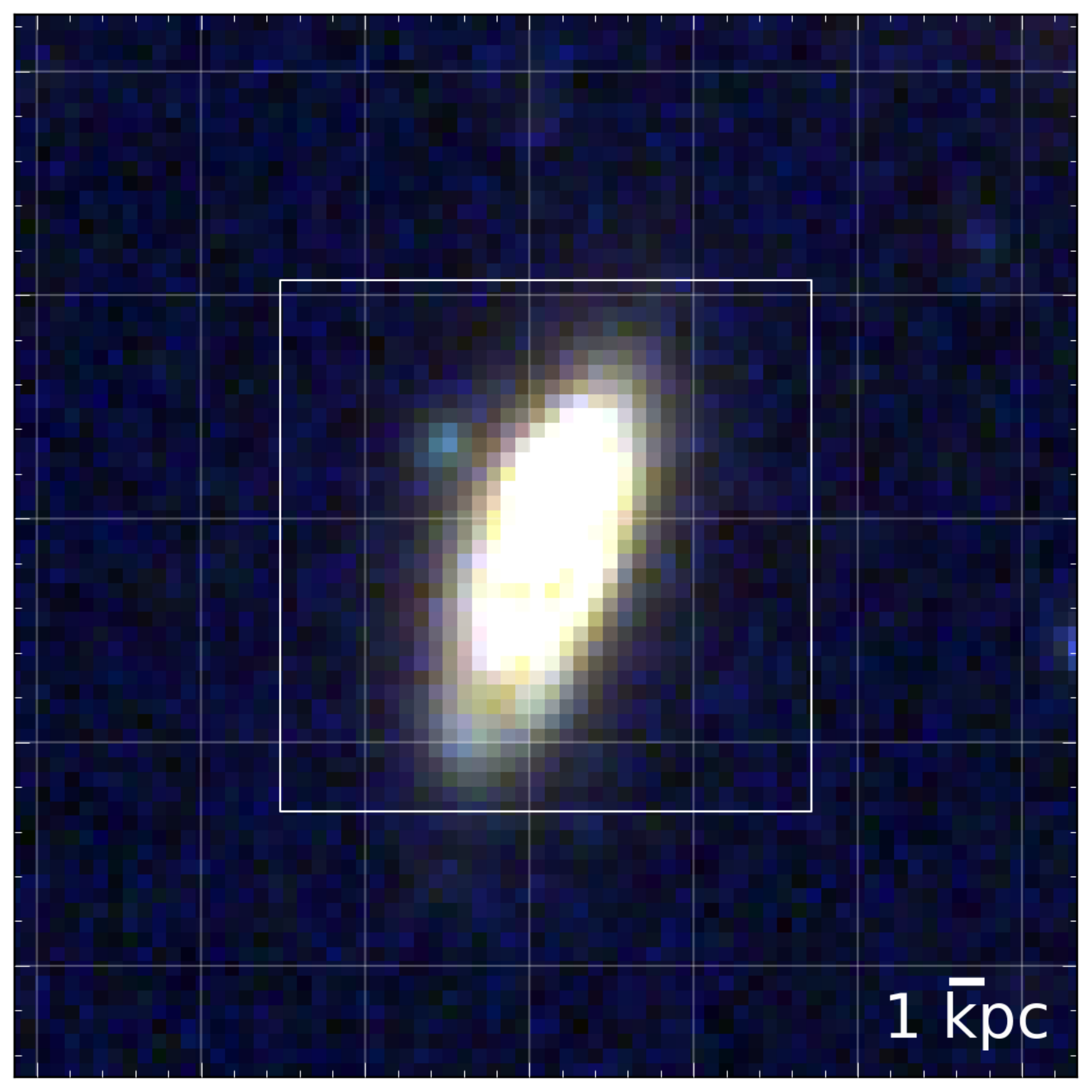}
\includegraphics[scale=0.14]{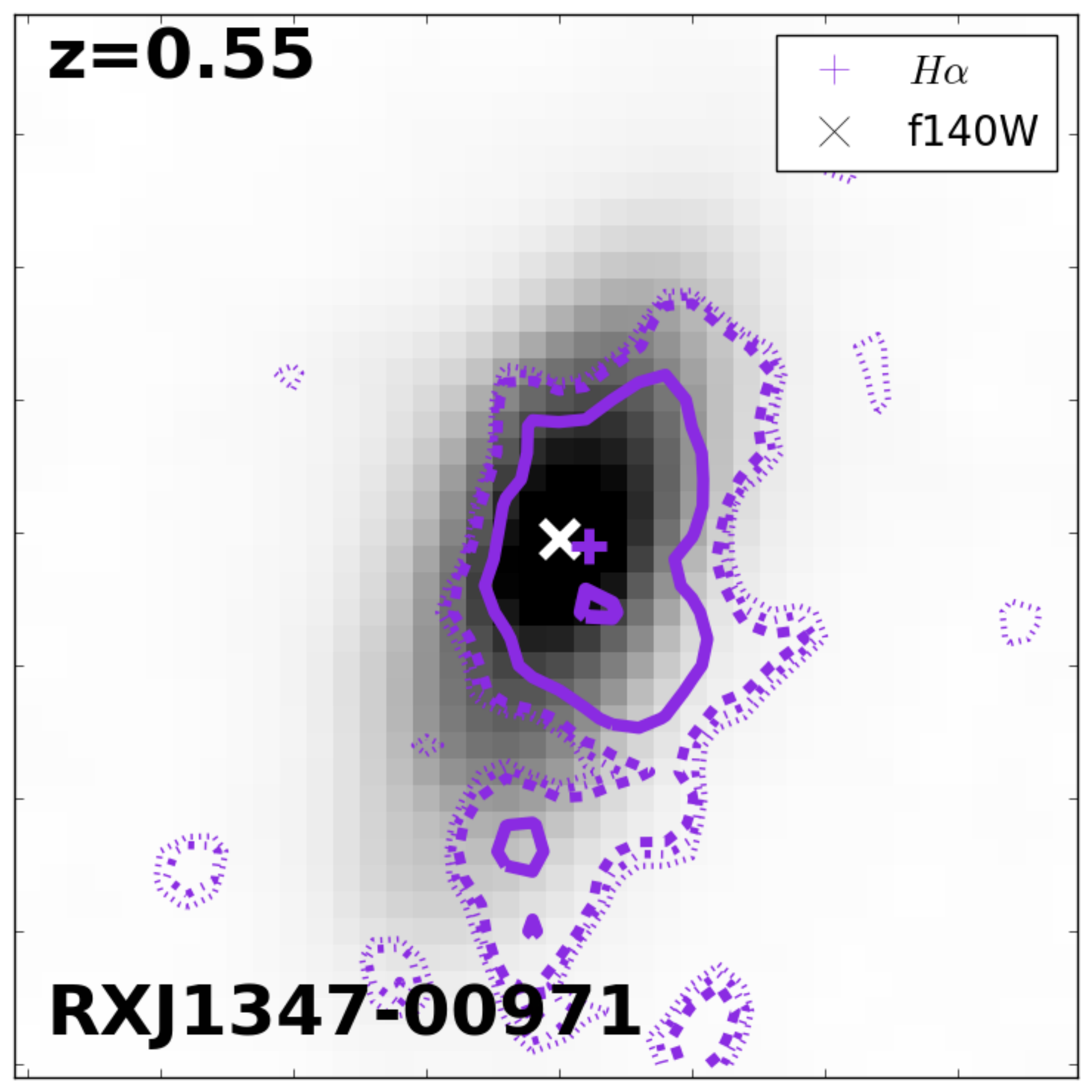}
\includegraphics[scale=0.14]{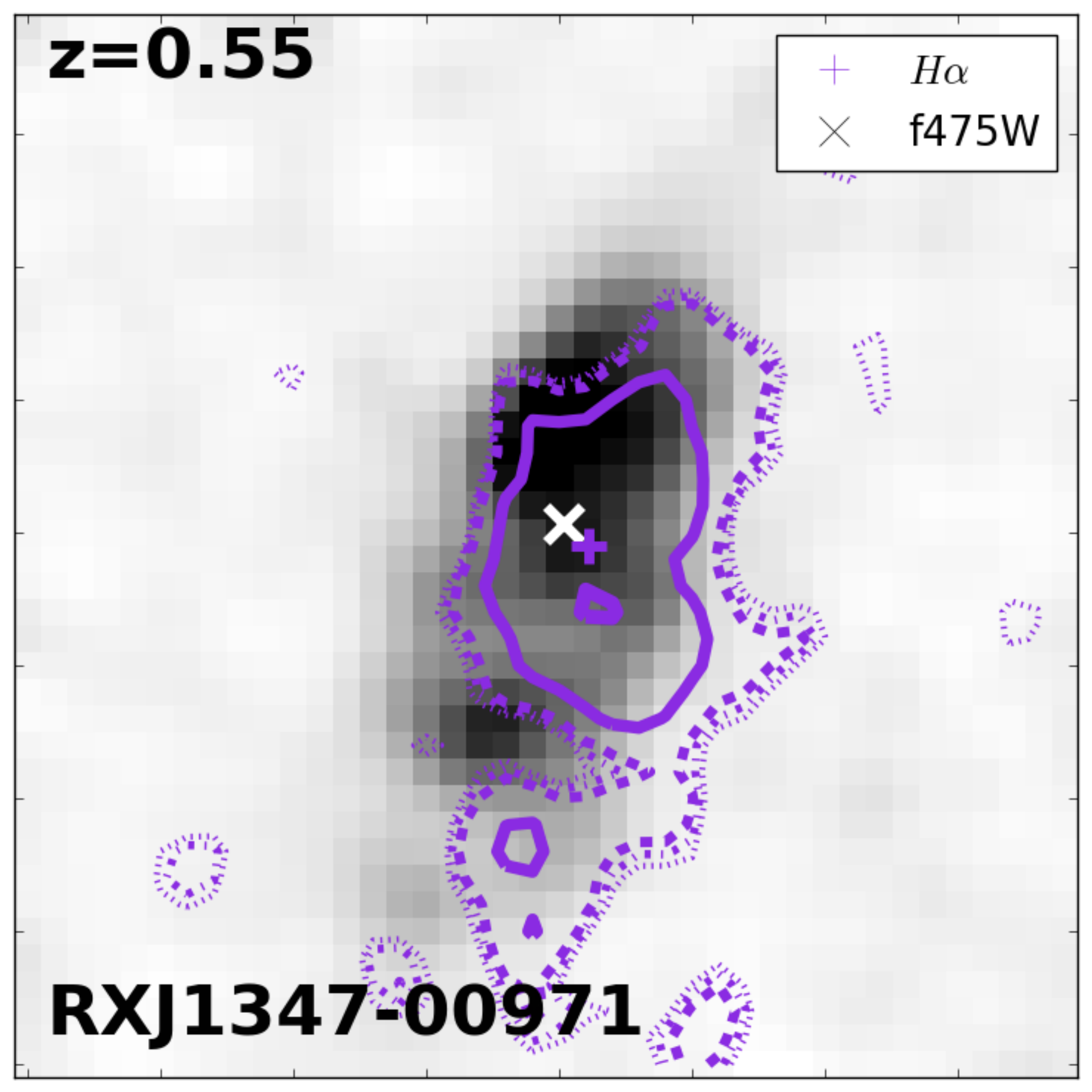}
\caption{Example of galaxies with spiral morphology and clumpy \Ha mainly due to minor mergers. Panels, 
colors, lines and labels are as in Fig. \ref{ell_ha}. 
 \label{spir_cl}}
\end{figure*}

\begin{figure*}
\centering
\includegraphics[scale=0.14]{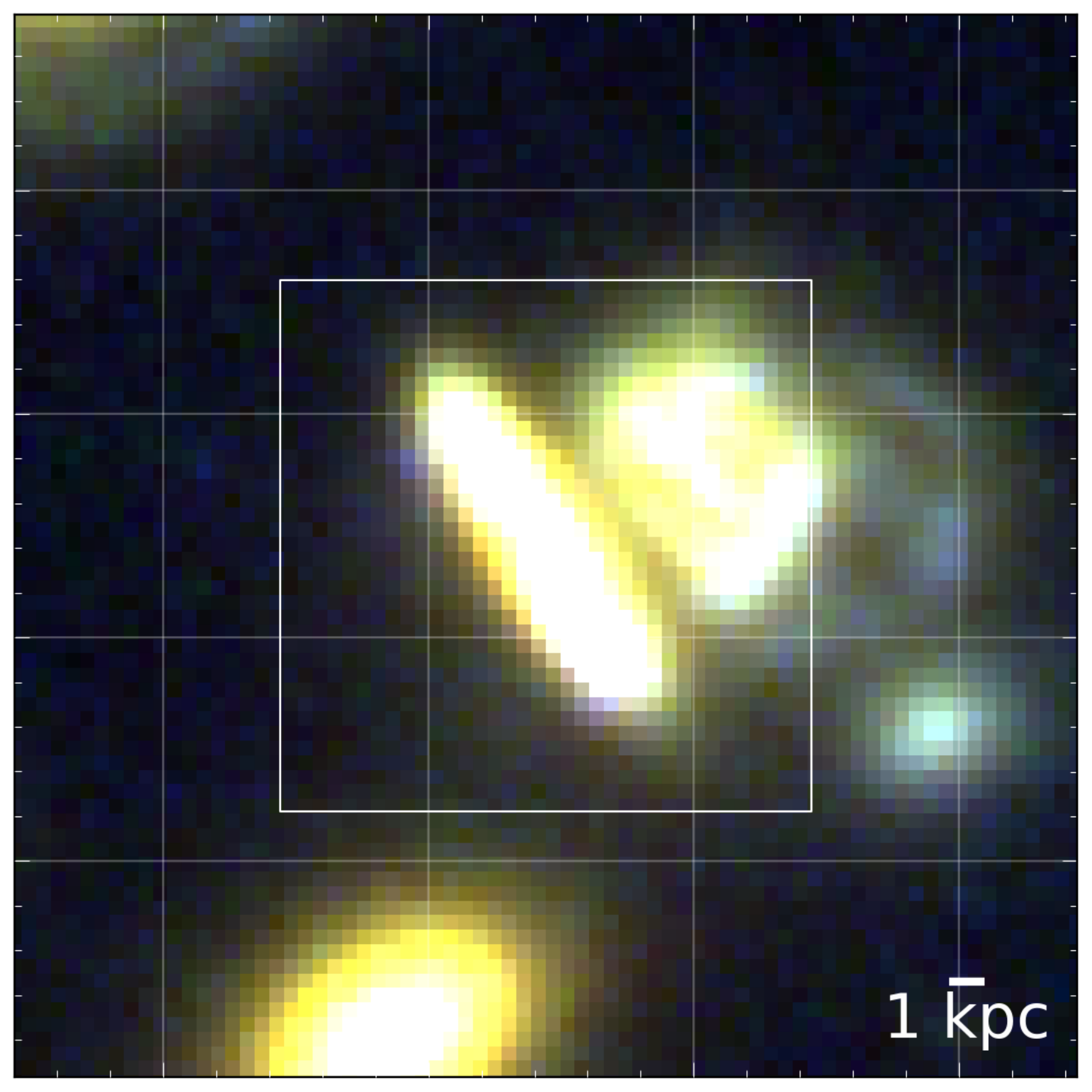}
\includegraphics[scale=0.14]{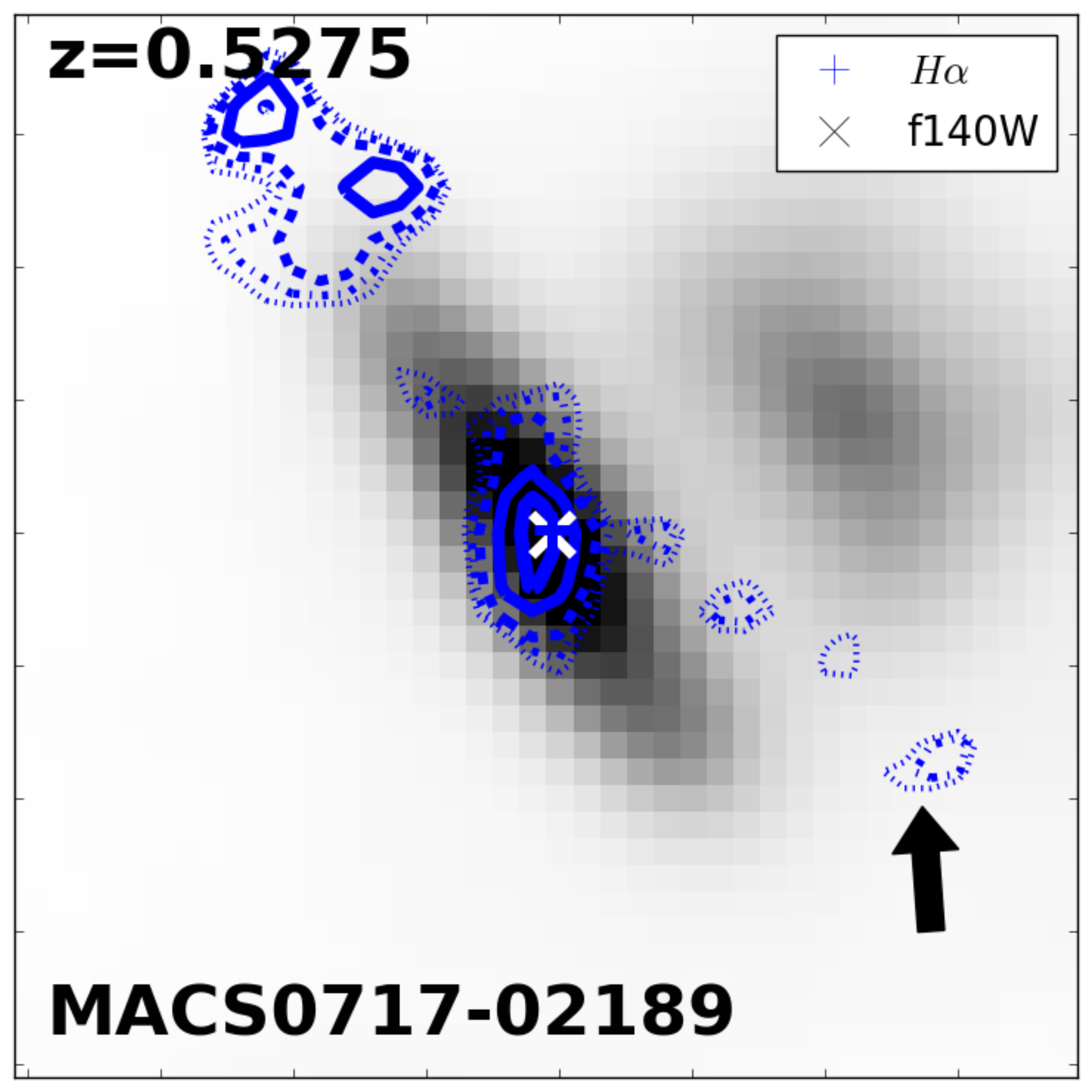}
\includegraphics[scale=0.14]{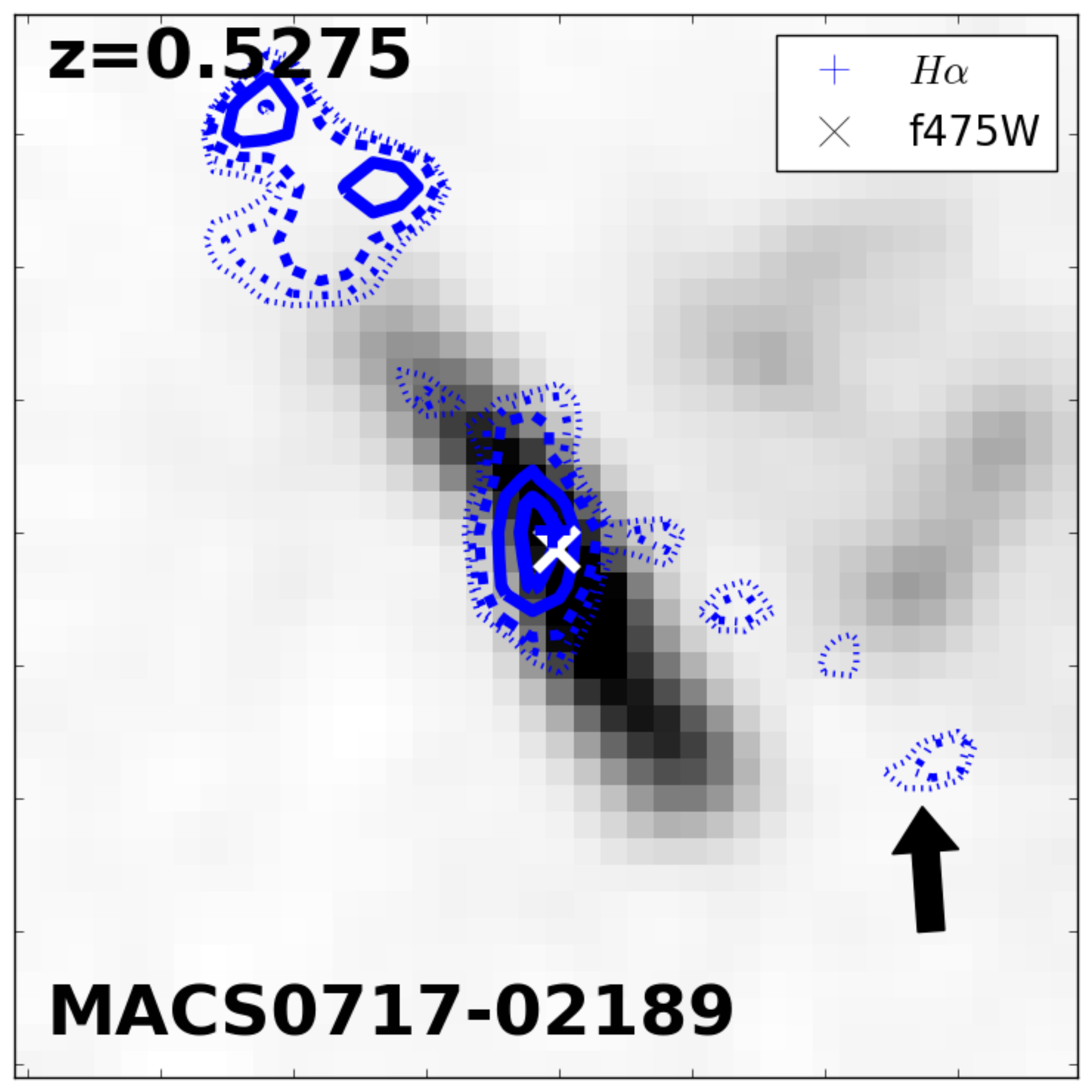}
\includegraphics[scale=0.14]{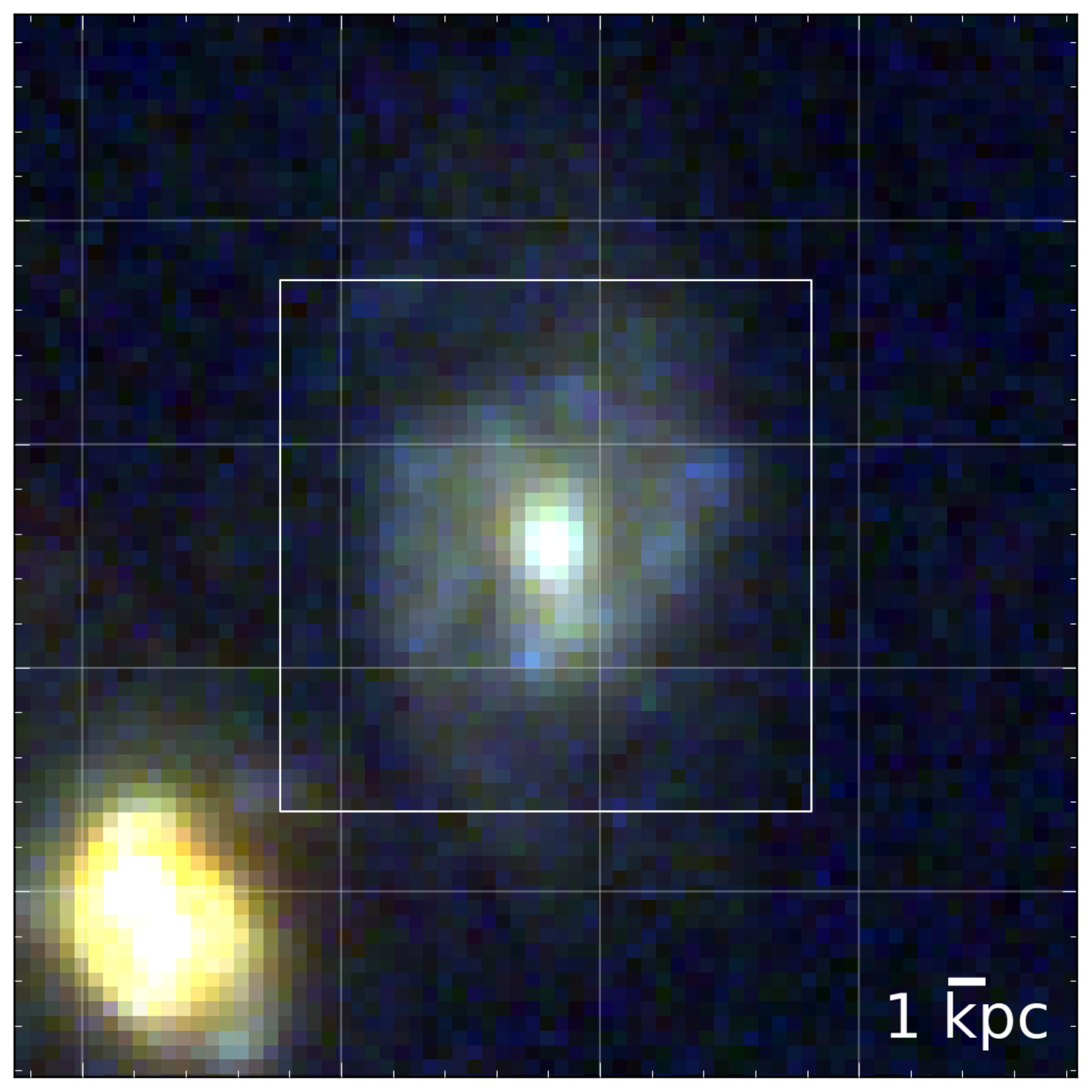}
\includegraphics[scale=0.14]{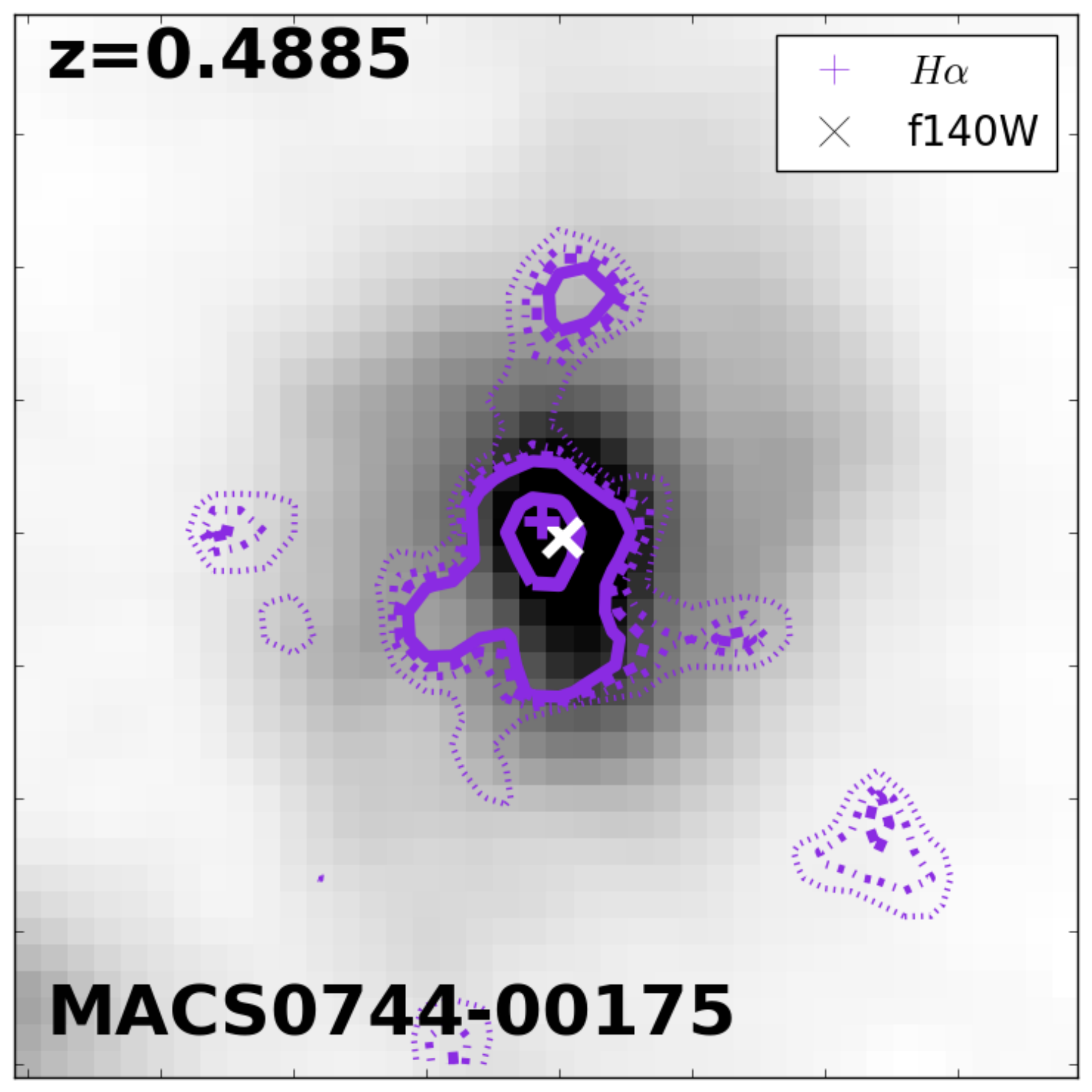}
\includegraphics[scale=0.14]{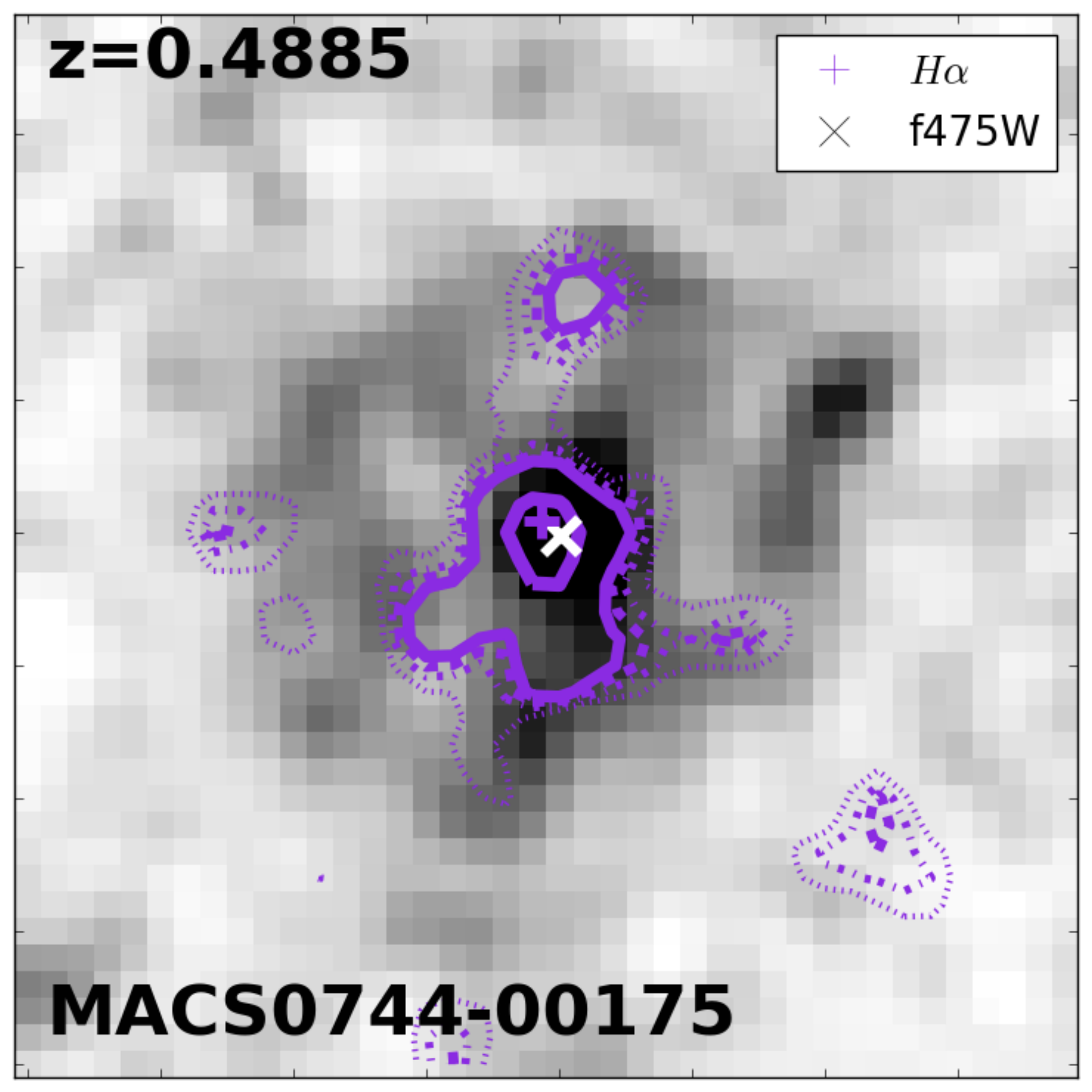}
\includegraphics[scale=0.14]{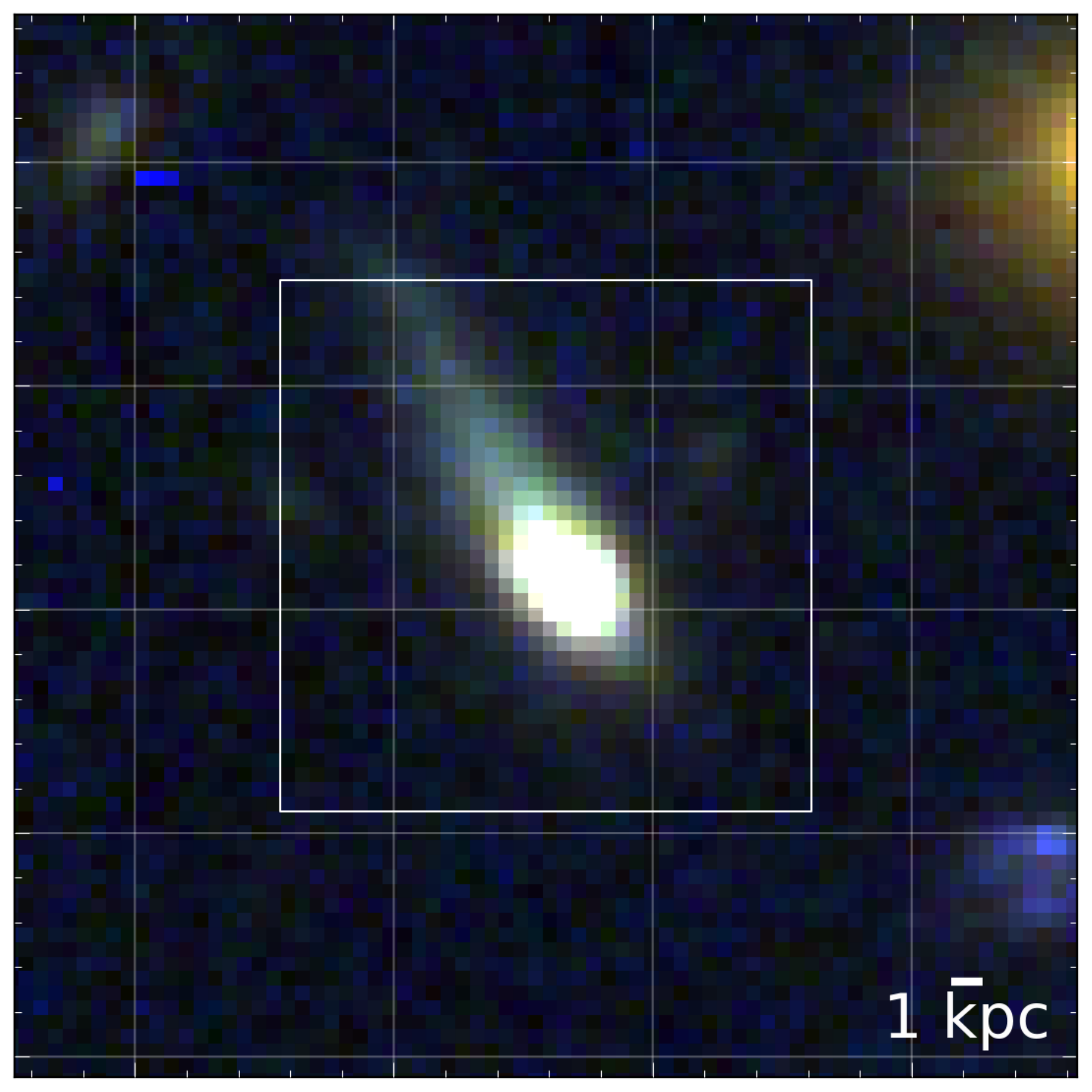}
\includegraphics[scale=0.14]{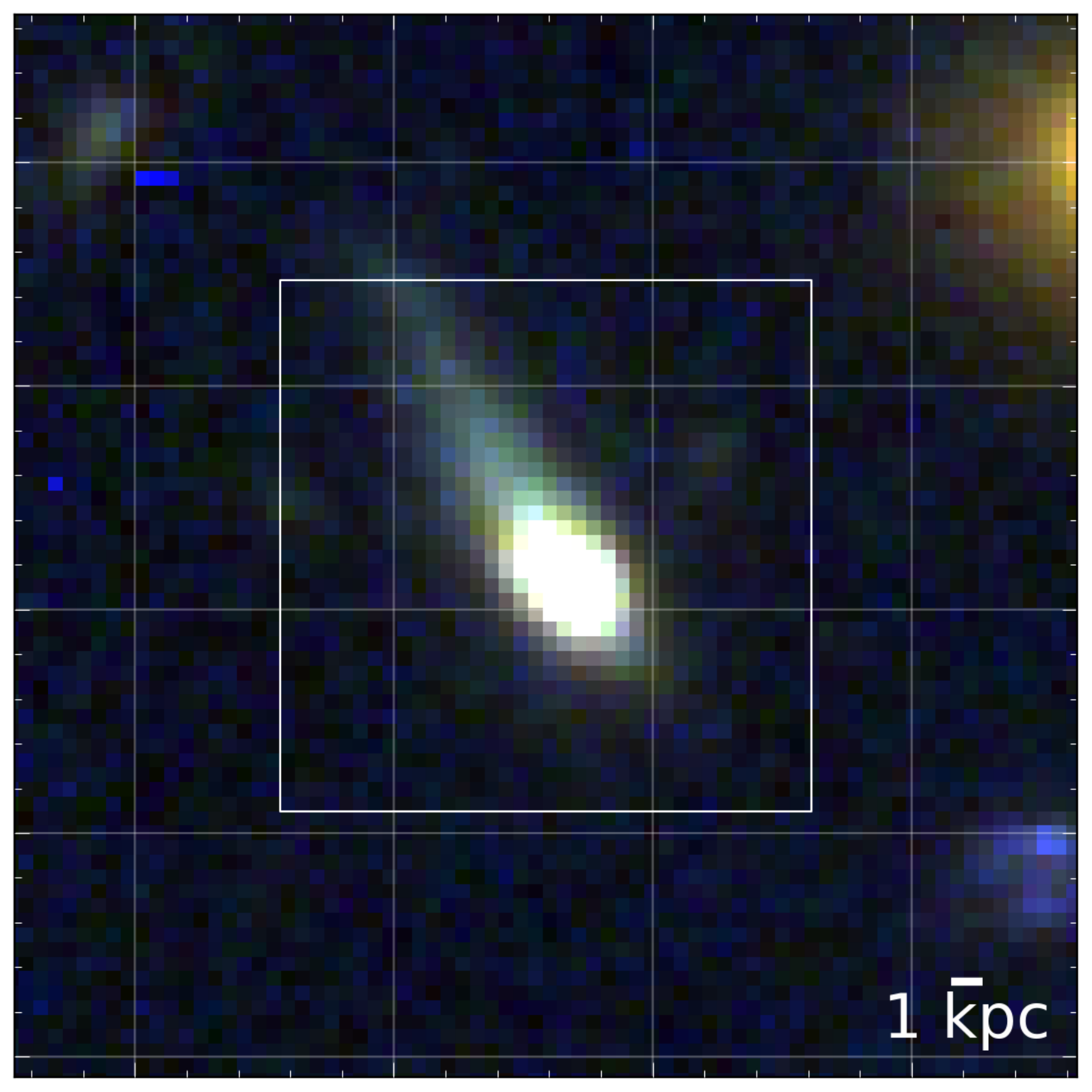}
\includegraphics[scale=0.14]{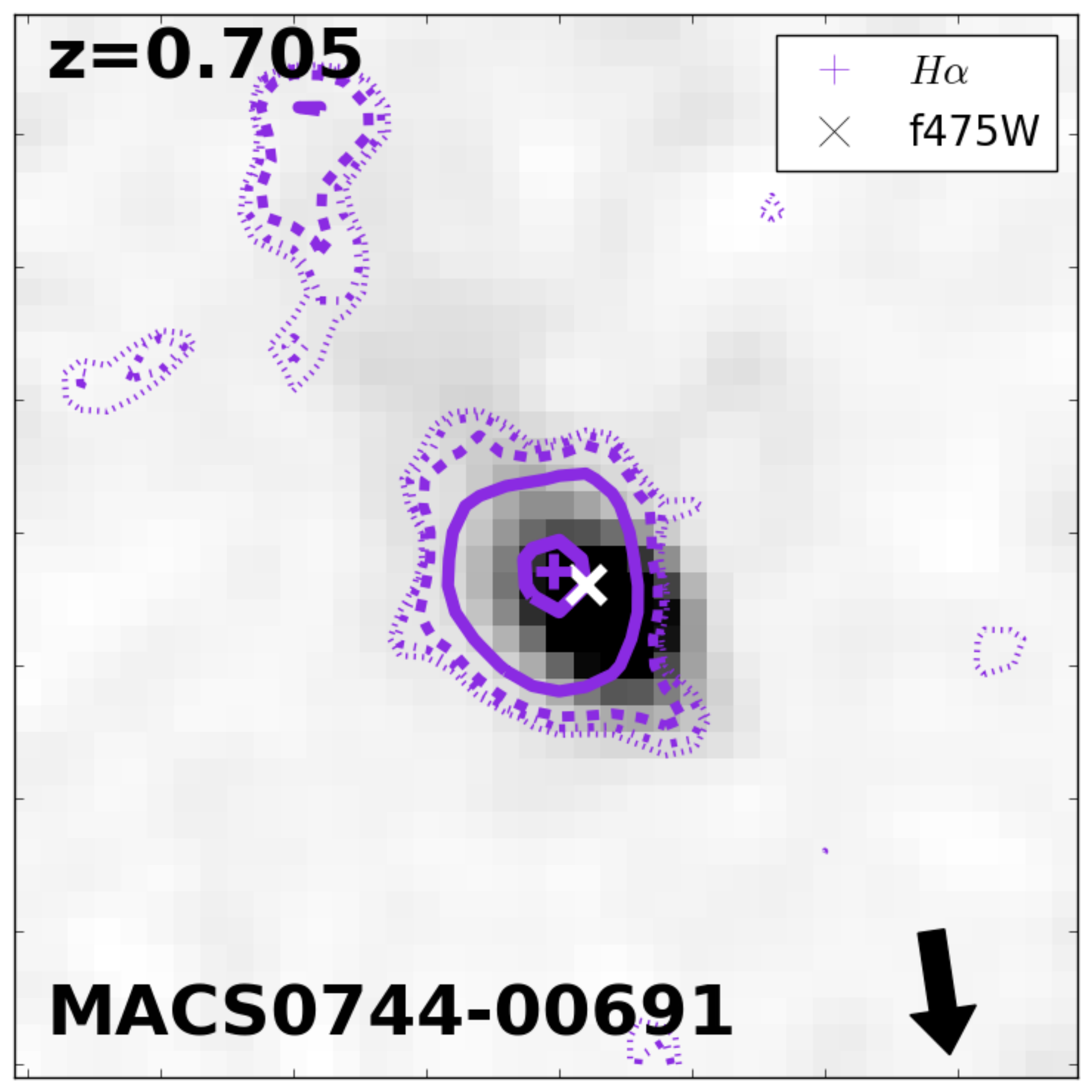}
\includegraphics[scale=0.14]{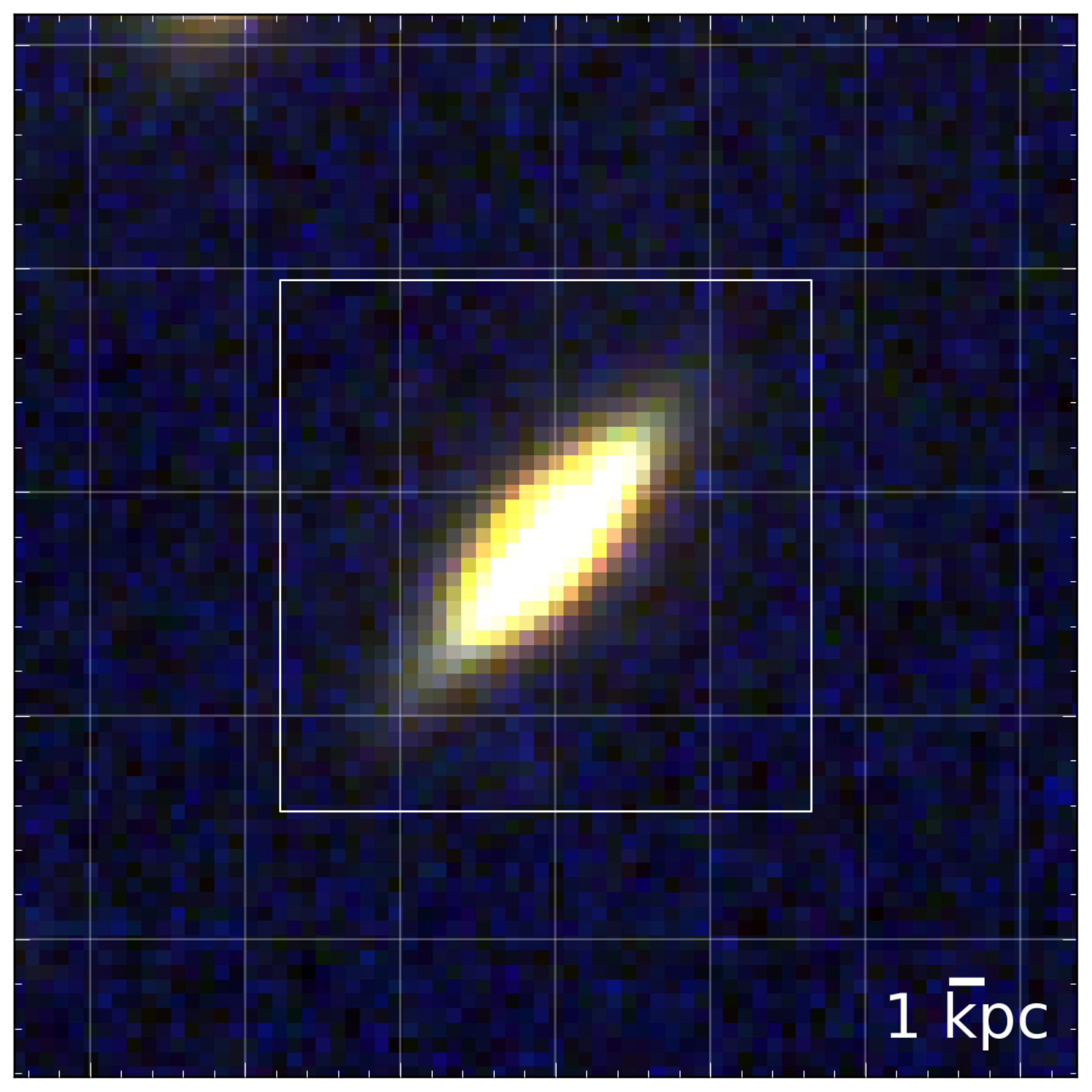}
\includegraphics[scale=0.14]{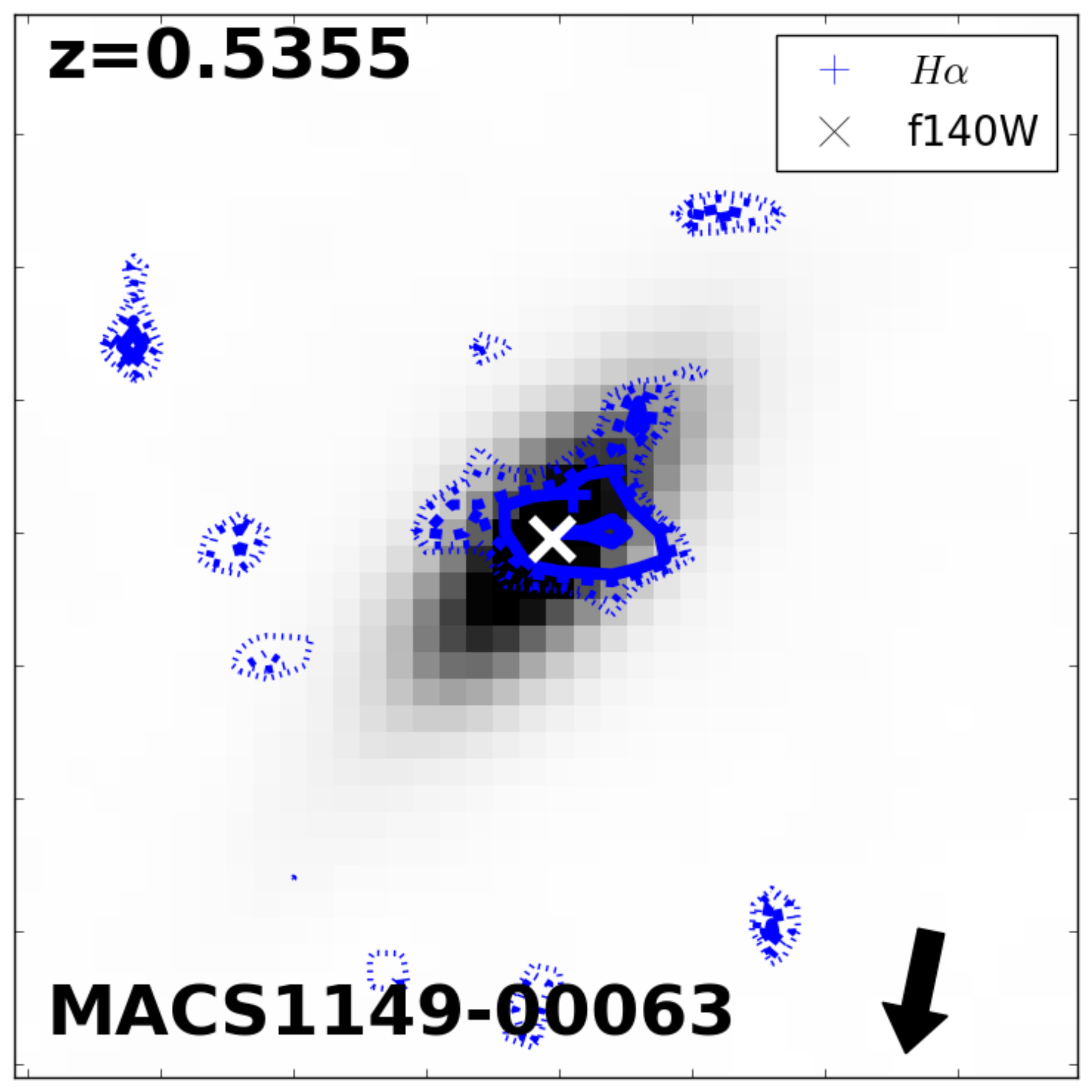}
\includegraphics[scale=0.14]{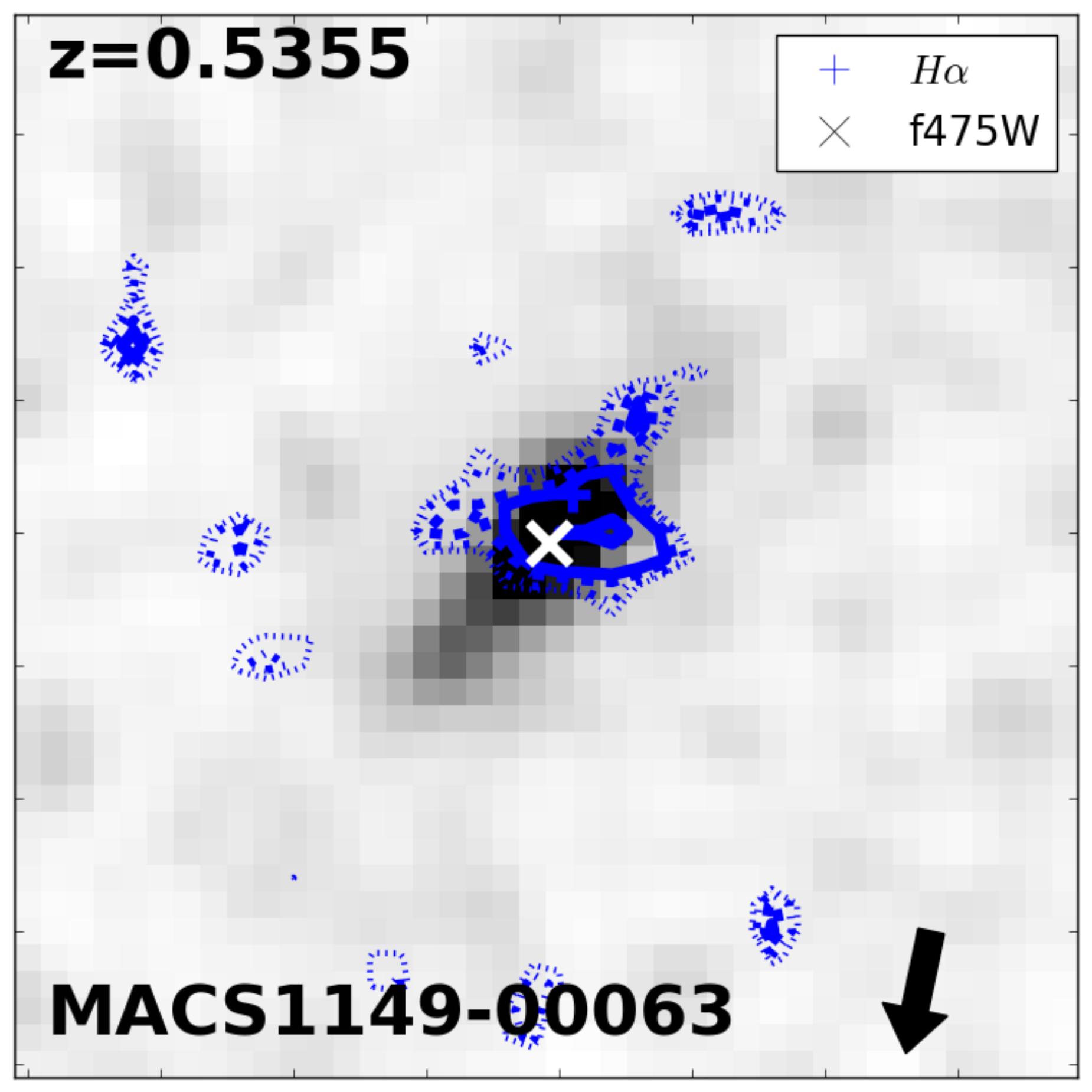}
\caption{Example of galaxies with compact centrally located \Ha. Panels, colors, lines and labels are as in Fig. \ref{ell_ha}. 
 \label{conc}}
\end{figure*}

Spiral galaxies are the most common class of \Ha emitters, in both environments. In clusters, $\sim 35\%$ of them show 
asymmetric \Ha maps, corresponding to  15\% of the total number of galaxies. In 
the field, 15\% of all spirals show asymmetric \Ha morphologies corresponding to 8\% of all field \Ha emitters. 
In clusters, in  65\% of the cases the asymmetry has been classified as consistent with ram pressure stripping, in the field this number is less than 30\%. 
Figure \ref{spir_as} shows some examples. Most of them can be characterized as jellyfish galaxies,  i.e. galaxies that
exhibit tentacles of material that appear to be stripped from the galaxy, and whose morphology is 
suggestive of non-gravitational removal mechanisms, such as ram pressure stripping 
\citep[e.g.,][]{fumagalli14, ebeling14}. Upon visual inspection, some of our galaxies really seem to be caught in the act of being 
stripped and are currently losing gas and material, producing a tail,
which extends in the opposite direction of the cluster center. The
bending of the \Ha disk (e.g. for MACS0717-02334) might suggest that
the galaxy is plunging into the ICM. We refer the
reader to Paper VIII for a more detailed investigation
of the relation between the \Ha morphology and cluster
properties. Interestingly, there are other galaxies
(e.g. RXJ2248-00104) that seem to have already lost most of their \Ha
disk, which appears to be truncated. RXJ2248-00104 is a good example
of multiple mechanisms as a merger, in addition to ram pressure
stripping, might be operating.

\subsubsection{Spiral galaxies with clumpy \Ha due to mergers}

In the field, almost 50\% of spiral galaxies present a clumpy \Ha
morphology, while in clusters this fraction is less than 20\%.  In the
field, in 80\% of the cases this \Ha morphology is attributed to the
occurrence of minor mergers. Figure \ref{spir_cl} shows some examples
in the field. For almost all of these galaxies the F475W filter shows
the presence of material infalling onto the main galaxies. This
material is not detected in F140W, suggesting that, though luminous,
it is not very massive. Therefore we classified it as minor mergers.

The median SFR of these galaxies is $\sim3 M_\sun \, yr^{-1}$, the median mass is $10^{9.9} M_\sun$.

\subsubsection{Galaxies with concentrated \Ha}

The last class of objects that we single-out for discussion is that made up of galaxies with very concentrated \Ha. Galaxies of all 
morphologies enter this category, both in clusters and in the field, except 
 that no S0s with concentrated \Ha have been identified in the field. In clusters (field), regular diffuse star formation seems to be ongoing for  35\% 
(45\%) of the objects. The \Ha disk seems fairly regular, though much smaller than the galaxy size, perhaps indicative of some other gentle process like strangulation, that is removing the most loosely bound  gas. 
In the other cases, both minor and major mergers seem to induce a 
concentrated \Ha morphology. Figure \ref{conc} shows some examples. It appears evident that some 
emission is more clumpy, some  very asymmetric. 

\subsection{Offsets between \Ha and the galaxy in the continuum}

\begin{figure*}
\centering
\includegraphics[scale=0.4]{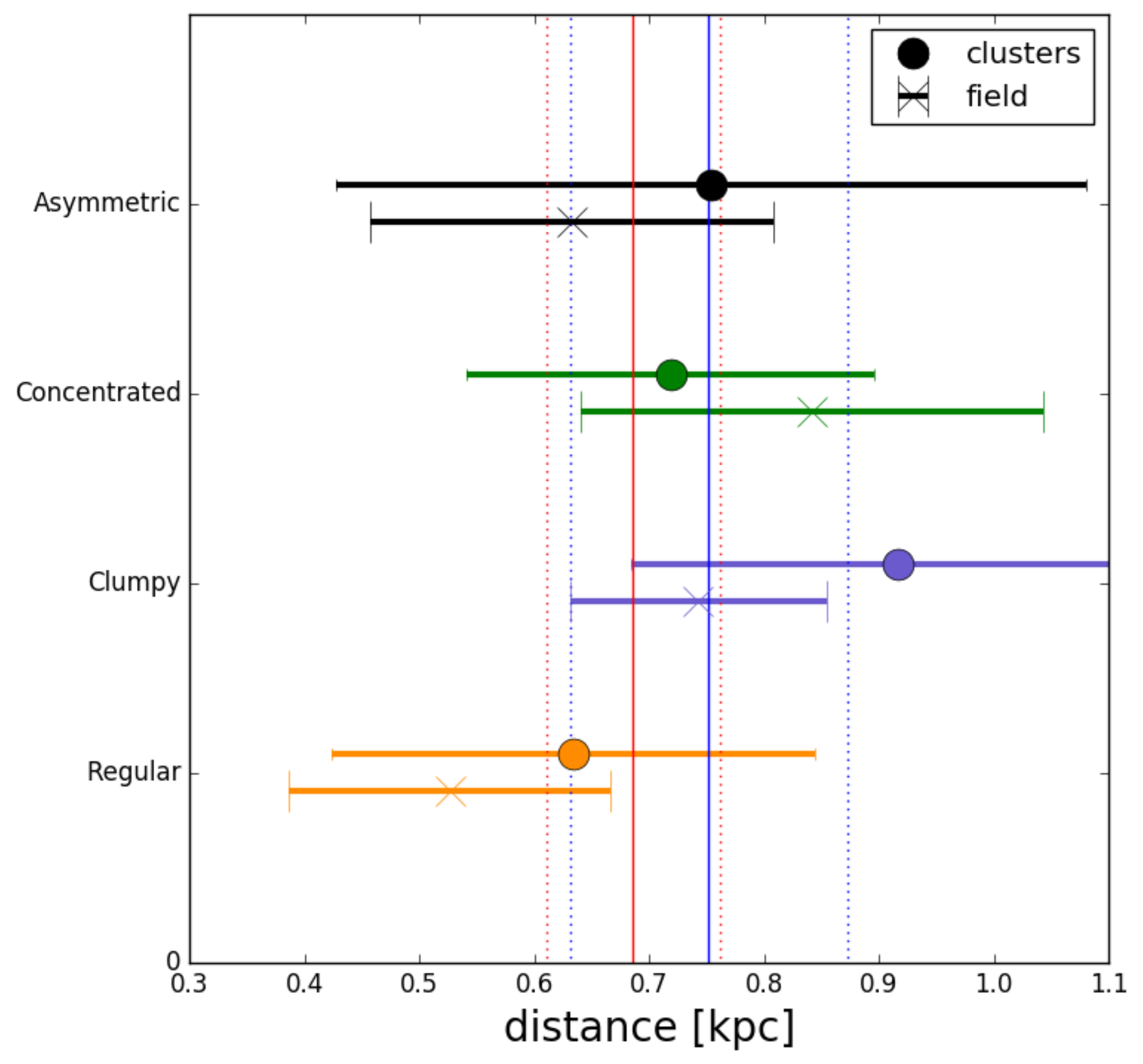}
\includegraphics[scale=0.4]{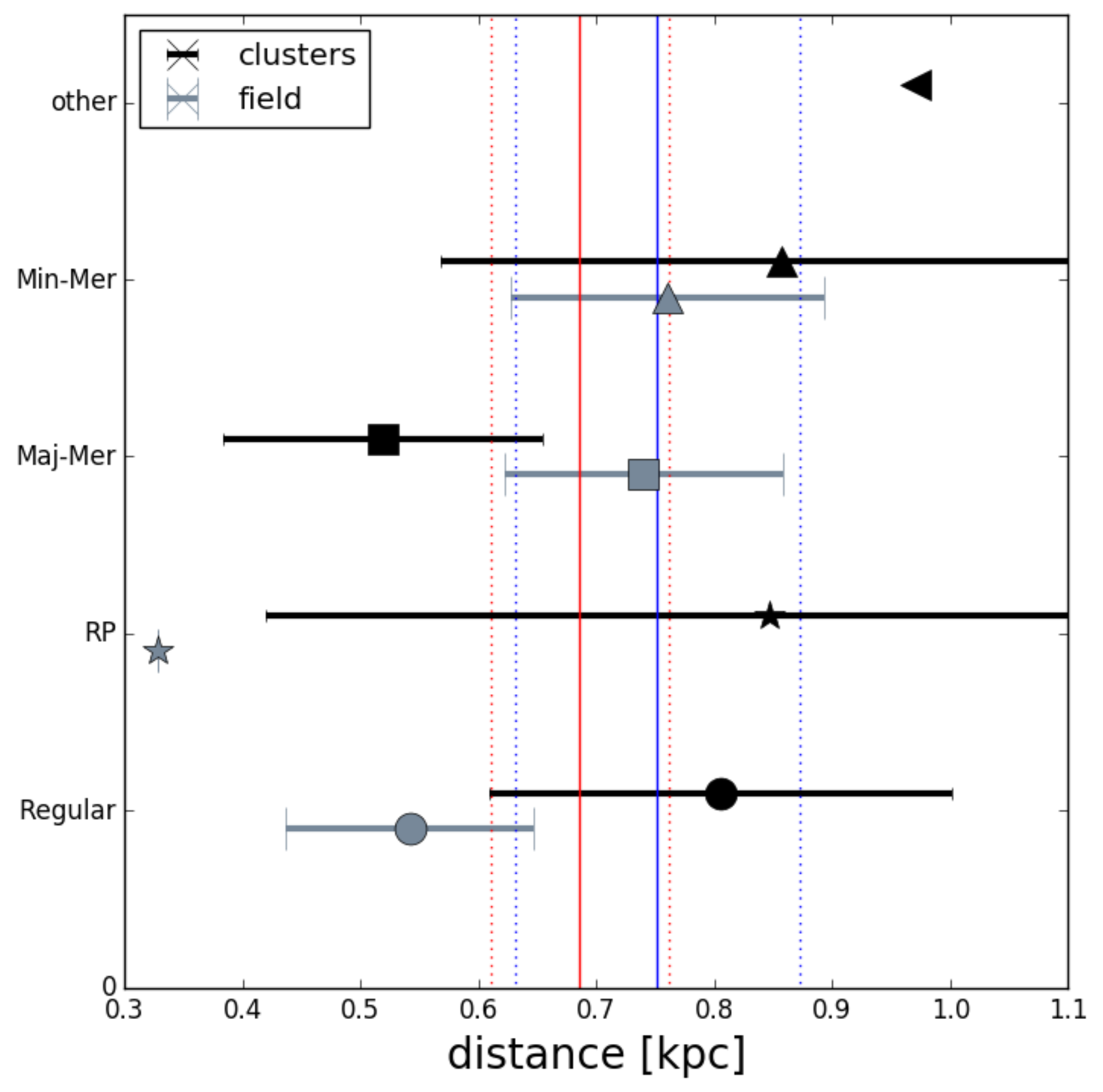}
\caption{Mean values along with errors of the distance between the peak of the \Ha emission and the continuum (F475W filter), for galaxies with both PAs. 
Left panel: galaxies in the field and clusters with different \Ha morphologies, as indicated in the label. Right panel:
galaxies in the field and clusters experiencing different processes, as indicated in the label.
Vertical solid and dotted lines represent the 
means with errors for the total populations (blue lines for clusters, red lines for the field). 
\label{fig:offset_Ha}}
\end{figure*}

In the analysis above, when comparing the \Ha emission to the image of
the galaxies in the continuum, we have neglected an important aspect:
in many cases there seems to be an offset between the peak of the \Ha
emission and the peak in the continuum.

Comparing the position of the peak of the \Ha
emission to that of the continuum, as traced by the
F475W,\footnote{For A2744 the F435W is used instead.}  
we found that in both environments, for most of
the galaxies the displacement is smaller than 1.5 kpc (plot not shown). The average
offset is $\sim 0.5$ kpc (typically 0.05$^{\prime\prime}$).  
We note that this offset is larger than any potential uncertainty in the astrometry for the different bands, 
which is of the order of a fraction of a pixel. 
For reference at the redshifts consider here, 1 pixel corresponds roughly to 0.2-0.3 kpc.
The existence of the offset
suggests that in most galaxies the bulk of the star formation is not
uniformly diffused. Past and current star formation are not co-located, as for example assumed 
by \cite{nelson12, nelson13, nelson15}.

These
results confirm our previous findings, based only on two clusters
\citep{vulcani15}.

Figure \ref{fig:offset_Ha} gives the mean values of the distance between the peak of the \Ha emission and the continuum (F475W filter), for galaxies
showing different \Ha morphologies (left) and possibly feeling different physical processes (right). Only galaxies with both PAs are considered. 
Similar results are obtained for the two offsets separately. 
Values are in agreement within uncertainties, but there are hints that e.g. galaxies with regular \Ha morphologies have smaller offset than galaxies
with a clumpy morphology (especially in clusters). In clusters, galaxies undergoing ram pressure events tend to have larger offset,  as also merging galaxies. 
Comparing the two environments, no robust systematic differences are detected. 

\subsection{Star formation rates}

\begin{figure}
\centering
\includegraphics[scale=0.45]{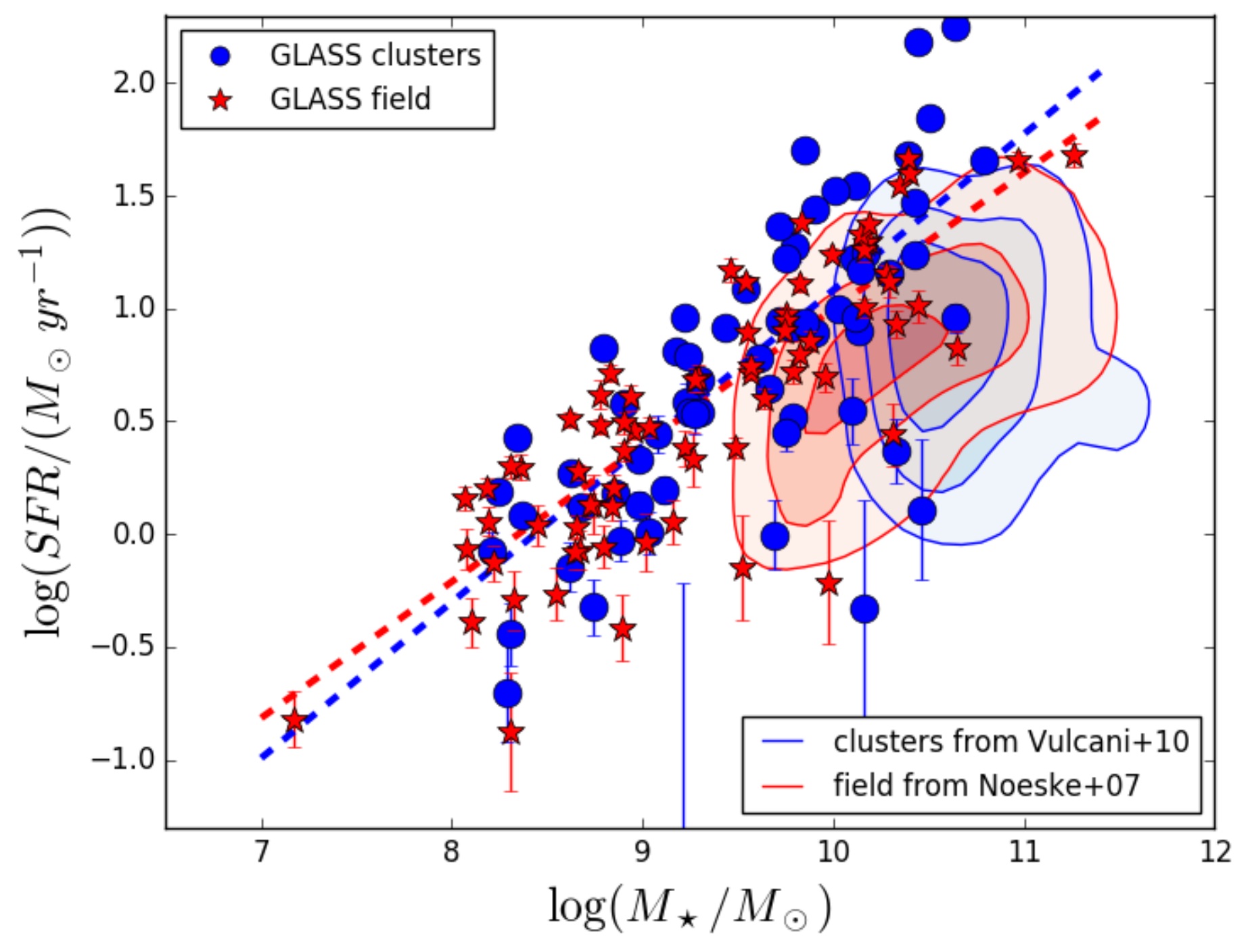}
\caption{
GLASS  SFR-mass relation over plotted to the field relation \citep[from][]
{noeske07} and the cluster relation at similar redshift  \citep[from][]{vulcani10}, both at similar redshift.
Blue filled circles and dashed blue line: GLASS cluster galaxies and fit; red filled stars and dashed red line: GLASS field galaxies and fit. The typical 
error on stellar masses is 0.2 dex and not shown for clarity.  Blue contours: EDisCS clusters, red 
contours: AEGIS field. For EDisCS and AEGIS, only blue emission line galaxies 
and galaxies detected at 24$\mu m$ above the mass and SFR completeness limits have been considered \citep[refer to][for details on the sample 
selection]{noeske07, vulcani10}. \label{fig:SFR_mass}}
\end{figure}

\begin{figure*}
\centering
\includegraphics[scale=0.48]{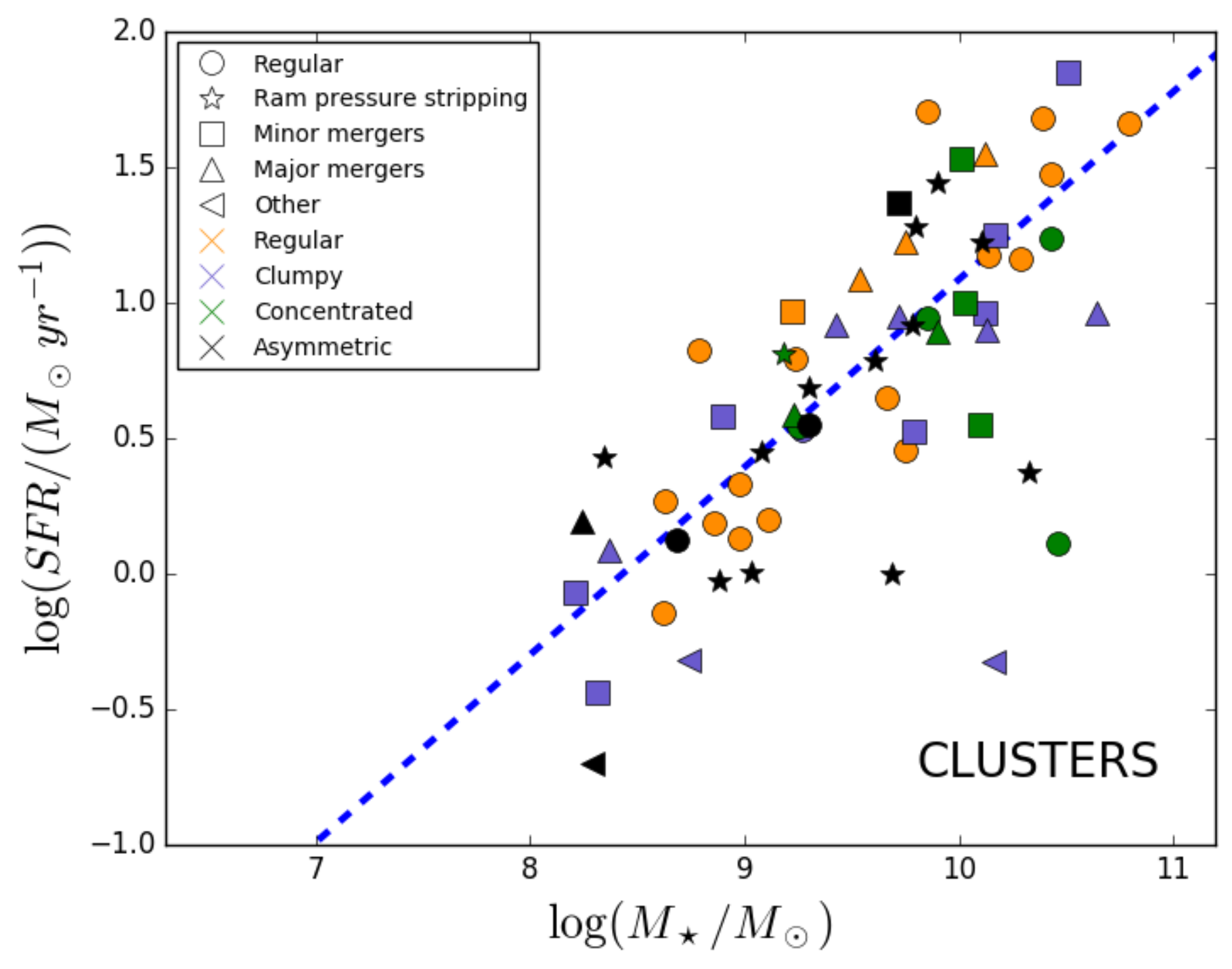}
\includegraphics[scale=0.48]{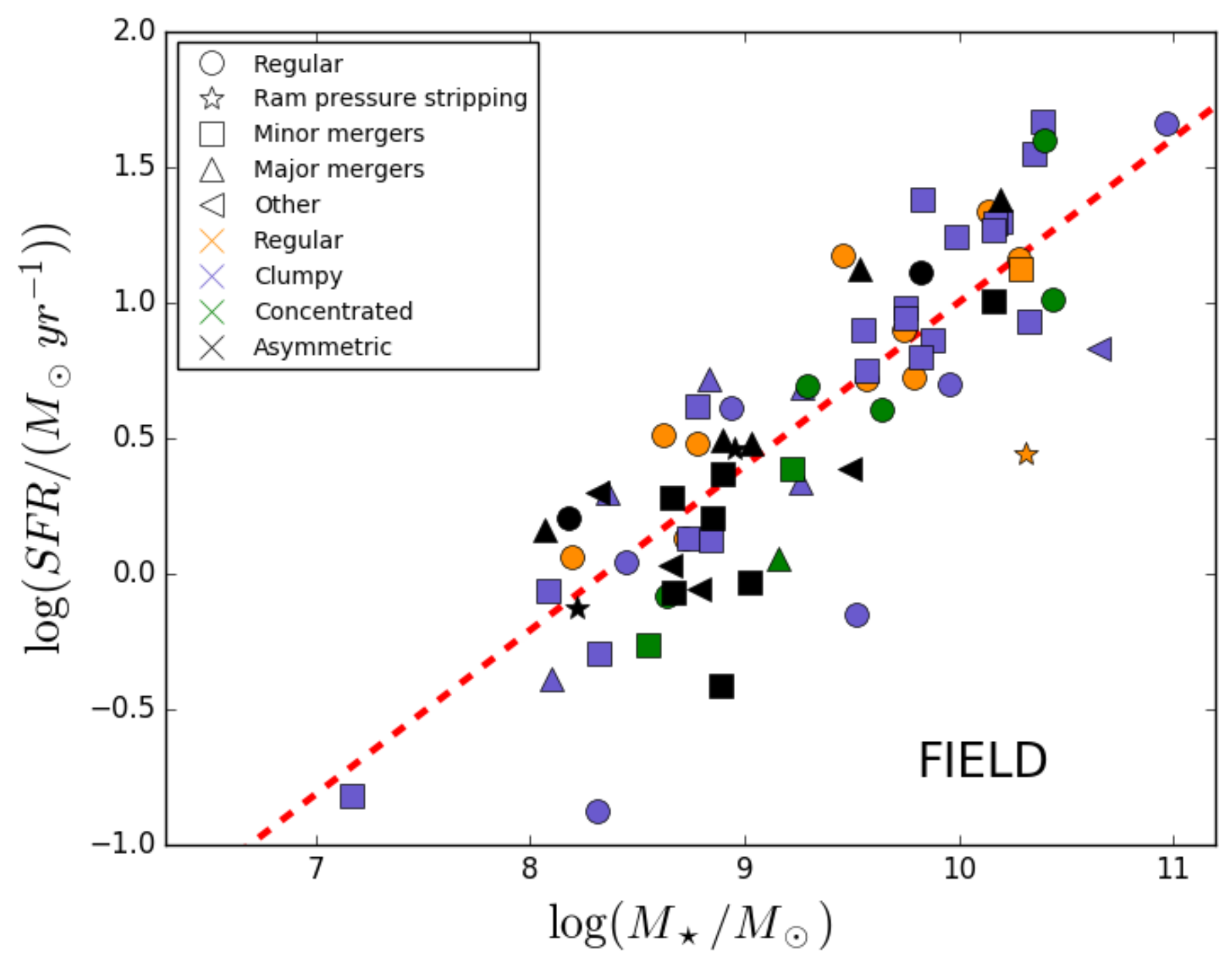}
\caption{
SFR-mass relation for galaxies of different
\Ha morphology and subject to different physical processes in clusters (left) and in the field (right). For the sake of clarity, error bars are not shown. The different symbols and colors are summarized in the label. 
Dashed lines represent the best fit relation for clusters (blue, left) and field (red, right). No clear trend between \Ha morphology and SFR: galaxies with comparable values of 
SFR can have very different \Ha morphologies.
\label{fig:SFR_mass_morphs}}
\end{figure*}

We can  relate the results obtained so far to the position of the galaxies on the SFR-mass plane, to 
investigate whether we can establish a link between the specific star formation rate (or the SFMS) and environmental processes.
Figure \ref{fig:SFR_mass} shows our cluster and field galaxies, overplotted the AEGIS \cite{noeske07} 
 field galaxies and  the EDisCS \cite{vulcani10}  
cluster galaxies, both  at $z\sim0.5$. 

The GLASS samples span a wide redshift range ($0.3<z<0.7$), therefore the spread is 
expectedly due to the evolution of the SFR-M$_\ast$ relation with $z$, but our sample is too small to investigate evolutionary trends. 

The  \cite{noeske07} sample is complete down to $\sim M\ast >10^{9.7}$ and $\log SFR>0$, the \cite{vulcani10} sample down to $\sim M\ast >10^{10.6}$ and similar SFRs. 
Therefore, the overlap between the GLASS and literature samples is limited. 
The GLASS galaxies tend to lay on the  SFR-mass relation of blue galaxies with 
emission lines or detected in the Infrared \citep[see][for details on their 
sample selection]{noeske07, vulcani10}, tracing the upper envelope. To some extent, this was expected 
given that our sample has been assembled by  selecting visually detected \Ha emitters, and thus preferentially rapidly star forming galaxies.  

Furthermore, it is important to note that the comparison with the
literature is not straightforward. \cite{noeske07, vulcani10} used long
slit spectroscopy, their SFR estimates are underestimated for extended
galaxies, while GLASS, making use of slitless grism spectroscopy,
gives a more reliable estimate for extended objects.

Our results  
show that at $0.3<z<0.7$ the vast majority of cluster galaxies can be as star forming as field galaxies of similar mass.  
In the literature, e.g.  \cite{patel09, vulcani10, koyama13, paccagnella16} have detected both at these redshift and at $z\sim0$ 
the existence of  a population of galaxies with a reduced SFR 
at fixed stellar mass  which is absent in the field. These galaxies are most likely in transition from being 
star forming to being passive. Probably due to our limited number statistic, we are not able to 
firmly single out this population, even though few galaxies in our clusters indeed show a reduced SFR 
given their mass.

In order to investigate the connection between process and star
formation, we then investigate whether galaxies with different \Ha
morphologies classified as experiencing different processes populate
specific regions of the SFR-mass plane.

From broad-band morphology (plot not shown), we find that the
different classes are almost normally distributed around the fit, with
elliptical galaxies presenting a large dispersion. Cluster galaxies in
the low SFR tail are spirals, while two of them are ellipticals.

Figure \ref{fig:SFR_mass_morphs} presents the SFR-mass relation for  galaxies of different
\Ha morphology  subject to different physical processes in the two environments separately 
We note that there is no clear trend between \Ha morphology and SFR: galaxies with comparable values of 
SFR can have very different \Ha morphologies. Nonetheless, some tentative trends emerge. 

Interestingly, cluster galaxies classified as undergoing ram pressure stripping (stars) seem to be located both above and below the SFR-mass relation, 
showing how star formation is either  enhanced in these systems \citep[in agreement with][]{poggianti16}, or suppressed, when 
we are witnessing a late stage of the phenomenon. As already mentioned, most of 
these galaxies show an asymmetric \Ha morphology and some of them look like jellyfish galaxies. 

Major mergers (triangles) seem to  induce an enhancement of the star formation, in clusters and also in the field. 
Overall, cluster galaxies in the tail of low SFR have \Ha concentrated and are produced by not identified (``other'') processes.  
In clusters, galaxies with \Ha concentrated (green symbols) are found only for $M_\ast>10^9 M_\ast$. Galaxies showing a regular \Ha morphology (circles) and experiencing regular processes (orange symbols) tend to be located above the best-fit of SFR-Mass relation. 

Even though we have used arguably one of the most sensitive diagnostic
tool to try and interpret the SFR-mass relation, we could not find any
clear trend within the SFR mass plane. It is possible that any trends
within the plane are obfuscated by uncertainties in our classification
scheme or perhaps our sample is not large enough. However, if the effects
on environmental processes on these integrated measures had been
dramatic we should have been able to see them. The lack of clear
trends is thus consistent with the idea that environmental process
play a secondary role in establishing the star formation rate for a
galaxy of a given mass.

\cite{nelson15} have also related the spatial extent of the \Ha distribution and stellar mass, by stacking \Ha images to reach deep surface brightness limits ($\sim 10^{-18} erg \,  s^{-1} \, cm^{-2}\, arcsec^{-2}$) for field galaxies at $z\sim1$. 
They mapped the \Ha distribution as a function of SFR(IR+UV) and found that above the main sequence \Ha is enhanced at all radii; below the main sequence \Ha is depressed at all radii. This suggests that at all masses the physical processes driving the enhancement or suppression of star formation act throughout the disks of galaxies. For $10^{10.5} < M_\ast/M_\sun < 10^{11}$, above the main sequence, they found that \Ha  is particularly enhanced in the center, indicating that gas is being funneled to the central regions of these galaxies to build bulges and/or supermassive black holes. Below the main sequence, the star forming disks are more compact.

In contrast, \cite{willett15} have analyzed  the local SFMS of disk galaxies and found that it is remarkably robust to the details 
of the spatial distribution of star formation within galaxies. They classified galaxies  in a wide range of morphological sub-types, i.e.  
number or pitch angle of spiral arms, presence of a large-scale bar; 
but did not detect no statistically significant difference in the relative position of these sub-types across the SFMS.
They concluded that  system which regulates star formation in galaxies is thus either not affected by the details of the spatial distribution of star formation, 
or its regulatory effect is so strong that it wipes out any such effect in a short time.

\subsection{Comparison between sizes in different bands for spiral galaxies}\label{sec:spirals}

\begin{figure*}
\centering
\includegraphics[scale=0.28]{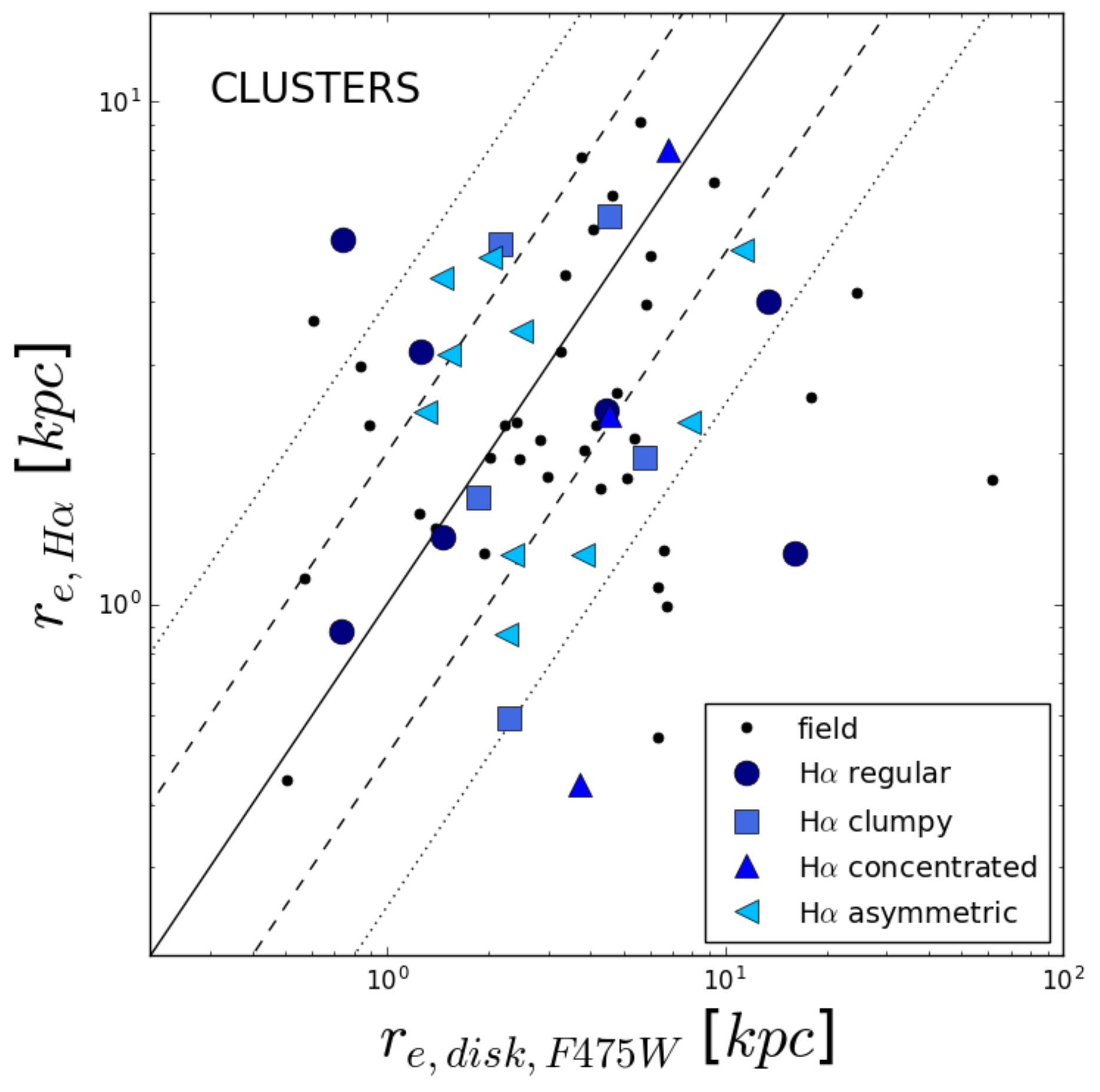}
\includegraphics[scale=0.28]{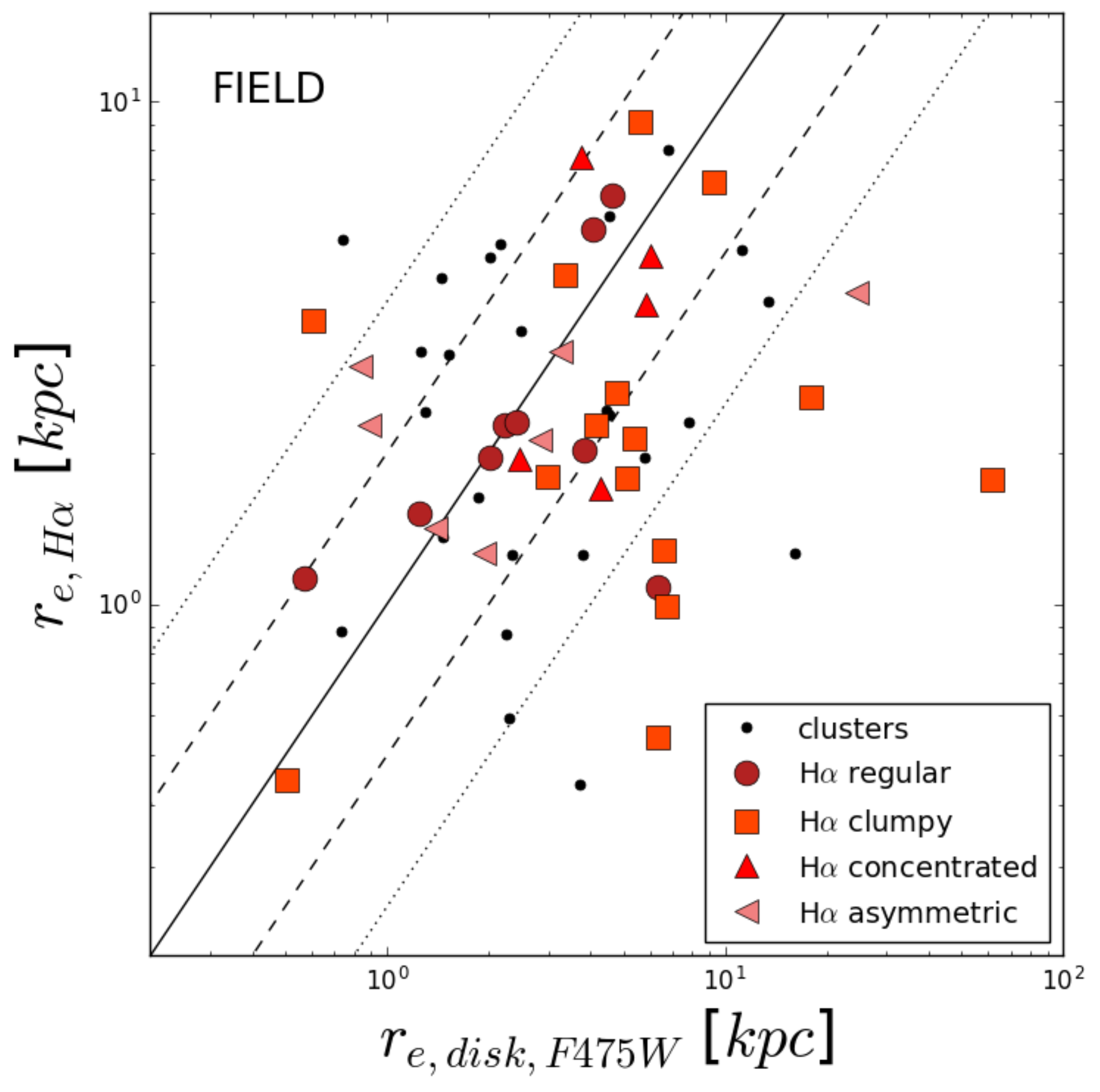}
\includegraphics[scale=0.28]{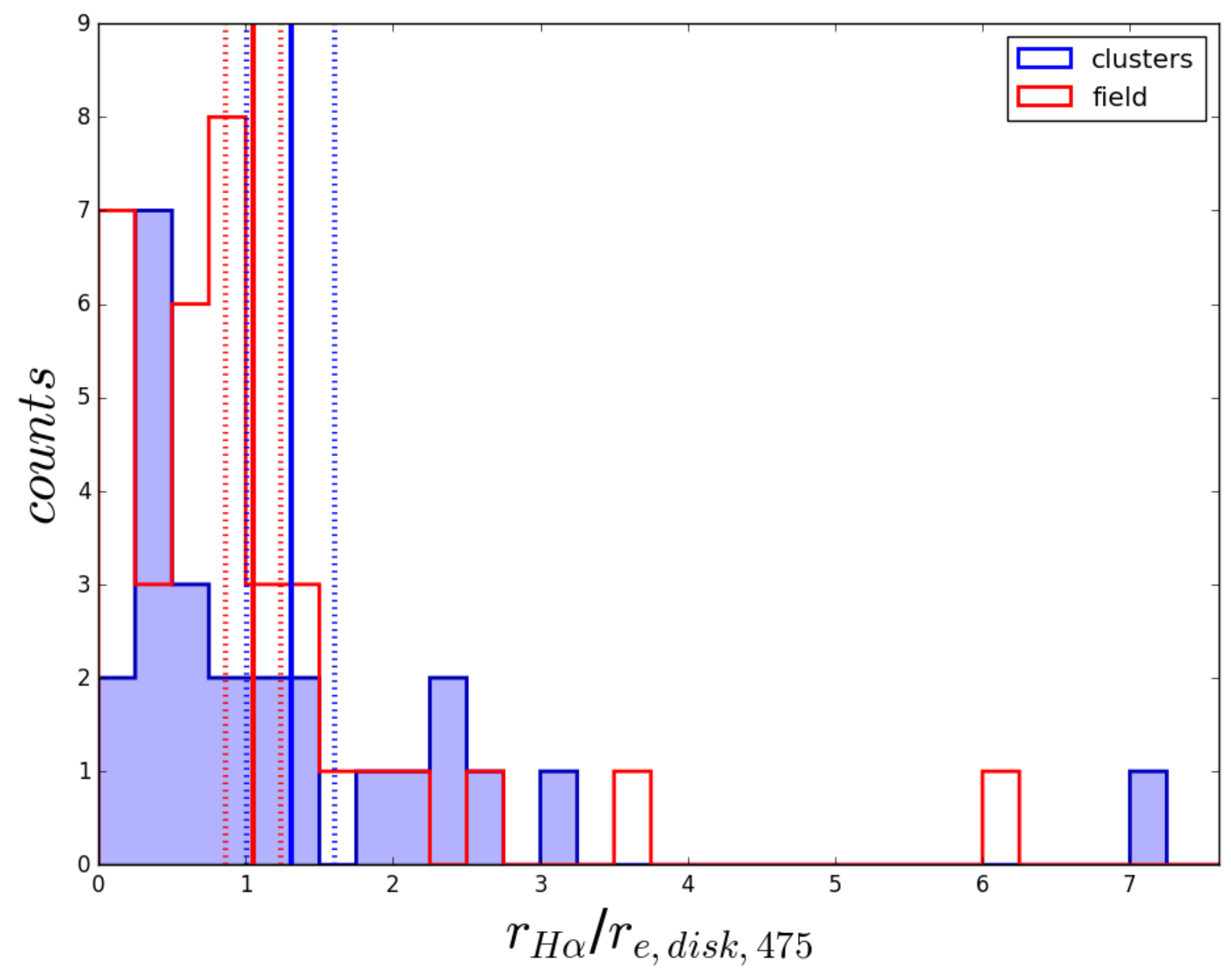}
\caption{Size comparisons for spiral galaxies in our sample, for which the assumption of the existence of 
the disk holds. Left panel: \Ha size vs. F475W size for cluster galaxies. Galaxies with different \Ha morphologies are highlighted, as indicated in the label. Lines show the 4:1, 2:1, 1:1, 1:2, 1:4 respectively. Small black dots are field galaxies, for comparison.   Central panel: \Ha size vs. F475W size for field galaxies. Galaxies with different \Ha morphologies are highlighted, as indicated in the label. Lines show the 4:1, 2:1, 1:1, 1:2, 1:4 respectively. Small black dots are cluster galaxies, for comparison.  Right panel:  distribution of the ratios of \Ha size to the F475W size, for galaxies in clusters (blue, filled histogram) and in the field (Red, empty histogram). Mean values along with 1-$\sigma$ errors are also indicated. The mean ratio between \Ha size and size in the continuum (F475W) is close to 1, in both environments, indicating that  \Ha traces the disk and suggesting that indeed star formation is mainly taking place in this component. Galaxies with different \Ha morphology are not strongly clustered in the size-size plane, even though concentrated objects tend to have smaller \Ha size, in both environments. In clusters, a population of galaxies with \Ha much larger than the size in the continuum tend to appear, maybe indication that this galaxies are currently ram pressure stripped. 
\label{fig:sizes_spirals}}
\end{figure*}

As discussed in Sec \ref{sec:sizes}, we have measured sizes for all the galaxies, assuming a double profile 
fit for the galaxies in the continuum, and a single profile fit with n=1 for \Ha 
maps. The former assumption is valid for galaxies with a regular morphology, such as ellipticals (which are 
supposed to have the second component negligible), S0s and spirals.
The latter assumption is valid under the hypothesis the \Ha is distributed across a disk. 
As we have seen so far, these assumptions are valid only for a fraction of galaxies in our sample, while the 
rest show non-canonical morphologies and host very irregular \Ha distributions. 

In what follows, we will consider only spiral galaxies and investigate whether in these objects \Ha actually 
follows the disk, indication that star formation is taking place in this 
component, or whether the two portions of the galaxies are uncorrelated. 
To do so, we compare the size of \Ha to the size of the disk of the galaxy, measured both from the F475W 
filter and the F140W filter. These two filters capture the light of two different 
stellar populations: the former is more sensitive to the younger population (SFR in the last million years), 
while the latter is more sensitive to the older stellar component. 
 
The left and central panels in Figure \ref{fig:sizes_spirals} compare the size of the \Ha disk to the size of the galaxy disk 
at  F475W for cluster and field galaxies separately. 
Galaxies with different \Ha morphology are also highlighted. The right panel  shows the distribution of the ratio of sizes
at the different wavelengths. 

Cluster and field galaxies occupy slightly different regions of the plane:  field galaxies seem to have systematically smaller \Ha sizes compared to the
continuum sizes, while in clusters there are  outliers with \Ha size much larger than the continuum size.
These cluster galaxies might be likely experiencing ram pressure stripping which is puffing up the \Ha disk (in agreement with our results presented in \citealt{vulcani15}).

Mean values
peak around one, indicating that  \Ha is approximately on the same scale of the disk, indicating that most of the star formation indeed is occurring in the disk as assumed.

Interestingly, the \Ha size is not strongly correlated to the \Ha morphology: while some systematic differences are detected (e.g. if \Ha is concentrated the \Ha size is smaller)
these are all within the observed scatter.

Using the F140W filter instead of the F475W, no strong differences are detected (plots not shown), except that sizes in the continuum are systematically smaller, as expected
given that we are looking at an older stellar population and the known color gradients in galaxies.
\begin{figure*}[!t]
\centering
\includegraphics[scale=0.45]{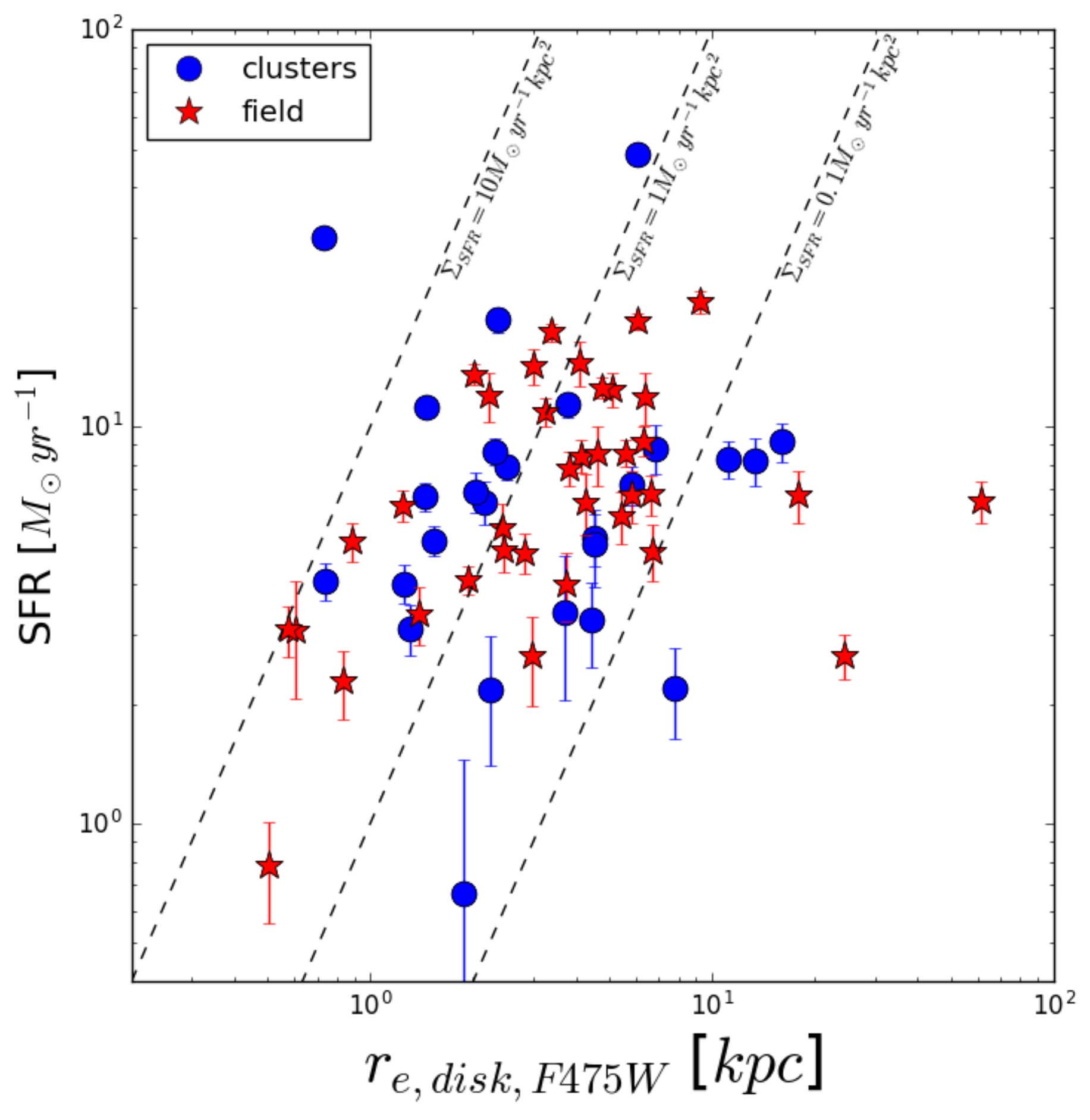}
\includegraphics[scale=0.45]{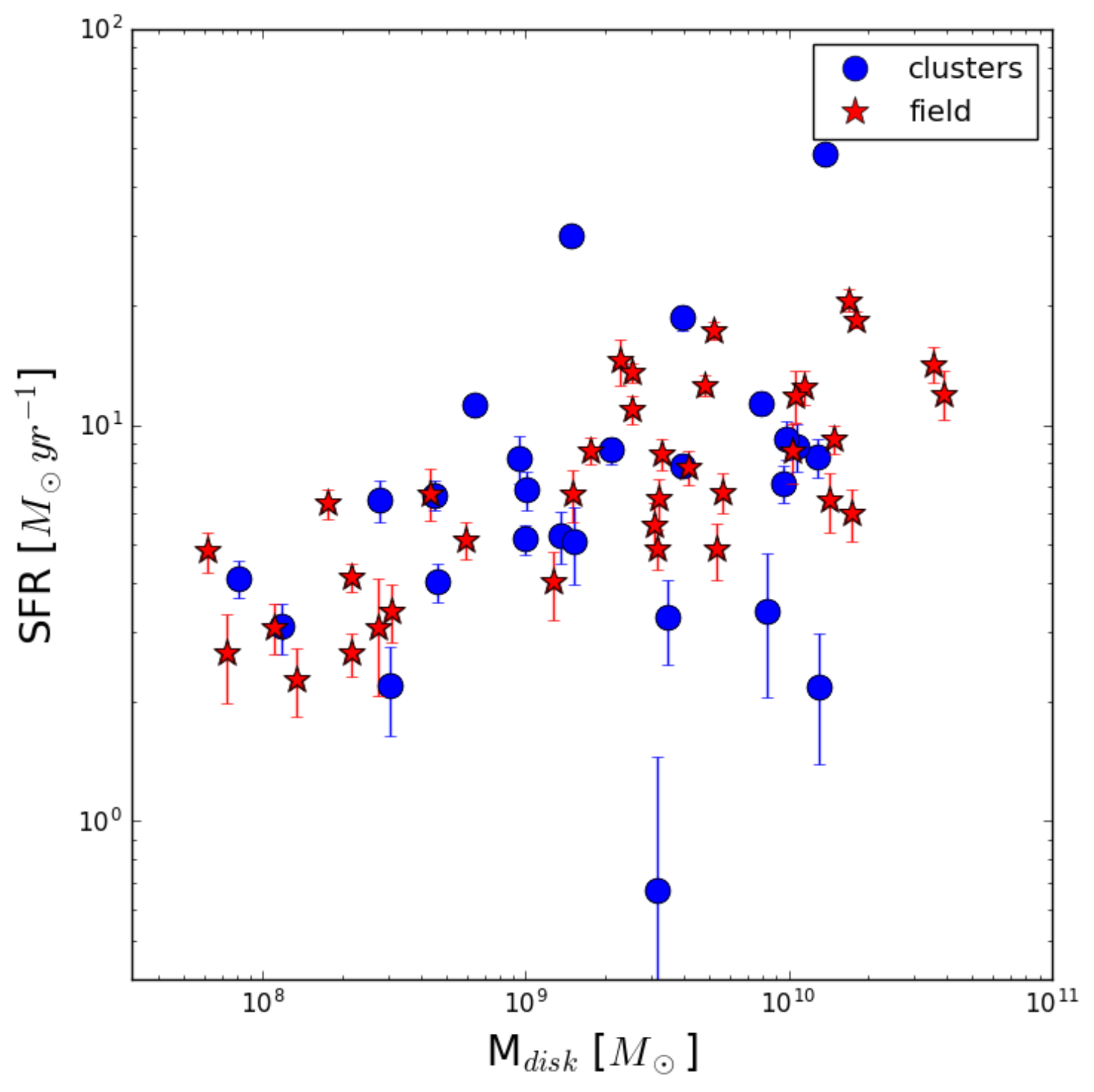}
\caption{Left: SFR-size disk relation for spiral galaxies in clusters (blue circles) and in the field (red stars). Dashed lines indicate the loci of constant SFR density. Right: SFR-Mass disk  relation for spiral galaxies in clusters (blue circles) and in the field (red stars).  \label{fig:SFR_spirals}}
\end{figure*}
Finally, for spiral galaxies, we can  investigate how the  SFR correlates with the size of the disk and its mass,  to investigate whether the \Ha disk and the galaxy disk coincide. 

The correlation between SFR and size (left panel of
Fig. \ref{fig:SFR_spirals}) is not very tight, in either environment.
Overplotted are also lines of constant star formation density
$\Sigma_{SFR}$. Our galaxies do not lie on any of these tracks,
spanning a range of almost two orders of magnitude in star formation
density. We interpret this as the result of a diversity of modes of
star formation, from different levels of star formation in disk with
different supplies of cold, to perhaps nuclear starbursts or
star formation associated with interactions and accretion.

Using the information on the B/T ratio, we can estimate the mass fo the disk in galaxies ($M_{disk}=(1-B/T)\times M_\ast$). Especially for field galaxies, this relation is much more tighter than the canonical SFR-Mass relation for the same galaxies (plot not shown), suggesting that indeed the bulk of the star formation is occurring in disks. 
 
\cite{abramson14} found that by normalizing galaxies by the stellar mass of the disk alone, 
the slope of the SFMS is consistent with only a linear trend (removing any dependence on mass). Although this correction to the disk stellar mass 
homogenizes the SFMS for disks with a range of B/T, the intrinsic dispersion ($\sigma$SFR) of the sequence must be a result of contributions by bars, 
disk dynamics, halo heating, AGN activity, environment and/or gas accretion history, among other factors \citep{dutton10}.
While the overall bulge strength does affect the position of a galaxy on the SFMS \citep{martig09, cheung12, fang13, kaviraj14, lang14, omand14}, 
the structure of the disk itself does not \citep{willett15}. These finding are consistent with recent models in which details of the feedback, which also relate strongly to the galaxy properties, have little 
effect on the SFMS \citep{hopkins14}. Alternatively, this also agrees with models in which the SFMS is the result of stochastic processes, rather than deterministic physics related to galaxy evolution \citep{kelson14}.

\section{Discussion and Conclusions}\label{sec:d_c}

Building on our pilot study presented in \cite{vulcani15}, we have
continued our exploration of the spatial distribution of star
formation in galaxies at $0.3<z<0.7$, as traced by the \Ha emission in
the field of view of the 10 GLASS clusters, detailing and
strengthening our previous results. We have produced \Ha maps taking
advantage of the WFC3-G102 and WFC3-G141 data at two orthogonal
PAs. We visually selected galaxies with \Ha in emission and, based on
their redshift, assigned their membership to the cluster. We used
galaxies in the foreground and background of the clusters to compile a
field sample at similar redshift. The cluster and field samples are
well matched in stellar mass and have very similar data quality
ensuring that any differences in the populations are not driven by the
stellar mass of the galaxies or instrumental selection effects.

We have visually classified both field and cluster galaxies, paying
particular attention to their broad-band morphology, the \Ha
morphology. We have introduced a new scheme to visually categorize
galaxies according to the main process that are affecting the mode of
star formation. Our is clearly a qualitative and approximate classification scheme, considering that multiple processes might be simultaneously at work, 
but we believe there is merit in categorizing in a self consistent manner the diversity of morphological features across environments. 
More quantitative tests on the ability of detecting  ram pressure stripping effect are currently underway (Paper VIII).

We have  correlated these quantities to the extent
of the \Ha emission and its position within the galaxy, in order to
present a complete characterization of the \Ha emitters in different
environments.

The main results of this analysis can be summarized as follows:

\begin{itemize}
\item Comparing the morphological distribution of \Ha emitters to that of a reference sample matched in mass and environment, we found systematic differences: among the \Ha emitters,  40\% of galaxies in clusters and 50\% of galaxies in the field present a spiral morphology. In both environments, the second most common morphological class is that of mergers, followed by ellipticals and S0s. In contrast, in the control sample, there is no dominant morphological type, with ellipticals and spirals the most represented ones. 

\item \Ha emitters can assume a variety of \Ha morphologies consistent with a diversity of physical processes. Nonetheless,  some patterns have been found. In the field \Ha emitters most likely present a clumpy \Ha morphology consistent with minor mergers or accretion (27\% of the galaxies), or a regular morphology where current star formation appears to be co-located with past star formation (18\%). Perhaps surprisingly, regular galaxies not affected by any strong process are the most common class in clusters (25\%), followed by asymmetric galaxies where clear sign of ram pressure stripping have been detected (18\%). 

The most common process label in clusters is ram pressure, while in the field it is mergers, mostly minor.
\item Comparing the position of the peak of the \Ha emission to that of the continuum, as traced by the F475W filter, we found that in both environments, for most of the galaxies the displacement is smaller than 1.5 kpc and the  average  offset is $\sim$0.5 kpc. The existence of the offset suggests that current star formation is not generally colocated with recent star formation, perhaps as the result of accretion of satellites or gas, or non gravitational interactions such as ram pressure stripping affecting the spatial distribution of the cold gas.

\item Overall, cluster and field galaxies share  a similar SFR-mass relation.
Galaxies with different \Ha morphologies produced by  different processes may  populate different regions of the SFR-mass plane, but due to our low number statistics we can not draw firm conclusions. Galaxies likely experiencing a ram pressure event are located either above the main sequence (SFR enhanced, maybe suggestive of ongoing stripping) or below (SFR suppressed, maybe suggestive of past stripping). Galaxies undergoing a major merger event tend to have SFR enhanced, both in clusters and in the field. In clusters, the tail at low SFR level is populated by clumpy or concentrated \Ha morphologies, due to some non identified process. 

\item For spiral galaxies, we compared the size of the current star formation (traced by \Ha) and the recent star formation (as traced by the disk in filter F475W). In general \Ha traces the disk, with the mean size ratio being close to unity, in both environments. Galaxies with different \Ha morphology are not strongly clustered in the size-size plane, even though concentrated objects tend to have smaller \Ha size, in both environments. In clusters, a population of galaxies with \Ha much larger than the size in the continuum tend to appear, consistent with their gas being currently being stripped or disturbed by environmental processes. 
\end{itemize}

 The emerging picture is that \Ha emitters are a very heterogeneous 
population, characterized by a range of  morphologies, sizes and  SFRs, in agreement with previous studies
conducted from the ground \citep[e.g.,][]{yang08, goncalves10, sobral13b, swinbank12, wisnioski15, stott16} and from space \citep{nelson12, nelson13, nelson15}.
 Therefore, a simple explanation can not describe our
observations. Even though we identified some small systematic
differences between galaxies in the field and in clusters, both
populations present very mixed morphologies and experience a variety
of processes.  

Non gravitational interactions such as ram pressure stripping seem to
play an important role in clusters while it is much less effective in
the field.  This is in agreement with previous studies that showed
how this phenomenon is expected to be important at the center of
massive clusters because of the large relative velocities and higher
densities of the ICM \citep{gunngott72, quilis01, font08,
bekki09}. \cite{bruggen08} showed that virtually all cluster galaxies
suffered weaker episodes of ram-pressure, suggesting that indeed this
physical process might have a significant role in shaping the observed
properties of the cluster galaxy population. They also found that
ram-pressure fluctuates strongly such that episodes of strong
ram-pressure are followed by two episodes of weaker ram pressure,
possibly allowing the gas reservoir to be replenished and intermittent
episodes of star formation to occur. In agreement with this, in our
sample we detected a broad range of ram pressure strengths affecting
the GLASS galaxies.  We also detected stripping in galaxies of
different broad-band morphology, consistent with the idea that gas
stripping does not directly and instantaneously affect galaxy
morphology.

We found that galaxy mergers and more generally major galaxy-galaxy
interactions are frequently detected in field \Ha emitters, and less
in massive clusters, where the large velocity dispersions impede
encounters (e.g. \citealt{mihos96}).  As expected, we found that
mergers trigger star formation.  Interestingly, \cite{sobral11} found instead that
  the increase in mergers within \Ha emitters is progressive from field into groups and into clusters.
 Their analysis  revealed that non-merger driven star formation is strongly suppressed in both rich groups/cluster 
 environments and for high stellar masses, implying that once potential mergers are neglected, stellar mass and environment both play separate and important roles.
 
Broadly speaking, we conclude that the effects of cluster-specific
mechanisms on galaxy evolution are detectable in our unprecedented
data. However, they are both subtle and complex. They are subtle in
the sense that no dramatic trend is found between the morphology of
the current star formation and the environment or other properties of
the galaxy. Every trend that we have found is weak and there are
always exceptions. This is consistent with previous work based on
spatially unresolved data that has concluded that the differences are
small, once one controls for stellar mass and other parameters
directly related to the time of initial collapse of the halo in which
the galaxy is found \citep{morishita16}. They are complex in the sense that the
richness of morphologies and sizes and relationships between current,
recent and past star formation cannot be easily reduced to a small
number of clear cut categories. This complexity limits the extent to
which data of this quality can be interpreted in the absence of full
blown quantitative calculations of these effects. The dynamical range,
resolution, and physical complexities that needs to be rendered in
order to carry out a detailed comparison between theory and data is
stupendous. However, given recent progress in hydrodynamical numerical
simulations \citep[e.g., Illustris,][]{vogelsberger14} it seems that detailed comparisons
between the kind of data derived here and simulated maps would be an
interesting exercise and might provide a way forward.

Another important issue that we have not addressed in this paper is
how the properties of galaxies depend on the detailed properties of
their host clusters. As discussed by \cite{treu15}, GLASS includes a
variety of clusters with different morphologies. The correlation
between the properties of the \Ha emitters and the hosting structure
is the subject of Paper VIII, where
we investigate trends with the clustercentric distance, the hot gas
density as traced by the X-ray emission, the surface mass density as
inferred from gravitational lens models and the galaxy local density,
in order to investigate in detail the role of the cluster environment
in shutting down star formation.

\section*{Acknowledgments}

We thank the anonymous referee for her/his constructive and helpful comments. 
Support for GLASS (HST-GO-13459) was provided by NASA through a grant
from the Space Telescope Science Institute, which is operated by the
Association of Universities for Research in Astronomy, Inc., under
NASA contract NAS 5-26555. We are very grateful to the staff of the
Space Telescope for their assistance in planning, scheduling and
executing the observations.  B.V. acknowledges the support from an
Australian Research Council Discovery Early Career Researcher Award
(PD0028506).
T.M. acknowledges support from a Japanese Ministry of Education, Culture, Sports, Science and 
Technology Grant-in-Aid for Scientific Research (26- 3871), and from a Japan Society for the 
Promotion of Science research fellowship for young scientists. 

\begin{appendix}
\section{Galaxies showing signs of ram pressure stripping} \label{RP}
\begin{figure*}
\centering
\includegraphics[scale=0.14]{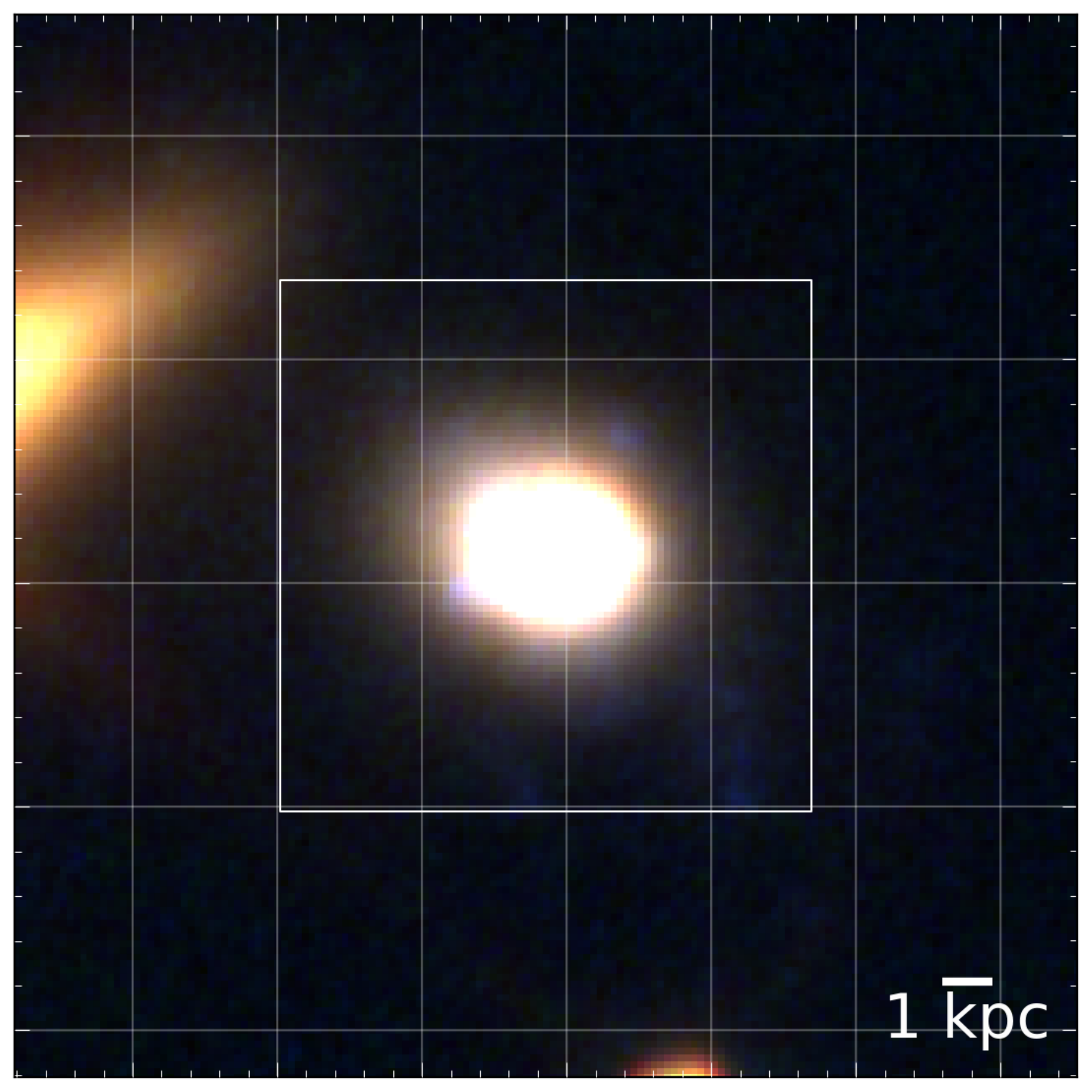}
\includegraphics[scale=0.14]{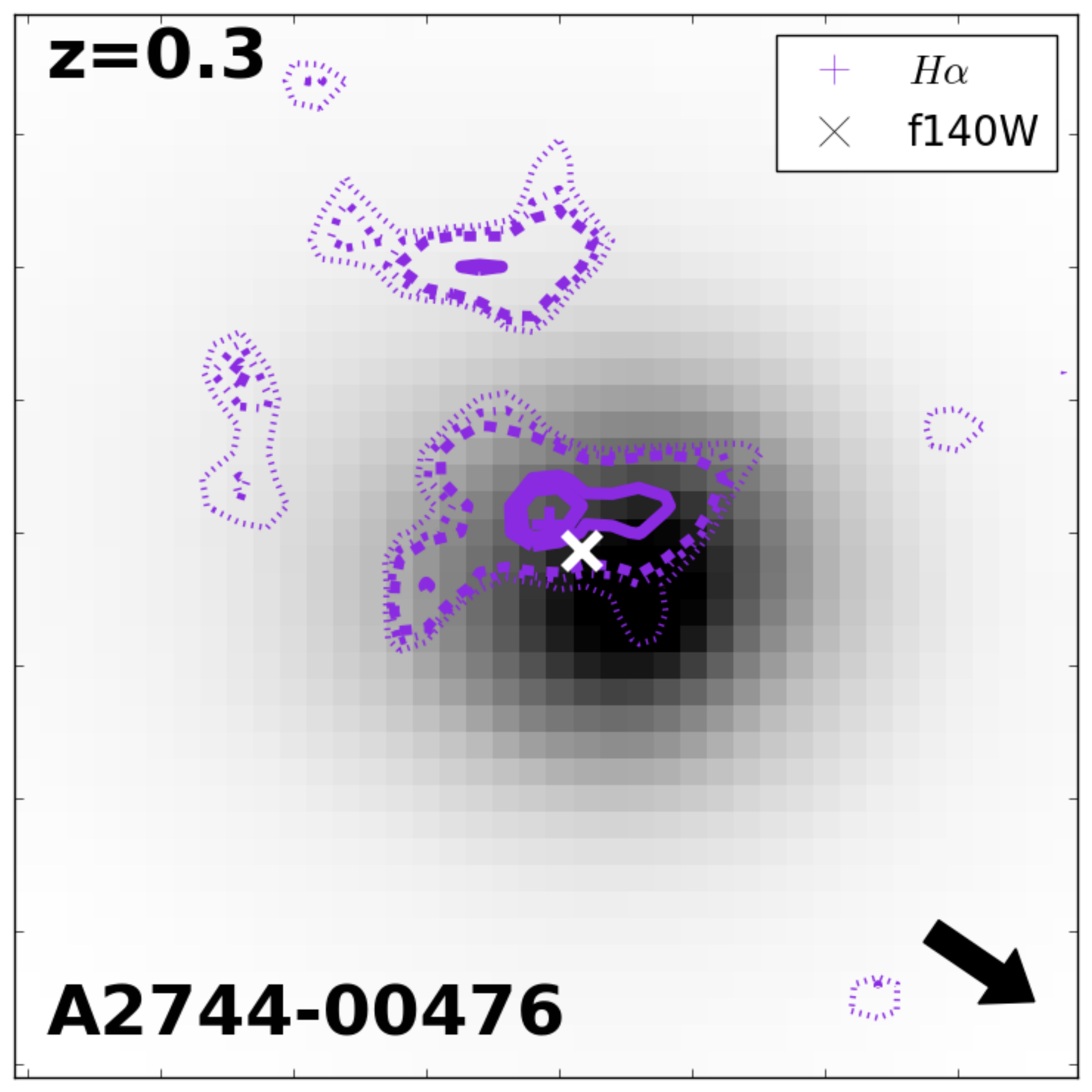}
\includegraphics[scale=0.14]{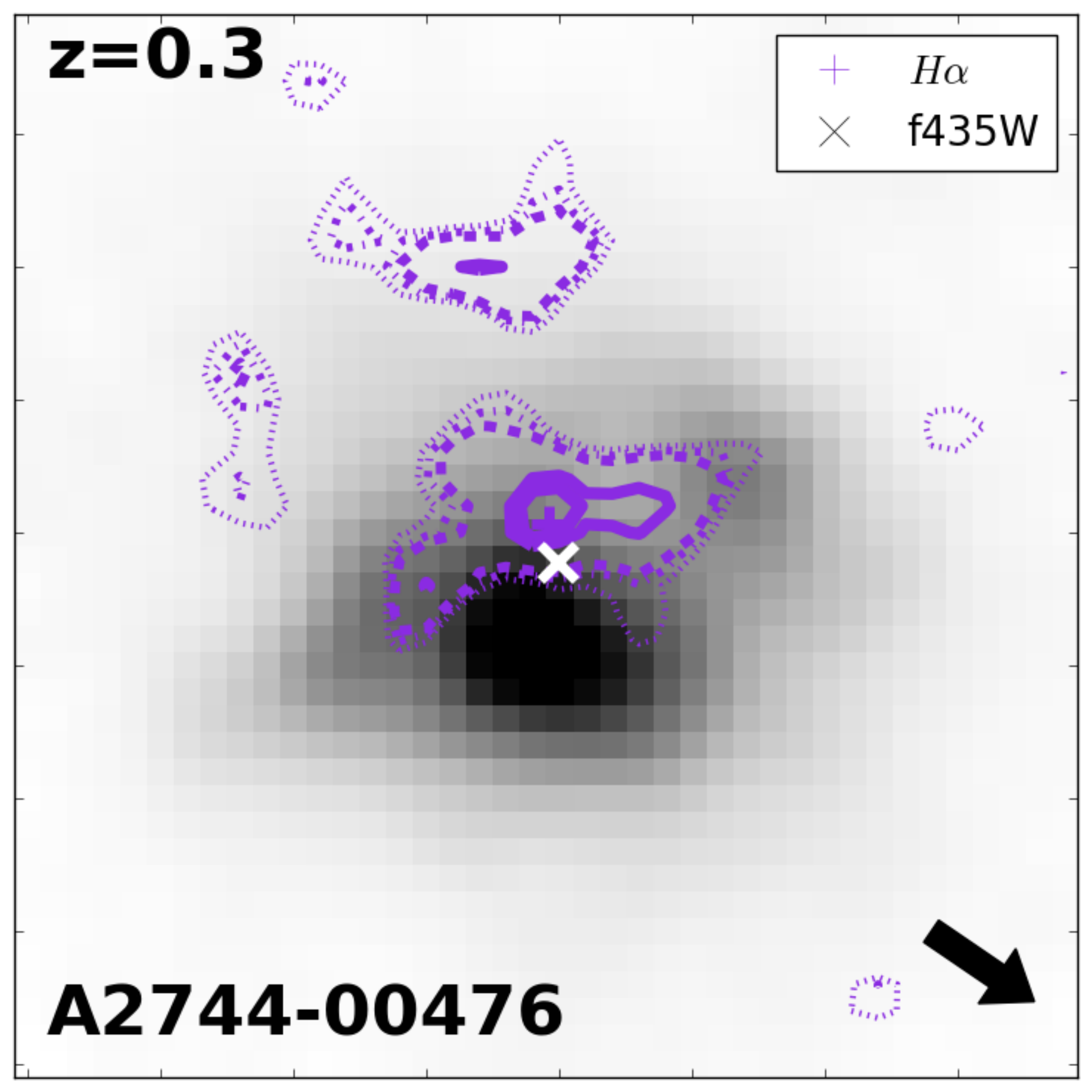}
\includegraphics[scale=0.14]{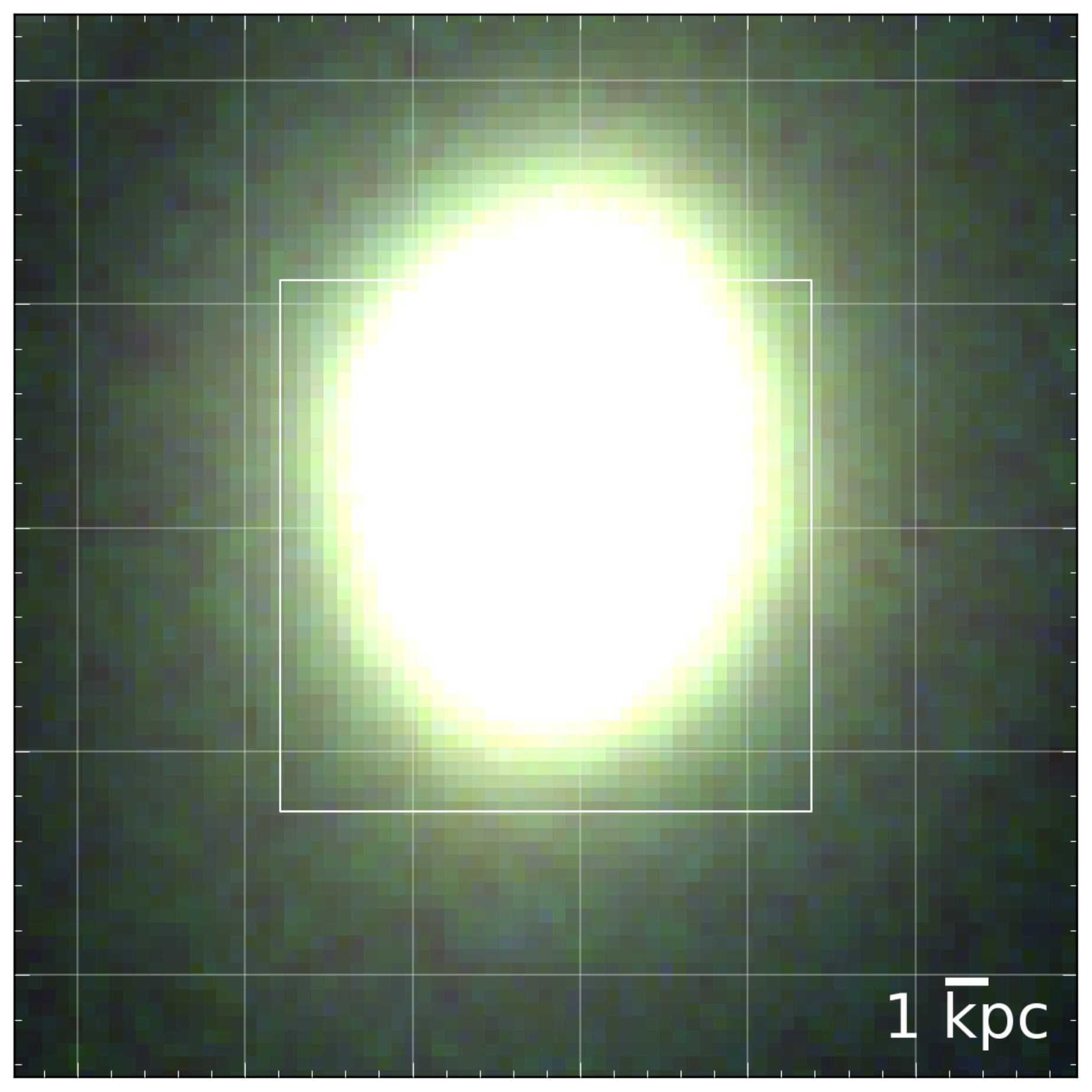}
\includegraphics[scale=0.14]{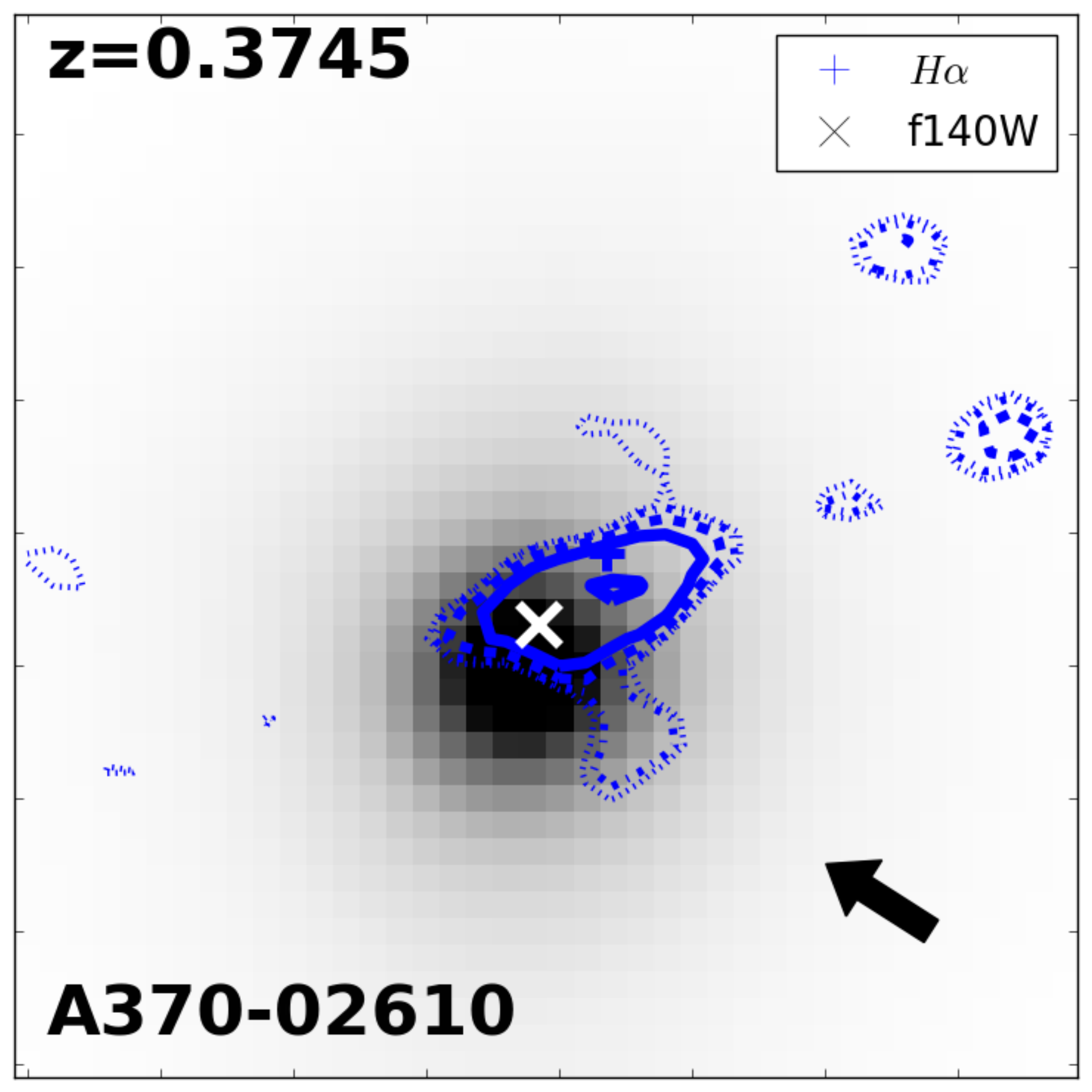}
\includegraphics[scale=0.14]{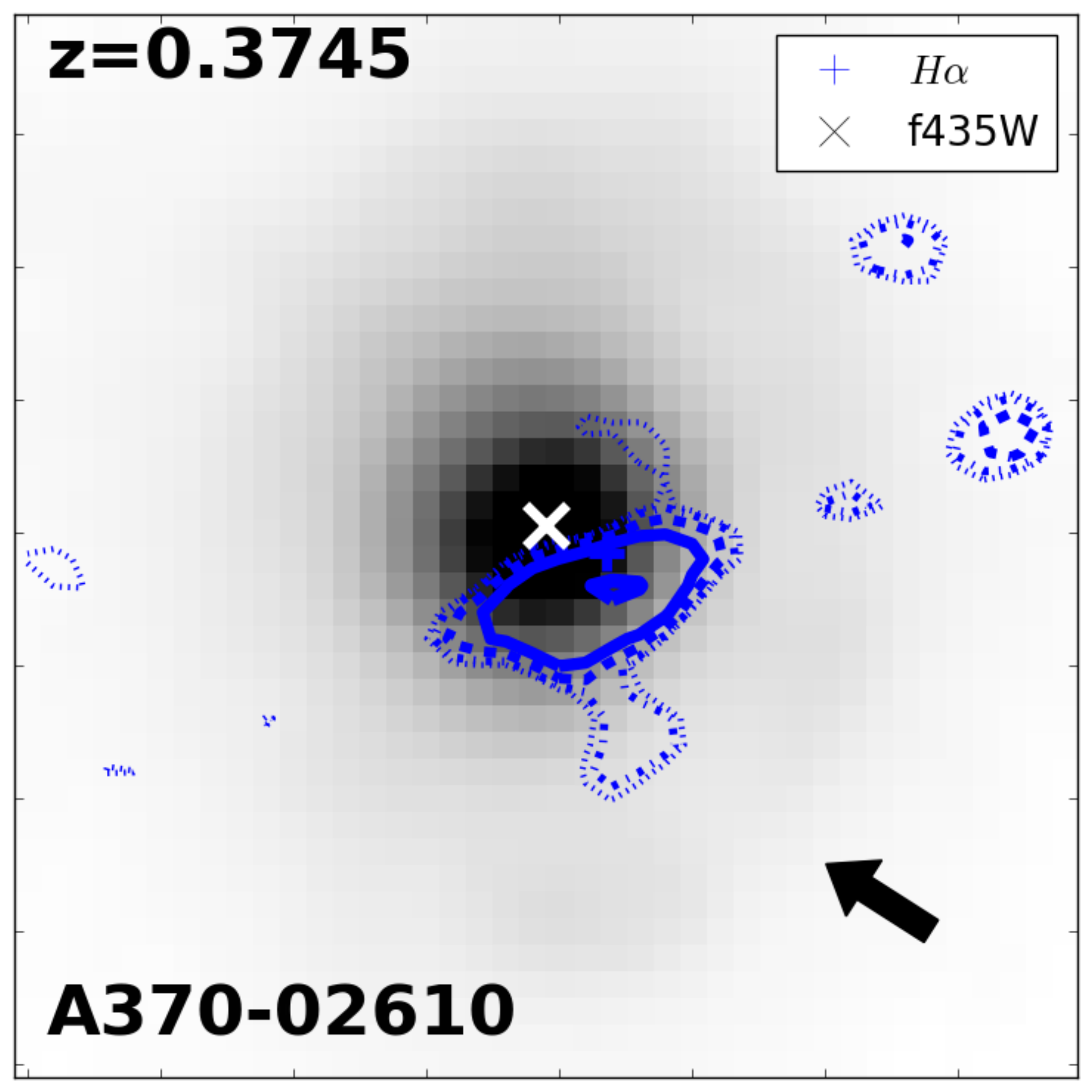}
\includegraphics[scale=0.14]{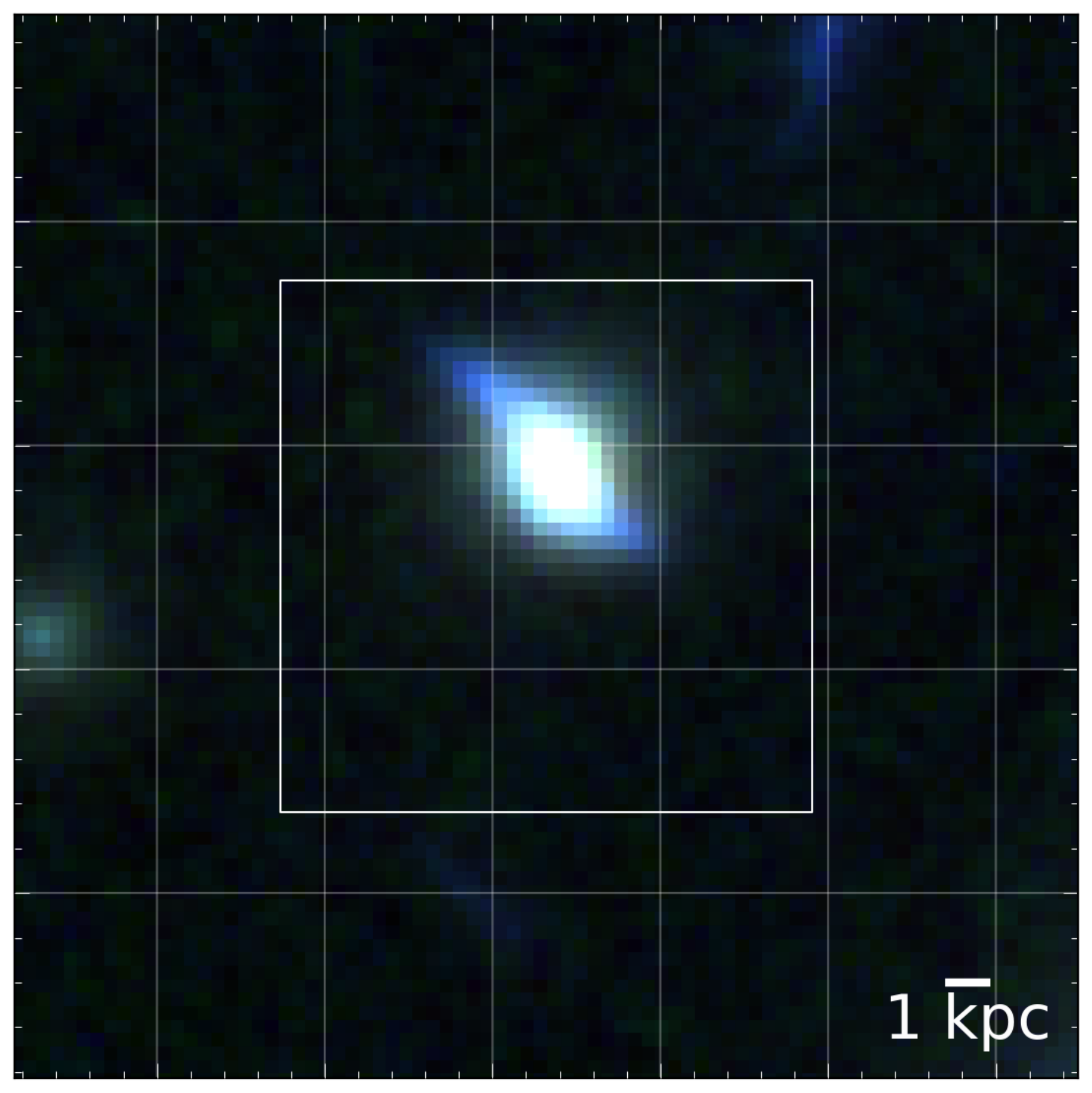}
\includegraphics[scale=0.14]{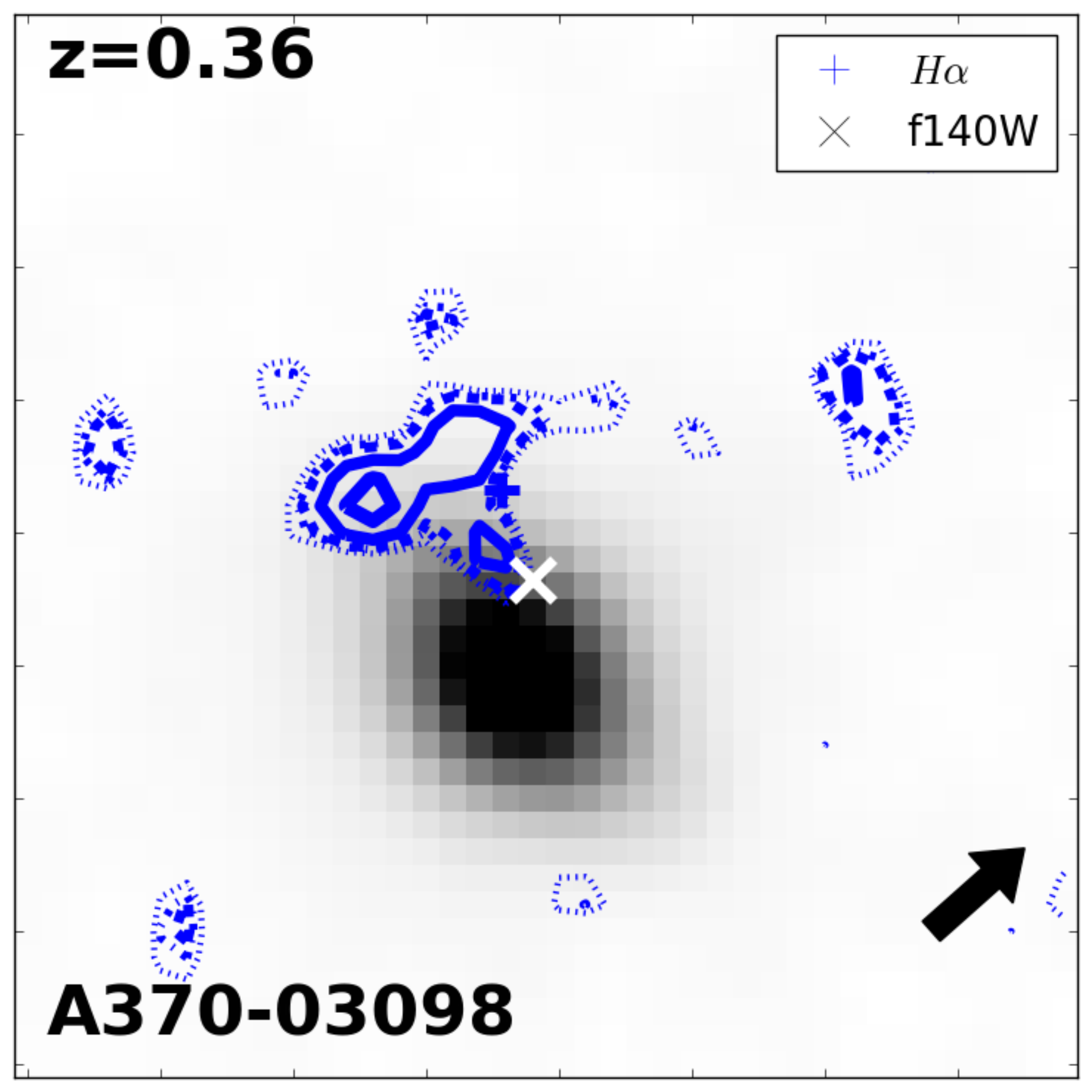}
\includegraphics[scale=0.14]{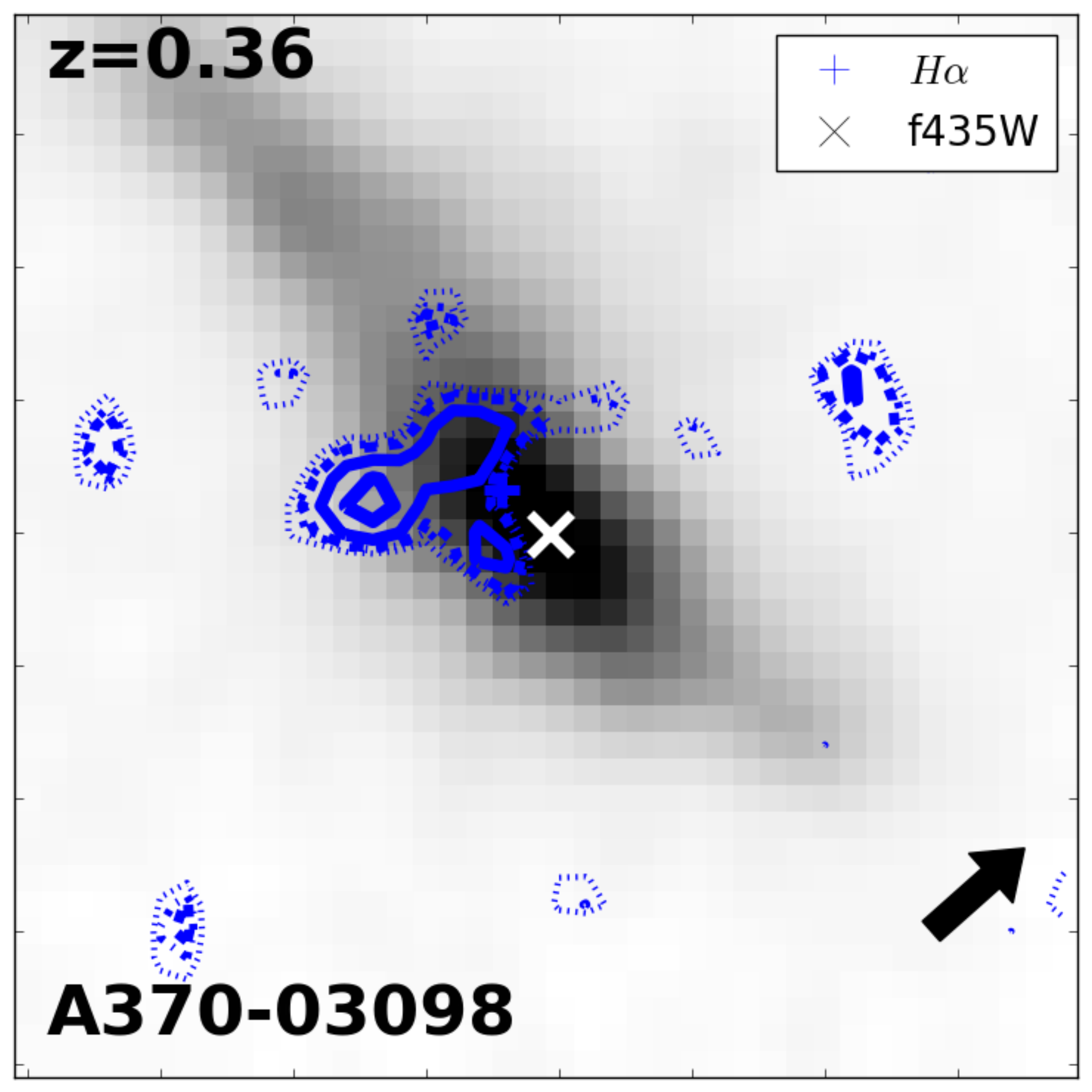}
\includegraphics[scale=0.14]{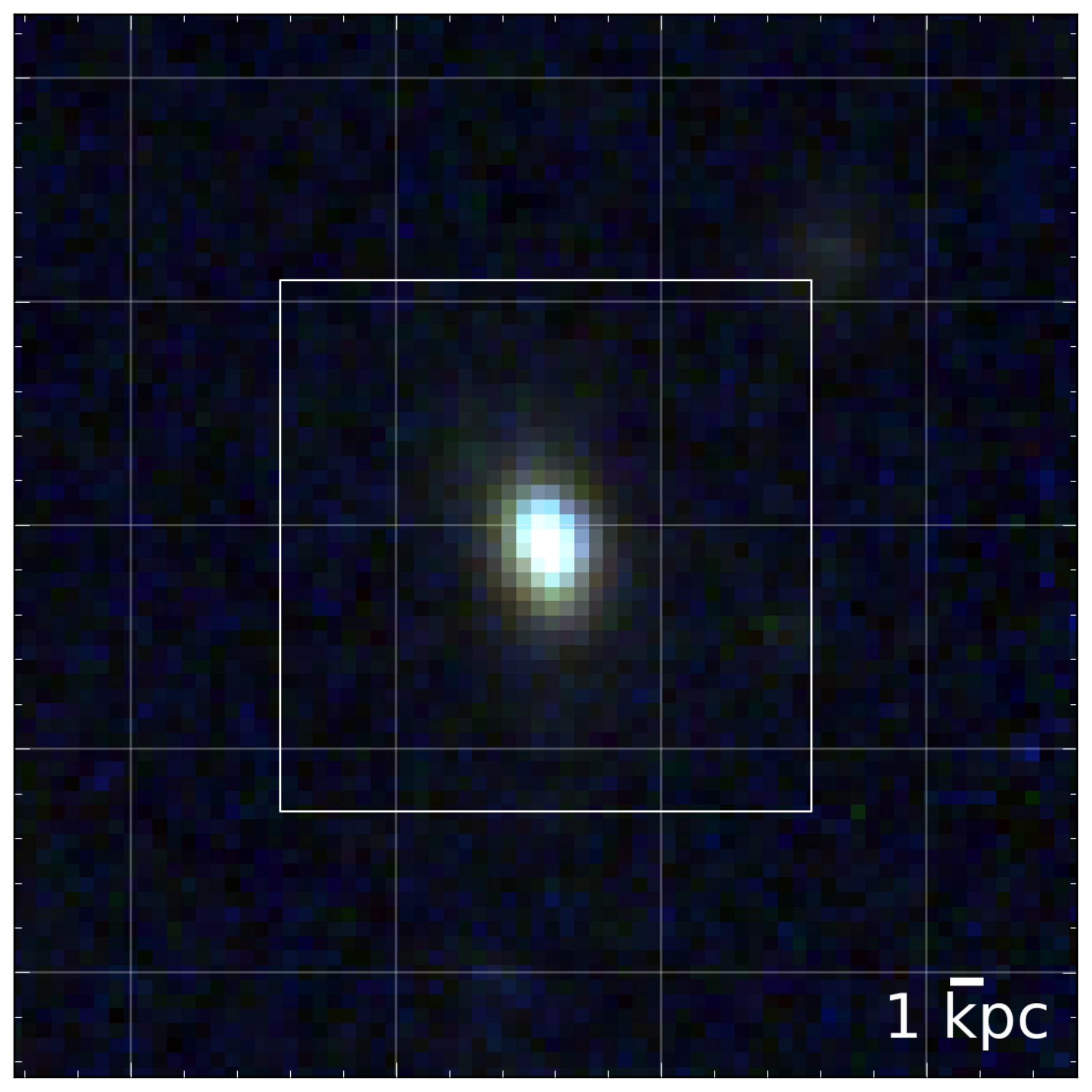}
\includegraphics[scale=0.14]{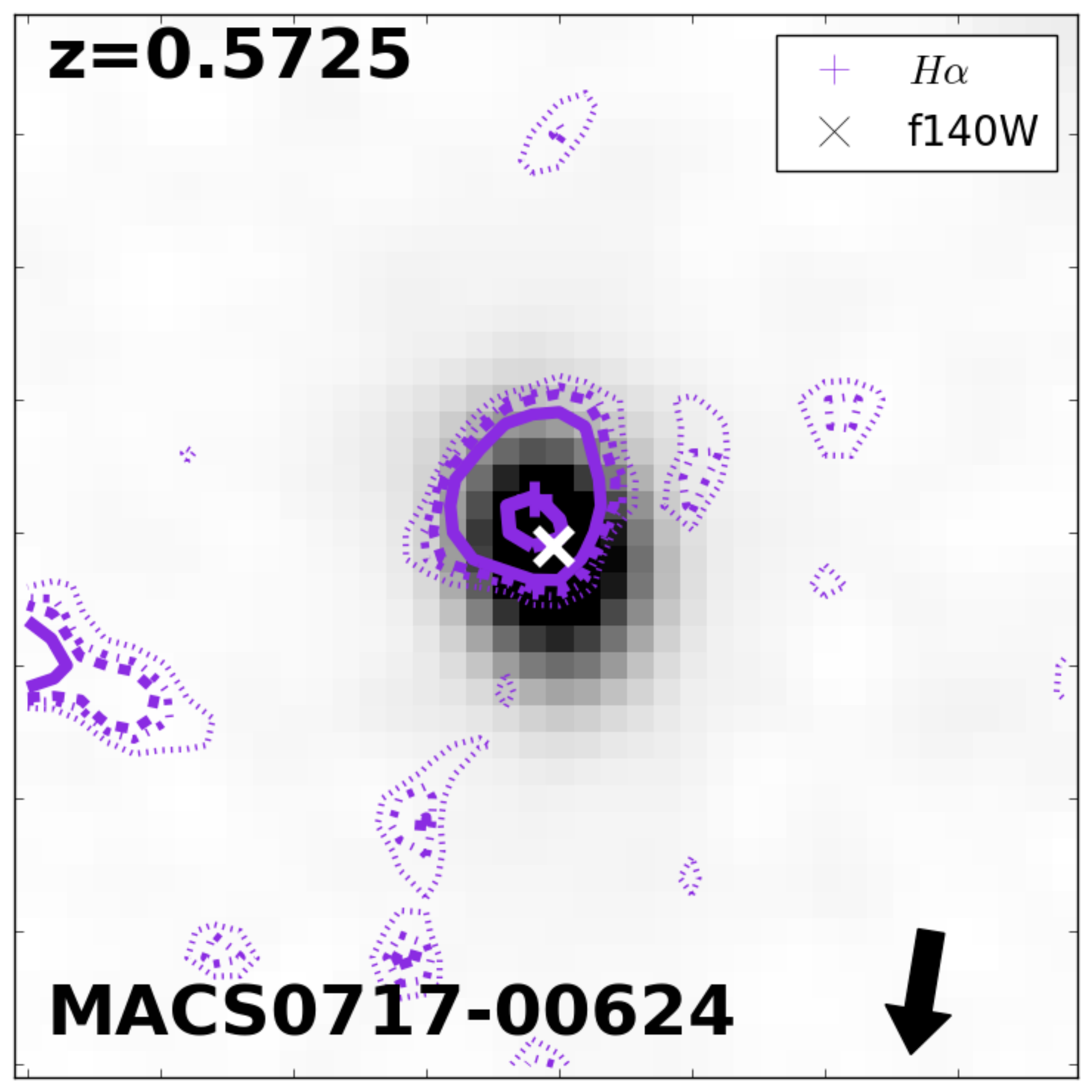}
\includegraphics[scale=0.14]{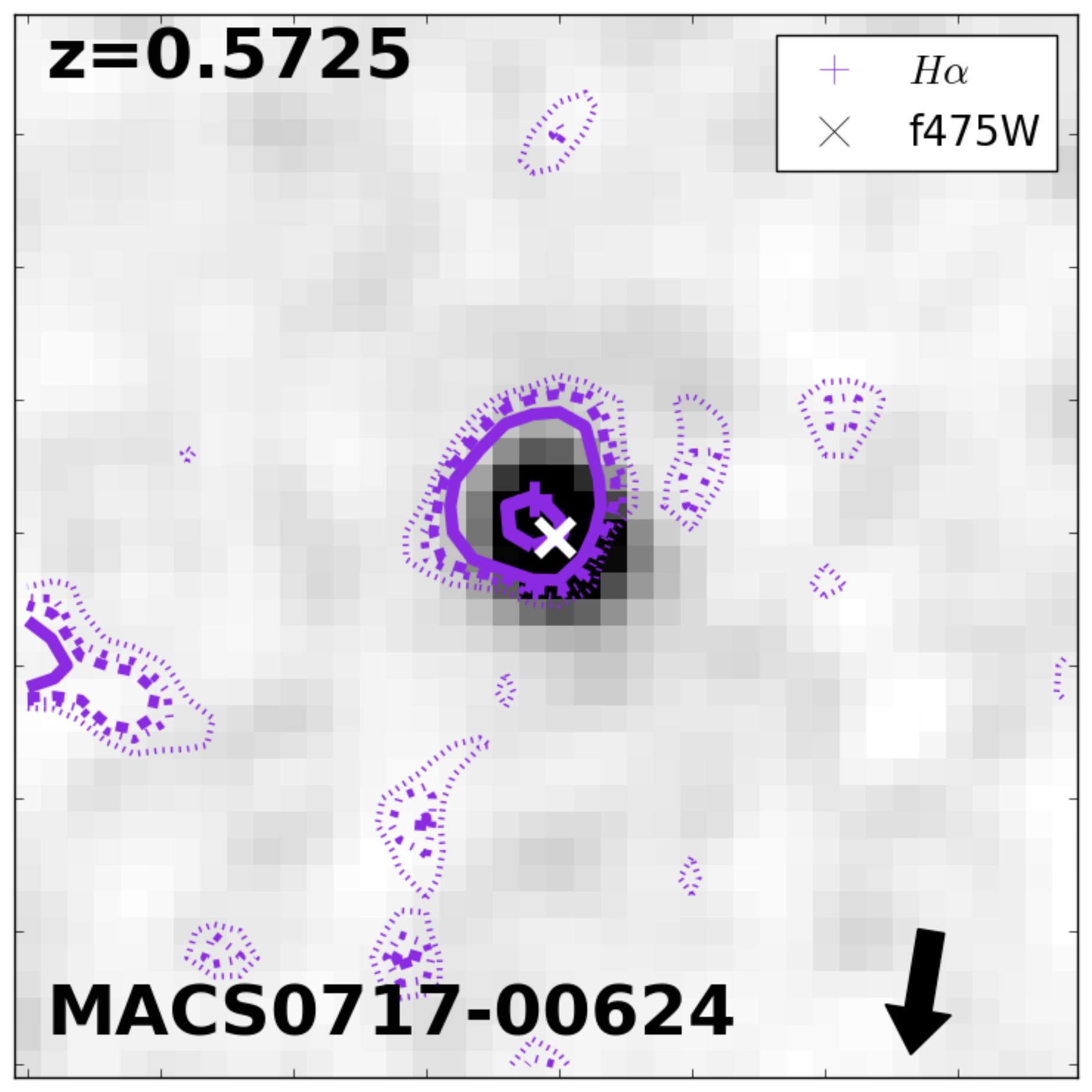}
\includegraphics[scale=0.14]{MACS0744_00691_rgb.pdf}
\includegraphics[scale=0.14]{MACS0744_00691_f140w_ha.pdf}
\includegraphics[scale=0.14]{MACS0744_00691_f475w_ha.pdf}
\includegraphics[scale=0.14]{MACS1149_01832_rgb.pdf}
\includegraphics[scale=0.14]{MACS1149_01832_f140w_ha.pdf}
\includegraphics[scale=0.14]{MACS1149_01832_f475w_ha.pdf}
\includegraphics[scale=0.14]{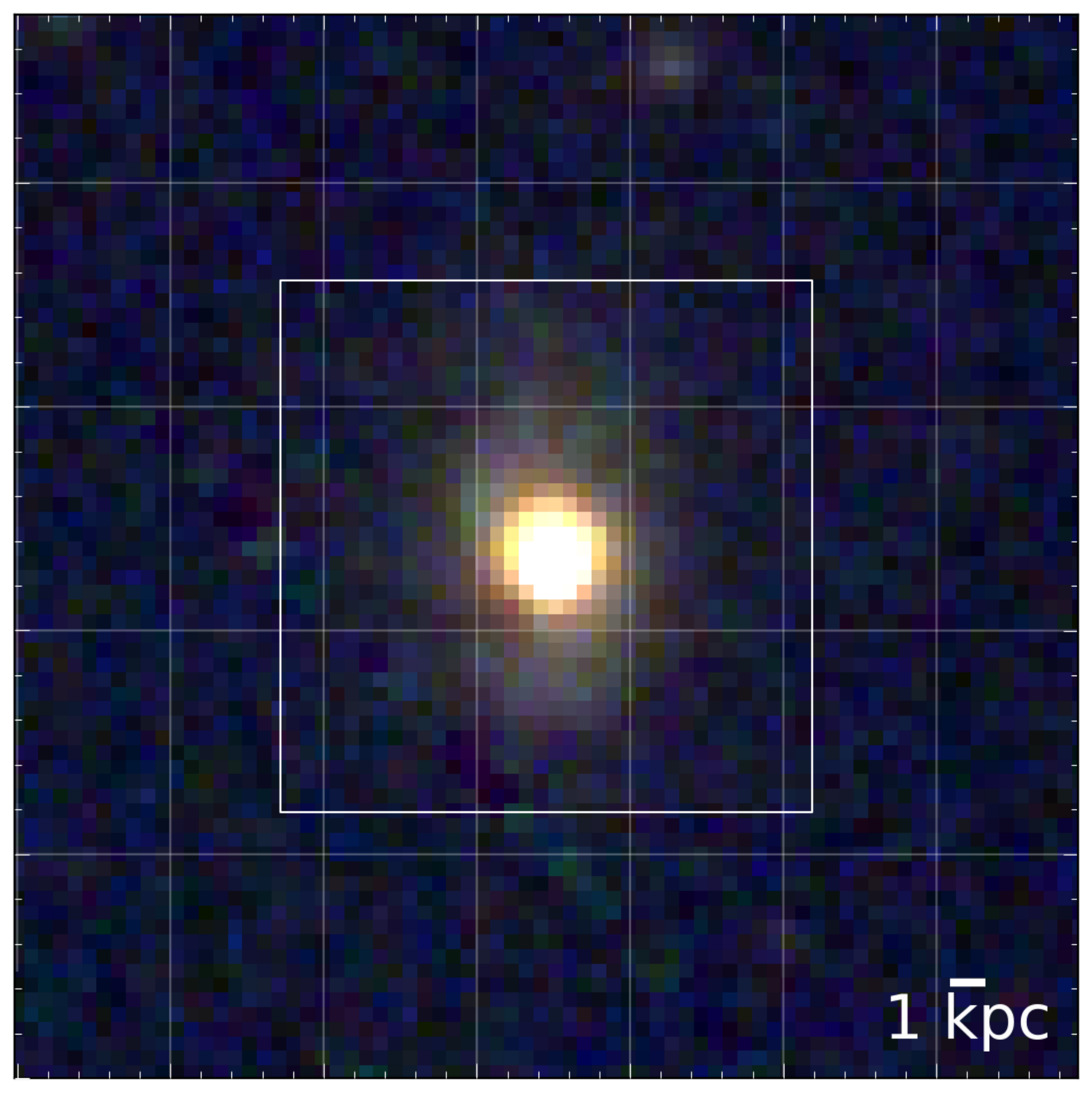}
\includegraphics[scale=0.14]{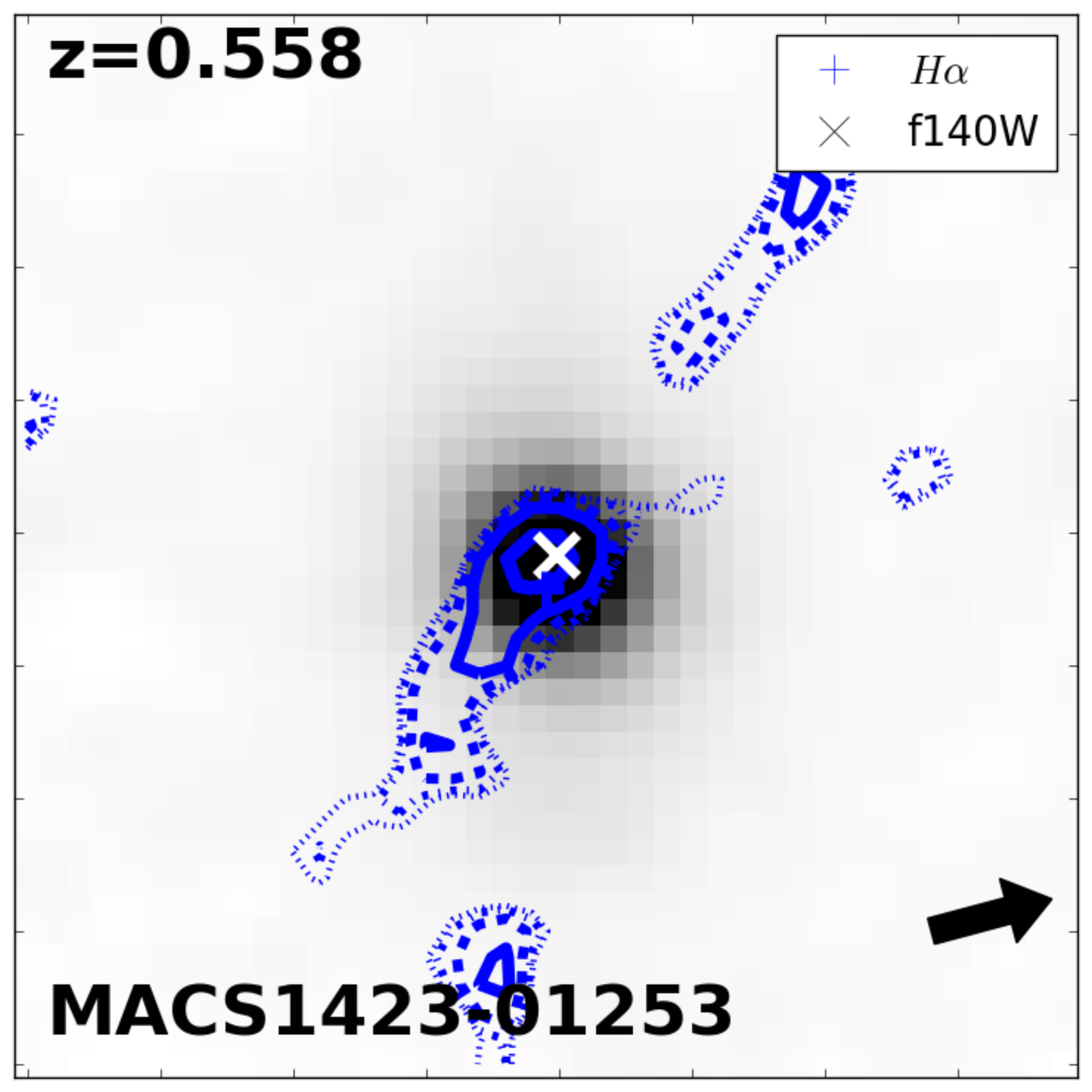}
\includegraphics[scale=0.14]{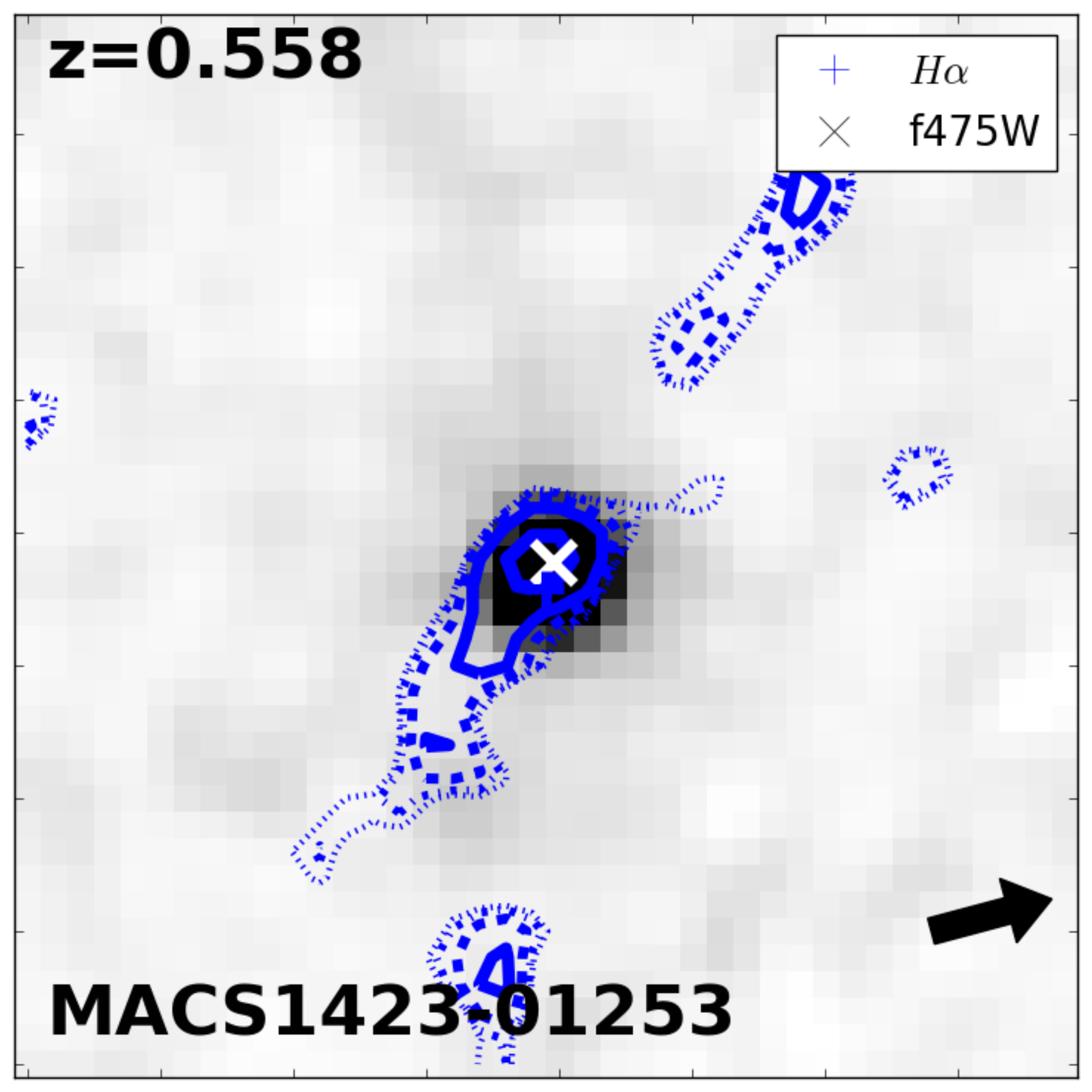}
\includegraphics[scale=0.14]{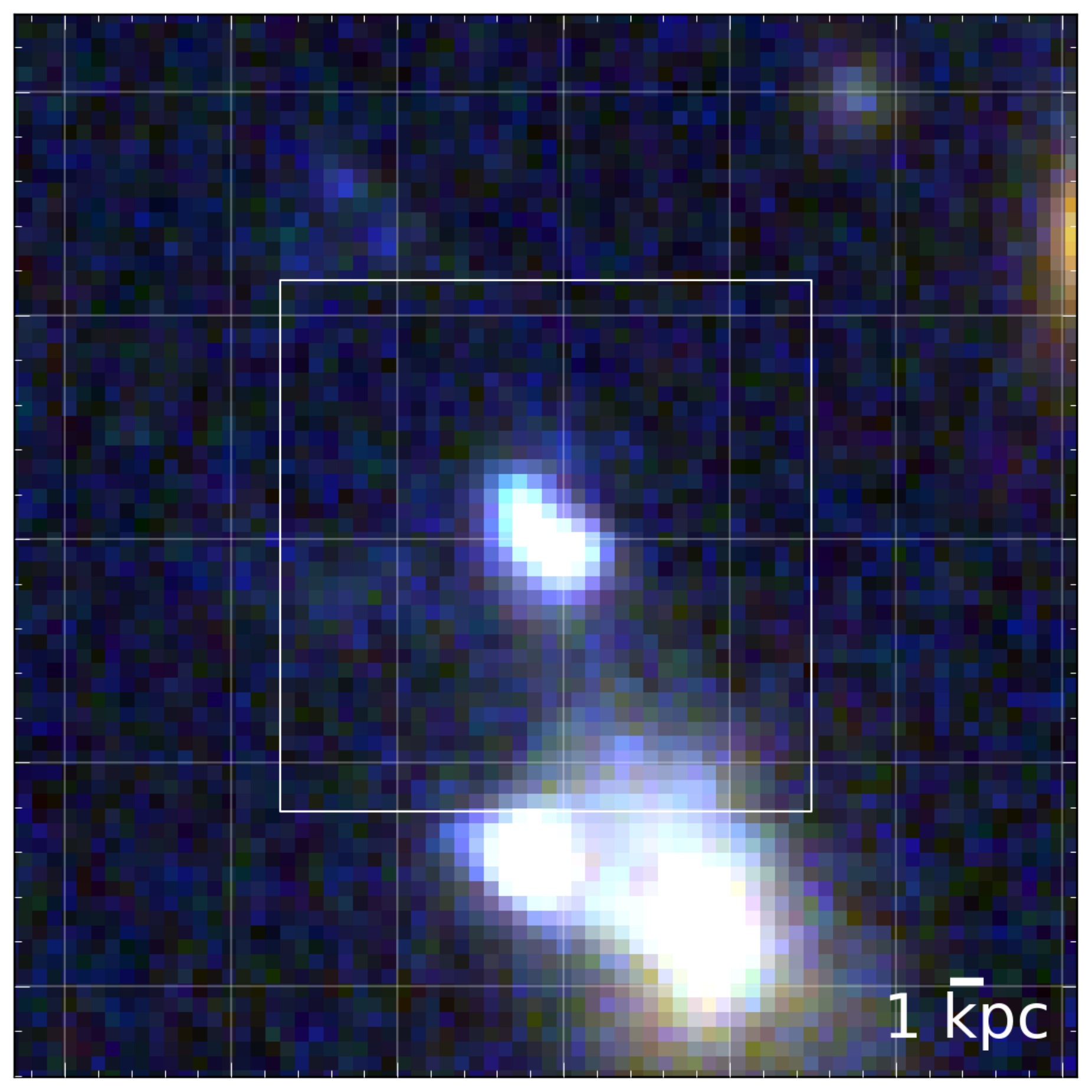}
\includegraphics[scale=0.14]{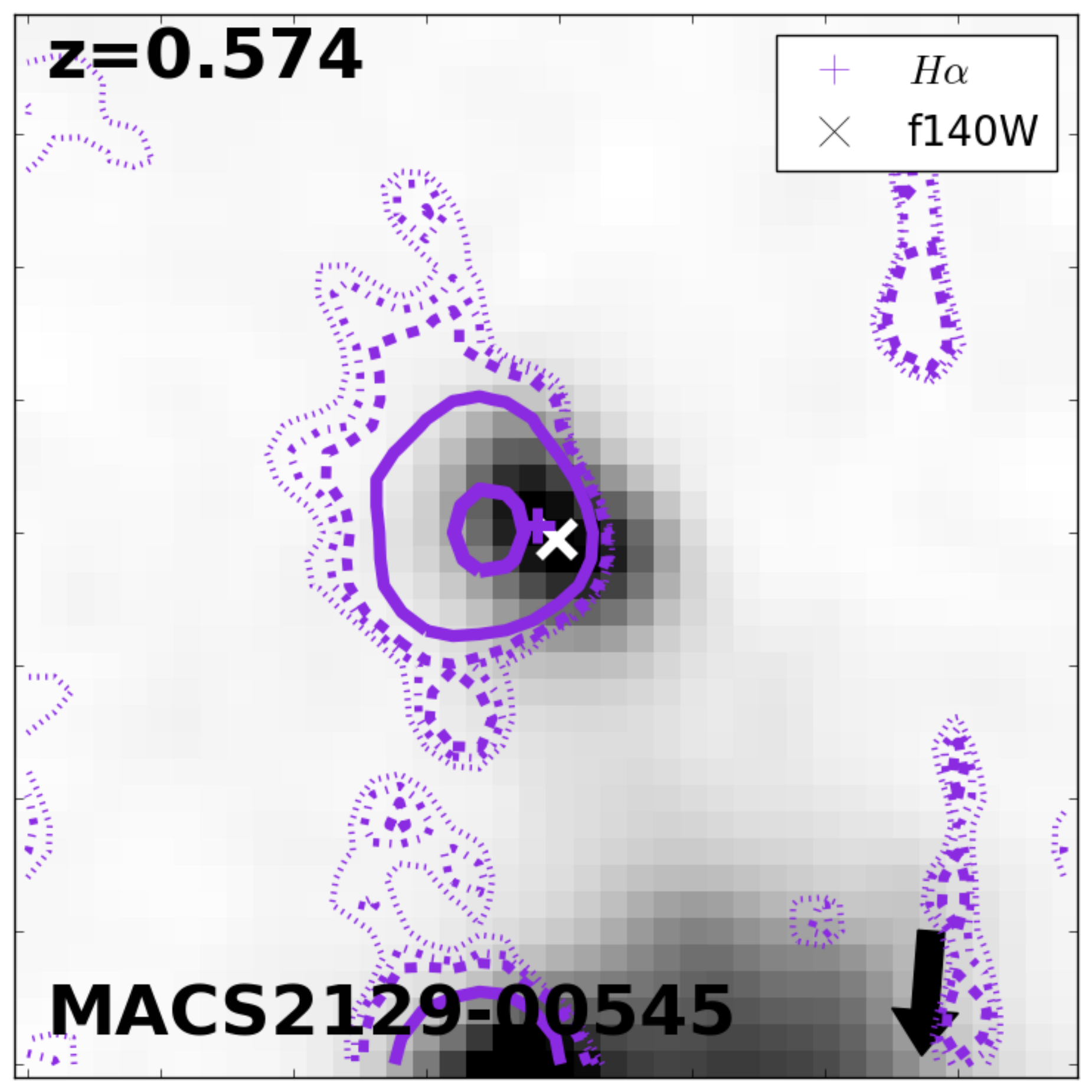}
\includegraphics[scale=0.14]{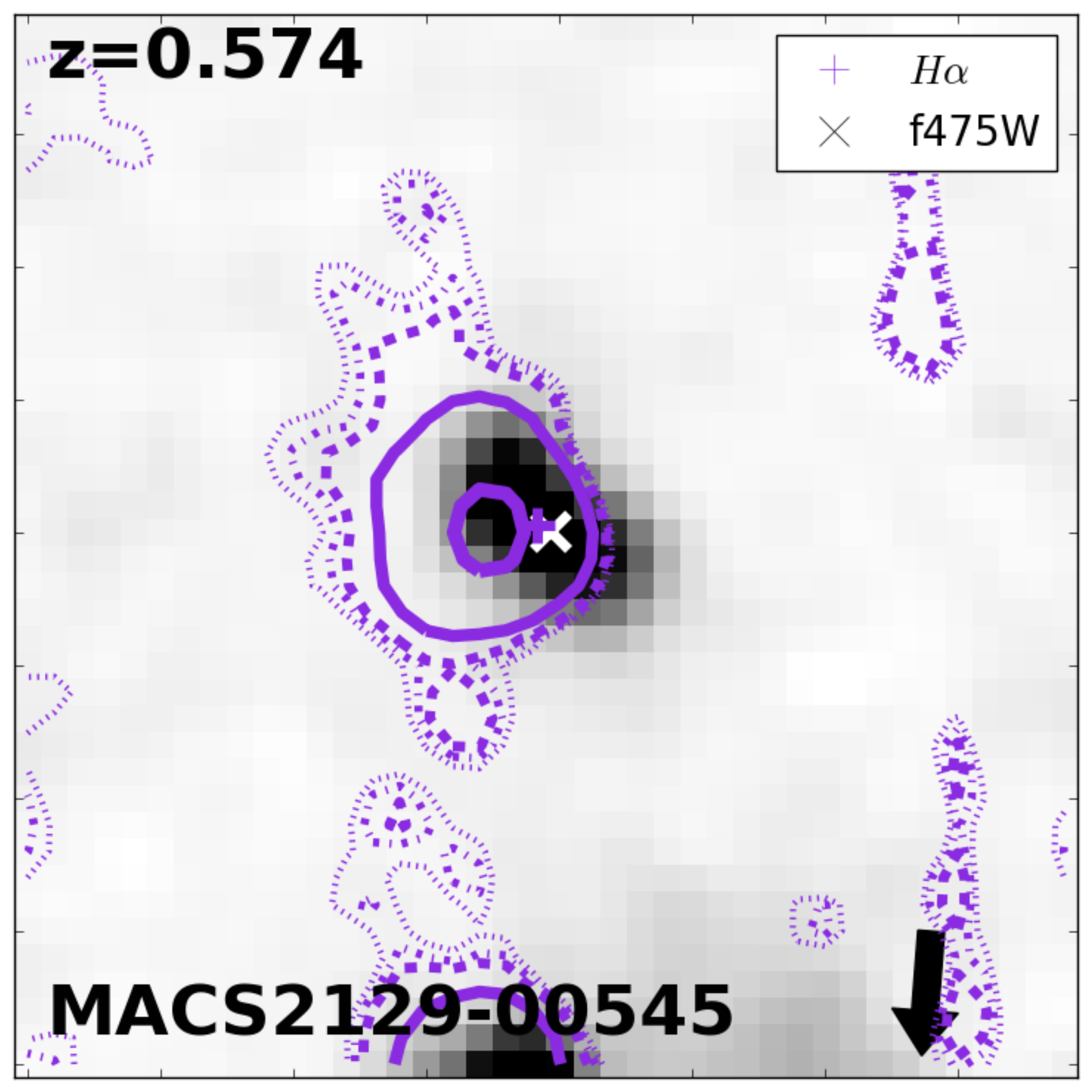}
\caption{Cluster galaxies labeled as ram pressure stripping candidates according to our classification scheme (\S\ref{sec:visinsp}). For each galaxy, the 
left panel shows the color composite image of the galaxy based on the 
CLASH \citep{postman12} or HFF \citep{lotz16} HST data. The blue channel is composed by the F435W, F475W, F555W, 
F606W, and F625W filters, the green by the F775W, F814W, F850lp, F105W, 
F110W filters, and the red by the F125W, F140W, F160W filters. The central panel 
shows the \Ha map superimposed on the image of the galaxy in the F140W filter 
and the right panel shows the \Ha map superimposed on the image of the galaxy in the F475W filter. 
Contour levels represent the 35$^{th}$, 50$^{th}$, 65$^{th}$, 80$^{th}$, 95$^{th}
$ percentiles of the light distribution, respectively. Blue contours indicate that \Ha maps are obtained  
from one spectrum, purple contours indicate that \Ha maps are obtained  
from  two orthogonal spectra, and green contours indicate that \Ha maps are obtained combining both the 
G102 and G141 grisms (only for $z>0.67$). In the color composite image, 
the Field of View is twice as big as the single band images. A smoothing filter has been applied to the 
maps and an arbitrary stretch  to the images for display purposes.  Arrows on the bottom right corner indicate the direction of the cluster center. The redshift of the 
galaxy is indicated on the top left corner.
 \label{fig:RP_clusters}}
\end{figure*}

\begin{figure*}
\centering
\includegraphics[scale=0.14]{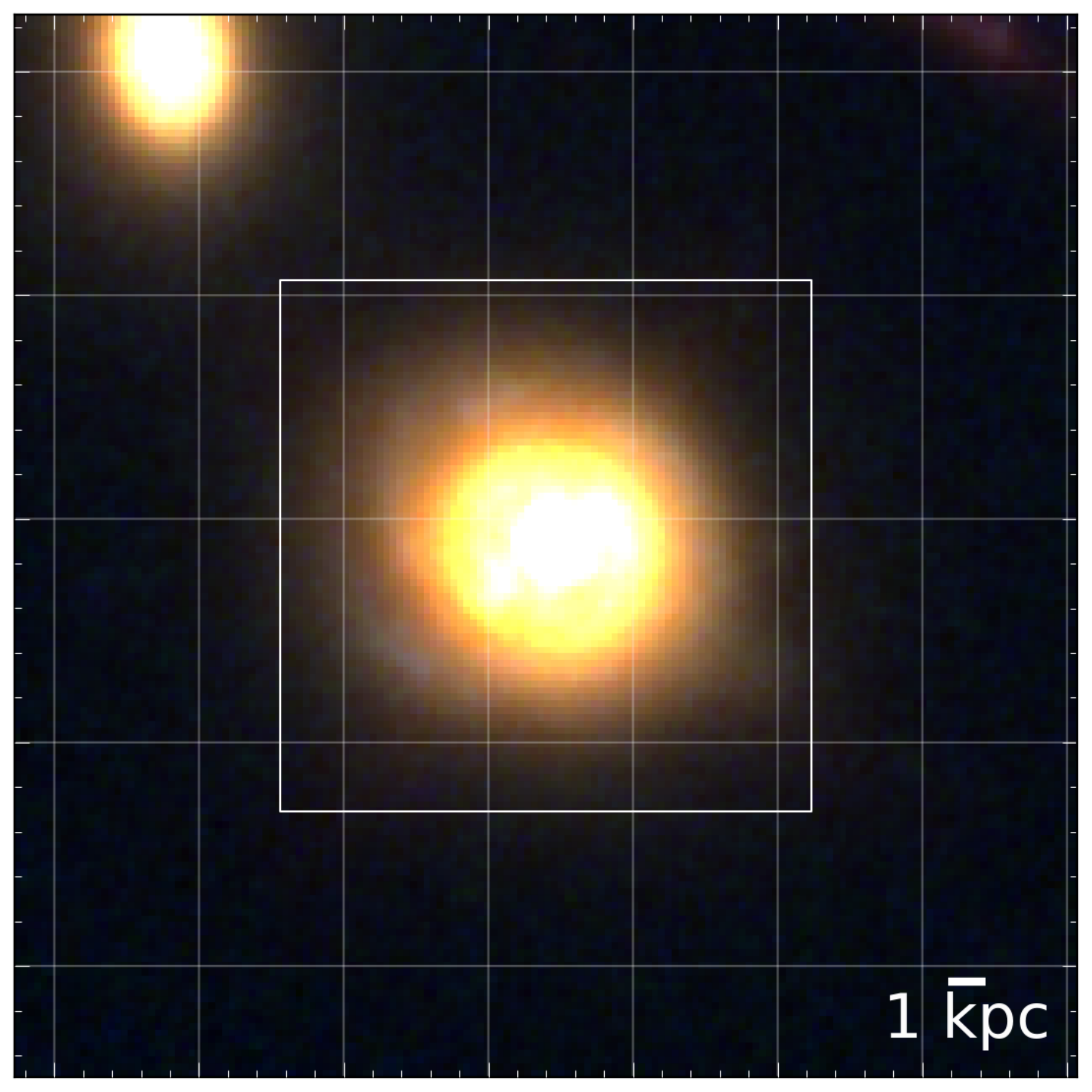}
\includegraphics[scale=0.14]{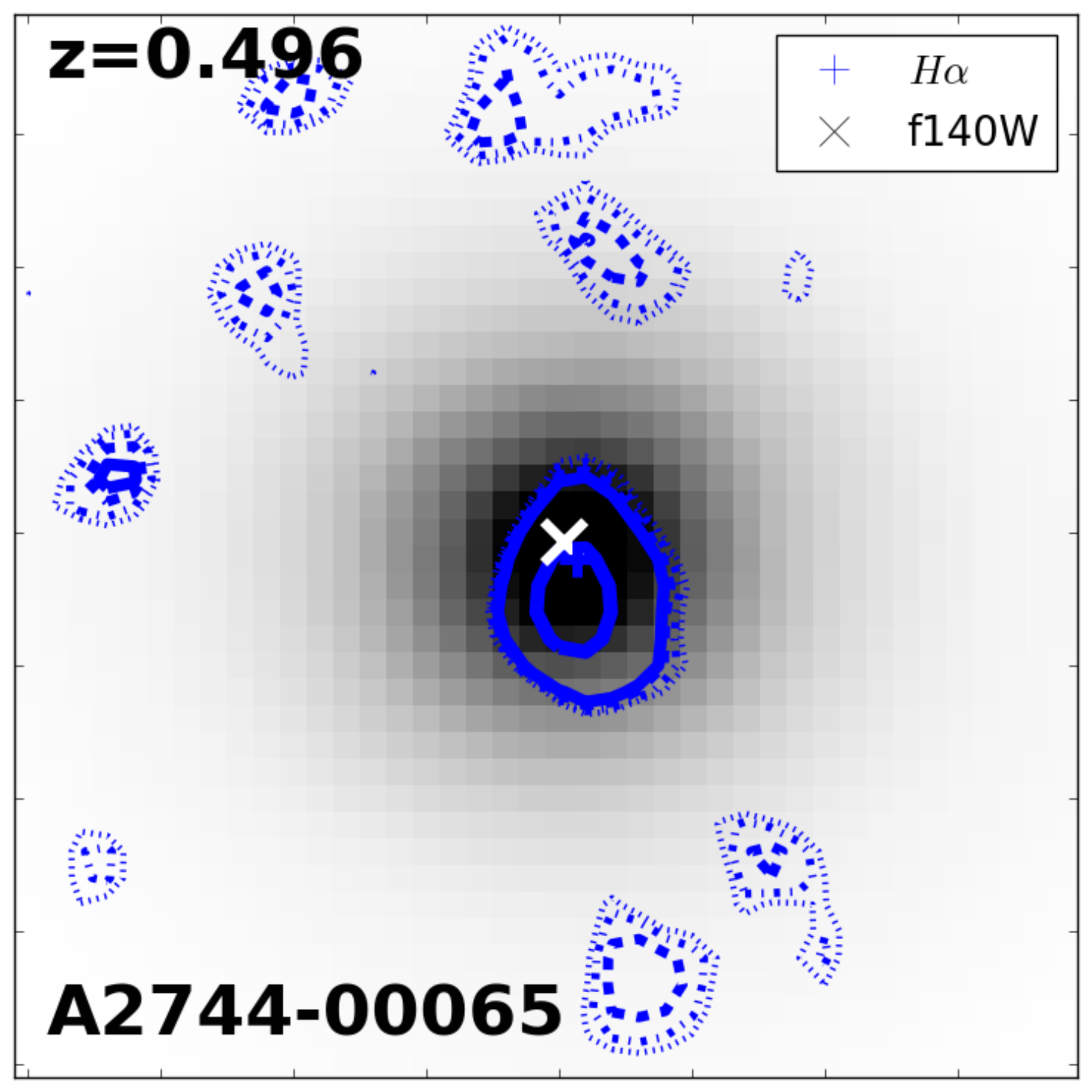}
\includegraphics[scale=0.14]{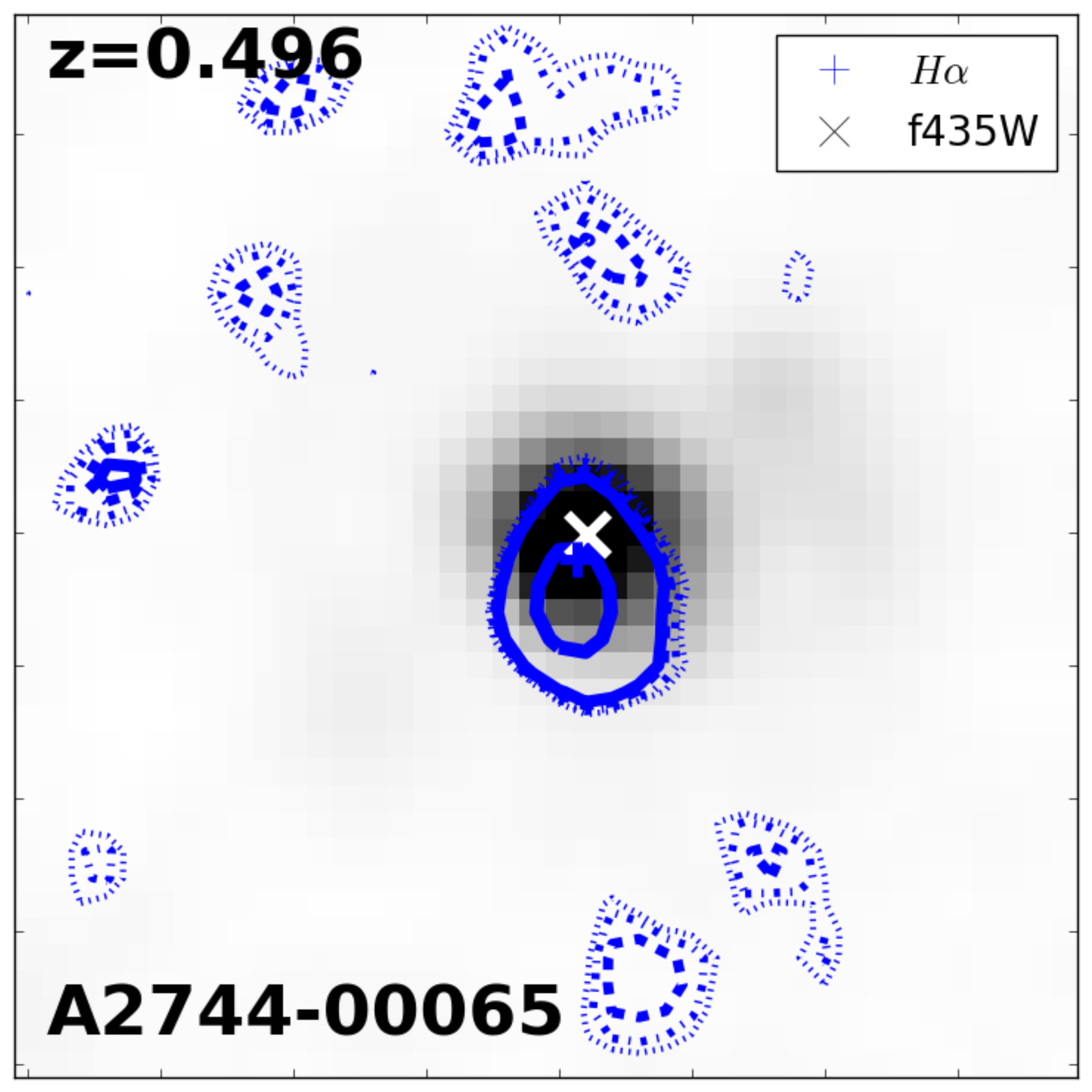}
\includegraphics[scale=0.14]{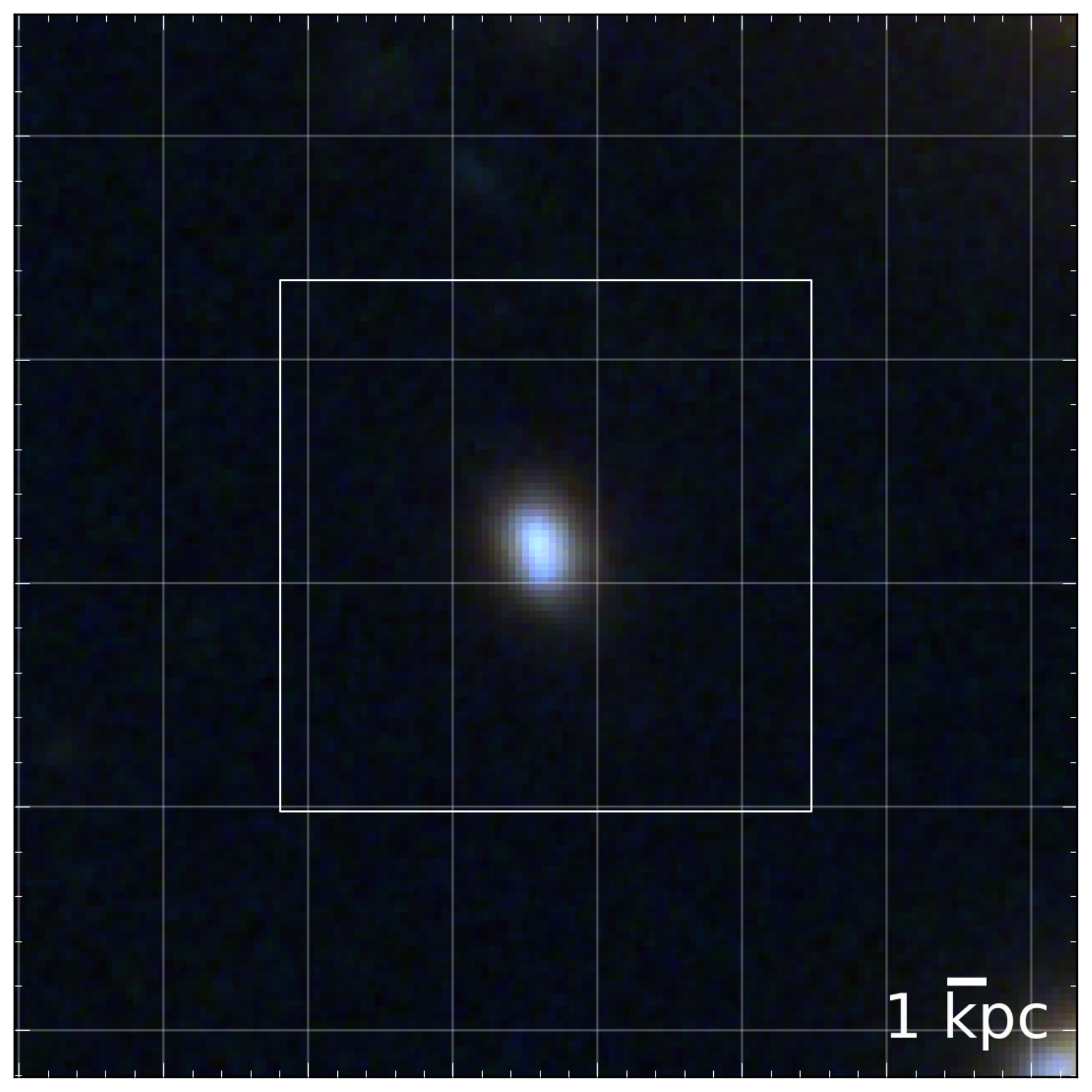}
\includegraphics[scale=0.14]{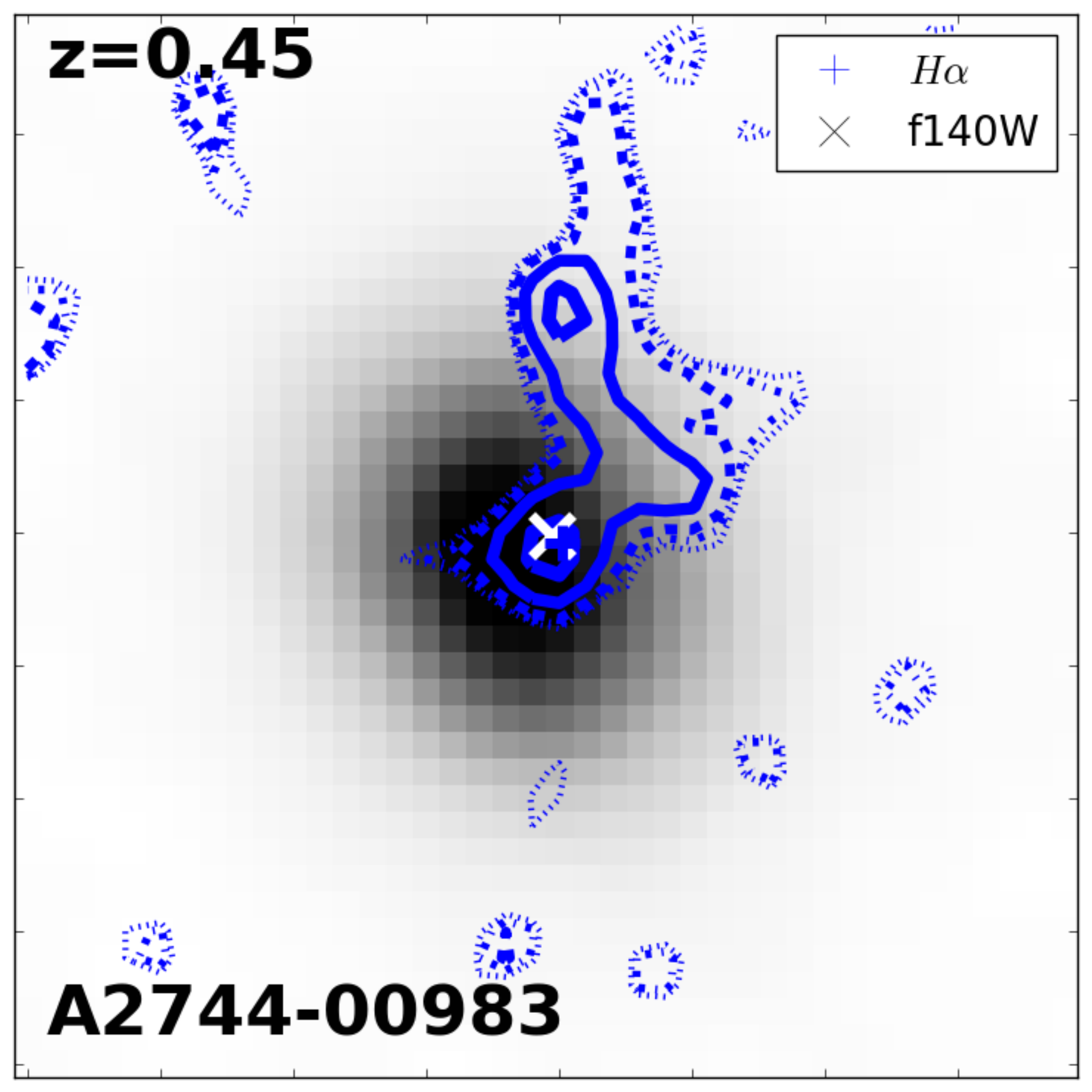}
\includegraphics[scale=0.14]{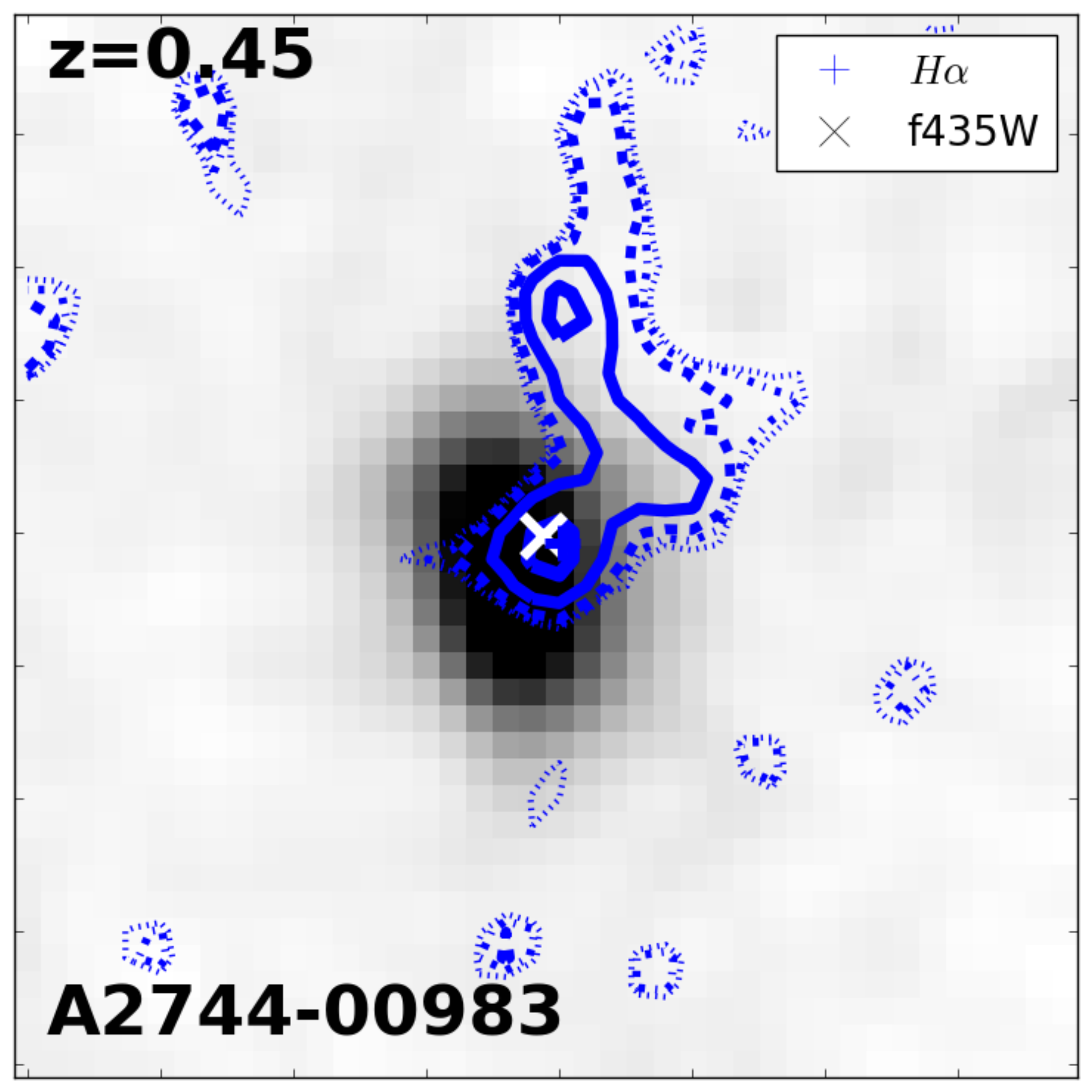}
\caption{Field galaxies labeled as ram pressure stripping candidates according to our classification scheme (\S\ref{sec:visinsp}). For each galaxy, the 
left panel shows the color composite image of the galaxy based on the 
CLASH \citep{postman12} or HFF \citep{lotz16} HST data. The blue channel is composed by the F435W, F475W, F555W, 
F606W, and F625W filters, the green by the F775W, F814W, F850lp, F105W, 
F110W filters, and the red by the F125W, F140W, F160W filters. The central panel 
shows the \Ha map superimposed on the image of the galaxy in the F140W filter 
and the right panel shows the \Ha map superimposed on the image of the galaxy in the F475W filter. 
Contour levels represent the 35$^{th}$, 50$^{th}$, 65$^{th}$, 80$^{th}$, 95$^{th}
$ percentiles of the light distribution, respectively. Blue contours indicate that \Ha maps are obtained  
from one spectrum, purple contours indicate that \Ha maps are obtained  
from  two orthogonal spectra, and green contours indicate that \Ha maps are obtained combining both the 
G102 and G141 grisms (only for $z>0.67$). In the color composite image, 
the Field of View is twice as big as the single band images. A smoothing filter has been applied to the 
maps and an arbitrary stretch  to the images for display purposes. The redshift of the 
galaxy is indicated on the top left corner.
 \label{fig:RP_field}}
\end{figure*}

In this Appendix we show all the galaxies that, upon visual inspection, were classified as probable ram pressure stripping candidates, both in clusters (Fig. \ref{fig:RP_clusters}) and in the field (Fig. \ref{fig:RP_field})
and that were not shown in the main text.

\section{The GLASS inspection GUI for morphologies (G\lowercase{i}G\lowercase{m})} \label{GIGm}

As described in the main text the GLASS inspection GUI for morphologies (GiGm) was developed to visually inspect the broad-band continuum morphology from ancillary imaging of the cluster and field galaxies in both the H$\alpha$ sample and the non-H$\alpha$  sample.
Similar to GiG and GiG$z$ presented in the Appendix of \cite{treu15}, GiGm is a \verb+Python+ based software available for download at \url{https://github.com/kasperschmidt/GLASSinspectionGUIs}.
GiGm is included in the most recent version of the self-contained script \verb+visualinspection.py+ that also contains GiG and GiG$z$. 
This means that an already functioning installation of GiG and GiG$z$ is trivially extended to also include GiGm by updating \verb+visualinspection.py+.
In this appendix we describe the basics of GiGm, but a more detailed description can be found in the \href{https://github.com/kasperschmidt/GLASSinspectionGUIs/blob/master/README.pdf}{GiG\_README}  available at \url{https://github.com/kasperschmidt/GLASSinspectionGUIs}.

\begin{figure*}
\begin{center}
\includegraphics[width=0.95\textwidth]{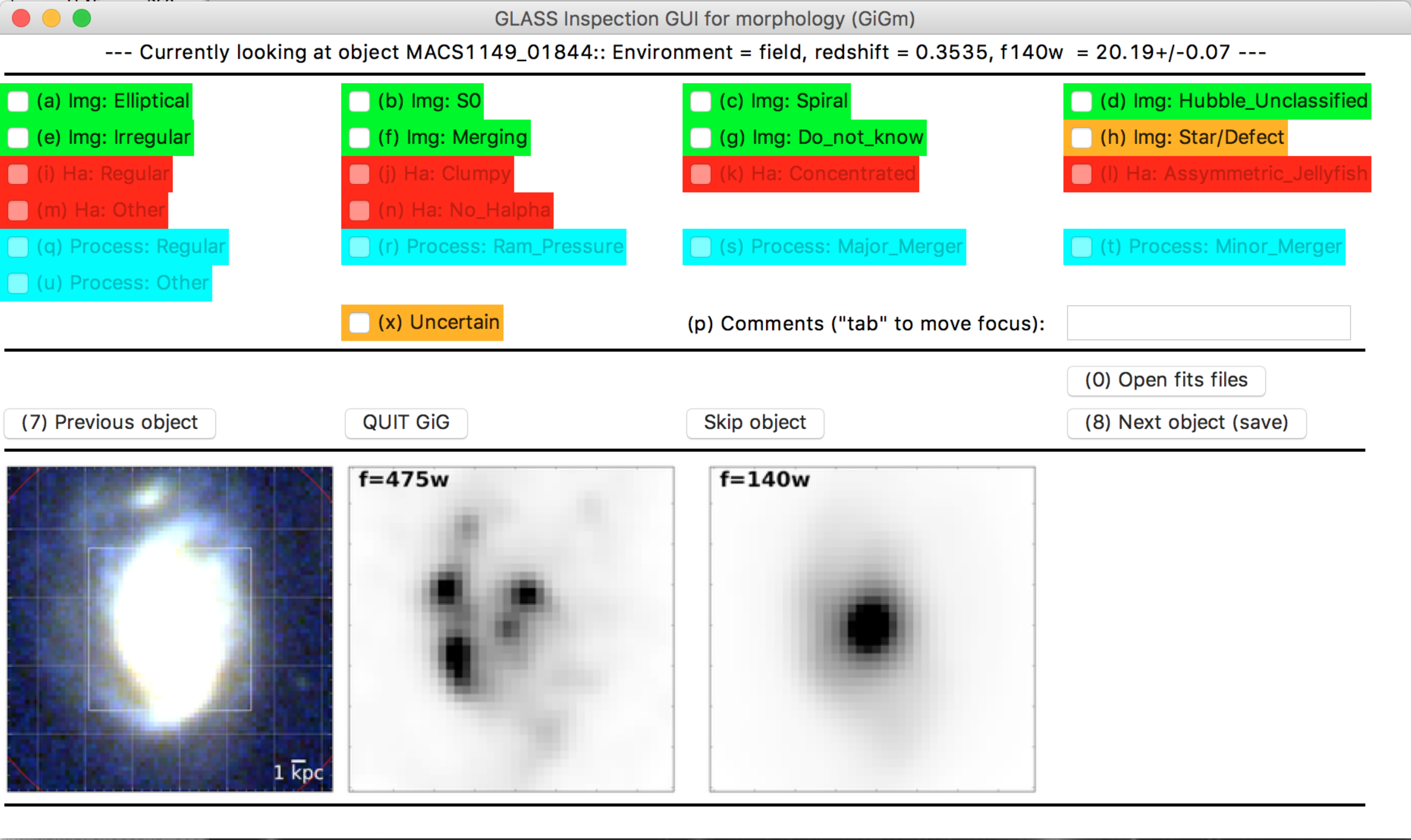}
\includegraphics[width=0.95\textwidth]{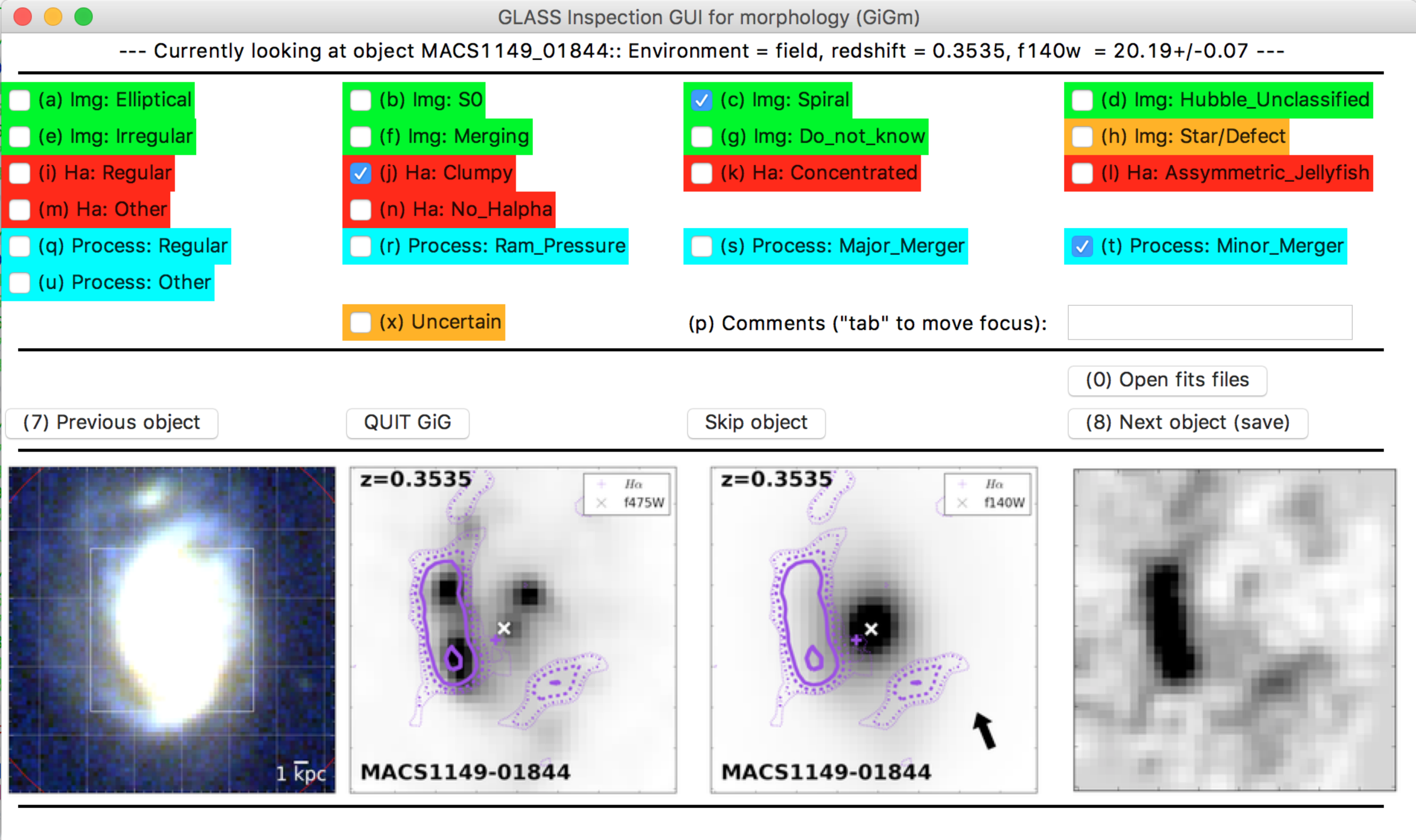}
\caption{Overview of the GiGm interface for the clumpy H$\alpha$ emitter MACS1149\_01844 shown in Figure~\ref{spir_cl}.
The top panel shows the initial inspection window which is used for all galaxies (H$\alpha$ emitters as well as non-H$\alpha$ emitters) classifying the morphology from direct image broad-band postage stamps (green check-boxes). 
A color composite and images in F475W and F140W are shown here.
If available, the fits-files of these can be opened with the "Open fits files" button to be able to manually adjust the scale and stretch of the images.
If the direct images of an H$\alpha$ emitter was inspected, after completing the morphological classification the H$\alpha$ morphology (red check-boxes) and star formation process (cyan check-boxes) classification become active, and the H$\alpha$ map postage stamp is shown and H$\alpha$ contours are over-layed on the direct image postage stamps and the \Ha map is shown as illustrated in the bottom panel.
If the direct image morphology inspection was performed on an object from the non-H$\alpha$ control sample GiGm simply advances to the next object.
This ensures truly independent inspections of the broad-band morphology and the H$\alpha$ morphology.
GiGm is publicly available at \url{https://github.com/kasperschmidt/GLASSinspectionGUIs}.
}
\label{fig:GiGm}
\end{center}
\end{figure*}

GiGm is run on a separate data directory containing prepared png (and fits) images, H$\alpha$ png (and fits) maps and versions of the png postage stamps with H$\alpha$ contours over-layed. If the fits images are located these can be opened with DS9 from within GiGm.
A general overview of the interface of GiGm is shown in Figure~\ref{fig:GiGm}.
The top panel depicts the inspection of the broad-band continuum morphology of the objects in the H$\alpha$ and non-\Ha control sample.
After completion of this inspection the GiGm interface will be added the H$\alpha$ map and H$\alpha$ contours on the broad band images for H$\alpha$ emitters as shown in the bottom panel of Figure~\ref{fig:GiGm}.
For non emitters GiGm will skip this step and simply advance to the next object in the data directory (or list of objects provided) when the broad-band morphological classification is completed. 
This ensures that the inspections of the broad-band morphology and the H$\alpha$ morphology are truly independent.
GiGm determines whether or not to enable the H$\alpha$ inspection based on the files available, i.e., if  an H$\alpha$ map is present for the current object, the H$\alpha$ classification will be enabled.
The details of the options for the morphological and H$\alpha$ inspections seen in Figure~\ref{fig:GiGm} are described in Section~\ref{sec:visinsp}.
To display the object information seen at the top of the GiGm windows in Figure~\ref{fig:GiGm} an information file is provide containing the object id, cluster name, redshift, magnitude, magnitude error, the name of the band the magnitude was measured in, and the environment (field or cluster) for each individual galaxy.
The output of the visual inspections with GiGm is and \verb+ascii+ file where 1 indicates that a check-box was set and 0 marks a check-box which was not set. As noted in Section~\ref{sec:visinsp} we combined these outputs and adopted the most common classification for each object among the four independent inspections. In case of broad disagreement, the galaxies were re-inspected and discussed to reach agreement.

\end{appendix}

\bibliographystyle{apj}
\bibliography{biblio_SFR2}

\begin{thebibliography}{}

\bibitem[{Abramson} {\it et al.}\ (2016)]{abramson16}
{Abramson}, L.~E. {\it et al.}\  2016, ArXiv e-prints.

\bibitem[{Abramson} {\it et al.}\ (2014)]{abramson14}
{Abramson}, L.~E. {\it et al.}\  2014, \apjl, 785, L36.

\bibitem[Atek {\it et al.}\ (2010)]{atek10}
Atek, H. {\it et al.}\  2010, The Astrophysical Journal, 723, 104.

\bibitem[{Balogh}, {Navarro} \& {Morris}(2000)]{balogh00}
{Balogh}, M.~L., {Navarro}, J.~F., and {Morris}, S.~L. 2000, \apj, 540, 113.

\bibitem[{Bekki}(1999)]{bekki99}
{Bekki}, K. 1999, \apjl, 510, L15.

\bibitem[{Bekki}(2009)]{bekki09}
{Bekki}, K. 2009, \mnras, 399, 2221.

\bibitem[{Bekki} \& {Couch}(2003)]{bekkicouch03}
{Bekki}, K. and {Couch}, W.~J. 2003, \apjl, 596, L13.

\bibitem[Bertin \& Arnouts(1996)]{bertin96}
Bertin, E. and Arnouts, S. 1996, Astronomy and Astrophysics Supplement, 117,
  393.

\bibitem[Brammer {\it et al.}\ (2014)]{brammer14}
Brammer, G.~B. {\it et al.}\  2014, STScI IRS.

\bibitem[Brammer {\it et al.}\ (2012)]{brammer12}
Brammer, G.~B. {\it et al.}\  2012, The Astrophysical Journal Supplement, 200,
  13.

\bibitem[{Br{\"u}ggen} \& {De Lucia}(2008)]{bruggen08}
{Br{\"u}ggen}, M. and {De Lucia}, G. 2008, \mnras, 383, 1336.

\bibitem[{Bruzual} \& {Charlot}(2003)]{bc03}
{Bruzual}, G. and {Charlot}, S. 2003, \mnras, 344, 1000.

\bibitem[{Butcher} \& {Oemler}(1984)]{butcher84}
{Butcher}, H. and {Oemler}, Jr., A. 1984, \apj, 285, 426.

\bibitem[{Calzetti} {\it et al.}\ (2000)]{calzetti00}
{Calzetti}, D. {\it et al.}\  2000, \apj, 533, 682.

\bibitem[{Castellano} {\it et al.}\ (2014)]{castellano14}
{Castellano}, M. {\it et al.}\  2014, \aap, 566, A19.

\bibitem[{Chabrier}(2003)]{chabrier03}
{Chabrier}, G. 2003, \pasp, 115, 763.

\bibitem[{Cheung} {\it et al.}\ (2012)]{cheung12}
{Cheung}, E. {\it et al.}\  2012, \apj, 760, 131.

\bibitem[{Coppin} {\it et al.}\ (2012)]{coppin12}
{Coppin}, K.~E.~K. {\it et al.}\  2012, \apjl, 749, L43.

\bibitem[{Darvish} {\it et al.}\ (2016)]{darvish16}
{Darvish}, B. {\it et al.}\  2016, \apj, 825, 113.

\bibitem[{Darvish} {\it et al.}\ (2014)]{darvish14}
{Darvish}, B. {\it et al.}\  2014, \apj, 796, 51.

\bibitem[{Dom{\'{\i}}nguez} {\it et al.}\ (2013)]{dominguez13}
{Dom{\'{\i}}nguez}, A. {\it et al.}\  2013, \apj, 763, 145.

\bibitem[{Dressler}(1980)]{dressler80}
{Dressler}, A. 1980, \apj, 236, 351.

\bibitem[{Dressler} {\it et al.}\ (1997)]{dressler97}
{Dressler}, A. {\it et al.}\  1997, \apj, 490, 577.

\bibitem[{Dutton}, {van den Bosch} \& {Dekel}(2010)]{dutton10}
{Dutton}, A.~A., {van den Bosch}, F.~C., and {Dekel}, A. 2010, \mnras, 405,
  1690.

\bibitem[{Ebeling}, {Ma} \& {Barrett}(2014)]{ebeling14}
{Ebeling}, H., {Ma}, C.-J., and {Barrett}, E. 2014, \apjs, 211, 21.

\bibitem[{Ellis} {\it et al.}\ (1997)]{ellis97}
{Ellis}, R.~S. {\it et al.}\  1997, \apj, 483, 582.

\bibitem[{Fang} {\it et al.}\ (2013)]{fang13}
{Fang}, J.~J. {\it et al.}\  2013, \apj, 776, 63.

\bibitem[{Font} {\it et al.}\ (2008)]{font08}
{Font}, A.~S. {\it et al.}\  2008, \mnras, 389, 1619.

\bibitem[{Fossati} {\it et al.}\ (2016)]{fossati16}
{Fossati}, M. {\it et al.}\  2016, \mnras, 455, 2028.

\bibitem[{Fumagalli} {\it et al.}\ (2014)]{fumagalli14}
{Fumagalli}, M. {\it et al.}\  2014, \mnras, 445, 4335.

\bibitem[{Garn} \& {Best}(2010)]{garnbest10}
{Garn}, T. and {Best}, P.~N. 2010, \mnras, 409, 421.

\bibitem[{Garn} {\it et al.}\ (2010)]{garn10}
{Garn}, T. {\it et al.}\  2010, \mnras, 402, 2017.

\bibitem[{Gehrels}(1986)]{gehrels86}
{Gehrels}, N. 1986, \apj, 303, 336.

\bibitem[{G{\'o}mez} {\it et al.}\ (2003)]{gomez03}
{G{\'o}mez}, P.~L. {\it et al.}\  2003, \apj, 584, 210.

\bibitem[{Gon{\c c}alves} {\it et al.}\ (2010)]{goncalves10}
{Gon{\c c}alves}, T.~S. {\it et al.}\  2010, \apj, 724, 1373.

\bibitem[{Goto} {\it et al.}\ (2003)]{goto03}
{Goto}, T. {\it et al.}\  2003, \mnras, 346, 601.

\bibitem[{Graham} \& {Driver}(2005)]{graham05}
{Graham}, A.~W. and {Driver}, S.~P. 2005, PASA, 22, 118.

\bibitem[{Gr{\"u}tzbauch} {\it et al.}\ (2011)]{grutzbauch11}
{Gr{\"u}tzbauch}, R. {\it et al.}\  2011, \mnras, 412, 2361.

\bibitem[{Gunn} \& {Gott}(1972)]{gunngott72}
{Gunn}, J.~E. and {Gott}, III, J.~R. 1972, \apj, 176, 1.

\bibitem[{Haines} {\it et al.}\ (2013)]{haines13}
{Haines}, C.~P. {\it et al.}\  2013, \apj, 775, 126.

\bibitem[{Hopkins} {\it et al.}\ (2014)]{hopkins14}
{Hopkins}, P.~F. {\it et al.}\  2014, \mnras, 445, 581.

\bibitem[{James} {\it et al.}\ (2005)]{james05}
{James}, P.~A. {\it et al.}\  2005, \aap, 429, 851.

\bibitem[{Kauffmann} {\it et al.}\ (2004)]{kauffmann04}
{Kauffmann}, G. {\it et al.}\  2004, \mnras, 353, 713.

\bibitem[{Kaviraj}(2014)]{kaviraj14}
{Kaviraj}, S. 2014, \mnras, 440, 2944.

\bibitem[{Kelson}(2014)]{kelson14}
{Kelson}, D.~D. 2014, ArXiv e-prints.

\bibitem[Kennicutt(1998)]{kennicutt98}
Kennicutt, R.~C. 1998, Astrophysical Journal v.498, 498, 541.

\bibitem[Kennicutt, Tamblyn \& Congdon(1994)]{kennicutt94}
Kennicutt, R.~C., Tamblyn, P., and Congdon, C.~E. 1994, Astrophysical Journal,
  435, 22.

\bibitem[{Kodama} {\it et al.}\ (2001)]{kodama01}
{Kodama}, T. {\it et al.}\  2001, \apjl, 562, L9.

\bibitem[{Koyama} {\it et al.}\ (2013)]{koyama13}
{Koyama}, Y. {\it et al.}\  2013, \mnras, 428, 1551.

\bibitem[{Kriek} {\it et al.}\ (2009)]{kriek09}
{Kriek}, M. {\it et al.}\  2009, \apj, 700, 221.

\bibitem[{Kroupa}(2001)]{kr01}
{Kroupa}, P. 2001, \mnras, 322, 231.

\bibitem[{Lang} {\it et al.}\ (2014)]{lang14}
{Lang}, P. {\it et al.}\  2014, \apj, 788, 11.

\bibitem[{Larson}, {Tinsley} \& {Caldwell}(1980)]{larson80}
{Larson}, R.~B., {Tinsley}, B.~M., and {Caldwell}, C.~N. 1980, \apj, 237, 692.

\bibitem[{Lewis} {\it et al.}\ (2002)]{lewis02}
{Lewis}, I. {\it et al.}\  2002, \mnras, 334, 673.

\bibitem[{Lilly} \& {Carollo}(2016)]{lilly16}
{Lilly}, S.~J. and {Carollo}, C.~M. 2016, ArXiv e-prints.

\bibitem[{Livermore} {\it et al.}\ (2012)]{livermore12}
{Livermore}, R.~C. {\it et al.}\  2012, \mnras, 427, 688.

\bibitem[{Lotz} {\it et al.}\ (2016)]{lotz16}
{Lotz}, J.~M. {\it et al.}\  2016, ArXiv e-prints.

\bibitem[{Madau}, {Pozzetti} \& {Dickinson}(1998)]{madau98}
{Madau}, P., {Pozzetti}, L., and {Dickinson}, M. 1998, \apj, 498, 106.

\bibitem[{Mantz} {\it et al.}\ (2010)]{mantz10}
{Mantz}, A. {\it et al.}\  2010, \mnras, 406, 1773.

\bibitem[{Martig} {\it et al.}\ (2009)]{martig09}
{Martig}, M. {\it et al.}\  2009, \apj, 707, 250.

\bibitem[{McGee}, {Bower} \& {Balogh}(2014)]{mcgee14}
{McGee}, S.~L., {Bower}, R.~G., and {Balogh}, M.~L. 2014, \mnras, 442, L105.

\bibitem[{Meert}, {Vikram} \& {Bernardi}(2015)]{meert15}
{Meert}, A., {Vikram}, V., and {Bernardi}, M. 2015, \mnras, 446, 3943.

\bibitem[{Menanteau} {\it et al.}\ (2005)]{menanteau05b}
{Menanteau}, F. {\it et al.}\  2005, in { American Astronomical Society Meeting
  Abstracts}, volume~37 of { Bulletin of the American Astronomical Society},
  1228.

\bibitem[{Merluzzi} {\it et al.}\ (2016)]{merluzzi16}
{Merluzzi}, P. {\it et al.}\  2016, \mnras, 460, 3345.

\bibitem[{Merluzzi} {\it et al.}\ (2013)]{merluzzi13}
{Merluzzi}, P. {\it et al.}\  2013, \mnras, 429, 1747.

\bibitem[{Mihos} \& {Hernquist}(1996)]{mihos96}
{Mihos}, J.~C. and {Hernquist}, L. 1996, \apj, 464, 641.

\bibitem[{Momcheva} {\it et al.}\ (2015)]{momcheva15}
{Momcheva}, I.~G. {\it et al.}\  2015, ArXiv e-prints.

\bibitem[{Moore} {\it et al.}\ (1996)]{moore96}
{Moore}, B. {\it et al.}\  1996, \nat, 379, 613.

\bibitem[{Moran} {\it et al.}\ (2005)]{moran05}
{Moran}, S.~M. {\it et al.}\  2005, \apj, 634, 977.

\bibitem[{Morishita} {\it et al.}\ (2016)]{morishita16}
{Morishita}, T. {\it et al.}\  2016, ArXiv e-prints.

\bibitem[{Muzzin} {\it et al.}\ (2012)]{muzzin12}
{Muzzin}, A. {\it et al.}\  2012, \apj, 746, 188.

\bibitem[Nelson {\it et al.}\ (2012)]{nelson12}
Nelson, E.~J. {\it et al.}\  2012, The Astrophysical Journal Letters, 747, L28.

\bibitem[{Nelson} {\it et al.}\ (2015)]{nelson15}
{Nelson}, E.~J. {\it et al.}\  2015, ArXiv e-prints.

\bibitem[Nelson {\it et al.}\ (2013)]{nelson13}
Nelson, E.~J. {\it et al.}\  2013, The Astrophysical Journal Letters, 763, L16.

\bibitem[{Noeske} {\it et al.}\ (2007)]{noeske07}
{Noeske}, K.~G. {\it et al.}\  2007, \apjl, 660, L43.

\bibitem[{Omand}, {Balogh} \& {Poggianti}(2014)]{omand14}
{Omand}, C.~M.~B., {Balogh}, M.~L., and {Poggianti}, B.~M. 2014, \mnras, 440,
  843.

\bibitem[{Paccagnella} {\it et al.}\ (2016)]{paccagnella16}
{Paccagnella}, A. {\it et al.}\  2016, \apjl, 816, L25.

\bibitem[{Patel} {\it et al.}\ (2009)]{patel09}
{Patel}, S.~G. {\it et al.}\  2009, \apjl, 705, L67.

\bibitem[{Peng} {\it et al.}\ (2002)]{peng02}
{Peng}, C.~Y. {\it et al.}\  2002, \aj, 124, 266.

\bibitem[{Poggianti} {\it et al.}\ (2009)]{poggianti09}
{Poggianti}, B.~M. {\it et al.}\  2009, \apjl, 697, L137.

\bibitem[{Poggianti} {\it et al.}\ (2016)]{poggianti16}
{Poggianti}, B.~M. {\it et al.}\  2016, \aj, 151, 78.

\bibitem[{Poggianti} {\it et al.}\ (1999)]{poggianti99}
{Poggianti}, B.~M. {\it et al.}\  1999, \apj, 518, 576.

\bibitem[{Porter} \& {Raychaudhury}(2007)]{porter07}
{Porter}, S.~C. and {Raychaudhury}, S. 2007, \mnras, 375, 1409.

\bibitem[{Porter} {\it et al.}\ (2008)]{porter08}
{Porter}, S.~C. {\it et al.}\  2008, \mnras, 388, 1152.

\bibitem[{Postman} {\it et al.}\ (2012)]{postman12}
{Postman}, M. {\it et al.}\  2012, \apjs, 199, 25.

\bibitem[{Postman} {\it et al.}\ (2005)]{postman05}
{Postman}, M. {\it et al.}\  2005, \apj, 623, 721.

\bibitem[{Price} {\it et al.}\ (2014)]{price14}
{Price}, S.~H. {\it et al.}\  2014, \apj, 788, 86.

\bibitem[{Quilis}, {Bower} \& {Balogh}(2001)]{quilis01}
{Quilis}, V., {Bower}, R.~G., and {Balogh}, M.~L. 2001, \mnras, 328, 1091.

\bibitem[{Schawinski} {\it et al.}\ (2009)]{schawinski09}
{Schawinski}, K. {\it et al.}\  2009, \mnras, 396, 818.

\bibitem[{Schmidt} {\it et al.}\ (2013)]{schmidt13}
{Schmidt}, K.~B. {\it et al.}\  2013, \mnras, 432, 285.

\bibitem[{Schmidt} {\it et al.}\ (2014)]{schmidt14}
{Schmidt}, K.~B. {\it et al.}\  2014, \apjl, 782, L36.

\bibitem[{Sobral} {\it et al.}\ (2012)]{sobral12}
{Sobral}, D. {\it et al.}\  2012, \mnras, 420, 1926.

\bibitem[{Sobral} {\it et al.}\ (2011)]{sobral11}
{Sobral}, D. {\it et al.}\  2011, \mnras, 411, 675.

\bibitem[{Sobral} {\it et al.}\ (2015)]{sobral15}
{Sobral}, D. {\it et al.}\  2015, \mnras, 450, 630.

\bibitem[{Sobral} {\it et al.}\ (2013)]{sobral13b}
{Sobral}, D. {\it et al.}\  2013, \apj, 779, 139.

\bibitem[{Stott} {\it et al.}\ (2016)]{stott16}
{Stott}, J.~P. {\it et al.}\  2016, \mnras, 457, 1888.

\bibitem[Straughn {\it et al.}\ (2011)]{straughn11}
Straughn, A.~N. {\it et al.}\  2011, The Astronomical Journal, 141, 14.

\bibitem[{Stroe} {\it et al.}\ (2015)]{stroe15}
{Stroe}, A. {\it et al.}\  2015, \mnras, 450, 646.

\bibitem[{Stroe} {\it et al.}\ (2014)]{stroe14}
{Stroe}, A. {\it et al.}\  2014, \mnras, 438, 1377.

\bibitem[{Swinbank} {\it et al.}\ (2012)]{swinbank12}
{Swinbank}, A.~M. {\it et al.}\  2012, \apj, 760, 130.

\bibitem[{Treu} {\it et al.}\ (2003)]{treu03}
{Treu}, T. {\it et al.}\  2003, \apj, 591, 53.

\bibitem[{Treu} {\it et al.}\ (2005)]{treu05}
{Treu}, T. {\it et al.}\  2005, \apj, 633, 174.

\bibitem[{Treu} {\it et al.}\ (2015)]{treu15}
{Treu}, T. {\it et al.}\  2015, \apj, 812, 114.

\bibitem[{van Dokkum} {\it et al.}\ (2011)]{vandokkum11}
{van Dokkum}, P.~G. {\it et al.}\  2011, \apjl, 743, L15.

\bibitem[{Vogelsberger} {\it et al.}\ (2014)]{vogelsberger14}
{Vogelsberger}, M. {\it et al.}\  2014, \mnras, 444, 1518.

\bibitem[{von der Linden} {\it et al.}\ (2010)]{vonderlinden10}
{von der Linden}, A. {\it et al.}\  2010, \mnras, 404, 1231.

\bibitem[{Vulcani} {\it et al.}\ (2010)]{vulcani10}
{Vulcani}, B. {\it et al.}\  2010, \apjl, 710, L1.

\bibitem[{Vulcani} {\it et al.}\ (2013)]{vulcani13}
{Vulcani}, B. {\it et al.}\  2013, \aap, 550, A58.

\bibitem[{Vulcani} {\it et al.}\ (2015)]{vulcani15}
{Vulcani}, B. {\it et al.}\  2015, \apj, 814, 161.

\bibitem[{Willett} {\it et al.}\ (2015)]{willett15}
{Willett}, K.~W. {\it et al.}\  2015, \mnras, 449, 820.

\bibitem[{Wisnioski} {\it et al.}\ (2015)]{wisnioski15}
{Wisnioski}, E. {\it et al.}\  2015, \apj, 799, 209.

\bibitem[{Wuyts} {\it et al.}\ (2012)]{wuyts12}
{Wuyts}, S. {\it et al.}\  2012, \apj, 753, 114.

\bibitem[{Wuyts} {\it et al.}\ (2013)]{wuyts13}
{Wuyts}, S. {\it et al.}\  2013, \apj, 779, 135.

\bibitem[{Yagi} {\it et al.}\ (2015)]{yagi15}
{Yagi}, M. {\it et al.}\  2015, \aj, 149, 36.

\bibitem[{Yang} {\it et al.}\ (2008)]{yang08}
{Yang}, Y. {\it et al.}\  2008, \aap, 477, 789.

\end{thebibliography}

\end{document}